\documentclass[11pt]{article}
\pdfoutput=1
\usepackage{jheppub}
\usepackage{amsmath}
\usepackage{bm}
\usepackage{slashed}
\usepackage{graphicx}

\newcommand{\tr}{\mathrm{tr}\,}
\newcommand{\Tr}{\mathrm{Tr}\,}
\newlength{\dummysp}
\newcommand{\beq}{\begin{eqnarray}}
\newcommand{\eeq}{\end{eqnarray}}

\newcommand{\e}{{\epsilon}}

\newcommand{\gappeq}{\mathrel{\rlap {\raise.5ex\hbox{$>$}}
{\lower.5ex\hbox{$\sim$}}}}
\newcommand{\lappeq}{\mathrel{\rlap{\raise.5ex\hbox{$<$}}
{\lower.5ex\hbox{$\sim$}}}}

\newcommand{\ben}{\begin{enumerate}}
\newcommand{\een}{\end{enumerate}}

\newcommand{\bit}{\begin{itemize}}
\newcommand{\eit}{\end{itemize}}

\newcommand{\eps}{\e}

\def\[{\left [}
\def\]{\right ]}
\def\({\left (}
\def\){\right )}
\def\R{{\mathbb R}}

\def\Z{{\mathbb Z}}
\def\T{{\mathbb T}}

\title{Mass-deformed Super Yang-Mills theory on $\mathbb T^4$:  sum over twisted sectors, $\mathbf{\theta}$-angle, and CP violation}
   
 \author[a]{Mohamed M. Anber,}\author[b]{Erich Poppitz} 
  
\affiliation[a]{Centre for Particle Theory, Department of Mathematical Sciences, Durham University, South Road, Durham DH1 3LE, UK}
\affiliation[b]{Department of Physics,   University of Toronto, 60 St George St., 
Toronto, ON M5S 1A7, Canada}
\emailAdd{mohamed.anber@durham.ac.uk}\emailAdd{poppitz@physics.utoronto.ca}    

\abstract{

{\flushleft{W}}e study $SU(N)$ super Yang-Mills theory with a small gaugino mass $m$ and vacuum angle $\theta$ on the four-torus $\mathbb{T}^4$ with 't Hooft twisted boundary conditions. Introducing a detuning parameter $\Delta$, which measures the deviation from an exactly self-dual $\mathbb{T}^4$, and working in the limits  $mLN \ll \Lambda LN \ll 1$ and $ \frac{(N-1) m^2 L^2}{4 \pi} \ll \Delta \ll 1$,
where $L$ is the torus size and $\Lambda$ the strong-coupling scale, we compute the scalar and pseudo-scalar condensates to leading order in $m^2L^2/\Delta$.  
The twists generate fractional-charge instantons, and we show that summing over all such contributions is crucial for reproducing the correct physical observables in the decompactified strong-coupling regime. From a Hamiltonian perspective, the sum over twisted sectors, already at small torus size, projects in the $m=0$ limit onto a definite superselection sector of the $\mathbb{R}^4$ theory. In the massless limit, we recover the exact value of the gaugino condensate  $|\langle \lambda \lambda \rangle| = 16\pi^2 \Lambda^3$, and demonstrate how a spurious $U(1)$ symmetry eliminates all ${\cal C}{\cal P}$-violating effects. Our results are directly testable in lattice simulations, and our method extends naturally to non-supersymmetric gauge theories.}

\begin{document}

\maketitle

\flushbottom

\section{Introduction}

The study of strongly coupled $4$-D Yang-Mills theories remains one of the most formidable challenges in physics. Lattice field methods have been instrumental in addressing this class of theories, offering crucial insights. However, the notorious sign problem severely limits the reach of lattice computations, particularly when exploring strongly coupled phenomena in the presence of a $\theta$ angle or finite density. Consequently, the development of alternative methods to probe these regimes is of great importance.

One method is to place the theory on a compact manifold smaller than the inverse strong-coupling scale $\Lambda$. Assuming that the setup preserves the global symmetries of the original theory, this approach can provide a portal to studying the theory in a weakly coupled and controlled regime.  A particularly natural choice is a four-dimensional torus, $\mathbb{T}^4$, as it aligns with lattice field computations. However, this introduces the challenge of selecting appropriate boundary conditions (BCs) for the gauge and matter fields. In the limit where $\mathbb{T}^4$ is large compared to $\Lambda^{-1}$, these BCs should not influence infrared observables. Yet, when $\mathbb T^4$ is small, where reliable calculations become feasible, different choices of BCs may lead to distinct physical outcomes. Then, the critical question is: what BCs should one use in the small-size limit that yields the correct result in the decompactified limit strong-coupling regime?

In the early 1980s, 't Hooft proposed that twisted boundary conditions on $\mathbb{T}^4$, which give rise to fractional instantons, could shed light on confinement in Yang-Mills theory \cite{tHooft:1980kjq,tHooft:1981nnx}. Shortly thereafter, Cohen and G\'omez recognized that these twisted boundary conditions also yield the correct number of fermion zero modes required to saturate the gaugino condensates in super Yang-Mills theory \cite{Cohen:1983fd}. However, the exact computation of the numerical coefficient of the condensate was only performed recently in \cite{Anber:2022qsz,Anber:2024mco}. Such computations were only possible after the development of the notion of higher-form symmetries \cite{Gaiotto:2014kfa} and their 't Hooft anomalies \cite{Gaiotto:2017yup}, their Hamiltonian-approach interpretation \cite{Cox:2021vsa}, as well as the detailed understanding of the moduli space of (multi-) fractional instantons \cite{Anber:2023sjn}. 

The $SU(N)$ gaugino condensates computations in \cite{Anber:2024mco} on a small $\mathbb T^4$ align with the results obtained through direct supersymmetric methods on $\mathbb R^4$: the bilinear condensate is given by\footnote{The definition of the strongly coupled scale $\Lambda$ used in \cite{Anber:2022qsz,Anber:2024mco} is as in \cite{Davies:2000nw}. We will not cite the numerous papers on the calculation of the gaugino condensate on $\R^4$, for a review see Shifman's textbook \cite{Shifman:2012zz}.} $\langle\mbox{tr}[\lambda\lambda] \rangle=16\pi^2\Lambda^3$. Moreover, the  small $\mathbb T^4$ calculational approach permits the computations of higher-order condensates, which are notoriously more difficult on $\mathbb R^4$ and can only be obtained via the power of the ADHM construction \cite{Dorey:2002ik}. 

While supersymmetry plays a crucial role in justifying the continuity of the small-$\mathbb{T}^4$ calculations to the large-volume (strong coupling) regime limit, one can still leverage semiclassical methods to perform calculations on small $\mathbb{T}^4$ even in the absence of supersymmetry.  In this work, we pave the ground for a systematic approach of semi-classical computations on $\mathbb T^4$ with twisted BCs beyond supersymmetry. The hope is that insights gained from these computations may offer valuable lessons about the strong coupling limit. To perform the computations, we break supersymmetry softly by adding a gaugino mass term, considering mass-deformed SYM (or SYM$^*$, as we denote it) and turning on a $\theta$ angle.   Several motivations make this calculation particularly valuable:

\begin{enumerate}

\item We have previously obtained exact results for the gaugino condensates on the twisted $\mathbb{T}^4$ in the massless limit. Extending this understanding to both the condensate and pseudo-scalar condensate as a small mass is introduced is an appealing challenge, especially as the calculations remain tractable in this regime. Direct soft supersymmetry-breaking methods on $\mathbb R^4$ provide an independent check on our calculations \cite{Konishi:1996iz,Evans:1996hi}. In addition to condensates, our method also allows the computations of correllators that might not be directly accessible via supersymmetric methods.

\item Previous attempts to compute CP-odd observables in the presence of the $\theta$-angle on the lattice have yielded inconclusive results; see \cite{Abramczyk:2017oxr} for a review. Performing an analytic calculation of such observables on a small $\mathbb{T}^4$ in the semi-classical regime can provide a useful benchmark or even guidance for expected lattice results. Our calculations of various observables in the mass-deformed SYM can be checked against lattice calculations. 

\item  The SYM$^*$ computations serve as prototype semi-classical calculations on $\mathbb T^4$, which can be adapted to more applications, such as determining the electric dipole moment in QED or in the Standard Model.

\end{enumerate}

\subsection{Setup: the partition function $Z^T$ as sum over twisted sectors}
\label{Setup}

We analyze mass-deformed $SU(N)$ SYM  on a small $\mathbb{T}^4$ with supersymmetry-preserving twisted boundary conditions imposed on both the gauge fields and gauginos. We also turn on a $\theta$ angle. Let $m$ and $V$ be the gaugino mass and the $\mathbb T^4$ volume, respectively.\footnote{In what follows, we find it convenient, sometimes, to define $V^{1\over 4} = L$, keeping in mind that not all sides of the torus have equal lengths. Throughout the paper, we will use $V$ and $L$ interchangeably.} 

We shall work in the limit $|m| L N\ll 1$ as well as $\Lambda L N \ll 1$.  Without loss of generality, we introduce twists along the 1-2 and 3-4 planes of $\mathbb T^4$, which naturally encompasses cases where twists are applied to only one of these planes or omitted altogether. The twists give rise to instantons with fractional topological charges $Q=-\frac{n_{12}n_{34}}{N}$ mod $1$, where $n_{12}$ and $n_{34}$ are the twists in the 1-2 and 3-4 planes, respectively \cite{vanBaal:1982ag}. The instantons must be self-dual to avoid instabilities (negative modes of gauge-field fluctuations in the instanton background). Thus, the action associated with such instantons is $S_I=\frac{8\pi^2 |Q|}{g^2}$. 

 The $\T^4$ partition function, $Z_{Q}$,  in a sector of  topological charge $Q$ is given by a Euclidean path-integral over the gauge field $A_\mu$ and gaugino $\lambda$:
\begin{eqnarray}\label{Zq}
Z_{Q}[\eta,\bar\eta]=   \int [DA_\mu][D\lambda][D\bar\lambda] e^{ - S_{\text{SYM}^*} + i \theta Q  - \int_{\mathbb T^4}(\eta\lambda+\bar\lambda\bar\eta) } \sim e^{- {8 \pi^2 |Q|\over g^2} + i \theta Q} \,,   
\end{eqnarray}
where $g^2 = g^2(L^{-1}) \ll 1$ and $\eta$ and $\bar\eta$ are external sources\footnote{
The action $S_{\text{SYM}^*}$ of SYM$^*$  with soft mass $m$, is written explicitly further below in (\ref{symaction2}) and  $\lambda, \bar\lambda$ are the gaugino fields in a two-component spinor notation.} and we have indicated that the partition function in the sector of charge $Q \ne 0$ is semiclassically suppressed at small $L$. The expectation values of the physical observables can be obtained from $Z_{Q}[\eta,\bar\eta]$ by taking derivatives with respect to the external sources. Yet, such expectation values are, in general, divergent and need to be regularized. To remedy this problem, we define a regularized version of $Z_{Q}[\eta,\bar\eta]$ by dividing by the path integral in the background of the instanton, but now incorporating the Pauli-Villars regulator;  this effectively corresponds to dividing by the determinants of the fluctuations around the instanton after introducing a Pauli-Villars mass. The softly broken SUSY provides a slick way to compute the determinants to the leading order in $mLN$. 
 Thus, we have
\begin{eqnarray}\label{ztotal 1}
Z_{Q}^{\scriptsize\mbox{Reg}}[\eta,\bar\eta]\equiv \frac{Z_{Q}[\eta,\bar\eta]}{\left[\int [DA_\mu][D\lambda][D\bar\lambda]e^{-S_{\scriptsize\mbox{SYM*}}}\right]_{Q, \scriptsize \mbox{PV}}}\,.
\end{eqnarray}

An instanton background explicitly breaks CP symmetry, necessitating the inclusion of the contribution from the anti-instanton, which carries the opposite topological charge. Thus, we recognize $Z_{Q}^{\scriptsize\mbox{Reg}}[\eta,\bar\eta]$ as a pre-partition function and define the total partition function $Z^{\scriptsize \mbox{T}}$ as the sum over topological sectors carrying charges $\pm Q$. As we shall argue, a given scalar or pseudo-scalar fermion correlator, as the ones in eqn.~(\ref{cond1}) below, will receive contributions from the topological sectors with charges $\pm \frac{k}{N}$, where $k$ is an integer including $k=0$. Therefore, it is natural to define the total partition function as a sum over all fractional and integer topological sectors:
\begin{eqnarray}\label{FULL PF}
Z^{\scriptsize T}[\eta,\bar\eta]=\sum_{Q=0,\pm\frac{1}{N},\pm\frac{2}{N},... } Z_{Q}^{\scriptsize\mbox{Reg}}[\eta,\bar\eta]\,.
\end{eqnarray}

As we will show below, the sum over topological sectors $Q = 0, \pm \tfrac{1}{N},\pm  \tfrac{2}{N}, \ldots$ can be realized on $\mathbb{T}^4$ by imposing a fixed twist $n_{34} = 1$ in the $34$-plane, while summing over all twists $n_{12} = 0, 1,  2, \ldots, N-1$ in the $12$-plane. For every value of the twist, one also adds arbitrary integers to the fractional topological charge determined by the twist. We explain this in more detail in  section \ref{sec:hamiltonian} on the Hamiltonian interpretation where we also show that, in SYM, the sum we define projects, already at finite volume, on one of the infinite-volume limit superselection sectors of the theory. 

The regularized expectation value of a gauge-invariant operator ${\cal O}$ is then given by
\begin{eqnarray}\label{Reg O}
\langle {\cal O}\rangle^{\scriptsize\mbox{Reg}}(x_1,x_2,..,x_n)=\left[(-1)^n\frac{\partial^n Z^{\scriptsize T}[\eta,\bar\eta] }{\partial\eta(x_1)\partial\eta(x_2)....\partial\bar\eta\partial....\bar\eta(x_n)}/{Z^{\scriptsize T}[\eta,\bar\eta]}\right]_{\eta=\bar\eta=0}\,,
\end{eqnarray}
where $n$ is the number of $\eta$ and $\bar\eta$ derivatives.

 We shall show that the leading-order semiclassical contributions to the bilinear correlators arise from the topological sectors with charges $\pm\frac{1}{N}$, while contributions from higher-charge instantons are subleading. We shall use our method to compute the scalar condensate\footnote{The relation between the four-component Majorana spinors $\bar\Psi, \Psi$, used in lattice simulations of SYM, as in \cite{Butti:2022sgy,Bonanno:2024bqg,Bonanno:2024onr}, and the two component spinors $\lambda, \bar\lambda$, as well as the action of parity and charge conjugation are spelled out in detail in appendix \ref{Discrete symmetries}.} $\tr \bar\Psi \Psi = \text{tr}(\lambda  \lambda +\bar\lambda  \bar\lambda)$, the pseudo-scalar condensate $-i \tr  \bar\Psi \gamma_5 \Psi =  i\text{tr}(\lambda  \lambda-\bar\lambda  \bar\lambda)$, and the fermion bilinears $\text{tr}(\lambda(x)\prod_\mu {\cal W}_\mu(x)  \lambda(0){\cal W}_\mu^\dagger(x) )$, and  $\text{tr}(\bar\lambda(x)\prod_\mu {\cal W}_\mu(x) \bar\lambda(0)   {\cal W}_\mu^\dagger(x))$, where ${\cal W}_\mu(x) $ are Wilson-lines insertions. These correlators are prototype examples of relevant quantities that can be cross-checked in lattice simulations in the strong-coupling regime. 
 For example, 
 \begin{eqnarray}
\label{corr 123}
 \langle \mbox{tr}\left[\lambda(x)\prod_{\mu=1}^4{\cal W}_\mu(x)\Gamma_{\nu_1\nu_2..} \lambda(0) {\cal W}^\dagger_\mu(x) \right]\rangle  \equiv {\sum\limits_{Q}  \langle \mbox{tr}\left[\lambda(x)\prod_{\mu=1}^4{\cal W}_\mu(x)\Gamma_{\nu_1\nu_2..} \lambda(0) {\cal W}^\dagger_\mu(x)
 \right] \rangle_{Q, unnorm.}  \over Z_T}\,.  \nonumber \\
\end{eqnarray}

Before we continue, we stress that defining the sum over sectors with arbitrary fractional topological charges requires specifying more data, namely the twists in the various $2$-planes of the $\T^4$, as already alluded  to above. As will become clear shortly, the sum we use is not equivalent to defining a $PSU(N)_0$ theory. Rather, our sum over twisted sectors is defined such that it projects to one of the $N$  degenerate flux sectors in Hilbert space. These sectors,  in the presence of an $n_{34}=1$ twist, are exactly degenerate  \cite{Cox:2021vsa}  at any finite volume  because of the mixed chiral-center anomaly. In effect, the anomaly in the presence of a twist allows us to project to one of the infinite-volume superselection sectors of the theory  already at arbitrarily small volume. This projection is what made possible the calculation of the infinite volume value of the gaugino condensate in the small-$\T^4$ theory. This is explained below, in section \ref{sec:hamiltonian}. 

\subsection{Structure of this paper}

There are three main threads that we follow in our study of SYM$^*$ in this paper. Here, we briefly review them in turn and point to the relevant sections, hoping to guide the reader through this rather long paper.

{\bf{\flushleft{Semiclassics on the small, twisted, and detuned $\T^4$:}}}
We  introduce SYM$^*$ theory  in section \ref{Mass-deformed SYM}. We discuss  its formulation on a twisted $\T^4$ in section \ref{sec:twistedbc}. We explain, in sections \ref{Twisting and fractional-instantons}, \ref{all the Correlators}, \ref{Determinants and regularization}, \ref{sec:observables}, the  steps involved in the semiclassical calculations of correlation functions  on the small-$\T^4$  in SYM$^*$, via  the partition function $Z^T$ summing over twisted sectors. Many important details are relegated to appendices \ref{Systematics of the propagator}-\ref{zeta function regularization}. The main technical effort is devoted to the calculation of the fermion propagators in SYM$^*$ in the $Q=k/N$ fractional instanton background to leading order in the $\T^4$ detuning parameter. 

Our main results and the scalar and pseudo-scalar condensates in SYM$^*$ are summarized in section \ref{intro:summary} of the Introduction.

{\bf{\flushleft{The Hamiltonian interpretation:}}} The partition function $Z^T$ as a sum over twisted sectors is interpreted in the Hamiltonian picture in section \ref{sec:hamiltonian}. We explain that the sum over twists projects,  already at finite volume, on a single superselection sector of the infinite-volume SYM theory.\footnote{The need to sum over twists  was also advocated for,  in two-dimensional QCD(adj), in \cite{Karthik:2024dhx}; see also the most recent \cite{Narayanan:2025cwl} for a discussion of additional subtleties of the 2d case.} The small soft-mass expansion is considered in the Hamiltonian framework in appendix \ref{sec:smallmhamiltonian}. The results are in qualitative agreement with those of the semiclassical calculation, summarized below in section \ref{intro:summary} of the Introduction.  

{\bf{\flushleft{Semiclassics on $\R \times \T^3$:}}} Somewhat outside of the main thrust of this paper---the  small-$\T^4$ semiclassical calculation---we discuss, in section \ref{sec:rtimest3}, SYM$^*$ in a different semiclassical, yet not analytically calculable,\footnote{This unwieldy phrase serves to point out that despite the weakly-coupled nature of the small-$\T^3$ theory, calculability has not yet been achieved, due to the limited  analytic understanding of the relevant fractional instantons.} limit:  that of $\R \times \T^3$ with a small twisted $\T^3$. This is of interest because it provides a semiclassical route to the semi-infinite volume limit. It also bears close relation to earlier studies of pure Yang-Mills theory \cite{RTN:1993ilw,Gonzalez-Arroyo:1995ynx}.

{\flushleft{W}}e begin  by  summarizing the main results of this paper.

\subsection{Summary of results and lessons}
\label{intro:summary} 
Here, we first summarize the results of our semiclassical calculations in SYM$^*$ on the small twisted $\T^4$. More general comments about the lessons learned are made in the end of this section.

We use the partition function 
$Z^T$ defined as a sum over twisted sectors in (\ref{FULL PF}).
The expectation values $\langle ... \rangle^{Reg.}$ shown below are defined as in (\ref{corr 123}), with appropriate subtractions discussed in the bulk of the paper, see sections \ref{Determinants and regularization} and \ref{sec:observables}.
Below, we give results of the semiclassical calculation for the scalar and pseudo-scalar condensates in SYM$^*$. As $m \rightarrow 0$, these reproduce well-known results obtained in SYM via holomorphicity. We stress, however, that our formalism allows for the calculation of more general correlators, not governed by the power of supersymmetry, see comments at the end of this section. 

We begin with the scalar condensate, $\tr \bar\Psi \Psi = \text{tr}(\lambda  \lambda +\bar\lambda  \bar\lambda)$,    to leading order in the semiclassical and small-$m$ expansion given by:
\begin{eqnarray} \label{scalar condensate intro}
 &&\langle\mbox{tr}\left[\bar\Psi\Psi\right] \rangle^{Reg.}\bigg\vert_{|m| LN  \ll 1}  \\
 && ~~~~~~~~~~~~\simeq    32\pi^2\Lambda^3  \left(1+  {|m|^2 L^2 \over c \Delta} \right)\cos\left(\frac{\theta}{N}\right)   + 16\pi^2\Lambda^3 \left( {L^2 m^{* \; 2}  \over c \Delta} e^{- i {\theta \over N}}+ { L^2 m^2  \over c \Delta} e^{i {\theta\over N}}\right)  ~. \nonumber
 \end{eqnarray}
 The pseudo-scalar condensate, $-i \tr  \bar\Psi \gamma_5 \Psi =  i\text{tr}(\lambda  \lambda-\bar\lambda  \bar\lambda)$,  is
\begin{eqnarray} \label{pseudo scalar condensate intro}
 &&-i\langle\mbox{tr}\left[\bar\Psi\gamma_5\Psi\right] \rangle^{Reg.}\bigg\vert_{|m| LN  \ll 1}  \\
&&~~~~~~~~~~~~ \simeq  -  32\pi^2\Lambda^3  \left(1+ {|m|^2 L^2   \over c \Delta}\right)\sin\left(\frac{\theta}{N}\right)
 +i 16 \pi^2\Lambda^3      \left( {L^2 m^{*\;2} \over c \Delta} e^{- i {\theta \over N}} -{L^2 m^{2} \over c \Delta} e^{  i {\theta \over N}}\right)   \,.\nonumber 
\end{eqnarray}
Here, $\Lambda^3 = \mu^3 e^{- {8 \pi^2 \over N g^2(\mu)}}/g^2(\mu)$  is the holomorphic strong-coupling scale expressed in terms of the canonical coupling $g^2(\mu)$ \cite{Shifman:1986zi,Arkani-Hamed:1997qui} and $L = V^{1\over 4}$ denotes the overall size of $\T^4$. 

{\flushleft{L}}et us now comment on the features of the scalar and pseudo-scalar condensates:
\begin{enumerate}
\item  Our current state of understanding of fractional instantons on $\T^4$ (see section \ref{Twisting and fractional-instantons}) only allows the semiclassical calculation to  be performed as an expansion in an additional  small parameter, the detuning parameter of the torus $0 < \Delta \ll 1$.\footnote{A good qualitative agreement between numerical (multi-) fractional instantons and the approximate analytic solutions obtained via the $\Delta$-expansion has been seen to hold for detuning parameters as large as $0.1$ or $0.2$, see \cite{GarciaPerez:2000aiw,Anber:2025yub}.} The $\Delta$ parameter entering (\ref{scalar condensate intro}, \ref{pseudo scalar condensate intro})  is:
\begin{eqnarray}\label{delta intro}
0 < \Delta \equiv {({N-1})L_3 L_4 - L_1 L_2  \over \sqrt{V}} \ll 1~, \end{eqnarray} as per eqn.~(\ref{def of Delta}) with $k=1$.
\item The weak-coupling semiclassical results for the scalar and pseudoscalar condensate hold in the small-$|m|$ and small-$\T^4$ limits, explicitly:\footnote{For readers interested in the large-$N$ limit, we note that the definitions of  both the scale and the condensate have to be modified, see \cite{Dorey:2002ik,Armoni:2003yv}. We also note that the parameter $c$ depends on $k$, the topological charge. The value given in (\ref{limit intro}) is the one appropriate for $k=1$, see eqn.~(\ref{energyshifts1}) in appendix \ref{sec:liftingzeromodes}.} 
\begin{eqnarray}\label{limit intro}
|m| L N \ll \Lambda L N \ll 1, ~~\text{and} ~~ {|m|^2 L^2 \over c} \ll  \Delta \ll 1, ~\text{with} ~ c \equiv {4 \pi \over  N-1}.
\end{eqnarray}
The additional $|m| L \ll \sqrt{c \Delta}$ limit is due to the nature of the small-$\Delta$ expansion. We refer to appendix  \ref{sec:liftingzeromodes} for detailed discussion.
 
Here we only briefly note that the factors ${|m|^2 L^2 \over c \Delta}$ are the leading contributions, at $\Delta \ll 1$, of  nonzero-mode fluctuations in the fractional instanton background to the determinants and fermion propagators. These   are subject to further additive  (to $|m|^2 L^2 /c \Delta$) corrections, proportional to $|m|^2 L^2$. These are in principle calculable, albeit with significantly more effort. However, in the limit (\ref{limit intro}), they are suppressed compared to the order-$|m|^2$ corrections shown. 
\item Both the scalar and pseudoscalar condensate are covariant\footnote{The covariance of the condensates is easiest to see from the condensates expressed in terms of two-component spinors, eqns.~(\ref{small mV C 1}) and (\ref{bar small mV C 1}).} under a spurious $U(1)$ ``symmetry,''  which also acts on the parameters of the theory  (see section \ref{Mass-deformed SYM}):
\begin{eqnarray}\label{spuriousU1 intro}\nonumber
U(1)_{spurious}:   \Psi &\rightarrow& e^{- i \alpha \gamma_5} \Psi,  \bar\Psi \rightarrow \bar\Psi e^{- i \alpha \gamma_5}  ~ (\text{or} ~ \lambda  \rightarrow  e^{i \alpha} \lambda,~ \bar\lambda  \rightarrow  e^{- i \alpha} \bar\lambda),\\
m  &\rightarrow& e^{- i 2 \alpha} m,~
m^*   \rightarrow   e^{i 2 \alpha} m^*,  ~
\theta  \rightarrow  \theta + 2N \alpha.
\end{eqnarray}
All correlation functions  are either $U(1)_{spurious}$ invariant, if they carry no $U(1)_{spurious}$ charge, or covariant, if charged under $U(1)_{spurious}$. 
\item When $|m| \rightarrow 0$, the small-$\T^4$ results for the condensate computed via $Z^T$ smoothly match the values of the condensates computed in {\it one} of the $N$ vacua of SYM theory on $\R^4$, as already seen in the $|m|=0$ calculation of \cite{Anber:2022qsz,Anber:2024mco}. 

This is due to the sum over twisted sectors in  $Z^T$.  As we explain in section \ref{sec:hamiltonian} on the Hamiltonian interpretation, the sum over twisted sectors  projects on one of the $\R^4$ superselection sectors of the $|m|=0$ theory already at finite volume. 

\item
At $|m|=0$, one finds that the pseudoscalar condensate (\ref{pseudo scalar condensate intro}) does not vanish (e.g. at $\theta \ne 0$). This  apparent ${\cal C} {\cal P}$-violation  is not physical and is due to the unavoidable ambiguity of a choice of field basis due to  (\ref{spuriousU1 intro}). Clearly, the $|m|=0$  ``${\cal C} {\cal P}$-violating'' condensate can be removed by rephasing the fields. Related ambiguities have been discussed on the lattice \cite{Abramczyk:2017oxr}.
\item As a byproduct, we can also calculate the quantity $\delta {\cal E} \equiv -  m\langle \mbox{tr}\left[\lambda\lambda\right] \rangle ^{Reg.} - m^*\langle \mbox{tr}\left[\bar\lambda\bar\lambda\right] \rangle ^{Reg.}$, obtaining (from (\ref{scalar condensate intro}, \ref{pseudo scalar condensate intro}))
\begin{eqnarray}\label{energy1 intro} 
\delta {\cal E}\big\vert_{|m| LN  \ll 1} \simeq  - 32\pi^2 \Lambda^3  |m| (1+ 2 {|m|^2  L^2   \over c \Delta})\cos\left(\frac{\theta_{\scriptsize \mbox{eff}}}{N}\right) ,
\end{eqnarray}
where $\theta_{\scriptsize \mbox{eff}}\equiv \theta + N {\rm arg}(m)$. This quantity is clearly invariant under $U(1)_{spurious}$ (\ref{spuriousU1 intro}).
The label $\delta {\cal E}$ is suggestive of the fact that,  in the infinite volume limit, $\delta {\cal E}$ would be the contribution of the condensates to the vacuum energy of SYM$^*$ for $|m| \ll \Lambda$: $\delta {\cal E}=- 32\pi^2 \Lambda^3 |m| \cos\left(\frac{\theta_{\scriptsize \mbox{eff}}}{N}\right)$, to linear order in $|m|$ (however, in the small $\T^4$ where our equation (\ref{energy1 intro}) was obtained, all energy eigenstates contribute).\footnote{We note that our $\delta{\cal{E}}$  has the same $\theta_{\scriptsize \mbox{eff}}$ dependence as the results obtained on $\R^4$ in \cite{Konishi:1996iz,Evans:1996hi} and that, to leading order in $|m|$, our small-$\T^4$ expression (\ref{energy1 intro}) exactly matches the one obtained by Konishi \cite{Konishi:1996iz} via soft-breaking technology  in one of the vacua of  the $\R^4$ theory.}
\end{enumerate} 
\smallskip

{\flushleft{O}}n the conceptual level, the main lesson we learned here is that the sum over twisted sectors, as defined in the paragraph after eqn.~(\ref{FULL PF}), or in the Hamiltonian framework in section \ref{sec:hamiltonian}, is needed  in order to obtain, already at small volume, expectation values which smoothly go to those calculated in one of the superselection sectors of the $\R^4$ theory. The sum over twisted sectors was implicitly used---but was not  explicitly stated---in our earlier gaugino condensate calculations \cite{Anber:2022qsz,Anber:2024mco}. Ultimately, the sum over twists is responsible, along with holomorphicity,  for the exact agreement between the small-$\T^4$ and $\R^4$ determinations of the gaugino condensate. 

In general non-supersymmetric theories with no mixed zero-form/one-form center symmetry anomalies, there is no exact degeneracy of electric flux sectors  in the finite volume Hilbert space.\footnote{It is the anomaly which allows to interpret the sum over twists as a projection to one of the degenerate sectors, see section \ref{sec:hamiltonian}.} However the sum over twisted sectors (for theories that permit twists)  might still be useful to isolate physical ${\cal CP}$-violating effects from the ${\cal CP}$-violation due to the twisted boundary conditions. 

At the technical level, we finally  stress that our expressions for propagators in the $Q=k/N$ fractional instanton background (section \ref{sec:propagatorkovern}) allow the calculation of more general correlators, for example the ones of section \ref{Correlators in the higher Q sector}, such as  $\langle \text{tr}(\lambda(x)\prod_\mu {\cal W}_\mu(x)  \lambda(0){\cal W}_\mu^\dagger(x) )\rangle$. While the $m \rightarrow 0$ limit of these is not governed by the power of supersymmetry, they could be checked in lattice simulations. In addition, our equations of section \ref{the2pointfunctionsection} also permit the calculation of further terms in the semiclassical expansion, see eqns.~(\ref{Tone1}, \ref{Tone2}, \ref{Tone3}, \ref{Ttwo}), suppressed w.r.t. the leading order ones presented here. 

\subsection{Mass-deformed SYM}
\label{Mass-deformed SYM}

We consider $SU(N)$  SYM theory on $\mathbb T^4$, and we break the SUSY softly by adding a gaugino mass term. In the following, we will study this theory in the presence of a nonvanishing $\theta$ angle. The Euclidean action is:
\begin{eqnarray}
\label{symaction2}
S_{\scriptsize\mbox{SYM}^*} = \frac{1}{g^2} \int\limits_{\T^4} \tr_\Box \left[ \frac{1}{2} F_{\mu\nu} F_{\mu\nu}  - 2\bar\lambda_{\dot\alpha}D_\mu\bar\sigma_\mu^{\dot\alpha \alpha}\lambda_\alpha  +m\lambda^\alpha\lambda_\alpha+m^*\bar\lambda_{\dot\alpha}\bar\lambda^{\dot\alpha} \right],\,
\end{eqnarray}
where we omit the topological term, $-i \theta Q$, with  $Q$ the topological charge of the configuration considered.
For generality, we consider a complex fermion mass $m=|m|e^{i\xi}$, and we take  $0\leq |m|\ll \Lambda$, where $\Lambda$ is the strong scale.
Here $A_\mu = A_\mu^a T^a$, where $a=1,2,...,N^2-1$, is the $SU(N)$ gauge field with hermitian Lie-algebra generators obeying $\tr_\Box \left(T^a T^b\right) =  \delta^{ab}$, $\lambda_\alpha = \lambda_\alpha^a T^a$ is the adjoint fermion (gaugino), and the field strength is $F_{\mu\nu}=-i[D_\mu,D_\nu]=\partial_\mu A_\nu-\partial_\nu A_\mu +i[A_\mu, A_\nu]$, where $D_\mu=\partial_\mu+iA_\mu$.  The symbol $\Box$ denotes the defining (fundamental) representation, with the normalization $\tr_\Box \left(T^a T^b\right) =  \delta^{ab}$ chosen to ensure that the simple roots satisfy $\bm \alpha^2=2$. The adjoint gaugino field is represented by $\bar\lambda_{\dot\alpha} =\bar\lambda_{\dot\alpha}^a T^a$ and $\lambda_\alpha = \lambda_\alpha^a T^a$, independent complex Grassmann variables.
The equations of motion are given by
\begin{eqnarray}\label{EOM}
(D_\mu F_{\mu\nu})^a=-i\bar \lambda \bar\sigma_\nu[T^a, \lambda]\,,\quad\bar \sigma_\mu^{\dot\alpha\alpha}D_\mu\lambda_\alpha=m^*\bar \lambda^{\dot\alpha}\,, \quad(\sigma_\mu)_{\alpha\dot\alpha}D_\mu\bar\lambda^{\dot\alpha}=m\lambda_\alpha\,,
\end{eqnarray}
where the covariant derivative is $D_\mu=\partial_\mu+i\left[A_\mu,\,\, \right]$. Here, $\sigma_\mu \equiv(i\vec\sigma,1)$, $\bar\sigma_\mu \equiv(-i\vec\sigma,1)$, $\vec \sigma$ are the Pauli matrices which determine the $\mu={1,2,3}$ components of the four-vectors $\sigma_\mu, \bar\sigma_\mu$. In addition, for any spinor, $\eta^\alpha = \epsilon^{\alpha \beta} \eta_\beta$, with $\epsilon^{12} = \epsilon_{21} = 1$, and likewise for the dotted ones. In addition, 
$
\bar\sigma_\mu^{\dot\alpha \alpha} = \epsilon^{\dot\alpha\dot\beta} \epsilon^{\alpha\beta} \sigma_{\mu \; \beta \dot\beta}$, $\sigma_{\mu \; \beta \dot\beta} = \epsilon_{\beta\alpha} \epsilon_{\dot\beta \dot\alpha} \bar\sigma_\mu^{\dot\alpha \alpha}$. All our notation is that of \cite{Dorey:2002ik}, except that we use Hermitean gauge fields.

For further use below, we note that the path integral with the action (\ref{symaction2}) has a $U(1)_{spurious}$  ``symmetry,'' which   rephases the gauginos and under which  the mass and the $\theta$-angle  transform as spurions:
\begin{eqnarray}\label{spuriousU1}\nonumber
U(1)_{spurious}:  \lambda &\rightarrow& e^{i \alpha} \lambda,~ \bar\lambda  \rightarrow  e^{- i \alpha} \lambda, \\
m  &\rightarrow& e^{- i 2 \alpha} m,~
m^*   \rightarrow   e^{i 2 \alpha} m^*, \nonumber \\
\theta &\rightarrow& \theta + 2N \alpha.
\end{eqnarray}
The anomaly-free $\Z_{2N}$ chiral symmetry of the $m=0$ SYM theory is contained in (\ref{spuriousU1}) and is generated by $\alpha = {2\pi \over 2N}$, the $U(1)_{spurious}$ transformation that shifts $\theta$ by $2 \pi$.  As (\ref{spuriousU1}) is also respected by the regulated theory, all our results for various expectation values, e.g.~$\langle \lambda\lambda\rangle$, $\langle \bar\lambda \bar\lambda\rangle$, etc., will be found  to  transform covariantly under (\ref{spuriousU1}), whereas $U(1)_{spurious}$-invariant quantities, such as $m \langle \lambda\lambda \rangle$, are invariant under (\ref{spuriousU1}).\footnote{Our results are covariant (eqns.~(\ref{small mV C 1}), (\ref{bar small mV C 1}))  or invariant (eqn.~(\ref{energy1})) under (\ref{spuriousU1}).}

We shall also use the $4$-component  Majorana spinors $\Psi$ and $\bar\Psi$, which are defined in terms of the Weyl fermions $\lambda$ and $\bar \lambda$ (suppressing the color index) as:\footnote{See Appendix \ref{Discrete symmetries} for a detailed discussion of the ${\cal{C}}$ and ${\cal{P}}$ transformations in both two-component and four-component spinor notation, both in infinite volume and in finite volume with twists. Also,  under $U(1)_{spurious}$ of  (\ref{spuriousU1}), we have $\Psi \rightarrow e^{- i \alpha \gamma_5} \Psi$, $\bar\Psi \rightarrow \bar\Psi e^{- i \alpha \gamma_5}$.}%
\begin{eqnarray}\label{Majorana def}
\Psi=\left[\begin{array}{cc}\lambda_{\alpha}\\\bar\lambda^{\dot\alpha} \end{array}\right]\,,\quad \bar\Psi=\left[\begin{array}{cc} \lambda^{\alpha}\bar\lambda_{\dot\alpha}\end{array}\right]\,.
\end{eqnarray}
The Majorana spinors are handy when computing  ${\cal{CP}}$-odd correlators and are also more convenient to compare with lattice results, e.g.~\cite{Butti:2022sgy,Bonanno:2024bqg,Bonanno:2024onr}. The use of the $4$-component spinor also necessitates the use of the Euclidean $\gamma$ matrices, defined in (\ref{gamma def1}) of Appendix \ref{Discrete symmetries}. There, we also write the fermionic terms in the action using the spinors (\ref{Majorana def}) and work out gauge-invariant operators. Among these, we are particularly interested in the scalar  and pseudo-scalar condensate operators \begin{eqnarray}\label{cond1}\mbox{tr}\left[\bar \Psi\Psi(x)\right]=\mbox{tr}\left[\lambda\lambda+\bar\lambda\bar\lambda\right](x)\,,\quad \mbox{tr}\left[\bar\Psi\gamma_5 \Psi(x)\right]=\mbox{tr}\left[-\lambda\lambda+\bar\lambda\bar\lambda\right](x)\,.\end{eqnarray} 
In the presence of $\theta$-angle, $\theta\neq 0$ or $\pi$, the theory explicitly breaks ${\cal{CP}}$ symmetry.  The transformation laws of the condensate and the pseudo-scalar condensate  under ${\cal{CP}}$ is %
\begin{eqnarray}\label{CPofcondensates}
\mbox{tr}[\bar \Psi\Psi] \xrightarrow[]{{\cal C}{\cal P}}\mbox{tr}[\bar \Psi\Psi]\,,\quad
\mbox{tr}[\bar \Psi\gamma_5\Psi ]\xrightarrow[]{{\cal C}{\cal P}} -\mbox{tr}[\bar \Psi\gamma_5\Psi]\,.
\end{eqnarray}
Thus, $\mbox{tr}[\bar \Psi\Psi]$ behaves as a scalar, while $\mbox{tr}[\bar \Psi\gamma_5\Psi ]$ is a pseudo-scalar. 

\subsection{Twisted boundary conditions on $\T^4$}

\label{sec:twistedbc}

We study the SYM$^*$ theory (\ref{symaction2}) on a 4-D torus $\mathbb T^4$. 
We take the torus to have periods of length $L_\mu$, $\mu=1,2,3,4$, where  $\mu,\nu$ runs over the spacetime dimensions. The gauge fields $A_\mu$  obey the boundary conditions
\begin{eqnarray}
A_\nu(x+L_\mu \hat e_\mu)=\Omega_\mu(x) A_\nu(x) \Omega_\mu^{-1}(x)-i \Omega_\mu(x) \partial_\nu \Omega_\mu^{-1}(x)\,,
\label{conditions on gauge field}
\end{eqnarray}
as we traverse $\mathbb T^4$ in each direction. The boundary conditions ensure that local gauge invariant quantities are periodic functions of $x$, with periods equal to the periods of $\T^4$.

 Here, $\Omega_\mu$ are
the transition functions (or twist matrices), $N\times N$ unitary matrices, and  $\hat e_\nu$ are unit vectors in the $x_\nu$ direction. The transition functions satisfy the cocycle conditions:
\begin{eqnarray}\label{cocycle}
\Omega_\mu (x + \hat{e}_\nu L_\nu) \; \Omega_\nu (x) = e^{i {2 \pi \over N} n_{\mu\nu}}\; \Omega_\nu (x+ \hat{e}_\mu L_\mu) \; \Omega_\mu (x)\,,
\end{eqnarray}
where the exponent $e^{i {2 \pi \over N} n_{\mu\nu}}$, with integers $n_{\mu\nu}=-n_{\nu\mu}$, is in the $\mathbb Z_N$ center of $SU(N)$.
The nonvanishing twists that we shall consider in this paper are of the form
\begin{equation}\label{twists1}
n_{12} = - n_{21} = - k,~ n_{34} = - n_{43}= 1,
\end{equation}
and are chosen so   a Yang-Mills configuration  obeying (\ref{conditions on gauge field})  carries   fractional topological charge \cite{tHooft:1979rtg,tHooft:1981sps,vanBaal:1982ag}:
\begin{eqnarray} \label{Q of n main}
Q= -\frac{n_{12}n_{34}}{N} ~({\rm mod} \; 1) = {k \over N} ~({\rm mod} \; 1)\,,
\end{eqnarray}
for $k \in {1,\ldots , N-1}$. To give an interpretation of the twists (\ref{twists1}), the partition function (\ref{FULL PF}) and expectation values (\ref{Reg O}), we next discuss the Hamiltonian formulation.

\subsection{Hamiltonian interpretation of the partition function $Z^T$ and correlators}
\label{sec:hamiltonian}

As mentioned above, to facilitate the interpretation of our results, it is desirable to have a Hamiltonian interpretation of the $\T^4$ partition function defined  in eqn.~(\ref{FULL PF}). To more precisely define (\ref{FULL PF}), we take the sum over topological sectors to correspond to taking the $3$-$4$ plane twist $n_{34}=1$, while summing over all values of the $1$-$2$ plane twists $n_{12}$.
Summing over all $n_{12}$, in view of (\ref{Q of n main}) readily reproduces the sum (\ref{FULL PF}) over all fractional and integer  topological charges defining our $Z^T$. Explicitly, for the Hamiltonian interpretation, we take the spatial directions to be $x_{2,3,4}$ and the Euclidean time direction, of extent $L_1$, to be $x_1$.

Thus our partition function of eqn.~(\ref{FULL PF}), with sources set to zero, is defined via a Hilbert space trace as
\begin{equation}
\label{full pf hamiltonian}
Z^T  \equiv {1\over N} \sum\limits_{k =0}^{N-1} \tr_{{\cal{H}}^{SYM}_{n_{34}=1}}\left( (-1)^F e^{- L_1 \hat H_{SYM} - L_1 \hat H_m} \; \hat T_2^{k} \right)~, 
\end{equation}
where the factor of $N$ is  inserted for future convenience; it cancels out in the computation of expectation values.
Expectation values, the counterpart of eqn.~(\ref{Reg O}), are computed by inserting $\hat{\cal{O}}$ in the partition function (\ref{full pf hamiltonian}):\footnote{If the observable $\hat{\cal{O}}$   involves operators taken at different ``times,'' e.g. $x_1=0$ and $x_1' \ne 0$, one has to split the evolution operator $e^{- L_1 \hat H}$; for brevity, we do not explicitly indicate this. In writing the above, we also assumed that  $\hat{\cal{O}}$  does not wind around the $\T^3$, i.e. commutes with $\hat T_2$.}
\begin{equation}\label{vevhamiltonian}
\langle {\cal{O}}(x_1=0) \rangle \equiv {1 \over N Z^T} \sum\limits_{k =0}^{N-1}\tr_{{\cal{H}}^{SYM}_{n_{34}=1}}\left( (-1)^F e^{- L_1 \hat H_{SYM} - L_1 \hat H_m} \; \hat T_2^{k}\; \hat{\cal{O}}(x_1=0) \right)~.
\end{equation}
We now explain the salient points in eqn.~(\ref{full pf hamiltonian}). The discussion below relies on ref.~\cite{Cox:2021vsa}, which also  contains a self-contained introduction to canonical quantization on $\T^3$ with twists, where the crucial relation,   the anomalous commutator of eqn.~(\ref{anomaly1}) below, is derived.

In (\ref{full pf hamiltonian}, \ref{vevhamiltonian}), $\hat H_{SYM}$ is the SYM hamiltonian, obtained from the action (\ref{symaction2}) with $m=0$,  while $\hat H_m$ is 
the $m, m^*$-dependent part of the SYM$^*$ Hamiltonian, $\hat H_m = \int_{\T^3} d^3 x( - m (\hat \lambda)^2 - m^* (\hat \lambda^\dagger)^2)$, where $\hat\lambda$ and $\hat\lambda^\dagger$ are canonically conjugated variables.  We separate the soft-breaking term, since we treat $m L N$ as small in what follows. 

The partition function is defined as a trace over the  physical Hilbert space of SYM on the $\T^3$ spanned by $x_{2,3,4}$, with spatial boundary conditions twisted by $n_{34}=1$. We denote this Hilbert space by  ${\cal{H}}^{SYM}_{n_{34}=1}$. 
The physical Hilbert space basis consists of states annihilated by Gauss' law. In addition, physical states are eigenstates of large gauge transformations, labelled by $\pi_3(SU(N))$, with eigenvalue $e^{i \theta}$ for a unit-winding transform. We work in a given $\theta$ sector. Using the physical states,  the $\theta$-vacuum  ensures that a sum over arbitrary integer topological charges is implicit\footnote{\label{foot1}In addition, see below, to the fractional charge $- {k \over N}$ due to the twist of boundary conditions in the time direction induced by the insertion of $\hat T_2^k$.} in each of the $N$ partition functions contributing to the sum in (\ref{full pf hamiltonian}). Finally,  in our discussion of the Hamiltonian formalism, we put the $\theta$ angle in the Hamiltonian (this is accomplished by a unitary transformation in the physical Hilbert space). 

It is well known, already from \cite{Witten:1982df}, that the energy spectrum of SYM on $\T^3$ with $n_{34}=1$ is gapped, for both gauge bosons and fermions, with the gap being of order $1/(LN)$ (see  Appendix \ref{sec:singletwistpropagator} for explicit expressions). There is only a discrete degeneracy left, as we  describe further below.

The partition function (\ref{full pf hamiltonian}) involves insertions of $\hat T_2$, the center symmetry generator in the spatial direction orthogonal to the $3$-$4$ plane of the $n_{34}=1$ twist. It plays a special role among the center symmetry generators $\hat T_2, \hat T_3, \hat T_4$, as already explained by 't Hooft \cite{tHooft:1980kjq,tHooft:1981nnx}.  Inserting $\hat T_2^k$  in the partition function twists the boundary conditions in the $L_1$ (time) direction by a  
center-symmetry transformation in the $x_2$ spatial direction, thus imposing a nonzero space-time twist $n_{12}$. The consequence of this twist, together with $n_{34}=1$, is that  the contribution to the partition function for given $k$ now includes a sum over topological charges $Q=-{k\over N} +  n$, for all $n \in \Z$, as per eqn.~(\ref{Q of n main}) (and as already alluded to in footnote \ref{foot1}).

Since center symmetry commutes with the Hamiltonian, $[\hat T_2, \hat H_{SYM}]=0$, every energy eigenstate with energy $E$ is also an eigenstate of $\hat T_2$, labeled by the discrete quantum number, the  ``electric flux'' $e_2 \in \Z\; ({\rm{mod}}\; N)$:\footnote{We ignore the similar flux labels $e_3, e_4$ in the $x_3, x_4$ spatial directions as they play no role in the discussion of the anomaly below. This is because $\hat T_3$ and $\hat T_4$ commute with $\hat X$  in the  $n_{34} \ne 0 ({\rm mod} N)$ background, as per \cite{Cox:2021vsa}.  For completeness, we also note that while $e_{3,4}$ also label energy eigenstates, as $\hat T_{3,4}$   commute with the Hamiltonian as well,  they do not label  degenerate states. In fact, in each of the $N$ degenerate sectors labelled by $e_2$, there are $N^2$  sectors labelled by $e_{3,4}$, whose degeneracy is lifted perturbatively. Briefly, this is because adjoint-field operators expanded as in (\ref{exp Jp}) carry $e_{3,4}$ flux quantum numbers determined by $p_{3,4}$. See \cite{GonzalezArroyo:1987ycm,Daniel:1990iz} for details and perturbative calculations of flux-splitting in pure Yang-Mills theory.}
\begin{equation}\label{T2action}
\hat T_2 |E, e_2\rangle = |E, e_2 \rangle e^{i {2 \pi \over N} e_2}~.
\end{equation}
Further, because of the mixed chiral-center anomaly, in addition to the supersymmetry degeneracies, all states in ${\cal{H}}^{SYM}_{n_{34}=1}$ have an exact $N$-fold degeneracy, as we now  review. The degeneracy follows from the realization that, with $n_{34}=1$, the  $\Z_{2N}$ chiral symmetry\footnote{Recall that the anomaly free chiral symmetry is the $\Z_{2N}$ subgroup of the $U(1)_{spurious}$ of eqn.~(\ref{spuriousU1}).} generator $\hat X$ does not commute with the center symmetry generator $\hat T_2$, the one generating spatial center symmetry transformations in the direction orthogonal to  the plane of the twist $n_{34}$s:
\begin{eqnarray} \label{anomaly1} 
\hat T_2\; \hat X \; \hat T_2^{-1} = e^{- i {2 \pi \over N}} \hat X~, ~{\rm{where}} ~ \hat X \; \hat \lambda \;\hat X^{-1} = e^{i {2 \pi \over 2 N}} \;\hat \lambda.
\end{eqnarray}
Because $\hat X$ is also a symmetry, the anomaly---the first equation in (\ref{anomaly1})---then implies that upon acting on an energy eigenstate, chiral symmetry lowers the electric flux $e_2$ by one unit:\footnote{Setting the undetermined phase factor to unity.} 
\begin{eqnarray}\label{chiral1}
\hat X |E, e_2\rangle = |E, e_2 -1\rangle~.
\end{eqnarray}Thus, the anomaly implies that all energy eigenstates on the $\T^3$ with  $n_{34}=1$ are $N$-fold degenerate, with  the  electric flux index $e_2 \in \Z$ (mod $N$) labeling the degenerate states. (Equivalently, the algebra of $\hat T_2$ and $\hat X$ obeying (\ref{anomaly1}) has $N$-dimensional nontrivial irreducible representations.)

In the infinite volume limit of SYM theory ($m=0$), the $N$ lowest-energy electric flux eigenstates $|E=0, e_2\rangle$ (all of zero energy in SYM) become the $N$ ground states of $\R^4$ SYM. These are interchanged by the discrete chiral symmetry $\hat X$, as per (\ref{chiral1}). For $m \ne 0$, the degeneracy is lifted (as the $\Z_{2N}$ chiral symmetry is explicitly broken), but for small enough $m$ the ground state of the deformed theory is expected to remain close to one of the SYM ground states.

From the above discussion, we can equivalently write the partition function (\ref{full pf hamiltonian}) as a sum over simultaneous $\hat H_{SYM}$ and $\hat T_2$ eigenstates $|E, e_2\rangle$:\footnote{As discussed above, we do not show the implicit sum over electric fluxes in the $x_3$ and $x_4$ directions, $e_3, e_4$.}
\begin{eqnarray}\label{ZTagain}
Z^T = \sum_{E, e_2} \langle E, e_2| (-1)^F e^{- L_1 \hat{H}_{SYM} - L_1 \hat{H}_m} \left( {1 \over N} \sum\limits_{k=0}^{N-1} \hat T_2^k \right)|E, e_2\rangle~,
\end{eqnarray}
Here, we have written the sum over $\hat T_2^k$ insertions in a form that explicitly shows that the sum over $k$ performs  a projection---as a simple consequence of (\ref{T2action})---on $e_2=0$ states.\footnote{It is trivial to modify the projector to select any $e_2 \ne 0$ by including appropriate phases (as in section \ref{sec:rtimest3}).}  

One consequence of   this projection is that in the infinite volume limit, the $\T^4$ twisted partition function $Z^T$ will go over to the partition function in one of the superselection sectors of SYM theory, rather than perform an average over all such sectors. This projection on a single superselection sector implies that  correlators in SYM theory computed via  (\ref{vevhamiltonian}) should obey cluster decomposition in the infinite volume limit.\footnote{That the sum over twisted sectors projects on a single superselection sector is expected to also hold if the supersymmetry-breaking mass is small enough, when the ground state of the deformed theory  remains close to one of the ground states of SYM.  Note that the perturbation $\hat H_m$ does not have off-diagonal matrix elements between different $e_2$ states: as it does not wind around the $x_2$ direction, it cannot change the flux. In addition, in SYM$^*$, which is expected to be in the universality class of pure YM theory, it is believed that there are different  superselection sectors only  at $\theta = \pi$. }

In the rest of this paper, we  
  compute and interpret the leading---at small $m LN$---contribution to the partition function $Z^T$, as well as to various expectation values (\ref{vevhamiltonian}), in the semiclassical approximation valid at a small $\T^4$, relying on our improved analytic understanding of multifractional instantons.

But before we embark on this, we note that the scaling of the results with $m$ and their $\theta$-angle dependence can be inferred form the Hamiltonian interpretation of $Z^T$ and the expectation values. The precise coefficients, however, can only be obtained in the path integral formalism.

Taking into account the properties of ${\cal{H}}^{SYM}_{n_{34}=1}$ described above, and recalling that $\hat H_m = \int_{\T^3} d^3 x( - m (\hat \lambda)^2 - m^* (\hat \lambda^\dagger)^2)$, we expand $e^{- L_1 \hat H_m}$ to leading  and subleading order  in $|m|$ to find  expressions for $Z^T$ and various fermion correlators computed via (\ref{vevhamiltonian}), in the leading semiclassical approximation. The details of this  combined small-$m$ and semiclassical expansion (valid at small $V$) are presented in Appendix \ref{sec:smallmhamiltonian}. Here we only note that the derivation   relies, in an essential way, on the selection rules for expectation values following from  chiral-center anomaly (\ref{anomaly1}, \ref{chiral1}) and the representation of the Hilbert space trace as a sum over degenerate $e_2$ flux sectors.  

The result for $Z^T$---where, for each term in  the small-$|m|$ expansion, we keep the leading semiclassical contribution (further contributions can be evaluated by parameterizing them by introducing further unknown constants)---can be formally\footnote{Because the definitions of the various constants in (\ref{ZTsmallmL main text}, \ref{bilinearhamiltonian main text}) require a discussion of regularization and renormalization conditions.} written as \begin{eqnarray}
Z^T &\simeq&  (1 + |m|^2 L^2 c_0) + c \; ( L m e^{- {8 \pi^2 \over N g^2}} e^{i {\theta \over N}} + L m^* {e^{-{8 \pi^2 \over N g^2}}} e^{ - i {\theta \over N}}) +{\cal{O}}(|m|^3, e^{- {16 \pi^2 \over N g^2}}) \nonumber \\
&=&  (1 + |m|^2 L^2 c_0) + c' L^4 |m| \Lambda^3 \; \cos {\theta + N \text{arg}\, m \over N}+ {\cal{O}}(|m|^3, e^{- {16 \pi^2 \over N g^2}}). \label{ZTsmallmL main text}
  \end{eqnarray}
On the second line we rewrote the result in terms of the strong coupling scale. We note that the second term, proportional to the space-time volume $L^4$,  in the partition function  obtained by summing over all appropriate twisted sectors, reproduces the well-known infinite-volume result for the $\theta$-dependence of the vacuum energy in one of the $\R^4$ vacua of SYM with soft breaking (obtained in e.g.~\cite{Konishi:1996iz}, via studying the soft-breaking in Seiberg-Witten theory).

The calculation of the bilinear gaugino condensates\footnote{We also note that (\ref{ZTsmallmL main text}) and (\ref{bilinearhamiltonian main text}) are invariant and covariant, respectively, under the $U(1)_{spurious}$ of eqn.~(\ref{spuriousU1}) and that  expansions for higher-order condensates and other correlation functions similar to (\ref{bilinearhamiltonian main text}) can also be obtained from the Hamiltonian formulation via an expansion in small-$m$, at the cost of introducing more unknown constants.
} via (\ref{vevhamiltonian}), in the same small-$m$, leading semiclassical approximation as the one leading to (\ref{ZTsmallmL main text}), gives:
\begin{eqnarray} \label{bilinearhamiltonian main text}
Z_T \; \langle  \hat\lambda^2 \rangle &=& 16 \pi^2 \Lambda^3(1 + c_1 |m|^2 L^2) +c_2  \;  {m^* \over L^2}+ c_3\;  (m^* L)^2  ~ \Lambda^3 e^{- i {\theta \over N}},\nonumber \\
Z_T \; \langle  (\hat\lambda^\dagger)^2  \rangle &=&  16 \pi^2 \Lambda^3(1 + c_1 |m|^2 L^2)  +c_2 \;    {m  \over L^2}+ c_3 \;  (m  L)^2  ~ \Lambda^3 e^{  i {\theta \over N}}~,
\end{eqnarray}
As for $Z^T$ above, we note that as $m \rightarrow 0$, one obtains the well-known $\R^4$ result already at finite volume. 
It is the sum over twisted sectors which allowed the computation of the infinite-volume gaugino condensate in one of the $\R^4$ vacua of SYM already from the small $\T^4$ \cite{Anber:2022qsz,Anber:2024mco}.\footnote{We also note that, as opposed to the results of the actual semiclassical calculations quoted in section  \ref{intro:summary}, there is no detuning parameter ($\Delta$) dependence here: the formal small-mass  expansion in the Hamiltonian is not aware of the need to detune the $\T^4$ to perform semiclassical calculation.}

The Hamiltonian formalism is not well-suited to performing actual calculations, as renormalization and regularization are most easily done in a path-integral framework and  are needed to compute the dimensionless $c_0$, $c'$, $c_{1,2,3}$.  Thus, we now return to the path integral formulation and the semiclassical calculations on small $\T^4$.

\section{The semiclassical path integral: twisting, fractional instantons, and fermions}
\label{Twisting and fractional-instantons}

To study the path integral formulation of $Z^T$ (\ref{FULL PF}) which allows us to perform actual semiclassical calculations we need two conditions. First, the weak-coupling approximation should be valid. Second, we should have analytic control over the instantons giving the leading contribution to the semiclassical path integral. To achieve the first condition it suffices to  take the spatial torus, $\T^3$ in the $x_{2,3,4}$ directions, small, such that $L N \Lambda \ll 1$. At this point, one can leave the $x_2$ time direction be infinite, effectively considering the $\R \times \T^3$ spacetime. However, we have no analytic understanding on the relevant saddle points.\footnote{Thus, we call this the  ``semiclassical, yet not calculable, limit.'' We discuss this $\R \times \T^3$ limit for SYM$^*$ in section \ref{sec:rtimest3}, where we also point its close relation  to older studies in pure YM theory \cite{RTN:1993ilw,Gonzalez-Arroyo:1995ynx}.} In order to satisfy the  condition of analytical calculability, we take the $L_1$ time direction to also be small and consider the small-$\T^4$ limit. This is the regime where the analytical $\Delta$-expansion, as we discuss below, gives us control over the nonperturbative saddle points.

Thus, to  activate fractional instantons, we put the theory of a 4-D torus $\mathbb T^4$ and impose general twists $n_{12}$ and $n_{34}$ as described in (\ref{cocycle}, \ref{twists1}, \ref{Q of n main}). As already explained, the various terms in the partition function $Z^T$ of eqn.~(\ref{FULL PF}), or (\ref{full pf hamiltonian}), correspond to summing over $n_{12}$ at fixed $n_{34}$.

't Hooft \cite{tHooft:1981nnx} found a solution to the cocycle conditions (\ref{cocycle}), giving rise to the fractional $Q$ in (\ref{Q of n main}). This was achieved by embedding the $SU(N)$ transition functions $\Omega_\mu(x)$ in $SU(k)\times SU(\ell)\times U(1)\subset SU(N)$, such that $N=k+\ell$.  To present the solution, we use the same notation followed in \cite{Anber:2023sjn}: we take primed upper-case Latin letters to denote elements of $k\times k$ matrices: $C', D'=1,2,...,k$, and the unprimed upper-case Latin letters to denote $\ell\times \ell$ matrices: $C,D=1,2,..,\ell$.   We also introduce the matrices $P_k$ and $Q_k$ (similarly the matrices $P_\ell$ and $Q_\ell$), the $k\times k$ (similarly $\ell\times\ell$) shift and clock matrices satisfying the relation 
\begin{equation}
\label{clockshift}
P_kQ_k=e^{i\frac{2\pi}{k}}Q_kP_k .
\end{equation} Explicitly, we have that  $(P_k)_{B'C'}=\gamma_k\delta_{B',C'-1 \; (\text{mod}\, k)}$ and $(Q_k)_{C'B'}=\gamma_k \; e^{i2\pi \frac{C'-1}{k}}\delta_{C'B'}$, for the matrix elements of $P_k$ and $Q_k$, where the coefficient $\gamma_k=e^{i\frac{\pi(1-k)}{k}}$ is chosen to ensure that $\mbox{Det}(P_k)=\mbox{Det}(Q_k)=1$. The matrix $\omega$ is the $U(1)$ generator:
\begin{eqnarray}\label{omega}
\omega=2\pi\mbox{diag}(\underbrace{\ell, \ell,...,\ell}_{k\, \mbox{times}},\underbrace{ -k,-k,...,-k}_{\ell\,\mbox{times}})\,,
\end{eqnarray}
commuting with $P_k,P_\ell, Q_k,Q_\ell$. 

The explicit form of the transition functions $\Omega_\mu$ obeying (\ref{cocycle}) with $n_{\mu\nu}$ of (\ref{twists1}):
 \begin{eqnarray}\label{the set of transition functions for Q equal r over N, general solution}
\nonumber
\Omega_1&=& (-1)^{k-1}I_k \oplus I_\ell e^{i \omega \frac{ x_2}{N  L_2}} = \left[\begin{array}{cc}(-1)^{k-1}I_k e^{i2\pi \ell   \frac{x_2}{N  L_2}}&0\\0& e^{-i 2\pi k\frac{x_2}{NL_2}}I_\ell\end{array}\right]\,,\\ 
\Omega_2&=&Q_k\oplus I_\ell = \left[\begin{array}{cc}Q_k&0\\0& I_\ell\end{array}\right],\\
\Omega_3&=&I_k\oplus P_\ell e^{i \omega \frac{x_4}{N\ell L_4}} = \left[\begin{array}{cc} e^{i2\pi  \frac{x_4}{N L_4}} I_k&0\\0& e^{-i 2\pi k\frac{x_4}{N \ell L_4}}P_\ell\end{array}\right]\,,\quad
\Omega_4=I_k\oplus Q_\ell = \left[\begin{array}{cc}I_k&0\\0& Q_\ell\end{array}\right], \nonumber
\end{eqnarray}
and $I_k$ ($I_\ell$) is the $k\times k$ ($\ell\times \ell$) unit matrix, reminding the reader that $\ell = N-k$. The reader can easily check that they obey the correct cocycle conditions, eqns.~(\ref{cocycle}, \ref{twists1}).

The moduli-independent part of the solution $A_\mu$ of self-dual instantons\footnote{As we shall be using anti-self-dual instantons as well, it is worth mentioning that they are obtained, for each given $k$, by replacing $n_{12} = -k \rightarrow n_{12} = + k$. This change of the cocycle condition can be implemented by only replacing $\Omega_1 \rightarrow \Omega_1^\dagger$ and keeping the other transition functions as in (\ref{the set of transition functions for Q equal r over N, general solution}). The effect on (\ref{naked expressions}, \ref{abelian F to leading}) is to change the signs of $A_2$ and $F_{12}$ only, keeping $A_4$ and $F_{34}$ the same. This charge $Q= - k/N$ background is anti-self-dual when  (\ref{abelian F to leading}) is self-dual, i.e. at the same values of $L_\mu$ where (\ref{selfdualtorus1}) holds. }  that satisfy the cocycle conditions is given in terms of $\omega$ of (\ref{omega}) by
 \begin{eqnarray}\label{naked expressions}
 A_1&=&0\,, \quad A_2=-\omega\frac{   x_1}{N  L_1L_2}\,, \quad  A_3=0\,, ~\quad A_4=-\omega \frac{x_3}{N\ell L_3L_4}\,.
 \end{eqnarray}
 The corresponding field strength  is constant  on $\mathbb T^4$:
\begin{eqnarray}\label{abelian F to leading}
F_{12}&=&-\omega\frac{1}{N  L_1L_2}\,,\quad F_{34}=-\omega\frac{1}{N\ell L_3L_4}\,.
\end{eqnarray}
 The reader can verify that the topological charge of this solution is $Q={k \over N}$.

There are $4$ bosonic translational moduli denoted by $z_\mu$. In addition, there are $4(k-1)$ moduli, denoted by $z_\mu$ and $a_\mu^{1},a_\mu^{1},..a_\mu^{k-1}$. These are the holonomies along the $SU(k)$ Cartan generators in each spacetime direction.  The matrix components of the moduli, denoted by $\delta A_\mu$, can be written using the Cartan generators $\bm H_k$ of $SU(k)$, embedded in $SU(N)$ by adding zeros in their lower $\ell \times \ell$ block,  as
\begin{eqnarray}
\nonumber
\delta A_1&=&-\omega \frac{z_1}{L_1}+{2\pi \over L_1} \bm a_1\cdot \bm H_{(k)}\,,\quad \delta A_2=-\omega\frac{z_2}{L_2}+{2\pi \over L_2} \bm a_2\cdot \bm H_{(k)}\,,\\
\delta A_3&=&-\omega \frac{z_3}{L_3}+{2\pi \over L_3} \bm a_3\cdot \bm H_{(k)}\,,\quad  \delta A_4=-\omega \frac{z_4}{L_4}+{2\pi \over L_4} \bm a_4\cdot \bm H_{(k)}\,,
\label{r over N abelian sol}
\end{eqnarray}
where, e.g., $\bm a_\mu=(a^1_\mu,a^2_\mu,..,a^{k-1}_\mu)$. 
Here ${\bm H}_{(k)}\equiv (H_{(k)}^1,...,H_{(k)}^{k-1})$ are the $SU(k)$ Cartan generators obeying $\tr \left[{ H}_{(k)}^{a}  { H}_{(k)}^{b} \right]= \delta^{ab}$, $a,b=1,...,k-1$. They can be expressed as ${ H}_{(k)}^b$=
diag$({ \nu}^b_1,  { \nu}^b_2,..., { \nu}^b_k)$, where $\bm\nu_1,...,\bm\nu_k$ are the weights of the fundamental representation of $SU(k)$. These are $(k-1)$-dimensional vectors that obey $\bm \nu_{B'} \cdot \bm \nu_{C'} = \delta_{B'C'} - {1 \over k}$, where $B',C'=1,..,k$.

\subsection{Fermions on the twisted $\mathbb T^4$}
\label{Fermions on the twisted T4}

We now turn to the adjoint fermions (gauginos), which obey the boundary conditions (\ref{conditions on gauge field}) without the inhomogeneous term
\begin{eqnarray} \label{boundaryconditions}
\lambda(x+L_\mu \hat e_\mu)=\Omega_{\mu}\lambda(x)\Omega_{\mu}^{-1}\,,
\end{eqnarray}
with $\Omega_\mu$ from (\ref{the set of transition functions for Q equal r over N, general solution}).  Omitting the spinor index, we write the gaugino field, an $N \times N$ traceless matrix (this is $\lambda_{i_1}^{i_2}$ and $\sum_{i_1=1}^{N}\lambda_{i_1}^{i_1}=0$), as a block of $k \times k$, $k \times \ell$, $\ell \times k$ and $\ell \times \ell$ matrices (recall $N = k + \ell$):
\begin{eqnarray} \label{blockform}
\lambda &=& \left[\begin{array}{cc}||\lambda_{C'B'}|| & ||\lambda_{C'B}||\\  ||\lambda_{CB'}||& ||\lambda_{CB}||\end{array}\right]~, ~ C',B' \in \{1,...k\}, ~ C,B  \in \{1,...\ell\}~, 
\end{eqnarray}
obeying the tracelessness condition
\begin{eqnarray}\label{tracelesslambda}
\sum_{i_1=1}^{N}\lambda_{i_1}^{i_1}= \sum\limits_{C'=1}^{k} \lambda_{C'C'} +  \sum\limits_{C=1}^{\ell} \lambda_{CC} =0\,.
\end{eqnarray}

The explicit form of the boundary conditions follows from  (\ref{boundaryconditions}) and (\ref{blockform}).  For $\lambda_{C'B'}$, they are
\begin{eqnarray}
\nonumber
\lambda_{C' B'}(x+L_1\hat e_1)&=&\lambda_{C' \; B'}(x)\,,\quad
\lambda_{C' B'}(x+L_2\hat e_2)= e^{i2\pi\frac{C'-B'}{k}}\lambda_{C' B'}(x)\,,\\
\lambda_{C' B'}(x+L_3\hat e_3)&=&\lambda_{C' B'}(x)\,,\quad \lambda_{C' B'}(x+L_4\hat e_4)=\lambda_{C' B'}(x)\,,
\label{BCS lambda A}
\end{eqnarray}
while  $\lambda_{CB}$ obeys\footnote{Here and below, $[C+1]_\ell \equiv C+1$ for $C = 1,..., \ell-1$ and $[C+1]_\ell = 1$ for $C = \ell$.}
\begin{eqnarray}\nonumber
\lambda_{CB}(x+L_1\hat e_1)&=&\lambda_{C B}(x)\,,\quad \lambda_{CB}(x+L_2\hat e_2)=\lambda_{CB}(x)\,,\\
\lambda_{CB}(x+L_3\hat e_3)&=&\lambda_{[C+1]_\ell \; [B+1]_\ell}(x)\,,\quad 
\lambda_{CB}(x+L_4\hat e_4)= e^{i2\pi\frac{C-B}{\ell}}\; \lambda_{CB}(x)\,,\label{BCS lambda a}
\end{eqnarray}
and  $\lambda_{C'B}$:
\begin{eqnarray}
\nonumber
\lambda_{C' B}(x+L_1\hat e_1)&=&\gamma_k^{-k}e^{i2\pi \frac{x_2}{L_2}} \;\lambda_{C'\;B}(x)\,,\quad 
\lambda_{C' B}(x+L_2\hat e_2)=\gamma_k e^{i2\pi\frac{(C'-1)}{k}} \;\lambda_{C'  B}(x)\,,\\\nonumber
\lambda_{C' B}(x+L_3\hat e_3)&=&\gamma_\ell^{-1}e^{i2\pi \frac{x_4}{\ell L_4}}\; \lambda_{C' [B+1]_\ell}(x)\,,\quad 
\lambda_{C' B}(x+L_4\hat e_4)=\gamma_\ell^{-1} e^{-i2\pi\frac{(B-1)}{\ell}}\; \lambda_{C' B}(x)\,.\\
\label{BCS lambda beta}
\end{eqnarray}
We also note that $\lambda_{C B'}$ obeys the h.c. conditions to (\ref{BCS lambda beta}).
In addition,  the dotted fermions $\bar\lambda$ obey boundary conditions equal to the ones given above, written in terms of a decomposition of $\bar\lambda$ in terms of $\bar\lambda_{C'B'}$, $\bar\lambda_{C'B}$, $\bar\lambda_{CB}$ and $\bar\lambda_{CB'}$,  identical to the one in (\ref{blockform}). 

It is clear from this treatment that we should distinguish between the fermions that live in the $U(1)\times SU(k)$, $SU(\ell)$, and the off-diagonal $k\times \ell$ sectors as they satisfy distinct boundary conditions. This will play an essential role in our subsequent discussions.

\subsection{Self-duality and fermions on the tuned $\mathbb T^4$}
\label{Self-duality and fermions on the tuned}

A self-dual fractional instanton must satisfy the relation $F_{12}=F_{34}$, from which we find that  the ratio
of the torus sides have to be tuned to
\begin{eqnarray}\label{selfdualtorus1}
\mbox{self-dual}\, \mathbb T^4:\quad \frac{L_1L_2}{L_3L_4}=N-k \,.
\end{eqnarray}
A torus with periods that satisfy the above relation is said to be a self-dual torus. The action of the self-dual solution is
\begin{eqnarray}\label{action of fractional instanton}
S_{0}=\frac{1}{2g^2}\int_{\mathbb T^4}\mbox{tr}\left[F_{\mu\nu}F_{\mu\nu}\right]=\frac{8\pi^2|Q|}{g^2}=\frac{8\pi^2 k}{N g^2}\,.
\end{eqnarray}

Let us examine the fermion Lagrangian on the tuned $\mathbb T^4$ in the background of the abelian self-dual instanton:
\begin{eqnarray}\label{detaild L f}\nonumber
g^2{\cal L}_f&=&-2 \bar\lambda_{\dot\alpha C'B'}\left[D_\mu \bar\sigma_{\mu}^{\dot\alpha\alpha}\right]_{B'D'}\lambda_{\alpha D'C'}-2 \bar\lambda_{\dot\alpha CB}\left[D_\mu\bar\sigma_{\mu}^{\dot\alpha\alpha}\right]_{BD}\lambda_{\alpha DC}\\\nonumber
&&-2 \bar\lambda_{\dot\alpha C'B}\left[D_\mu \bar\sigma_{\mu}^{\dot\alpha\alpha}\right]_{BD}\lambda_{\alpha DC'}-2 \bar\lambda_{\dot\alpha CB'}\left[D_\mu\bar\sigma_{\mu}^{\dot\alpha\alpha}\right]_{B'D'}\lambda_{\alpha D'C}\\
\nonumber
&&+m\lambda^{\alpha}_{CB}\lambda_{\alpha BC}+m\lambda^{\alpha}_{C'B'}\lambda_{\alpha B'C'}+m\lambda^{\alpha}_{CB'}\lambda_{\alpha B'C}+m\lambda^{\alpha}_{C'B}\lambda_{\alpha BC'}\\
&&+m^*\bar\lambda_{\dot\alpha CB}\bar \lambda_{ BC}^{\dot\alpha}+m^*\bar \lambda_{\dot\alpha C'B'}\bar \lambda^{\dot\alpha}_{ B'C'}+m^*\bar \lambda^{}_{\dot\alpha CB'}\bar\lambda_{ B'C}^{\dot\alpha}+m^*\bar\lambda^{}_{ \dot\alpha C'B}\bar\lambda_{ BC'}^{\dot\alpha}\,.
\end{eqnarray}
In this equation, $[D_\mu]_{BD}$ denotes the $B,D$ component of the adjoint covariant derivative acting on $\lambda$ represented in the block form of eqn.~(\ref{blockform}), and likewise for the other components. 

One  then observes that the $k \times k$ and $\ell \times \ell$ components of (\ref{blockform}) do not couple to the abelian background.
Further, because the background field lies along the $U(1)$ generator $\omega$, it is easy to see $\left[D_\mu \bar\sigma_{\mu}^{\dot\alpha\alpha}\right]_{B'D'}\propto\delta_{B'D'}$ and $\left[D_\mu\bar\sigma_{\mu}^{\dot\alpha\alpha}\right]_{BD}\propto \delta_{BD}$. Then, the fermion Lagrangian ${\cal L}_f$ takes the simple form:
\begin{eqnarray}\label{detaild L f simple}\nonumber
g^2{\cal L}_f&=&-2 \bar\lambda_{\dot\alpha C'B'}\partial_\mu \bar\sigma_{\mu}^{\dot\alpha\alpha}\lambda_{\alpha B'C'}-2 \bar\lambda_{\dot\alpha CB}\partial_\mu\bar\sigma_{\mu}^{\dot\alpha\alpha}\lambda_{\alpha BC}\\\nonumber
&&-2 \bar\lambda_{\dot\alpha C'B}\left[D_\mu \bar\sigma_{\mu}^{\dot\alpha\alpha}\right]_{BB}\lambda_{\alpha BC'}-2 \bar\lambda_{\dot\alpha CB'}\left[D_\mu\bar\sigma_{\mu}^{\dot\alpha\alpha}\right]_{B'B'}\lambda_{\alpha B'C}\\
\nonumber
&&+m\lambda^{\alpha}_{CB}\lambda_{\alpha BC}+m\lambda^{\alpha}_{C'B'}\lambda_{\alpha B'C'}+m\lambda^{\alpha}_{CB'}\lambda_{\alpha B'C}+m\lambda^{\alpha}_{C'B}\lambda_{\alpha BC'}\\
&&+m^*\bar\lambda_{\dot\alpha CB}\bar \lambda_{ BC}^{\dot\alpha}+m^*\bar \lambda_{\dot\alpha C'B'}\bar \lambda^{\dot\alpha}_{ B'C'}+m^*\bar \lambda^{}_{\dot\alpha CB'}\bar\lambda_{ B'C}^{\dot\alpha}+m^*\bar\lambda^{}_{ \dot\alpha C'B}\bar\lambda_{ BC'}^{\dot\alpha}\,,
\end{eqnarray}
and $[D_\mu]_{BB}$ and $[D_\mu]_{B'B'}$ represent the action of the covariant derivative in the background   (\ref{naked expressions}), including the holonomies (\ref{r over N abelian sol}), on the $\ell \times k$ and $k \times \ell$ components of $\lambda$ of (\ref{blockform}).

In the massless limit, it was observed in \cite{Anber:2022qsz,Anber:2024mco} that the fractional abelian self-dual instanton with topological charge $Q=\frac{1}{N}$ supports more fermion zero modes than needed to saturate the gaugino condensate. This is because both Dirac operators $D = \sigma_\mu D_\mu$ and $\bar D = \bar\sigma_\mu D_\mu$ have non-empty kernels in the constant field strength instanton background (this, however, does not contradict the index theorem, as the index is $I= \mbox{ker}\,\bar D -\mbox{ker}\,D$).   While we believe that this is a technical issue, its detailed resolution on the tuned $\T^4$ requires further work. With the extra fermion zero modes, the equations of motion for the bosonic fields in (\ref{EOM}) acquire a nonzero r.h.s., making the background inconsistent, as already remarked in \cite{Anber:2022qsz}. This should be resolvable by appropriately deforming the gauge field background, but the relevant calculations have not yet been performed in the tuned $\T^4$.\footnote{\label{piljinfootnote}A similar problem---the appearance of a nonzero r.h.s. of the bosonic equations of motion due to fermion zero modes---arises in the study of moduli-space dynamics of magnetic monopoles coupled to adjoint fermions, see Sec.~8 in \cite{Weinberg:2006rq}, where related calculations are discussed. We thank Piljin Yi for discussions of this.}

\subsection{Detuned $\mathbb T^4$: nonabelian self-dual instantons, fermion zero modes, and fermion propagator}
\label{Detuned T4 nonabelian self-dual instantons and fermions}

To cure the problem of the extra zero modes, in \cite{Anber:2022qsz,Anber:2023sjn,Anber:2024mco} we chose to deviate from the tuned $\mathbb T^4$ by relaxing the condition (\ref{selfdualtorus1}) and seeking a nonabelian self-dual instanton solution, which was constructed as an expansion in a small detuning parameter $\Delta$. One wonders whether the deviation from the exact massless limit can cure the problems one encounters in the tuned $\mathbb T^4$ without seeking the nonabelian solutions. As we shall show in later discussion,  the issues we encounter on the tuned $\mathbb T^4$ haunt us also at nonzero $m$. 

Therefore, even as we give the gauginos a small mass, we will introduce a detuning parameter\footnote{\label{foot parameterize}For use below, working in a fixed sector with $Q=k/N$, one can parametrize the periods of the detuned $\mathbb T^4$ as 
  \begin{eqnarray}
 L_1=L(1+\xi_1 \Delta)p_1^2\,,\quad  L_2=L(1+\xi_2 \Delta)p_2^2\,, \quad  L_3=L(1+\xi_3 \Delta)p_3^2\,, \quad  L_4=L(1+\xi_4 \Delta)p_4^2\,.
 \end{eqnarray}
 
These periods respect the relation (\ref{def of Delta}), with error ${\cal O}(\Delta^2)$, provided that the parameters 
 \begin{eqnarray}
 \frac{p_1p_2}{p_3p_4}=\frac{1}{k(\xi_3+\xi_4-\xi_1-\xi_2)}=\sqrt{N-k}\,.
 \end{eqnarray}
  } 
\begin{eqnarray}\label{def of Delta}
\Delta\equiv\frac{k (N-k) L_3L_4-k L_1L_2}{\sqrt V}
\end{eqnarray}
 and seek nonabelian self-dual fractional instanton solutions as expansion in $\Delta$ on the detuned $\mathbb T^4$, an approach pioneered in \cite{GarciaPerez:2000aiw} and further developed in \cite{Gonzalez-Arroyo:2004fmv,Gonzalez-Arroyo:2019wpu} and \cite{Anber:2023sjn}. 
 
Before we continue, let us  address the following question.  
Suppose we tune the sides of the torus so that, for $k=1$, we have $\Delta_{k=1} \ll 1$. For these fixed torus sides, then, for how many values of $k>1$ do we still have $\Delta_k \ll 1$---so that we can use the $\Delta$-expansion for more than one fractional charge instanton  sector?  The parameterization given in  footnote \ref{foot parameterize},  or a direct use of eqn.~(\ref{def of Delta}), allows us to provide an answer.  It is most easily stated in the limit  $\Delta_{k=1} \rightarrow 0$, where one finds that $|\Delta_k|= {k (k-1) \over \sqrt{N-1}}$. Thus, the $\Delta$-expansion can be arranged to work for a range of topological sectors from $1$ to $k$ such that $k(k-1) \ll \sqrt{N-1}$.\footnote{\label{footnoteDelta}One can use the more precise expression, $\Delta_{k} \simeq {k (1 - k) \over \sqrt{N-1}} + {(2 N -k -1) k \over 2 (N-1)} \Delta_{k=1} + {\cal{O}}(\Delta_{k=1}^2)$, to check that the indicated range of $k$  does not significantly change  upon varying $0 \le \Delta_{k=1} \le 1/k$ (in  \cite{Anber:2025yub}, a  comparison of the $\Delta$-expansion with ``exact'' $SU(3)$ lattice fractional instantons showed that it works well up to $\Delta \sim 0.2$).}
 
 The procedure of finding a nonabelian multi-fractional instanton as an expansion in $\Delta$ was thoroughly discussed in \cite{Anber:2023sjn}, so we do not repeat it here. We found that a nonabelian self-dual fractional instanton with topological charge $Q = \frac{k}{N}$ on the detuned $ \mathbb{T}^4 $ can be understood as a liquid of instantons, consisting of $k$ overlapping lumps, each carrying  two fermion zero modes. To the leading order in ${\cal O}(\Delta)$ we have
 \begin{eqnarray}\label{perturbed A}
 A_\mu(x)=\hat A_\mu+\left[\begin{array}{cc} \Delta{\cal S}_\mu^{(\Delta) (k)} & \sqrt{\Delta}{w}_\mu^{(\sqrt \Delta)}\\ \sqrt{\Delta} {w}_\mu^{\dagger(\sqrt \Delta)} & \Delta{\cal S}^{(\Delta)(\ell) }_\mu\end{array}\right]\,.
 \end{eqnarray}
 Here, $\hat A_\mu$ is the abelian self-dual solution provided in (\ref{naked expressions}) along with its holonomies (\ref{r over N abelian sol}). The superscript $k$ and $\ell=N-k$ over ${\cal S}$ denote the dimension of the matrix, while $\Delta$ denotes the order of approximation.  Notice that one needs to impose the condition $\mbox{tr}\left[{\cal S}_\mu^{(\Delta) (k)}+{\cal S}^{(\Delta)(\ell) }_\mu\right]=0$ since ${\cal S}$ is an $SU(N)$ matrix, and hence, traceless. 
 
 In \cite{Anber:2023sjn}, we were able to obtain an explicit expression of $w_\mu^{\sqrt\Delta}$ (order-$\sqrt{\Delta}$, $k\times \ell$ matrix), whose exact expression is not important for most of what follows.\footnote{The explicit form of $w_\mu^{\sqrt\Delta}$ will be used in Appendix \ref{sec:liftingzeromodes}, where references to the relevant equations from \cite{Anber:2023sjn} are given, see eqn.~(\ref{integral1}).} The ${\cal S}_\mu^{(\Delta) (k)}$ and ${\cal S}_\mu^{(\Delta) (\ell)} $ terms are more complex. While a systematic method exists for determining them, we did not pursue this endeavor. For consistency to order ${\cal O}(\Delta)$, the fermion Lagrangian (\ref{detaild L f simple}) must be supplemented with the additional terms
 \begin{eqnarray}\label{detuned coupling}
 g^2\delta {\cal L}_f&=&2 i \sqrt{\Delta} \sum_{C,C',D,D',F,F'} \left(\begin{array}{cc}\bar\lambda_{C'D'}\bar\sigma_\mu  & \bar\lambda_{C'D}\bar\sigma_\mu  \cr \bar\lambda_{C D'}\bar\sigma_\mu  & \bar\lambda_{CD}\bar\sigma_\mu  \end{array} \right)  \left\{ \left(\begin{array}{cc} 0 &   w_{\mu \; D'F}^{(\sqrt \Delta)}  \cr  w_{\mu \; DF'}^{\dagger (\sqrt \Delta) }  & 0\end{array} \right)  \left(\begin{array}{cc} \lambda_{F'C'} & \lambda_{F'C} \cr \lambda_{FC'} & \lambda_{FC} \end{array}\right) - ... \right\}, \nonumber \\
 \end{eqnarray}
where the second term in the curly brackets has the two matrices appear in opposite order, completing the commutator (its explicit form shall not be needed in what follows).

\subsubsection{Fermion zero modes in the $Q=k/N$ background}

For $m=0$, the Dirac equation in the self-dual fractional instanton background has $2k$ undotted fermion zero modes. There are no fermion zero modes associated with the dotted fermions on the detuned torus. This is expected since the undotted zero modes shown above exactly saturate the index theorem, which requires the existence of $2k$ fermion zero modes in a $Q=\frac{k}{N}$ instanton background. 
Within the $\Delta$-expansion, explicit expressions for these were found in \cite{Anber:2023sjn}:
 \begin{eqnarray}
\label{lambdaseries}
\lambda_{B'C'} &=&  \lambda_{B'C'}^{(0)} + {\cal O}(\Delta)\,, \quad
\lambda_{BC } =   \lambda_{BC}^{(0)} + {\cal O}(\Delta)\,,\nonumber \\
\lambda_{C' B } &=&  \sqrt{\Delta} \; \lambda_{C'B }^{(\sqrt{\Delta})} + {\cal O}(\Delta^{3/2})\,, \quad
\lambda_{C B' } =  \sqrt{\Delta}\;  \lambda_{C B'}^{(\sqrt{\Delta})} + {\cal O}(\Delta^{3/2})\,.
 \end{eqnarray}
 The ${\cal O}(\Delta^0)$ contributions, $\lambda_{B'C'}^{(0)}$, are given by
\begin{eqnarray} \label{fermion1}
\lambda^{(0)}_{\alpha \; B'C'} &=& \delta_{B'C'}\; \theta_\alpha^{C'}\,,\quad
\lambda^{(0)}_{\alpha \; BC} = -  \;\frac{\delta_{BC}}{N-k} \sum_{C'=1}^{k} \theta_\alpha^{C'}, 
\end{eqnarray}
 where $\theta_\alpha^{C'}$ are constant spinors, and we momentarily restored the spinor index $\alpha=1,2$. 
The ${\cal O}(\sqrt\Delta)$ contributions are given by 
\begin{eqnarray}\label{lambdazeromode11}\nonumber
\lambda_{1 \; C'B}^{(\sqrt{\Delta})} &= & \eta_2^{C'} {\cal G}_{3 \; C'B}^{}(x)\,,\quad
\lambda_{2 \; C'B}^{(\sqrt{\Delta})} = 0\,,\\
\lambda_{1 \; CB'}^{(\sqrt{\Delta})} &= & 0\,,\quad
\lambda_{2 \; BC'}^{(\sqrt{\Delta})} = \eta_1^{C'}   {\cal G}_{3 \; C'B}^{* \; }(x).
\end{eqnarray}
where
 \begin{equation}\label{detuned zero 1}
 \eta^{C'} \equiv    \theta^{C'} + \frac{1}{N-k} \sum\limits_{B'=0}^{k-1} \theta^{B'}\,,
 \end{equation} 
 and ${\cal G}_{3 \; C'D}^{}(x)$ are complicated functions on $\mathbb T^4$, whose explicit form is given in Appendix C in \cite{Anber:2023sjn} (they shall not play a role in this work).
 Thus, we have in total $2 k$ zero modes labeled by $\theta_{1,2}^{C'}$, with $C'=1,...,k$. The ${\cal O}(\Delta)$ terms are more complex, and while a systematic method exists for determining them, we will not pursue this. 
 
The bottom line is that to ${\cal O}(\Delta^0)$,  the fermion zero modes arise from fermions residing in the Cartan subalgebra of the $U(1) \times SU(k)$ sector. 
We now introduce the notation for these zero modes that we use further in the paper. We use the labels $p=1,...,k$ and $\beta=1,2$ to label the $2k$ different zero modes. As $C$-number functions 
(the Grassmann variables are attached to the $C$-number solutions of the Dirac equation when defining the path integral, see Appendix \ref{Systematics of the propagator}), the zero modes  are $SU(N)$ adjoint elements carrying an undotted-fermion index $\alpha$, i.e. their wavefunctions are denoted, in all generality $(\psi^{(0) }_{\alpha, p, \beta})^{ij}$. 

 At order $\Delta^0$, in order to describe these zero modes, we 
  we combine the $U(1)$ generator $\omega$ of (\ref{omega}) with the $SU(k)$ Cartan generators and define the new basis of $N\times N$ matrices  
 \begin{equation}
\label{Hbasisfirst}
\bm {\tilde H}\equiv \left(\frac{\omega}{2\pi   \sqrt{N k(N-k)}}, \bm H_{(k)}\right) = (\tilde H^1, \tilde H^2,... \tilde H^k), ~ \mbox{tr}\left[\tilde H^{b_1} \tilde H^{b_2}\right]=\delta_{b_1b_2}, ~ b_1, b_2 = 1,\ldots, k~,
\end{equation} where we denoted the generator proportional to $\omega$ by $\tilde H^1$ and the $SU(k)$ Cartan generators, embedded into $N\times N$ matrices (by filling in zeros), by $\tilde H^{b}$, $b=2,...,k$. The $2 k$ zero modes (\ref{fermion1}) then are then rewritten using this basis as
\begin{eqnarray}\label{zeromodesnormalized}
(\psi^{(0) }_{\alpha, p, \beta})^{ij} = {1 \over \sqrt{V}} \epsilon_{\alpha\beta} \; (\tilde H^p)^{ij}~, ~p=1,...,k, \; \beta={1,2}~,
\end{eqnarray}
 where we stress again that $p, \beta$ are indices used to label the $2k$ zero modes. The normalization factor is introduced so that, from (\ref{Hbasisfirst}),   the  zero-modes (\ref{zeromodesnormalized}) obey the normalization used in Appendix \ref{Systematics of the propagator}:
 \begin{eqnarray}\label{zeromodenorm}
 \int\limits_{\T^4} \tr  \psi^{(0) \; \alpha}_{ p,\beta}   \psi^{(0)}_{q,\beta', \alpha} = \delta_{pq} \epsilon_{\beta\beta'}.
 \end{eqnarray} 

\subsubsection{The fermion propagator in the self-dual $Q=k/N$ background}
\label{sec:propagatorkovern}

 The main tool used in our semiclassical calculations of the  fermion correlation functions in the fractional instanton background is the expression for the unnormalized propagator of the fermions in a general self-dual $Q= {k \over N}$ background.  The background is assumed ``generic,'' namely such that the covariant Laplacian acting on scalars in the adjoint representation\footnote{As follows from the well-known Weitzenb\" ock formulae (\ref{relations 1}), the absence of adjoint Laplacian zero modes is equivalent to the absence of dotted fermion  zero modes.} has no zero modes. Our $\Delta$-expansion background on the detuned $\T^4$ is an example of such a background.
 
 The propagator,   whose detailed derivation is given in  Appendix  \ref{Systematics of the propagator},  is determined by
 the eigenfunctions $\phi_n$  of the covariant Laplacian and the $2k$ undotted-fermion zero modes, $\psi^{(0)}_{\alpha, p, \beta}$ (given, to leading order in $\Delta$, in (\ref{zeromodesnormalized})). The Laplacian eigenfunctions  $\phi_n$ are Hermitean ($N\times N$) adjoint scalar fields obeying $D_\mu D_\mu \phi_n = -\omega_n^2 \phi_n$, where $D_\mu = \partial_\mu + i [A_\mu,..]$ is the adjoint covariant derivative. The  $\phi_n$ satisfy $\T^4$ boundary conditions twisted by $\Omega_\mu$ (i.e.~(\ref{conditions on gauge field}) without the non-homogeneous term) and are normalized as $\int\limits_{\T^4} \tr \phi_n \phi_m = \delta_{nm}$. 
 The wave functions of the $2k$ adjoint-fermion zero modes are $\psi^{(0)}_{\alpha, p, \beta}$ ($\alpha$ is the spinor index and $p=1,...,k$, $\beta=1,2$ label the $2k$ zero modes), subject to the normalization (\ref{zeromodenorm}) and omitting the adjoint indices.

 The unnormalized propagator in the self-dual $Q={k/N}$ background is given as  a sum over the fermion zero modes and the eigenvalues  and eigenfunctions of the adjoint Laplacian. It has the form of eqn.~(\ref{propagator 2}), which we reproduce here:
  \begin{eqnarray}\label{propagator 2 maintext}
&& \left(\begin{array}{cc} \langle \lambda_\alpha(x)  \otimes \lambda^\beta(y) \rangle &\langle  \lambda_\alpha(x) \otimes  \bar\lambda_{\dot\beta}(y) \rangle \cr \langle \bar\lambda^{\dot\alpha}(x) \otimes  \lambda^\beta(y) \rangle & \langle \bar\lambda^{\dot\alpha}(x) \otimes  \bar\lambda_{\dot\beta}(y) \rangle\end{array} \right)_{unnorm.}  \\
&&~~= {\cal{D}}_k^f(m){g^2 \over 2} \left\{ \left(\begin{array}{cc}  { m^* \over |m|^2}  \sum\limits_{p=1}^k \left( \psi^{(0)}_{\alpha p, 1}(x) \otimes  \psi^{(0) \beta}_{\;\;\; p,2}(y) - \psi^{(0)}_{\alpha p, 2}(x)  \otimes \psi^{(0) \beta}_{\;\;\; p,1}(y)\right) &~ 0\cr 0& ~0\end{array}\right)  \right.\nonumber  \\
&&\left. ~~~~~~~+  \sum_n \left(\begin{array}{cc}  {m^* \over \omega_n^2 + |m|^2 } \; (\sigma_\mu \bar\sigma_\nu)_\alpha^{\; \; \beta} \;  {D_\mu \phi_n(x)  \otimes  D_\nu \phi_n(y) \over \omega_n^2}&  {\sigma_{\mu \; \alpha \dot\beta} \over \omega_n^2 + |m|^2}\; {D_\mu \phi_n(x)  \otimes  \phi_n(y)}\cr - { \bar\sigma_{\nu}^{\dot\alpha \beta}  \over \omega_n^2 + |m|^2} \; { \phi_n(x)  \otimes D_\nu \phi_n(y)}&  {m \over \omega_n^2 + |m|^2} \; \delta^{\dot\alpha}_{\dot\beta}\; \phi_n(x)  \otimes \phi_n(y)  \end{array}\right) \right\}. \nonumber
\end{eqnarray}
Here $D_\mu$ is  adjoint representation covariant derivative in the $Q=k/N$ background and 
 \begin{eqnarray}
\label{dfermion maintext}
{\cal{D}}_k^{f}(m) =  \left({2 m \over g^2}\right)^k \prod_n \left({16\over g^4} (\omega_n^2 + |m|^2)\right) \prod_{p=1}^k \eps_p  \prod_{n} {\eps_n}.
\end{eqnarray}
is the massive fermion determinant in the same background, which includes a product over all eigenvalues of the Laplacian.\footnote{The parameters $\epsilon_p, \epsilon_n = \pm 1$ were introduced in (\ref{measure}) to  define the Grassmann integrals. We take $\epsilon_p=\epsilon_n =1$ in what follows.} We emphasize that this is the unnormalized propagator, while the problem of regularization will be attacked in section \ref{Determinants and regularization}.

All outer products of wave functions and their derivatives appearing in (\ref{propagator 2 maintext}) should be  explicitly  understood as
 \begin{eqnarray}\label{outer 11}
 \langle \lambda_\alpha(x) \otimes  \lambda^\beta(y) \rangle &\rightarrow& \langle \lambda_{ij \; \alpha}(x) \lambda_{kl}^\beta(y) \rangle, \nonumber \\
 D_\mu \phi_n(x) \otimes \phi_n(y) &\rightarrow& (D_\mu \phi_{n})_{ij}(x)\; \phi_{n \; kl}(y),~ \text{etc.}
 \end{eqnarray}
where $i,j,k,l = 1,...N$ are   adjoint indices. 

 As already noted, eqn.~(\ref{propagator 2 maintext}) is very generally valid: it holds in any exactly self-dual $Q=k/N$ background, assumed to be ``generic,'' i.e. such that the adjoint Laplacian $D_\mu D_\mu$ has no zero modes. No exact expression for such a background is known, much less expressions for the eigenvalues and eigenfunctions of the Laplacian. 
 What makes (\ref{propagator 2 maintext}) useful is the fact that, within the $\Delta$-expansion,  the self-dual background is given in  a series expansion in $\Delta$, with the first few terms shown in (\ref{perturbed A}). 
All quantities appearing   in (\ref{propagator 2 maintext}) should be understood via the same $\Delta$-expansion: the eigenvalues of the Laplacian $\omega_n$, its eigenfunctions $\phi_n$, the background field entering the covariant derivative $D_\mu$, and the zero mode wave functions $\psi_{\alpha p,i}^{(0)}$ are all given as an expansion in $\Delta$.
 
The calculation of (\ref{propagator 2 maintext})---and therefore, the determination of fermion correlators in the instanton background---is  feasible due to the fact that the order-$\Delta^0$  fractional instanton background (\ref{naked expressions}) is simply a $U(1) \in SU(N)$ constant field strength background along the generator $\omega\sim  {\rm{diag}} ({\ell I_k},{ -k I_\ell})$ of eqn.~(\ref{omega}). The adjoint Laplacian in this background   factorizes into Laplacians acting only in the $k\times k$, $\ell\times \ell$, as well as the $k \times \ell$ and $\ell \times k$ parts of  the $N \times N$ adjoint matrix.
The eigenvectors and eigenvalues of each of these Laplacians can then be separately determined and used in the calculating (\ref{propagator 2 maintext}). Each of these eigenvectors only has nonzero components in either the $k\times k$, $\ell \times \ell$, or in $k\times \ell$ and $\ell \times k$, thus their contribution to the fermion propagators factorizes. 

As already discussed near eqn.~(\ref{blockform}), the  $i,j,k,l = 1,...,N$ adjoint indices  
then naturally  split into $SU(\ell)$ $(C,B,..)$ and $SU(k)$ $(C',B',...)$ ones. This splitting is used to develop explicit expressions for $\phi_n$ and the propagators, by solving for the  Laplacian eigenfunctions within the $\Delta$-expansion. This is a task that we systematically undertake in various voluminous Appendices:  Appendix \ref{sec:diagonalsuktimesu1}  for the $SU(k) \times U(1)$ parts of the adjoint, Appendix \ref{appx:suellpropagator} for the $SU(\ell)$ parts of the adjoint matrix, and, finally, Appendix \ref{sec:offdiagonalktimesell} for the $k\times \ell$ and $\ell \times k$ off-diagonal parts of the $SU(N)$ adjoint. The one subtle point, studied in detail in Appendix \ref{sec:liftingzeromodes}, is the lifting of the order-$\Delta^0$ Laplacian zero mode at order $\Delta$. There,  we also discuss how this affects  the $\Delta$-expansion of the propagator (\ref{propagator 2 maintext}).

We now briefly summarize our findings for the fermion propagators, focusing on the $11$ and $22$ elements of (\ref{propagator 2 maintext}), the  $\langle \lambda \lambda\rangle$ and $\langle\bar\lambda \bar\lambda\rangle$ propagators, the ones that we focus on in the rest of the paper. The off-diagonal elements of (\ref{propagator 2 maintext}) can similarly be determined from the results of the Appendices, if needed in the calculation of   correlation functions other than the ones we compute here.

{\bf \flushleft{Fermions in $SU(k) \times U(1)$:}} We start with the fermions that live in the $U(1) \times SU(k)$ space.  It proves easier to use the Cartan-Weyl basis of $SU(k)$: these are the Cartan $\bm H_{(k)}$ and root $E_{\bm\beta_{B'C'}}$ generators. Here, $\bm\beta_{B'C'}$, $B'\neq C'=1,2,..,k$ are the $k^2-k$ distinct (positive and negative) roots. 
We begin with the propagators of the $SU(k) \times U(1)$ Cartan components of the gauginos.   Using the basis of $SU(k) \times U(1)$ Cartan generators already introduced in (\ref{Hbasisfirst}), we expand the fermion field in diagonal ($\lambda_b$) and off-diagonal components
\begin{eqnarray} 
\lambda(x) = \sum\limits_{b=1}^k \lambda_b(x) \tilde{H}_b + \text{off diagonal},\label{Hbasis old}
\end{eqnarray}
where the off-diagonal $SU(k)$ pieces are considered further below (see (\ref{lambdabarsquared2 main text})). The propagator of the undotted $SU(k) \times U(1)$ Cartan components of the fermions is given in (\ref{11element}) of Appendix \ref{sec:diagonalsuktimesu1} and has the form
\begin{eqnarray} \label{11element main text}
  \langle \lambda_{b\;\alpha}(x) \lambda_{b'}^\beta(y) \rangle_{unnorm.} &=&\delta_{bb'} \; \delta_\alpha^\beta \;{\cal{D}}_k^f(m) {g^2 \over 2 V}    \left({ m^* \over |m|^2}  +  \sum\limits_{p_\mu \in {2 \pi\over L_\mu} \Z}' {m^* \over p_\mu^2} e^{i p_\mu (x_\mu - y_\mu)}(1 + \ldots)\right), \nonumber \\
\end{eqnarray}
where the prime over the summation sign over $n_\mu$ ($p_\mu = {2 \pi n_\mu \over L_\mu}$)  excludes the point $n_1=n_2=n_3 = n_4 = 0$. 
The propagator of the dotted fermions in the Cartan of $SU(k)\times U(1)$, on the other hand, has the form given in (\ref{22element3}) of Appendix \ref{sec:liftingzeromodes}\footnote{We stress that in writing the value of $c$ below we made the simplifying assumption $k \ll N$, only to simplify the writing of the propagator. The values of $c$ for $b=1$ and $b=2,...,k$ slightly differ, due to the different energy shifts, see (\ref{energyshifts1}).}
\begin{eqnarray}\label{22element3 maintext}
 \langle \bar\lambda_b^{\dot\alpha}(x) \bar\lambda_{b' \dot\beta}(y) \rangle_{unnorm.}  &=&  \delta_{bb'}  {\cal{D}}_k^f(m) \; {g^2 \over 2 V} \; \delta_{\dot\beta}^{\dot\alpha} \; \left({m  L^2 \over c \Delta }(1 + \ldots) + \sum\limits_{p_\mu \in {2 \pi\over L_\mu} \Z}' { 
 m \; e^{i p_\mu (x_\mu - y_\mu)} \over p_\mu^2  }(1 + \ldots) \right), \nonumber \\
 \text{where} \; c &\equiv& {4 \pi \over kN}~.
\end{eqnarray} 
The order-$\Delta^0$ Laplacian has $k$ zero modes, with constant eigenfunctions n the $SU(k) \times U(1)$ Cartan directions (as already noted, the presence of these zero modes is in one to one correspondence with the presence of zero modes of the dotted fermions). These zero modes are lifted at order $\Delta$. This lifting is the reason for the appearance of the
 the $m L^2\over c \Delta$ term in the propagator of each of the $k$ Cartan components of the dotted fermions in (\ref{22element3 maintext}) (naturally, there are no such terms in the undotted fermion propagator (\ref{11element main text})).  
The constant $c$ is determined  in 
 Appendix \ref{sec:liftingzeromodes}, where  the lifting $k$ zero eigenvalues of the order $\Delta^0$ Laplacian   to $c \Delta \over L^2$ at order $\Delta^1$ is determined by a perturbative calculation, see (\ref{energyshifts1}) there.

In both (\ref{11element main text}) and (\ref{22element3 maintext}) the omitted terms denoted by $\ldots$ represent corrections that scale as
$\Delta \ll 1$, $(|m|L)^2 \ll 1$, or ${(|m|L)^2 \over c \Delta} \ll 1$. We stress that the small-$m$ small-$\Delta$ limit we consider is:
\begin{eqnarray}\label{smalllimit maintext}
{(|m|L)^2 \over c} \ll   \Delta \ll 1~, ~ c \equiv {4 \pi \over k N}~,
\end{eqnarray}
and that  the order of limits is motivated by the fact that the SYM theory results are  obtained at fixed  $\Delta \ll 1$ and $m =0$.

The remaining  propagator of $k\times k$ components of the fermions is the one of the off-diagonal $SU(k)$ components. This is derived in Appendix \ref{sec:noncartansuktimesu1} and is easiest to give using explicit index notation, where we stress that  $D' \ne E'$ and $F' \ne G'$:
 \begin{eqnarray}
\label{lambdabarsquared2 main text}
&&\langle \bar\lambda^{\dot\alpha}_{D'E'}(x) \bar\lambda_{\dot\beta \; F'G'}(y) \rangle_{unnorm.}~ \nonumber \\
&& = {g^2 \over 2V} {\cal{D}}_k^f(m)  \\
&& \times \sum\limits_{p_\mu = {2 \pi n_\mu \over L_\mu}}  
  {m\delta^{\dot\alpha}_{\dot\beta} \over |m|^2 +  (p_\mu  + \delta_{\mu 2} {2 \pi \over L_2}{D'-E'\over k})^2 }  
    e^{  i    x_\mu(p_\mu + \delta_{\mu,2} {2 \pi (D'-E')\over k L_2}) - i y_\mu(p_\mu - \delta_{\mu,2} {2 \pi (F'-G')\over k L_2}) } \delta_{D'G'}\delta_{E'F'}~.    \nonumber
\end{eqnarray}
The undotted propagator $\langle \lambda_{\alpha \; D'E'}(x) \lambda^\beta_{F'G'}(y)\rangle$ is given by a virtually identical expression, with the only replacement $m \rightarrow m^*$ and $ \delta^{\dot\alpha}_{\dot\beta} \rightarrow \delta_\alpha^\beta$, as per eqn.~(\ref{lambdasquared2}). 
As in (\ref{22element3 maintext}) and (\ref{11element main text}), for consistency with the small-$\Delta$ and small-$|m|L$  expansion, the $|m|^2$ term in the denominator in (\ref{lambdabarsquared2 main text}) above (and in (\ref{lambdasquared2})) should be omitted.

{\bf \flushleft{Fermions in $SU(\ell)$:}} We now turn to the propagators of the $SU(\ell)$ components of the fermions, simply quoting the result from Appendix \ref{appx:suellpropagator}, eqn.~(\ref{lambdabarsuell1}):
 \begin{eqnarray}\label{lambdabarsuell1 maintext}
&& \langle \bar\lambda^{\dot\beta}_{BC}(x)  \bar\lambda_{\dot\alpha \;DE}(y)\rangle_{unnorm.}  
  \\
 &&= \delta_{\dot\alpha}^{\dot\beta}  \; {\cal D}^f_k(m) \; {g^2 \over 2 \ell V}  \nonumber \\
 && \times \sum\limits_{k_\mu = {2 \pi n_\mu \over L_\mu}, n_\mu \in \Z} \sum\limits_{(p_3,p_4) \in \Z_\ell^2} {m \;  e^{- i (x_\mu - y_\mu) (k_\mu + \delta_{\mu 3} {2 \pi p_3 \over \ell L_3} + \delta_{\mu 4} {2 \pi p_3 \over \ell L_3})}  \over |m|^2 +  \sum\limits_{\mu = 1}^4 \left(k_\mu + \delta_{\mu 3} {2 \pi p_3 \over \ell L_3} + \delta_{\mu 4} {2 \pi p_3 \over \ell L_3}\right)^2}  \; (J_{p_3, p_4})_{BC}  (J_{- p_3, - p_4})_{DE}, \nonumber 
 \end{eqnarray}
 where $J_{\bm p} = e^{-i\frac{\pi p_3p_4}{\ell}}Q_\ell^{-p_3}P_\ell^{p_4}$, with $P_\ell$ and $Q_\ell$ the $\ell \times \ell$ shift and clock matrices, see (\ref{JPELL1}). The sum over $p_3, p_4$ does not include $p_3=p_4 =0$.
An  expression identical to (\ref{lambdabarsuell1 maintext}) is obtained for  $\langle \lambda_{\alpha \; BC}(x)  \lambda(x)^\beta_{DE }(y) \rangle$, with the replacement $m \rightarrow m^*$ and $\delta^{\dot\alpha}_{\dot\beta} \rightarrow \delta_\alpha^\beta$.

{\bf \flushleft{Fermions in the off diagonal $k \times \ell$ and $\ell \times k$ blocks:}}The final remaining nonvanishing propagators in the $Q=k/N$ background are for the off-diagonal, $k \times \ell$ and $\ell\times k$, components of the fermions. Finding these propagators is  the subject of Appendix \ref{sec:offdiagonalktimesell}, with the result given  in (\ref{barlambdaktimesellpropagator}, \ref{lambdaktimesellpropagator}):
\begin{eqnarray}\label{barlambdaktimesellpropagator maintext}
 \langle \bar\lambda^{\dot\alpha}_{C'C}(x) \bar\lambda_{\dot\beta \; DC'}(y) \rangle_{unnorm.} =   \delta^{\dot\alpha}_{\dot\beta} \; {\cal{D}}_k^f(m)\; {g^2 \over 2V} \sum\limits_{\ell_{(1)}, \ell_{(2)} = 0}^\infty {m \;  \varphi _{C'C \ell_{(1)} \ell_{(3)}}(x) \;  \varphi^*_{C'D \ell_{(1)} \ell_{(3)}}(y)  \over \omega_{\ell_{(1)}, \ell_{(2)} }^2 + |m|^2}  ~.
\end{eqnarray}
Here $ \omega_{\ell_{(1)}, \ell_{(2)} }^2 =  {4 \pi \over L_1 L_2} (\ell_{(1) }+ \ell_{(3) } + 1)$ are the Laplacian eigenvalues in the $k\times \ell$ and $\ell \times k$ subspace. The undotted propagator has a slightly different expression
\begin{eqnarray} \label{lambdaktimesellpropagator maintext}
 \langle \lambda_{\gamma \; C'C} (x) \lambda^\beta(y)_{D C'}\rangle_{unnorm.}  = {\cal{D}}_k^f(m)\; {g^2 \over 2V}  \sum\limits_{\ell_{(1)}, \ell_{(2)} = 0}^\infty {m^* \;   \sigma_{\mu \gamma \dot\gamma} D_\mu  \varphi _{C'C \ell_{(1)} \ell_{(3)}}(x) \;  \bar\sigma_\nu^{\dot\gamma \beta}  D_\nu^* \varphi^*_{C'D \ell_{(1)} \ell_{(3)}}(y)  \over \omega_{\ell_{(1)}, \ell_{(2)} }^2(\omega_{\ell_{(1)}, \ell_{(2)} }^2 + |m|^2)} ~.\nonumber\\
\end{eqnarray}
We note that the $k\times \ell$ and $\ell\times k$ propagators are the only ones that have dependence on the $4k$ moduli (\ref{r over N abelian sol}) of the fractional instanton (because these components are the only ones coupling to the background (\ref{naked expressions})). The moduli dependence enters through the eigenfunctions of the Laplacian, $\phi_{C'C \; \ell_{(1)} \ell_{(3)}}$. The expression for these is found in (\ref{finalvarphiBCS}) and is too bulky to quote here; we only note that these eigenfunctions are ultimately determined by the normalized eigenfunctions of simple harmonic oscillators and intricately depend on the twisted boundary conditions. 

{\flushleft{A}}s already noted, in all propagators above (\ref{lambdabarsquared2 main text}, \ref{lambdabarsuell1 maintext}, \ref{barlambdaktimesellpropagator maintext}, \ref{lambdaktimesellpropagator maintext}), the additive $|m|^2$ in the denominators should be dropped in the leading small-$|m|L$, small-$\Delta$ limit of (\ref{smalllimit maintext}), as was already done for the $SU(k) \times U(1)$ Cartan propagators in (\ref{11element main text}, \ref{22element3 maintext}).

\subsection{Gauge-invariant observables}
\label{Guage-invariant observables}
We will focus on gauge-invariant observables, which include condensates or gauge-invariant densities, Wilson loops, and spacetime-dependent correlators.

We begin with open Wilson  lines, which are used to construct gauge-invariant spacetime-dependent correlators. These Wilson lines are defined as:
\begin{eqnarray}\label{Wilson 1 main}
{\cal W}_\mu (x) \equiv e^{i\int_0^{x_\mu} \left(A_\mu(x) \right)}\,,
\end{eqnarray}
and can be decomposed into contributions from the $SU(k)$ and $SU(\ell)$ spaces. The explicit form of ${\cal W}_\mu$ in the fractional instanton background is given in Appendix  \ref{Expressions of Wilsons lines}. 
We are interested in gauge-invariant fermion bilinears. The adjoint fermion $\lambda(x)$ transforms as
\begin{eqnarray}\label{adjoint trans}
\lambda'(x)=U(x)\lambda(x)U^\dagger (x)\,.
\end{eqnarray}
Thus,  for the adjoint fermions, we can construct the gauge-invariant bilinear operators (the insertions $\Gamma_{\nu_1\nu_2...}$ are $\sigma_\mu$ or $\bar\sigma_\mu$ matrices):
\begin{eqnarray}
\mbox{tr} \left[\lambda(x)\Gamma_{\nu_1\nu_2...} \prod_{\mu=1}^4{\cal W}_\mu(x) \lambda(0) {\cal W}^\dagger_\mu(x)  \right]\,,
\end{eqnarray}
which simplifies greatly when the Wilson lines ${\cal W}_\mu$ are abelian, the case when we consider the computations to ${\cal O}(\Delta^0)$.

\section{Correlators}
\label{all the Correlators}

In this section, we study the expectation values of the $2$-point or higher-point fermion operators, or simply ``the correlators.'' We shall perform our study in the background of a sector carrying a general topological charge $\pm\frac{k}{N}$, where $k=1,.., N-1$. We begin, however, with a few comments about the partition function and correlators in the $Q=0$ sector.

\subsection{Comments on the partition function and the sector $Q=0$}
\label{Comments on the sector Q0}

As outlined in Section \ref{Setup}, the computation of a physical observable is performed separately in each topological sector, including the sector with $Q = 0$. The final result is obtained by summing the contributions from all sectors. Here, we focus on the partition function and correlator in the $Q = 0$ sector.  

A vanishing topological charge can be achieved by imposing no twists in any direction, i.e.~taking $n_{12} = n_{34} = 0$ and imposing periodic boundary conditions for all fields. However, in the massless gaugino limit, the absence of twists introduces subtleties into the calculations. Specifically, the partition function $ Z_{Q=0}[\eta=0,\bar{\eta}=0]$  appears to vanish  in this limit due to the presence of fermion zero modes and the necessity to integrate over them.\footnote{It is not known (to us) how to reconcile the path integral intuition with the nonzero value of the Witten index computed with periodic boundary conditions on $\T^3$ in the Hamiltonian formalism. We only note that there are related subtleties in this calculation, alluded to in \cite{Witten:2000nv}. }
 Furthermore, in the massless case, sectors with $Q \neq 0$ also support fermion zero modes. As a result, the total partition function, given by $ \sum_{Q} Z_Q[\eta=0,\bar{\eta}=0]$, vanishes identically, preventing the computation of physical observables via (\ref{Reg O}). 
 
  To overcome the vanishing of the total partition function, we follow the procedure we adopted in \cite{Anber:2023sjn}: the $Q=0$ is selected by imposing twists in only one two-plane, e.g., by taking $n_{12}=0, n_{34}=1$. This twist lifts all the continuous zero modes of the supersymmetric theory, leaving behind $N$ inequivalent gauge configurations with zero action. Such configurations are distinguished by the $N$ values of the Wilson line wrapping, for example, the $x_2$ direction (see  \cite{Witten:1982df}, and, for a discussion of the subtleties involved in path integral framework \cite{Anber:2024mco}). 
  Thus, we have in the massless-gaugino limit, after regularizing the theory and thanks to supersymmetry, which provides a direct way to obtain the determinants:%
\begin{eqnarray}\nonumber
Z^{\scriptsize T}[\eta=\bar\eta=0]_{m=0}&=&\sum_{Q=0,\frac{\pm 1}{N},\frac{\pm 2}{N},... } Z_{Q}^{\scriptsize\mbox{Reg}}[\eta=0,\bar\eta=0]=\underbrace{N}_{Q=0\, \mbox{sector}}+\underbrace{0+0+...}_{\mbox{higher Q sectors}}\\
&=&N\,. \label{ztotal 2}
\end{eqnarray}
Adding a gaugino mass modifies this result. In writing the expression for $Z^T$ at $m \ne 0$ below, we assume the validity of the semiclassical approximation on $\T^4$ and ignore higher than one-loop corrections:  
\begin{eqnarray}\nonumber  \label{grand ZT with mass}\nonumber
&&Z^{\scriptsize T}[\eta=\bar\eta=0]_{m\neq 0} = \sum_{Q=0,\frac{\pm 1}{N},\frac{\pm 2}{N},... } Z_{Q}^{\scriptsize\mbox{Reg}}[\eta=0,\bar\eta=0] \\\nonumber
&=& {\cal F}_0 (|m|^2, M_{PV})\left[N +\sum_{k>0}   e^{-\frac{8\pi^2 k}{Ng^2}}\mu_{B}^{(k)}\left({{\cal F}_k(m, M_{PV}) \over   {\cal F}_0(|m|^2, M_{PV}) }e^{i\frac{\theta k}{N}}+{{\cal F}_{-k}(m, M_{PV}) \over   {\cal F}_0(|m|^2, M_{PV}) }e^{-i\frac{\theta k}{N}} \right)\right]. \\ 
\end{eqnarray}  
Here ${\cal F}_0(|m|^2, M_{PV})$ and ${\cal F}_{\pm k}(m, M_{PV})$ denote the determinants of the fluctuations in the $Q=0$ (perturbative vacuum with $n_{34}=1$, $n_{12}=0$ twist) sector  and $Q=\pm\frac{k}{N}$ (multifractional instanton) sectors, respectively, after using the Pauli-Villars regulators, where $\mu_{B}^{(k)}$ is the volume of the bosonic moduli space in the sector $Q=k/N$ (see (\ref{volume of mub})). 
The reason we pulled out the overall factor of the $Q=0$ determinant is that, as is well known, this quantity is UV divergent in the softly-broken SYM theory, while as we will argue below the ratio ${{\cal F}_k(m, M_{PV})/{\cal F}_0(|m|^2, M_{PV}) }$ is UV finite.
The details of the regularization are discussed in section \ref{Determinants and regularization}.

We can also study the bilinear fermion correlators in the sector $Q=0$ subject to the twists $n_{12}=0, n_{34}=1$. The propagator is found in Appendix \ref{sec:singletwistpropagator}, where it is noted that it is similar to the $SU(\ell)$ propagator already found in (\ref{lambdabarsuell1 maintext}), see Appendix \ref{appx:suellpropagator} for derivation, upon replacing $\ell \rightarrow N$. Here we quote the final result, omitting the fermion determinant:
 \begin{eqnarray}\label{Q0 correlators}
&& \langle \lambda_{\alpha \; ij}(x)  \lambda^\beta_{kl}(y)\rangle
  \\
 &&= \delta_{\alpha}^{\beta}  \;  {g^2 \over 2 N V}   \sum\limits_{k_\mu = {2 \pi n_\mu \over L_\mu}, n_\mu \in \Z} \sum\limits_{{\bm p} \in \Z_N^2} {m^* \;  e^{- i (x_\mu - y_\mu) (k_\mu + \delta_{\mu 3} {2 \pi p_3 \over N L_3} + \delta_{\mu 4} {2 \pi p_3 \over N L_3})}  \over |m|^2 +  M_{\bm p,k}^2}  \; (J_{p_3, p_4})_{ij}  (J_{- p_3, - p_4})_{kl}, \nonumber  
 \end{eqnarray}
 where $J_{\bm p} = e^{-i\frac{\pi p_3p_4}{N}}Q_N^{-p_3}P_N^{p_4}$, with $P_N$ and $Q_N$ the $N \times N$ shift and clock matrices ($J_{\bm p}$ obey (\ref{J identity}) with $\ell \rightarrow N$).
Here, $i,j,k,l=1,2,..,N$ are the color indices.

\subsection{Correlators in the $Q=\frac{k}{N}\neq 0$ sector}
\label{Correlators in the higher Q sector}

Using the machinery introduced in Section \ref{Detuned T4 nonabelian self-dual instantons and fermions}, we proceed to calculate the $2$-point correlators in the background of a nonabelian self-dual instanton carrying topological charge $Q=\frac{k}{N}$ on the detuned $\mathbb T^4$. We shall perform our analysis to leading order in $\Delta$.  We are interested in computing gauge-invariant fermion bilinears, of the form:\begin{eqnarray}\label{the full things}\nonumber
&&\langle {\cal C}_{\nu_1\nu_2...}(x)\rangle_{\frac{k}{N}}\equiv\\\nonumber
&&\langle \mbox{tr}\left[\lambda(x)\prod_{\mu=1}^4{\cal W}_\mu(x)\Gamma_{\nu_1\nu_2..} \lambda(0) {\cal W}^\dagger_\mu(x) \right]\rangle=\langle \lambda_{i_1 i_2} (x)\prod_{\mu=1}^4\left[{\cal W}_{\mu}\right]_{i_2 i_3}(x)\left[{\cal W}^\dagger_{\mu}\right]_{i_4 i_1}(x) \Gamma_{\nu_1\nu_2..}\lambda_{i_3 i_4}(0)\rangle\,,\\
\end{eqnarray}
where the indices $i_1,i_2,..=1,2,..,N$ are the color indices\footnote{The fermions and Wilson lines are in the adjoint representation, transforming as $\lambda_{ij} \rightarrow U_{i k} \lambda_{kl} (U^\dagger)_{kj}$, and we are using Einstein's summation convention.} and the ${\cal W}_\mu$ are open Wilson  lines  in the $x_\mu$ direction, given by (\ref{abelian wilson gr1}, \ref{abelian wilson gr2}, \ref{Wilson lines in tilde H coor}). The insertions $\Gamma_{\nu_1\nu_2..}$ are contractions of $\sigma_\mu$ and $\bar\sigma_\mu$ matrices, with appropriate spinor indices to contract with the spinor indices of $\lambda$. Identical $2$-point correlators can also be constructed by replacing $\lambda(x)\lambda(0)$ with  $\bar\lambda(x)\bar\lambda(0)$ or $\bar \lambda(x)\lambda(0)$, with appropriate insertions of  $\sigma_\mu$ or $\bar\sigma_\mu$ matrices to contract the spinor indices. Decomposing the $SU(N)$ adjoint fermions into $U(1)\times SU(k)$, $SU(\ell)$, and $k\times \ell$ components, with $N=k+\ell$, taking into account that to the leading order in $\Delta$ the Wilson lines are abelian,  the correlator takes the form (temporarily removing the $\langle... \rangle_{\frac{k}{N}}$ brackets to avoid cluttering)
\begin{eqnarray}\label{general correlator}
{\cal C}_{\nu_1\nu_2...}(x)={\cal C}^{(1)}_{\nu_1\nu_2...}(x)+{\cal C}^{(2)}_{\nu_1\nu_2...}(x)+{\cal C}^{(3)}_{\nu_1\nu_2...}(x)\,,
\end{eqnarray} 
where the three kinds of correlators are, with repeated indices are summed over
\begin{eqnarray}\label{general correlator123}\nonumber
{\cal C}^{(1)}_{\nu_1\nu_2...}(x)=\lambda_{B_1B_2}(x)\prod_{\mu=1}^4{\cal W}_{\mu B_2B_3}(x)\Gamma_{\nu_1\nu_2..} \lambda_{B_3B_4}(0)  {\cal W}^\dagger_{\mu B_4B_1}(x)+(B_i\leftrightarrow B_i')\,,\\ \nonumber
{\cal C}^{(2)}_{\nu_1\nu_2...}(x)=\lambda_{B_1B_2'}(x)\prod_{\mu=1}^4{\cal W}_{\mu B_2'B_3'}(x)\Gamma_{\nu_1\nu_2..} \lambda_{B_3'B_4}(0)  {\cal W}^\dagger_{\mu B_4B_1}(x)+(B_i\leftrightarrow B_i')\,,\\
{\cal C}^{(3)}_{\nu_1\nu_2...}(x)=\lambda_{B_1B_2}(x)\prod_{\mu=1}^4{\cal W}_{\mu B_2B_3'}(x)\Gamma_{\nu_1\nu_2..} \lambda_{B_3'B_4}(0)  {\cal W}^\dagger_{\mu B_4B_1}(x)+(B_i\leftrightarrow B_i')\,,
\end{eqnarray}
and the range of primed and unprimed indices is, as per our convention adopted since (\ref{blockform}), $B_i=1,2,..,\ell=N-k$ and $B'_{i}=1,2,...,k$. In the following, it will be clear that only ${\cal C}^{(1)}_{\nu_1\nu_2...}$ and ${\cal C}^{(2)}_{\nu_1\nu_2...}$   contribute to the correlators to order $\Delta^0$. The correlator  ${\cal C}^{(3)}_{\nu_1\nu_2...}$ gives a contribution that is a higher order in $\Delta$, and its computation requires the explicit form of ${\cal S}^{(\Delta) k}$ and ${\cal S}^{(\Delta) \ell}$, information that is not available to us.  

The correlator ${\cal C}^{(1)}_{\nu_1\nu_2\ldots}(x)$ receives contributions exclusively from either the $U(1)\times SU(k)$ or $SU(\ell)$ sectors, as there are no mixed indices $BB'$ present.   The contributions to  ${\cal C}^{(1)}_{\nu_1\nu_2\ldots}(x)$ from each of these fermions will be denoted as ${\cal C}^{(1),( i)}_{\nu_1\nu_2\ldots}(x)$, ${\cal C}^{(1),(ii)}_{\nu_1\nu_2\ldots}(x)$, and ${\cal C}^{(1),(iii)}_{\nu_1\nu_2\ldots}(x)$, respectively.


We begin with the computation of ${\cal C}^{(1),( i)}_{\nu_1\nu_2\ldots}(x)$. Using the $\tilde{\bm H}$ basis (\ref{Hbasisfirst}) we readily find:
\begin{eqnarray}\label{C1 corr}\nonumber
\langle {\cal C}^{(1), (i)}_{\nu_1\nu_2...}(x)\rangle_{\frac{k}{N}}&=&\langle \tilde\lambda_{b_1}(x)\Gamma_{\nu_1\nu_2..} \tilde \lambda_{b_2}(0)\prod_{\mu=1}^4 {\cal W}_{\mu b_1b_2}(x) {\cal W}_{\mu b_2b_1}^{\dagger}(x)  \rangle_{\frac{k}{N}} \\
&=&\langle \tilde \lambda_{b_1}^\alpha(x) \tilde \lambda_{\beta b_2}(0)\rangle_{\frac{k}{N}}\Gamma_{\alpha\nu_1\nu_2..}^\beta\delta_{b_1b_2}\,,
\end{eqnarray}
and the sum is over $b_1,b_2=1,2,...,k$, the $U(k)$ Cartan components from (\ref{Hbasis old}). Notice that the Wilson  lines cancel out owing to their abelian nature to ${\cal O}(\Delta^0)$; see Eq. (\ref{Wilson lines in tilde H coor}). Similarly, repeating the analysis for the dotted fermions, we find:
\begin{eqnarray}\label{C1bar corr}\nonumber
\langle \bar{\cal C}^{(1), (i)}_{\nu_1\nu_2...}(x)\rangle_{\frac{k}{N}}&=&\langle \bar{\tilde\lambda}_{b_1}(x)\Gamma_{\nu_1\nu_2..} \bar{\tilde \lambda}_{b_2}(0)\prod_{\mu=1}^4 {\cal W}_{\mu b_1b_2}(x) {\cal W}_{\mu b_2b_1}^{\dagger}(x) \rangle_{\frac{k}{N}} \\
&=&\langle \bar{\tilde \lambda}_{b_1}^{\dot \alpha}(x) \bar{\tilde \lambda}_{\dot\beta b_2}(0)\rangle_{\frac{k}{N}}\Gamma_{\dot\alpha\nu_1\nu_2..}^{\dot\beta}\delta_{b_1b_2}\,,
\end{eqnarray}
In the $\tilde{\bm H}$  basis, the fermion propagators in the $Q={k\over N}$ background, to order $\Delta^0$ , are given by 
(\ref{11element main text}) for the undotted fermions and (\ref{22element3 maintext}) for the undotted fermions; these expressions should be substituted into (\ref{C1 corr}), (\ref{C1bar corr}).
 
In a background with a negative topological charge $Q=-\frac{k}{N}$, the unnormalized correlators
$\langle {\cal C}^{(1),(i)}_{\nu_1\nu_2...}(x)\rangle_{-\frac{k}{N}}$ and $\langle \bar{\cal C}^{(1),(i)}_{\nu_1\nu_2...}(x)\rangle_{-\frac{k}{N}}$ take the same structure of (\ref{C1 corr}) and (\ref{C1bar corr}). Now, however,  the dotted fermions, instead, acquire zero modes, and the propagators get switched from (\ref{11element main text}) and (\ref{22element3 maintext}). Explicitly, 
the undotted fermion propagator in the $Q=-{k/N}$ background acquires the form (\ref{22element3 maintext}) of the dotted fermion background in the $Q=k/N$ background, with the obvious replacement of spinor indices as well as $m \rightarrow m^*$, including in the $ ({2m \over g^2})^k$ prefactor of the fermion determinant (\ref{dfermion maintext}). On the other hand, the dotted fermion propagator in the $Q={- k/N}$ background has the form of the undotted propagator in the $Q=k/N$ background, eqn.~(\ref{11element main text}) with the $m \leftrightarrow m^*$, spinor indices, and fermion zero mode (undotted to dotted (\ref{zeromodesnormalized})) replacement.

We now move to the correlator ${\cal C}^{(1),(ii)}_{\nu_1\nu_2...}(x)$. It receives contributions from the fermions that live along the roots of $SU(k)$. Using the propagator (\ref{lambdabarsquared2 main text}), omitting the fermion determinant ${\cal D}_k^f(m)$,  recalling that $B_1' \ne B_2'$, and the Wilson lines (\ref{abelian wilson gr1}) we find
\begin{eqnarray}\label{C1ii}\nonumber
\langle {\cal C}^{(1),(ii)}_{\nu_1\nu_2...}(x)\rangle_{\frac{k}{N}}&=&\langle\lambda_{B_1'B_2'}(x)\Gamma_{\nu_1\nu_2..} \lambda_{B_2'B_1'}(0)\rangle_{\frac{k}{N}}  \prod_{\mu=1}^4{\cal W}_{\mu B_2'B_2'}(x) {\cal W}^\dagger_{\mu B_1'B_1'}(x)\\\nonumber
&=&\frac{g^2}{2V}  \; \Gamma^{\alpha}_{\alpha\,,\nu_1\nu_2..}\sum_{B_1'\neq B_2'=1}^k\sum_{p_\mu=\frac{2\pi \mathbb Z}{L_\mu}}\frac{m^* e^{-ix_\mu \left(p_\mu-2\pi\bm a_\mu\cdot(\bm \nu_{B_2'}-\bm \nu_{B_1'})\right)}e^{i2\pi (B_1'-B_2')\frac{x_2}{kL_2}} }{p_1^2+\left(p_2+\frac{2\pi(B_1'-B_2')}{ k L_2}\right)^2+p_3^2+p_4^2+|m|^2}\,,\\
\end{eqnarray}
with an identical expression of $\langle\bar  {\cal C}^{(1),(ii)}_{\nu_1\nu_2...}(x)\rangle_{\frac{k}{N}}$ after replacing $m^*$ with $m$. Also, identical expressions are obtained in the background of anti-instantons. 

The correlator ${\cal C}^{(1),(iii)}_{\nu_1\nu_2...}(x)$ receives contributions from the fermions in the $SU(\ell)$ sector.  Using the propagator (\ref{lambdabarsuell1 maintext}) without the fermion determinant and the expression of Wilson  lines in (\ref{abelian wilson gr2}), we note that  the Wilson  lines cancel each other out, to find, again not including ${\cal D}_k^f(m)$:
\begin{eqnarray}\label{C1iii}
\langle  {\cal C}^{(1),(iii)}_{\nu_1\nu_2...}(x)\rangle_{\frac{k}{N}}
&=&\frac{g^2}{2V}   \Gamma^\alpha_{\alpha\,,\nu_1\nu_2,..} \sum_{\bm p\in \mathbb Z_\ell^2\neq 0, k_\mu\in \frac{2\pi\mathbb Z}{L_\mu}}\frac{m^*e^{-i\left(k_\mu x_\mu+ \frac{p_3x_3}{\ell L_3}+\frac{p_4x_4}{\ell L_4}\right)}}{|m|^2+M_{(\ell) \bm p,k}^2}\,, \nonumber \\
\end{eqnarray}
where
\begin{eqnarray}\label{Mell pk}
M_{(\ell)\bm p,k}^2=\left[k_1^2+k_2^2+\left(k_3+\frac{2\pi p_3}{\ell L_3}\right)^2+\left(k_4+\frac{2\pi p_4}{\ell L_4}\right)^2\right]\,,
\end{eqnarray}
 Identical expression  of $\langle\bar  {\cal C}^{(1),(iii)}_{\nu_1\nu_2...}(x)\rangle_{\frac{k}{N}}$ follow after replacing $m^*$ with $m$. Also, identical expressions are obtained in the background of anti-instantons. 

Finally, we discuss the correlators $\langle{\cal C}^{(2)}_{\nu_1\nu_2...}(x)\rangle_{\frac{k}{N}}$, which receive contributions from the fermions that live in the off-diagonal space $k\times \ell$. Using the propagator (\ref{lambdaktimesellpropagator maintext}), we find
\begin{eqnarray}\label{C2form}\nonumber
\langle{\cal C}^{(2)}_{\nu_1\nu_2...}(x)\rangle_{\frac{k}{N} }&=&\frac{g^2}{2V}  \Gamma^{\alpha}_{\alpha \nu_1\nu_2..}\\\nonumber
&&\times \sum_{B_1=1}^\ell \sum_{B_2'=1}^k \sum_{\ell_{(1)},\ell_{(3)}=0}^{\infty}\frac{m^* D_\mu \varphi_{B_2'B_1}^{\ell_{(1)},\ell_{(3)}}(x,\hat\phi)\left(D_\mu\varphi_{B_2'B_1}^{\ell_{(1)},\ell_{(3)}}(0,\hat\phi)\right)^*}{\left(\frac{4\pi}{L_1L_2}\left(1+\ell_{(1)}+\ell_{(3)}\right)\right)\left(\frac{4\pi}{L_1L_2}\left(1+\ell_{(1)}+\ell_{(3)}\right)+|m|^2\right)}\,.\\
\end{eqnarray}
 Here, $\varphi_{B_2'B_1}^{\ell_{(1)},\ell_{(3)}}(x,\hat\phi)$ are the covariant Laplacian eigenvectors in the $k\times \ell$ sector, given by (\ref{finalvarphiBCS}). An almost identical expression can be given for $\langle\bar {\cal C}^{(2)}_{\nu_1\nu_2...}(x)\rangle_{\frac{k}{N}}$ using the propagator (\ref{barlambdaktimesellpropagator maintext}):
\begin{eqnarray}\nonumber
\langle\bar {\cal C}^{(2)}_{\nu_1\nu_2...}(x)\rangle_{\frac{k}{N}}&=&\frac{g^2}{2V}  \Gamma^{\dot\alpha}_{\dot\alpha \nu_1\nu_2..} \sum_{B_1=1}^\ell \sum_{B_2'=1}^k \sum_{\ell_{(1)},\ell_{(3)}=0}^{\infty}\frac{m \varphi_{B_2'B_1}^{*\ell_{(1)},\ell_{(3)}}(x,\hat\phi)\varphi_{B_2'B_1}^{\ell_{(1)},\ell_{(3)}}(0,\hat\phi)}{\frac{4\pi}{L_1L_2}\left(1+\ell_{(1)}+\ell_{(3)}\right)+|m|^2}\\
&&\times  \prod_{\mu=1}^4\left({\cal W}_{\mu B_2'B_2'}(x) {\cal W}^\dagger_{\mu B_1B_1}(x)+{\cal W}_{\mu B_1B_1}(x) {\cal W}^\dagger_{\mu B_2'B_2'}(x)\right)\,.
\end{eqnarray}
The propagators $\langle{\cal C}^{(2)}_{\nu_1\nu_2...}(x)\rangle_{-\frac{k}{N}}$ and $\langle\bar {\cal C}^{(2)}_{\nu_1\nu_2...}(x)\rangle_{-\frac{k}{N}}$ in the anti-instanton backgrounds can be constructed similarly.

In this section, we have ignored the contribution of the path-integral fluctuations (including the integral over the bosonic moduli), and we also ignored the Pauli-Villars regulators and the fermion determinants. Both issues will be dealt with in the next section.

\section{Determinants, regularization, and the bosonic moduli space}
\label{Determinants and regularization}

\subsection{Determinants and regularization}
\label{sec:detandreg}

In the limit of a massless gaugino, and thanks to supersymmetry, the bosonic and fermionic determinants of the nonzero modes cancel out, yielding unity. Introducing a gaugino mass term modifies this result; however, in the regime where $m LN \ll 1$, which we consider, supersymmetry still allows for a simple leading-order expansion for the determinants in $m LN$. In the following, we present our calculations in a background with a positive topological charge $Q=\frac{k}{N}$, which can be easily extended  negative charges or the trivial topological sector.  

The path integral in the  $Q = \frac{k}{N}$ instanton background, which we denote by  ${\cal E}_{\frac{k}{N}}(0)$ (where the argument ($0$) is only a reminder that this is not the regulator contribution) is given by the formal one-loop expression:\footnote{The expression on the second line is the one for $k \ge 0$. For $k < 0$, one replaces $m \rightarrow m^*$.}
\begin{eqnarray}\label{det0}
{\cal E}_{\frac{k}{N}}(0) &\equiv& \left[\int [DA_\mu][D\lambda][D\bar\lambda] e^{-S_{\scriptsize\mbox{SYM*}}}\right]_{\frac{k}{N}} ={{\cal{D}}_{k}^f(m) \over ({\rm{det}}' {\cal O}^{G}_{\,\mu\nu})^{1\over 2} \; ({\rm{det}} (- D^2))^{-{1 \over 2}}} \nonumber \\
&=&  { \left({ 2 m \over g^2 }\right)^k \prod_n \left({16 \over g^4}  (\omega_n^2 + |m|^2)\right)  \over \prod_n  \omega_n^2 }  
\end{eqnarray}
The above quantity ${\cal E}_{\frac{k}{N}}(0)$ represents the ratio of various determinants of fluctuations around the instanton.  The numerator corresponds to the fermionic sector, with determinant given by (\ref{dfermion maintext}). The denominator is the bosonic contribution of ghosts and gauge fields, with zero modes removed from the gauge field fluctuation determinant (${\rm{det}}' {\cal O}^{G}$) by introducing collective coordinates. 
We already used (\ref{dfermion maintext})  for 
 ${\cal{D}}_{k}^f(m)$  in the numerator to obtain the last line above.\footnote{The various $2/g^2$ factors in the numerator of (\ref{det0}) are left for easy comparison with (\ref{dfermion maintext}); they  cancel with the regulator contributions below.}  We recall that $\omega_n^2$ are the eigenvalues of the Laplacian, $-D_\mu D_\mu$, which has no zero modes in our self-dual background.
Finally, we also used the
  fact that the product of the bosonic fluctuation determinants in the denominator, with the zero modes omitted, equals the  product of eigenvalues of the adjoint Laplacian, as is well known, see e.g.~\cite{Vandoren:2008xg}. It should be understood that the expression (\ref{det0}) should be further integrated over the instanton moduli (recall our general semiclassical formula (\ref{grand ZT with mass}), where the moduli space integral is denoted $\mu_B^k$ and the ratio of determinants in each instanton sector by ${\cal F}_k$).

The expression for the ratios of determinants in the instanton background is regulated by multiplying each determinant by the  determinants  of  massive  Pauli-Villars (PV) fields  of alternating statistics, of mass $\sim M_{PV}$, larger than any physical mass scale. 
We introduce $R$ PV fields of masses $M_{i}^2$, $i=1,...,R$, and statistics $e_i$ ($e_i = \pm 1$), obeying 
\begin{eqnarray}
\label{PVstats}
\sum\limits_{i=1}^R e_i = -1, ~ \sum\limits_{i=1}^R e_i M_i^{2 q} = 0, ~ q = 1,..., R-1,
\end{eqnarray}
and, for use below, define the scale
\begin{eqnarray}\label{PVscale123}
  M_{PV} \equiv   \prod\limits_{i=1}^R M_i^{- e_i}.
\end{eqnarray}
Then the regulated fermion determinant becomes:
\begin{eqnarray}\label{fermion123}
{  m^k   } \prod_n   (\omega_n^2 + |m|^2) \rightarrow {m^k \over M_{PV}^k} \prod_n \prod_{i=0}^R (\omega_n^2 +  |m|^2 + M_i^2)^{e_i}~,
\end{eqnarray}
where we defined 
\begin{eqnarray}\label{requalzero}
e_0 \equiv 1, \; \text{and} \; M_0^2 \equiv 0.
\end{eqnarray}
The convergence of the product on the r.h.s. of (\ref{fermion123}) follows from (\ref{PVstats}), upon taking $R$ large enough.\footnote{The necessity of  the conditions (\ref{PVstats}) is easiest to see in the case of a free particle in the continuum (here, at least $R=3$ is required but a solution of (\ref{PVstats}) only exists for $R >3$):  taking the logarithm of the product, converting the sum to an integral, and demanding convergence at large $\omega$.} The overall factor of $M_{PV}$ arises from the zero modes of the regulator fermions.

Similarly, for the gauge contribution, we replace the denominator in (\ref{det0}) by
\begin{eqnarray}\label{gauge123}
({\rm{det}}' {\cal O}^{G}_{\,\mu\nu})^{1\over 2} \; ({\rm{det}} (- D^2))^{-{1 \over 2}} 
 \rightarrow {1 \over M_{PV}^{4k}} \prod_n \prod_{i=0}^R (\omega_n^2  + M_i^2)^{e_i}~,
\end{eqnarray}
where the factor $M_{PV}^{-4k}$ is the combination (\ref{PVscale123}) of $M_i^4$ factors from each regulator determinant, which has no zero modes and the same relations (\ref{PVstats}) apply.

Thus, combining (\ref{fermion123}, \ref{gauge123}), we obtain the finite version of (\ref{det0}): 
\begin{eqnarray}\label{det1}
{\cal E}_{\frac{k}{N}}(0) \over {\cal E}_{\frac{k}{N}}(M_{PV})  &=&    m^k M_{PV}^{3k}  \prod_n \prod\limits_{i=0}^R  \left(1 +  {|m|^2 + M_i^2 \over \omega_n^2} \over 1 +  { M_i^2 \over \omega_n^2}\right)^{e_i} \nonumber \\
&\equiv& {\cal F}_k(m, M_{PV}) \equiv m^k M_{PV}^{3k} {\cal D}_k(|m|^2),  \end{eqnarray}
where we recall (by (\ref{requalzero})) that $i=0$ corresponds to the physical field contribution (\ref{det0}) while $i=1,...R$ correspond to the regulators, and that convergence is assured by having the regulator masses obey (\ref{PVstats}).
This regulated determinant ${\cal E}_{\frac{k}{N}}(0)/{\cal E}_{\frac{k}{N}}(M_{PV})$  is precisely what is denoted ${\cal F}_k(m, M_{PV})$ in (\ref{grand ZT with mass}).  For convenience, we also defined the quantity
\begin{eqnarray}\label{defofDratio}
{\cal D}_k(|m|^2)\equiv  \prod_n (1 + {|m|^2\over \omega_n^2})\bigg\vert_{reg.} =   \prod_n \prod\limits_{i=0}^R  \left(1 +  {|m|^2 + M_i^2 \over \omega_n^2} \over 1 +  { M_i^2 \over \omega_n^2}\right)^{e_i}~.
\end{eqnarray}
In the supersymmetric limit $m \rightarrow 0$, the infinite product in (\ref{defofDratio})  equals unity.\footnote{In this limit, the overall factor of $m^k$ is cancelled by the $1/m$ factors from the zero-mode contributions of the fermion  propagator in (\ref{propagator 2 maintext}). Recall that the $Q=k/N$-instanton amplitude is zero without the insertion of $2k$ undotted adjoint fermions.} This is nothing but the usual supersymmetric cancellation of the quantum fluctuations of the nonzero modes in the instanton background, seen here at one loop. It is also clear that when $|m|^2 >0$, the regulator contributions do not cancel. Thus,  ${\cal D}_k(|m|^2) -1$ vanishes as $|m|L \rightarrow 0$.

In summary, the ratios of one-loop determinants in the $Q=k/N$ background reduce to the computation of  ${\cal D}_k(|m|^2)$. To compute it, we only need to know the eigenvalues of the adjoint Laplacian.
This work was undertaken in Section \ref{sec:propagatorkovern} where it was understood that the adjoint Laplacian  factorizes in the small-$m$ small-$\Delta$ limit of (\ref{smalllimit maintext}). 
Using this, we can immediately write the following  expressions for the functions ${\cal D}_k(|m|^2)$:
\begin{eqnarray}\label{Dratio}
 {\cal D}_k(|m|^2) &\equiv&  \prod_n (1 + {|m|^2\over \omega_n^2})\bigg\vert_{reg.} \\
&=& \underbrace{\prod_{n \in SU(k) \times U(1)} (1 + {|m|^2\over \omega_n^2})}_{ {\cal D}_k(|m|^2, SU(k)\times U(1))} \; \underbrace{\prod_{n \in SU(\ell) } (1 + {|m|^2\over \omega_n^2})}_{{\cal D}_k(|m|^2, SU(\ell))} \;\underbrace{\prod_{n \in k \times \ell (\ell \times k) } (1 + {|m|^2\over \omega_n^2})}_{{\cal D}_k(|m|^2,  (k \times \ell))  }\bigg\vert_{reg.}, ~ \nonumber  
 \end{eqnarray}
 where all products on the second line are also assumed to be regularized.\footnote{All expressions involving infinite products here and further below should be understood to be regularized, e.g.~as in  (\ref{defofDratio}); for brevity this is not explicitly indicated.}
 As shown above,  the product of eigenvalues factorizes into products of those that lie   in $SU(k)\times U(1)$ (with eigenvalues computed in Appendix \ref{sec:diagonalsuktimesu1}, \ref{sec:cartansuktimesu1}, \ref{sec:liftingzeromodes}, \ref{sec:noncartansuktimesu1}), those in $SU(\ell)$ (with eigenvalues computed in Appendix \ref{appx:suellpropagator}), as well as those in the $k\times \ell$ and $\ell \times k$ part of the adjoint (with eigenvalues computed in Appendix \ref{sec:offdiagonalktimesell}). 

We now consider each of the products in (\ref{Dratio}) in turn, beginning with the $SU(k) \times U(1)$ contributions (as usual, we denote $L = V^{1/4}$):
\begin{eqnarray}\label{ukdeterminant}
&&{\cal D}_k(|m|^2, SU(k)\times U(1)) \\
&&= \left(1 +{ |m|^2L^2 \over c \Delta}\right)^k \prod\limits_{k_\mu ={2 \pi n_\mu\over L_\mu}}^\prime \left(1 + {|m|^2 L^2 \over (L k_\mu)^2}\right)^k  \prod\limits_{D' \ne E' =1}^k \prod\limits_{k_\mu = {2 \pi n_\mu\over L_\mu}}\left(1 + {|m|^2 L^2 \over ( L k_\mu + \delta_{\mu 2} {L \over L_2} { D'-E' \over k })^2}\right),  \nonumber
\end{eqnarray}
where the first term is the contribution of the $k$ order-$\Delta$ eigenvalues of the Laplacian and the rest is due to the eigenvalues in the Cartan and off-diagonal elements of $SU(k) \times U(1)$, respectively and the product $\prod^\prime_{n_\mu \in \Z}$ excludes the term $n_1=n_2=n_3=n_4=0$.
The $SU(\ell)$ term reads:
\begin{eqnarray}\label{suelldeterminant}
&&{\cal D}_k(|m|^2, SU(\ell)) = \prod\limits_{p_3,p_4 = 0, (p_3, p_4) \ne (0,0)}^{\ell-1} \prod\limits_{k_\mu = {2 \pi n_\mu\over L_\mu}}\left(1 + {|m|^2 L^2 \over (L  k_\mu + \delta_{\mu 3} {L \over L_3} {2 \pi p_3 \over \ell}+\delta_{\mu 4} {L \over L_4} {2 \pi p_4 \over \ell})^2}\right),  \nonumber \\
\end{eqnarray}
while the off-diagonal term, $k \times \ell$, contribution is
\begin{eqnarray}\label{kbyelldeterminant}
{\cal D}_k(|m|^2, (k\times \ell)) =  \prod\limits_{\ell_{(1)}, \ell_{(3)} = 0}^\infty    \left(1 + {|m|^2 L^2   \over {4 \pi L^2 \over L_1 L_2} (\ell_{(1)}+ \ell_{(3)} + 1)}\right)^{2k}, \nonumber \\
\end{eqnarray} 
where we recall from Appendix \ref{sec:offdiagonalktimesell}, eqn.~(\ref{eigenvectors ktimesell hermitean}), that for any   $\ell_{(1)}, \ell_{(3)}$ there $2k$ Hermitean eigenvectors of the adjoint Laplacian.

We also recall that the same expression as (\ref{defofDratio}) also gives the result in the trivial topological sector contributing to the normalization factor $Z^T$ of (\ref{grand ZT with mass}):
 \begin{eqnarray}\label{f0sector}
{\cal F}_{0}&=& {\prod_n (1 + {|m|^2 \over \omega_n^2})} = \prod_{p_3,p_4 = 0, (p_3, p_4) \ne (0,0)}^{N-1} \prod\limits_{k_\mu = {2 \pi n_\mu\over L_\mu}} \left( 1 + {|m|^2 \over M_{\bm p,k}^2}\right) \,,
\end{eqnarray}
where $\omega_n^2 \rightarrow M_{\bm p,k}^2$ of eqn.~(\ref{valueof Mpks}) and the product over $n$ is replaced by a product over the values of ${\bm p = (p_3,p_4),k}$ discussed there (and we have omitted indicating the need for regularization). Finally, we recall from (\ref{grand ZT with mass}) that (\ref{defofDratio}) is divided by (\ref{f0sector}).

In the next two sections, we first (section \ref{sec:F0UV}) study the  UV divergences of ${\cal F}_{0}$, the one-loop fluctuation determinants in the $Q=0$ sector (equal, by (\ref{det1}), to (\ref{defofDratio}) with $k=0$). Then, in section \ref{sec:uvfiniteratio}, we  argue for the UV finiteness  of the ratio of (\ref{defofDratio}), the nonzero mode determinant in the $Q={k/N}$ background, to ${\cal F}_0$,
  \begin{eqnarray}\label{DtoF ratio}
{{\cal D}_k(|m|^2) \over {\cal F}_{0}}~.
\end{eqnarray}  
Recall that this is the expression that enters the contributions to the partition function $Z^T$, eqn.~(\ref{grand ZT with mass}), of the $Q=k/N$ sectors.
We give a qualitative argument and then present an analytic, non-rigorous but suggestive, argument for UV finiteness, based on our calculation of the Laplacian spectra in the fractional instanton background. 

\subsubsection{The UV divergence of the $Q=0$ sector determinant}
  
  \label{sec:F0UV}
  
  The explicit (unregulated) expression  for  ${\cal F}_{0}(|m|^2)$ determinant 
\begin{eqnarray}\label{f0sector 1}
{\cal F}_{0}(|m|^2)&=& {\prod_n (1 + {|m|^2 \over \omega_n^2})} = \prod \prod_{p_3,p_4 = 0, (p_3, p_4) \ne (0,0)} \prod\limits_{k_\mu = {2 \pi n_\mu\over L_\mu}} \left( 1 + {|m|^2 \over M_{\bm p,k}^2}\right) \,,
\end{eqnarray}
where the masses $M_{\bm p,k}$ in the $Q=0$ sector are given by
\begin{eqnarray}\label{valueof Mpks}
M_{\bm p,k}^2=\left[k_1^2+k_2^2+\left(k_3+\frac{2\pi p_3}{NL_3}\right)^2+\left(k_4+\frac{2\pi p_4}{NL_4}\right)^2\right]\,. 
\end{eqnarray}
We note that ${\cal F}_{\pm 0}(|m|^2, M_{PV})$ is simply given by   ${\cal D}^f_0(m)/{\cal D}_0^f(0)$, where ${\cal D}^f_0(m)$  the $k=0$ fermion determinant given by (\ref{dfermion maintext}) with $k=0$ and $\omega_n^2 \rightarrow  M_{\bm p,k}^2$ (and the product over eigenvalues is  over all $n_\mu$ and $p_{3,4}$). The division by ${\cal D}_0^f(0)$ is due to the gauge and ghost fluctuations.\footnote{As described in more detail in section \ref{sec:detandreg}. The regulated version via multiple Pauli-Villars (PV) regulators is also given there.} 

The point we want to make now is that ${\cal F}_{0}(|m|^2)$ is UV divergent, due to the quadratic and log-divergent contributions to the vacuum energy in SYM$^*$. In fact, we can infer from (\ref{f0sector 1}) that:
\begin{eqnarray}\label{f0sector 2}
&&\ln {\cal F}_{0}(|m|^2) \\
&&= \sum\limits_{k_\mu = {2 \pi n_\mu\over L_\mu}} \sum_{p_3,p_4 = 0, (p_3, p_4) \ne (0,0)} \ln{ |m|^2 + M_{\bm p,k}^2 \over M_{\bm p,k}^2} 
\simeq {V}  (N^2 -1) \int^{M_{PV}} {d^4 k \over (2 \pi)^4} \ln {|m|^2 + k^2 \over k^2} + \ldots \nonumber ,\end{eqnarray}
where, focusing on the UV divergent part, we replaced the sum by an integral. The dots denote finite contributions that depend on $|m|^2 L^2$ and on the boundary conditions.
 The integral in (\ref{f0sector 2}) is the standard expression for the one-loop vacuum energy in SYM$^*$, which has quadratically and logarithmically divergent pieces, proportional to $M_{PV}^2 |m|^2$ and $|m|^4 \ln M_{PV}$, respectively. The point is that these divergent pieces are independent of the volume and the boundary conditions and can be subtracted away by introducing a cosmological constant counterterm. This counterterm is the same in all topological sectors\footnote{As we argue later, the $Q={k/N}$ determinants  ${\cal F}_{\pm k}$, with $|k|>0$, are also UV divergent, but the ratio ${\cal F}_{\pm k}/{\cal F}_{\pm 0}$ is UV finite. Thus, the $Q={k/N}$ contributions to $Z^T$ are rendered finite by the same cosmological constant counterterm.} contributing to (\ref{grand ZT with mass}).
 
 The calculation of the finite pieces in (\ref{f0sector 2}) can be performed with the multiple PV regulators discussed above or, for example, using $\zeta$-function regularization, as discussed in Appendix \ref{zeta function regularization}.\footnote{However, regularization has to be consistent, i.e.~ideally the same for the calculation of all quantities; else, one is faced with finite renormalizations if  different regularization schemes for different parts of the instanton calculation are used. The precise  calculation of the finite parts presents a challenge which we will not address in this paper.} 
We define the subtraction of the divergent cosmological constant so that the renormalized value of ${\cal{F}}_0$---and thus of $Z^T$ (\ref{grand ZT with mass})---is:
 \begin{eqnarray}\label{f0sector2}
[{\cal F}_{0}(|m|^2)]_{Reg} = 1 + {\cal{O}}(|m| NL)^p~,
\end{eqnarray}
where $p>0$ and we assume $|m| NL \ll 1$, i.e. the mass is smaller than the mass gap in the $Q=0$ sector.
The computation of the finite ${\cal O} \left((m LN)^p\right)$ correction in the trivial sector, as defined by $\zeta$-function regularization, is in Appendix \ref{zeta function regularization} (see the discussion starting at (\ref{start discu of prod})), where we show that $p=2$.

\subsubsection{UV finiteness of the ratio of $Q=k/N$ sector to $Q=0$ sector determinants}  
\label{sec:uvfiniteratio}

The UV finiteness of this ratio  is a consequence of the cancellation of the nonzero mode fluctuations in the instanton background  in the SUSY limit $m=0$, where adjoint fermion and gauge nonzero mode fluctuations exactly cancel, as we now review.
Each of the two contributions to the ratio (\ref{DtoF ratio}), those of the adjoint fermion or gauge field, is separately UV divergent, even for $m=0$ (as shown in the classic computation \cite{tHooft:1976snw} on $\R^4$, precisely for the ratio of determinant in the instanton background to the determinant in the trivial background, as in (\ref{ratioUV}) below). There is a logarithmic divergence in each piece which is responsible for the nonzero-mode contribution to the running of the gauge coupling. The log-divergent pieces due to the gauge field and adjoint fermion exactly cancel for $m=0$. Adding a small gaugino mass does not affect the cancellation of divergences, but leads to small finite contributions (these have been calculated on $\R^4$, e.g.~\cite{Brown:1978bta,Carlitz:1978yj}). The log-divergent pieces are independent of volume and boundary conditions, hence this cancellation will persist in our geometry as well.

We can see the UV finiteness of (\ref{DtoF ratio}) explicitly, by 
focusing on the contribution of the large eigenvalues $\omega_n^2$ in each expression in (\ref{Dratio}).
We note that, except for ${\cal D}_k(|m|^2, (k\times \ell)), $\footnote{We  consider (\ref{kbyelldeterminant}) separately below, see discussion after (\ref{ratioUVkbyell}).} the high eigenvalues in (\ref{ukdeterminant},\ref{suelldeterminant}) are 
identical, scaling as $p_\mu^2$. We also note that, in the limit of large eigenvalues, ${\cal F}_{0}$ involves $N^2-1$ identical factors, since for $k_\mu \gg 1$ the $p_3, p_4$ factors are inessetial. Further, 
for large eigenvalues ${\cal D}_k(|m|^2, SU(k)\times U(1))$ has $k^2$ identical factors,  and ${\cal D}_k(|m|^2, SU(\ell))$ also has $\ell^2-1$ identical factors. 

Thus, consider the ratio of one of the $k^2$ terms in (\ref{ukdeterminant}), for definiteness the one with $D'-E' = 1$, to one of the $N^2$ terms in (\ref{f0sector}), to find 
\begin{eqnarray}
\ln {{\cal D}_k(|m|^2, SU(k)\times U(1))\vert_{one~ term} \over {\cal F}_{0}\vert_{one~ term} }= \sum\limits_{k_\mu = {2 \pi n_\mu \over L_\mu}, n_\mu \gg 1} \ln (1 + {|m|^2 \over (k_\mu  + \delta_{\mu 2} {1 \over k L_2})^2})  - \ln(1 + {|m|^2 \over M_{p, k}^2})   \nonumber \\
\end{eqnarray}with $M_{\bm p,k}^2$ of eqn.~(\ref{valueof Mpks}) for some fixed $\bm p$.
We next note that  the log is a slowly varying function of $n_\mu$. Thus, we  replace the sum by integral, ignoring the remainder terms of the Euler-Mclaurin formula. Then we find 
  \begin{eqnarray}\label{ratioUV}
&&\ln {{\cal D}_k(|m|^2, SU(k)\times U(1))\vert_{one~ term} \over {\cal F}_{0}\vert_{one~ term} } \simeq {V \over (2\pi)^4} \int d^4 k \left(\ln (1 + {|m|^2 \over (k_\mu  + \delta_{\mu 2} {1 \over k L_2})^2})  - \ln(1 + {|m|^2 \over M_{p, k}^2})\right)   \nonumber \\
&&={V \over (2\pi)^4} \int d^4 k \left(\ln(1 + {|m|^2 \over (k_\mu  + \delta_{\mu 2} {1 \over k L_2})^2})   - \ln(1 + {|m|^2 \over (k_\mu + \delta_{\mu 3} \frac{2\pi p_3}{NL_3} + \delta_{\mu 4}  \frac{2\pi p_4}{NL_4})^2}\right) 
\end{eqnarray}
  Then, clearly, this is UV finite (and, in fact, vanishes if the integral is understood to be over all $\R^4$ in $k$-space, as we can shift the variable of integration; which value of $p_3, p_4$, $D'-E'$ we picked is irrelevant). 
A similar argument goes through for the $SU(\ell)$ contribution, where we  consider the ratio ${{\cal D}_k(|m|^2, SU(k)\times U(1))\vert_{one~ term}\over {\cal F}_{0}\vert_{one~ term}  }$ instead and obtain a similar UV finite expression.

These arguments can be made more rigorous if one considers instead, the regulated versions of the two terms in the ratio, e.g.~as in the rightmost equation in ~(\ref{defofDratio}), as well as keeping track over the remainder of the Euler-Mclaurin formula.

For completeness, let us consider the $k\times \ell$ contribution (\ref{kbyelldeterminant}) as well, keeping the remaining $2 k\times \ell$ terms  (out of the $N^2-1$ ones, we already used $k^2 + \ell^2 -1$) in ${\cal F}_{0}$, choosing one fixed $p_3,p_4$-value in the integrand below:
\begin{eqnarray}\label{ratioUVkbyell}
\ln {{\cal D}_k(|m|^2, (k \times \ell))  \over {\cal F}_{0}\vert_{2 k \times \ell \;  terms} }=2 k   \sum\limits_{\ell_{(1)}, \ell_{(3)} = 0}^\infty    \ln\left(1 + {|m|^2     \over {4 \pi  \over L_1 L_2} (\ell_{(1)}+ \ell_{(3)} + 1)}\right) - 2 k \ell \sum\limits_{k_\mu = {2 \pi n_\mu \over L_\mu}} \ln(1 + {|m|^2 \over M_{p, k}^2})~.   \nonumber \\
\end{eqnarray}
Again, we replace the sums by integrals to obtain for the r.h.s. above, recalling that $\ell L_3 L_4 = L_1 L_2$ was used to obtain (\ref{kbyelldeterminant}):
\begin{eqnarray}\label{kbyell ratio1}
&&\ln {{\cal D}_k(|m|^2, (k \times \ell))  \over {\cal F}_{0}\vert_{2 k \times \ell \;  terms} }\\
&\simeq& 2 k \ell {V \over (4 \pi)^2} \int d q_1^2 d q_2^2 \; \ln (1 + {|m|^2     \over q_1^2 + q_2^2+1}) - 2 k \ell {V\over (2 \pi)^4} \int d^4 k  \ln (1 + {|m|^2 \over (k_\mu + \delta_{\mu 3} \frac{2\pi p_3}{NL_3} + \delta_{\mu 4}  \frac{2\pi p_4}{NL_4})^2}), \nonumber
\end{eqnarray}
where $q_{1}^2 = {4 \pi \over L_1 L_2 } \ell_{(1)}$, $q_2^2= {4 \pi \over \ell L_3 L_4 } \ell_{(3)}$ only take positive values. The following manipulations indicate that there is no UV divergence above, ignoring the $2 k \ell$ factors from now on. The second integral  in (\ref{kbyell ratio1}) is, going to spherical coordinates at large $k^2$:
\begin{eqnarray} \label{second integral}
{2 \pi^2 \over 2 (2 \pi)^4} \int d k^2 k^2 \ln (1 + {|m|^2 \over k^2}) = {1 \over 16 \pi^2}  \int d k^2 k^2 \ln (1 + {|m|^2 \over k^2})
\end{eqnarray}
while in the first integral in (\ref{kbyell ratio1}), integrating over the $q_1>0$, $q_2>0$ quadrant in polar coordinates $q_1 = r \cos \phi$, $q_2 = r \sin \phi$, we find
\begin{eqnarray}\label{first integral}
{4 \over (4 \pi)^2} \underbrace{\int\limits_{0}^{\pi\over 2} \cos \phi \sin\phi d \phi}_{= {1 \over 2}} \int dr r^3 \ln (1 + {|m|^2 \over r^2})  =  { 1\over 16  \pi^2}  \int dr^2 r^2 \ln (1 + {|m|^2 \over r^2}),  
\end{eqnarray}
i.e. the same expression as (\ref{second integral}), showing that the UV-divergent parts in (\ref{kbyell ratio1}) cancel. Admittedly, our manipulations are only suggestive and non-rigorous and should be repeated with regulated expressions. However, this will only be needed if the higher order terms, expected to be of order $(|m|L)^2$ are to be computed.  

{\flushleft{T}}he moral of the story is that, going back to (\ref{Dratio}), is that the ratio is, approximately
\begin{eqnarray}\label{Dratio last}
{ {\cal D}_k(|m|^2) \over {\cal F}_0} \simeq \left(1 + k { (|m|L)^2 \over c \Delta}\right)~,
\end{eqnarray}
i.e.~is given by the $k$ smallest eigenvalues of the adjoint Laplacian, the ones of order $\Delta$, lying in the $SU(k)\times U(1)$ subspace of the adjoint. The corrections would multiply (\ref{Dratio last}) by terms like, $1+ (|m|L)^2$ or $1+ \Delta$, and their computation is a challenging task (because the order-$\Delta$ shifts of the eigenvalues $\omega_n^2$ has to be computed as well).

Thus, in the small gaugino mass limit, $|m|NL\ll 1$, we obtain, recalling (\ref{grand ZT with mass}, \ref{f0sector2}) with $p=2$%
\begin{eqnarray}  \label{ZT with mass}\nonumber
Z^{\scriptsize T}[\eta=0,\bar\eta=0]_{m\neq 0} = \sum_{Q=0,\frac{\pm 1}{N},\frac{\pm 2}{N},... } Z_{Q}^{\scriptsize\mbox{Reg}}[\eta=\bar\eta=0] = N(1 + {\cal O} \left((m LN )^2\right))+{\cal O}(\Lambda^3 |m| L^4)\,.\\
\end{eqnarray}  
 The correction ${\cal O}(\Lambda^3 |m| L^4)$ comes from the sectors with topological charges $Q=\pm \frac{1}{N}$, and is computed in section \ref{the2pointfunctionsection}.   Both results match our computations using the Hamiltonian formalism of section \ref{sec:hamiltonian}, recall eqn.~(\ref{ZTsmallmL main text}) there.

{\bf{\flushleft{Comment on tuned vs. detuned $\T^4$:}}} At this stage, it is important to emphasize a key aspect of the theory mentioned at the outset of our construction. We previously argued that the theory must be placed on a detuned $\mathbb T^4$ to avoid the emergence of unwanted additional fermion zero modes. Let us now examine the consequences for the regularized theory when it is instead defined on a tuned $\mathbb T^4$, corresponding to setting the detuning parameter $\Delta=0$. To streamline our analysis, we restrict attention to the sector $Q=\frac{1}{N}$, noting that the same reasoning extends to sectors with higher topological charge. 

On a tuned $\mathbb T^4$, and in the strict limit $m=0$, the Dirac equation can be solved explicitly, revealing the presence of $4$ dotted and $2$ undotted zero modes. This outcome is consistent with the index theorem, which predicts the difference $\mbox{dim ker} \bar D - \mbox{dim ker} D = 4 - 2 = 2$. However, the surplus dotted zero modes are excessive for supporting bifermion gaugino condensates.
 When a small mass term is introduced, these would-be zero modes contribute to the fermion determinant, as previously discussed, and give  the result \begin{eqnarray} \left[\frac{{\cal E}_{\frac{1}{N}}(0)}{{\cal E}_{\frac{1}{N}}(M_{PV})}\right]_{\Delta=0}\cong m^3 M_{PV}\,. \end{eqnarray} The fully regularized bilinear condensate is obtained by replacing the operator ${\cal O}$ in (\ref{Reg O}) with $\mbox{tr}[\lambda\lambda(0)]$, after appropriate regularization (see also section \ref{sec:observables} below). This yields
$ \langle\mbox{tr}[\lambda\lambda(0)]\rangle^{\scriptsize \mbox{reg}} \propto m^2,
$
a result in clear contradiction with the expected behaviour of the condensate, which should remain finite as $m \rightarrow 0$. This clearly incorrect and unexpected outcome arises due to the proliferation of fermion zero modes on the tuned $\mathbb T^4$. As already noted, this issue---circumvented here and in our earlier work by detuning the $\T^4$---awaits further study---see also footnote \ref{piljinfootnote}.

\subsection{The bosonic moduli space}
\label{The bosonic moduli space}

The final piece of information needed to carry out the complete calculations of the regularized correlator is the determination of the moduli space and its measure. This exercise was performed in \cite{Anber:2023sjn}, and it was found that the measure  $d\mu_B$ over the moduli space in the $Q=\pm\frac{k}{N}$ sector is given, in terms of the collective coordinates $z_\mu$ and $a_\mu^b$  ($b=1...k-1$), by
\begin{eqnarray}\label{debeta}
d\mu_B^{(k)}=\frac{\prod_{\mu=1}^4 \prod_{b=1}^{k-1} d a_\mu^b dz_\mu \sqrt{\mbox{Det}\, {\cal U}_B^{(k)}}}{k!(\sqrt{2\pi})^{4k}}\,,
\end{eqnarray}
where  
\begin{eqnarray}\label{Det UB}
\sqrt{\mbox{Det}\, {\cal U}_B^{(k)} }=N^2 \left(\sqrt{k (N-k)}\right)^4\left(\frac{8\pi^2 \sqrt V}{g^2}\right)^{2k}\,.
\end{eqnarray}
The moduli space $\Gamma^{(k)}$ is defined via 
 \begin{eqnarray}\label{range of z and a bulk}
 \Gamma^{(k)}=\left\{\begin{array}{l} z_{2} \in [0,1), z_{1} \in [0, {1 \over N})\,,\\ 
z_{3,4} \in [0, {1 \over N-k})\,,\\
\bm a_\mu  \in \Gamma_{w}^{{SU(k)}}\; \text{for} \;  \mu=1,2,3,4\,,\end{array}\right.
\end{eqnarray}
and $ \Gamma_{w}^{{SU(k)}}$ is the fundamental domain of the $SU(k)$ weight lattice. The fact that the integration over the Euclidean time coordinate $z_1$ should be restricted to $[0,{1 \over N})$---corresponding to not counting the center symmetry images of the instanton in the Euclidean time direction---is explained in \cite{Anber:2024mco}, see Appendix G there. Integration of $d\mu_B^{(k)}$ over $\Gamma^{(k)}$ yields  the volume of the moduli space in the sector $Q=|k|/N$
\begin{eqnarray}\label{volume of mub}
\mu_B^{(k)}=\frac{N}{k!} \left(\frac{4\pi \sqrt V}{g^2}\right)^{2k}\,.
\end{eqnarray}  
%

\section{Putting things together: the observables}
\label{sec:observables}

In this section, we collect the different pieces needed to calculate the regularized observable $\langle {\cal O} \rangle^{\scriptsize\mbox{Reg}}$ given by (\ref{Reg O}). Here, ${\cal O}(x_1,..,x_n)$ is a multifermion operator at distinct points $x_1,..,x_n$ on the detuned $\mathbb T^4$. 
One can develop a formalism for general multifermion correlators using all we have assembled till now. 
However, from now on, we focus on the regularized expectation value of the $2$-point function 
\begin{eqnarray}\label{two point 1}
{\cal O}_{\nu_1\nu_2...}(x)=\mbox{tr} \left[\lambda(x)\Gamma_{\nu_1\nu_2...} \prod_{\mu=1}^4{\cal W}_\mu(x) \lambda(0) {\cal W}^\dagger_\mu(x)  \right].
\end{eqnarray}
whose unnormalized expectation values in the $Q=k/N$ sectors were computed in section \ref{all the Correlators}, starting from eqn.~(\ref{the full things}). We thus obtain the master formula for $\langle{\cal O}_{\nu_1\nu_2...}(x)\rangle^{Reg.}$ computed in the semiclassical approximation, using the notation for the determinants ${\cal F}_{\pm k}$, already introduced in (\ref{grand ZT with mass}), omitting the $\nu_1, \nu_2,...$ indices 
\begin{eqnarray}\label{grand master F} 
&&\langle {\cal O}(x) \rangle^{\scriptsize\mbox{Reg}} \\
&&=\frac{ {\cal F}_0\left[{N }  \langle  {\cal O} \rangle_{0} +\sum_{k=1}e^{-\frac{8\pi^2 k}{Ng^2}} \int_{\Gamma^{(k)}} d\mu_B^{(k)} \left(   e^{i\theta\frac{k}{N}}  { {\cal F}_k \over {\cal F}_0} {\langle {\cal O} \rangle_{\frac{k}{N}}  } + e^{-i\theta\frac{k}{N}}   { {\cal F}_{-k} \over {\cal F}_0} {\langle {\cal O} \rangle_{\frac{-k}{N}}}\right)\right] }{{\cal F}_0 \left[N+\sum_{k=1}e^{-\frac{8\pi^2 k}{Ng^2}}\mu_B^{(k)}\left({{\cal F}_{k} \over {\cal F}_0} e^{i\theta\frac{k}{N}} +{{\cal F}_{-k} \over {\cal F}_0}e^{-i\theta\frac{k}{N}} \right)\right]}\,,\nonumber 
\end{eqnarray}
where the denominator is simply $Z^{\scriptsize T}[\eta=\bar\eta=0]$ of eqn.~(\ref{grand ZT with mass}). The expectation values $\langle {\cal O}(x) \rangle_{\frac{k}{N}}$  in the instanton backgrounds are computed in section  \ref{Correlators in the higher Q sector} for $k \ne 0$  (and in section \ref{Comments on the sector Q0}, eqn.~(\ref{Q0 correlators}) for $k=0$).  The factors ${\cal F}_k \over {\cal F}_0$ are the regulated determinants of fluctuations from section \ref{Determinants and regularization}, eqn.~(\ref{det1}) for ${\cal F}_k$ (and noting that  ${\cal F}_{-k}$ is the same expression but with $m \rightarrow m^*$). 

For further use, we rewrite (\ref{grand master F}) substituting (\ref{det1}) for ${\cal F}_k$. We also keep in mind that ${\cal D}_{ k}(|m|)/{\cal F}_0$, denoted simply ${\cal D}_{k}/{\cal F}_0$ below, is, as discussed in section \ref{sec:uvfiniteratio}, to be substituted by the leading-order small-$\Delta$ expression  (\ref{Dratio last}):\footnote{Recall the limit (\ref{smalllimit maintext}).} 
\begin{eqnarray}\label{grand master F 1} 
&&\langle {\cal O}(x_1,..x_n) \rangle^{\scriptsize\mbox{Reg}}   \\
&&=\frac{ {\cal F}_0\left[N  \langle  {\cal O} \rangle_{0} +\sum_{k=1}e^{-\frac{8\pi^2 k}{Ng^2}}  M_{PV}^{3k} \int_{\Gamma^{(k)}} d\mu_B^{(k)}\left( m^k  e^{i\theta\frac{k}{N}} {{\cal D}_{k} \over {\cal F}_0} {\langle {\cal O} \rangle_{\frac{k}{N}}  } + (m^*)^k e^{-i\theta\frac{k}{N}} {{\cal D}_{k} \over {\cal F}_0} {\langle {\cal O} \rangle_{\frac{-k}{N}}}\right)\right] }{{\cal F}_0 \left[N+\sum_{k=1}e^{-\frac{8\pi^2 k}{Ng^2}} M_{PV}^{3k} \mu_B^{(k)}\left(m^k  {{\cal D}_{k} \over {\cal F}_0} e^{i\theta\frac{k}{N}} +(m^*)^k {{\cal D}_{k} \over {\cal F}_0}e^{-i\theta\frac{k}{N}} \right)\right]}\,. \nonumber 
\end{eqnarray}
 The  correlators $\langle {\cal O} \rangle_{\frac{k}{N}}$ and $\langle {\cal O} \rangle_{\frac{-k}{N}}$ typically include Wilson line insertions and depend on moduli parameters $z_\mu$ and $\bm{a}_\mu$. For each $k$ in the sum, the contribution from the $Q=\pm \frac{k}{N}$ sectors requires an integration over the corresponding moduli space $\Gamma^{(k)}$ from section \ref{The bosonic moduli space}.
Finally, after integrating over the moduli space, everything is expressed through the strong coupling scale using  $\Lambda^3\equiv M_{PV}^3e^{-\frac{8\pi^2}{Ng^2}}/g^2$.

Keeping the leading order in $|m|$ in each topological sector,  using   (\ref{det1}) with (\ref{Dratio last}), but omitting also the   $(|m|L)^2/\Delta$ terms,  we obtain the simplified result
\begin{eqnarray}\label{master F}\nonumber
&&\langle {\cal O}(x_1,..x_n) \rangle^{\scriptsize\mbox{Reg}}=\\
&&\frac{1}{N}\left\{N\langle  {\cal O} \rangle_{0}+\sum_{k=1}M_{PV}^{3k} e^{-\frac{8\pi^2 k}{Ng^2}} \int_{\Gamma^{(k)}} d\mu_B^{(k)}\left(m^ke^{i\theta\frac{k}{N}}\langle {\cal O} \rangle_{\frac{k}{N}}+m^{*k}e^{-i\theta\frac{k}{N}}\langle {\cal O} \rangle_{\frac{-k}{N}}\right) \right\}\,.
\end{eqnarray}
The prefactor $1/N$ corresponds to the total partition function $Z^T[\eta=\bar\eta=0]$, as defined in Eq.~(\ref{ZT with mass}) after omitting corrections of order ${\cal O}(|m|NL)$ and higher. 
 
\subsection{The $2$-point function}
\label{the2pointfunctionsection}

 Recalling the propagators (\ref{Q0 correlators}) in the trivial sector, the correlators (\ref{general correlator}, \ref{general correlator123}, \ref{C1 corr},  \ref{C1bar corr}, \ref{C1ii}, \ref{C1iii}, \ref{C2form}) in the background with topological charge $Q=k/N$ as well as the corresponding propagators in the anti-instanton background (of charge $Q=-k/N$), plugging the results into the master formula (\ref{grand master F 1}) and using $\Lambda^3\equiv M_{PV}^3e^{-\frac{8\pi^2}{Ng^2}}/g^2$, we obtain an explicit expression of the regularized two-point correlator (\ref{two point 1}). Although the resulting formula can be written, it is still quite cumbersome and will not be written out here in full. 

Thus, in what follows, we will instead focus on analyzing the behaviour of (\ref{two point 1}) in the coincidence limit, $x \rightarrow 0$ and for $\Gamma_{\nu_1 \nu_2 ...}$ taken to be the unit matrix. Thus,  we compute the ``gaugino condensate'' $\langle \tr \lambda\lambda(0) \rangle$, using the partition function defined by summing over all fractional topological sectors, as in (\ref{grand master F 1}). 
The quotation marks above serve to remind us that the composite operator $\tr \lambda\lambda(x)$ suffers, at $m\ne0$,
 an additive divergent renormalization\footnote{As opposed to the $m=0$ case.} and thus needs a UV definition.
 To define the composite operator at the level of our calculations, we subtract an infinite part proportional to the unit operator. We take this to be the infinite part of the normalized (i.e.~with ${\cal D}_0^f(|m|)$ divided out) propagator (\ref{Q0 correlators}) in the $Q=0$ sector in the coincident $x \rightarrow y$ limit.
 This definition can  be, after replacing the divergent sum in (\ref{Q0 correlators})  with an integral, as in (\ref{f0sector 2}), formally written as
 \begin{eqnarray}\label{lambdadefinition}
[\tr \lambda \lambda(x)]_{Reg} &=& \tr \lambda \lambda(x) - m^*  {g^2 \over (2\pi)^4} (N^2 - 1) \int^{M_{PV}} {d^4 k \over |m|^2 + k^2} \\
&=&   \tr \lambda \lambda(x) - m^*  g^2 f(M_{PV}, |m|), \nonumber 
\end{eqnarray}
where we defined, schematically\footnote{We stress, again, that the precise form of the subtraction depends on the way the divergent sum is regulated; for a consistent manner, this should be done using the same regulators as everywhere else, for example, with multiple PV fields.}
\begin{eqnarray}\label{defoff}
  f(M_{PV}, |m|) \equiv \int^{M_{PV}} {d^4 k  \; (N^2 - 1) \over (2 \pi)^4 \;  ( |m|^2 + k^2)} = {d \over d |m|^2}\left[ (N^2 -1) \int^{M_{PV}} {d^4 k \over (2 \pi)^4} \ln {|m|^2 + k^2 \over k^2}\right]. \nonumber \\ \end{eqnarray}
As noted above, the quantity $f(M_{PV},|m|)$  is the derivative w.r.t. $|m|^2$ of the divergent contribution to the cosmological constant from (\ref{f0sector 2}).
Clearly, the subtraction in (\ref{lambdadefinition})  is only formally defined;  in order to compute the finite part, one requires a precise definition of the subtraction in a given regularization scheme\footnote{However, we also note that when $m$ is real, the pseudo scalar condensate $-i\langle\mbox{tr}\left[\bar\Psi\gamma_5\Psi\right] \rangle \equiv  i \langle \tr \lambda\lambda(0) \rangle  - i  \langle \tr \bar\lambda\bar\lambda(0) \rangle^{\scriptsize\mbox{}}$ is UV finite, and there is no need to additional subtraction. For the general phase of $m$, a linear combination between scalar, $\langle\mbox{tr}\left[\bar\Psi\Psi\right] \rangle \equiv   \langle \tr \lambda\lambda(0) \rangle  +  \langle \tr \bar\lambda\bar\lambda(0) \rangle^{\scriptsize\mbox{}}$,  and pseudo scalar condensates is UV finite, e.g., for purely imaginary $m$, the scalar one is UV finite.}.

As for the case of the cosmological constant, due to the fact that the additive UV divergent contributions to the composite operator $\tr \lambda\lambda$  do not depend on the volume and the boundary conditions, the definition of $[\tr \lambda \lambda(x)]_{Reg.}$ of (\ref{lambdadefinition}) renders the expectation value finite for all nonzero $Q=k/N$ sectors.\footnote{That this is true follows from our discussions in section \ref{sec:uvfiniteratio} and the relation between the infinite contributions to the cosmological constant and to the composite operator from eqn.~(\ref{defoff}).}

This being said, we continue,  proceeding from (\ref{grand master F 1}) with ${\cal O} = [\tr \lambda\lambda]_{Reg.}$ to define 
\begin{eqnarray}\label{bigtracedef}
\langle \tr \lambda\lambda(0) \rangle^{\scriptsize\mbox{Reg}} \equiv  {N {\cal F}_0 \over Z^T}\left({\cal T}_0+{\cal T}_{1(i)}+{\cal T}_{1(ii)}+{\cal T}_{1(iii)}+{\cal T}_{2}\right)
\end{eqnarray}
where the expectation value is computed as in (\ref{grand master F}) and the quantities ${\cal T}_0, {\cal T}_{1(i)}, {\cal T}_{1(ii)}, {\cal T}_{1(iii)}, {\cal T}_{2}$  will be defined explicitly below: ${\cal T}_0$ corresponds to $Q=0$ sector propagator while the rest are the sums over all nontrivial topological sectors of the Cartan and non-Cartan components of $SU(k) \times U(1)$ (${\cal T}_{1(i)}$ and ${\cal T}_{1(ii)}$, respectively),  the $SU(\ell)$  ( ${\cal T}_{1(iii)}$), and the off-diagonal $k\times \ell$ and $\ell\times k$ components (${\cal T}_{2}$).
 
We begin with ${\cal T}_0$,  the contribution fro the trivial, $Q=0$, topological sector. Comparing (\ref{bigtracedef}) with (\ref{grand master F}), we conclude that ${\cal T}_0$ is simply the trace of the regulated propagator in the  trivial $Q=0$ sector, explicitly
\begin{eqnarray} \label{Tzero}
{\cal T}_0= \langle [\tr \lambda\lambda]_{Reg.} \rangle_{Q=0} =  \frac{g^2}{V} \sum_{\bm p \in \mathbb Z_N^2\neq 0, k_\mu\in \frac{2\pi\mathbb Z}{L_\mu}}\frac{m^*}{|m|^2+M_{\bm p,k}^2}\bigg\vert_{Reg.}\,,
\end{eqnarray}
where $M_{\bm p,k}^2$ is defined in (\ref{valueof Mpks}), and the subscript regulated on the divergent sum denotes the subtraction of the infinite part from eqn.~(\ref{lambdadefinition}). 

The following term, ${\cal T}_{1(i)}$, sums the contributions of all topological sectors to the expectation value of   the propagator along the Cartan directions of $U(1)\times SU(k)$.  In the background of an instanton with topological charge $Q=k/N$  for $k>0$, this is given in (\ref{11element main text}), while the c.c. of eqn.~(\ref{22element3 maintext}) holds in the background of an anti-instanton with $Q=-k/N$, $k>0$.\footnote{We have written a sum over $|k|=0,1,2....$ terms using the $\Delta$ expansion in each of them. However, the reader should keep in mind that discussion in section \ref{Detuned T4 nonabelian self-dual instantons and fermions} (near footnote \ref{footnoteDelta}) when the torus sides are taken so that $\Delta_{k=1}$ is small, only a finite number of $\Delta_{k>1}$ are small enough for the $\Delta$ expansion to apply. Thus, while all $|k|>1$ terms are semiclassical at small $LN\Lambda$, if $|k|$ is large enough, they are semiclassically suppressed yet not analytically calculable.} The integration of the Cartan propagators over the moduli space is trivial in all terms \begin{eqnarray}\nonumber \label{Tone1}
{\cal T}_{1(i)}&=&  \sum_{k=1}\frac{(16\pi^2\Lambda^3)^k}{(k-1)!} {{\cal D}_k \over {\cal F}_0} e^{i\frac{k\theta}{N}}\left(\frac{mV}{g^2}\right)^{k-1}\\
&&+\sum_{k=1}\frac{(16\pi^2\Lambda^3)^k}{(k-1)!}{{\cal D}_k \over {\cal F}_0}e^{i\frac{k\theta}{N}}\left(\frac{mV}{g^2}\right)^{k-1}\sum_{k_\mu\in \frac{2\pi \mathbb Z }{L_\mu}, k_\mu k_\mu \neq 0}\frac{|m|^2}{ k_\mu k_\mu}\bigg\vert_{Reg.}\\
&&+\sum_{k=1}\frac{(16\pi^2\Lambda^3)^k}{(k-1)!} {{\cal D}_k \over {\cal F}_0} e^{-i\frac{k\theta}{N}}\left(\frac{m^*V}{g^2}\right)^{k-1}\left({(m^* L)^2 \over c \Delta} + \sum_{k_\mu\in \frac{2\pi \mathbb Z}{L_\mu},  k_\mu k_\mu \neq 0}\frac{m^{*2}}{ k_\mu k_\mu}\bigg\vert_{Reg.}\right)\,, \nonumber 
\end{eqnarray}
and ${\cal D}_k/{\cal F}_0$  is understood to be  (\ref{Dratio last}). 
The first term is the would-be zero mode contribution, coming from all the $Q=\frac{k}{N}$ sectors, while the second and third terms represent the contribution from the non-zero modes of all the $Q=k/N$ and $Q=-k/N$ sectors, with $k\geq 1$.  

Similarly, the term ${\cal T}_{1(ii)}$ correspond to the propagator (\ref{C1ii}) defined along the non-zero roots of $SU(k)$:
\begin{eqnarray} \label{Tone2}\nonumber
{\cal T}_{1(ii)}&=&\sum_{k=1}\frac{(16\pi^2\Lambda^3)^k}{k!}{{\cal D}_{k} \over {\cal F}_0} \left\{e^{i\frac{k\theta}{N}}\left(\frac{mV}{g^2}\right)^{k-1}|m|^2+
e^{-i\frac{k\theta}{N}}\left(\frac{m^*V}{g^2}\right)^{k-1}m^{*2}\right\}\\
&&\times\sum_{B_1'\neq B_2'=1}^k\sum_{p_\mu=\frac{2\pi \mathbb Z}{L_\mu}}\frac{1}{p_1^2+\left(p_2+\frac{2\pi(B_1'-B_2')}{ k L_2}\right)^2+p_3^2+p_4^2+|m|^2}\bigg\vert_{Reg.}\,.
\end{eqnarray}
We also have the term  ${\cal T}_{1(iii)}$ that corresponds to the propagator (\ref{C1iii}) of the $SU(\ell)$ sector:
\begin{eqnarray}\label{Tone3}\nonumber
{\cal T}_{1(iii)}&=&\sum_{k=1}\frac{(16\pi^2\Lambda^3)^k}{k!}{{\cal D}_{k} \over {\cal F}_0}\left\{e^{i\frac{k\theta}{N}}\left(\frac{mV}{g^2}\right)^{k-1}|m|^2+
e^{-i\frac{k\theta}{N}}\left(\frac{m^*V}{g^2}\right)^{k-1}m^{*2}\right\}\\
&&\times \sum_{\bm p\in \mathbb Z_\ell^2\neq 0, k_\mu\in \frac{2\pi\mathbb Z}{L_\mu}}\frac{1}{|m|^2+M_{(\ell) \bm p,k}^2}\bigg\vert_{Reg.}\,,
\end{eqnarray}
and $M_{(\ell) \bm p,k}^2$  is defined in (\ref{Mell pk}).

In evaluating ${\cal T}_{1(i)}$, ${\cal T}_{1(ii)}$, and ${\cal T}_{1(iii)}$ in the coincidence limit $x = 0$, the integration over the moduli space was trivial, as the corresponding propagators are independent of the moduli-space coordinates in this limit. In contrast, the propagator required for computing ${\cal T}_2$ retains a nontrivial dependence on the moduli-space coordinates, as is clear from the propagator (\ref{C2form}) in the instanton with charge $Q=k/N$ as well as the counter propagator in the anti-instanton background, necessitating a more involved analysis. To this end, we use two identities (no summation over $\ell_{(1)}, \ell_{(3)}$)

\begin{eqnarray}\label{phisum1}
\int_{\Gamma^{(k)}}\sum_{C'=1}^k \sum_{C=1}^{N-k}(\varphi^{\ell_{(1)}, \ell_{(3)}}_{C'C}(x,\hat\phi))^* \; \varphi^{\ell_{(1)}, \ell_{(3)}}_{C'C}(x,\hat\phi)&=&\left({8 \pi^2 \sqrt{V} \over g^2}\right)^{2k} {N   \over (k-1)! (2\pi)^{2k}} ~,
\end{eqnarray}
which can be directly proved from the explicit form of $\varphi^{\ell_{(1)}, \ell_{(3)}}_{C'C}(x,\hat\phi)$ in (\ref{finalvarphiBCS}). Similarly, one obtains 
\begin{eqnarray}\label{phisum2}
\int_{\Gamma^{(k)}}\sum_{C'=1}^k \sum_{C=1}^{N-k} D_\mu \varphi_{C'C}^{\ell_{(1)},\ell_{(3)}}(x,\hat\phi)\left(D_\mu\varphi_{C'C}^{\ell_{(1)},\ell_{(3)}}(0,\hat\phi)\right)^* =\omega^2_{\ell_{(1)},\ell_{(3)}}  \left({8 \pi^2 \sqrt{V} \over g^2}\right)^{2k} {N   \over (k-1)! (2\pi)^{2k}} ,\nonumber \\
\end{eqnarray}
where $\omega^2_{\ell_{(1)},\ell_{(3)}}$ is given just after (\ref{barlambdaktimesellpropagator maintext}) (as well as  in (\ref{EVSktimesell})). With this information, we readily obtain
\begin{eqnarray}\label{Ttwo} \nonumber
{\cal T}_{2}&=&\sum_{k=1}  \frac{(16\pi^2\Lambda^3)^k}{(k-1)!}{{\cal D}_{k} \over {\cal F}_0} \left\{e^{i\frac{k\theta}{N}}\left(\frac{mV}{g^2}\right)^{k-1}|m|^2+
e^{-i\frac{k\theta}{N}}\left(\frac{m^*V}{g^2}\right)^{k-1}m^{*2}\right\}\\
&&\times\sum_{\ell_{(1)},\ell_{(3)}=0} \frac{1}{\frac{4\pi}{L_1L_2}\left(1+\ell_{(1)}+\ell_{(3)}\right)+|m|^2}\bigg\vert_{Reg.}\,.
\end{eqnarray}
In accordance with our $\Delta$-expansion and the limit (\ref{smalllimit maintext}), the $|m|^2$ terms in the denominators of (\ref{Tone2}, \ref{Tone3}, \ref{Ttwo}) should be omitted.

{\flushleft{\bf The zero mass limit:}}
In the strict massless limit $m=0$, we have $Z^T=N$ and ${{\cal D}_k \over {\cal F}_0} = 1$, we find that only the sector $Q=1/N$ contributes to the $2$-point function via the term ${\cal T}_{1(i)}$, leading to  
\begin{eqnarray}\label{massless limit}
\mbox{lim}_{m= 0}\,\langle \tr \lambda\lambda(0) \rangle^{\scriptsize\mbox{Reg}} =  16\pi^2\Lambda^3 e^{i\frac{\theta}{N}}\,,
\end{eqnarray}
a position-independent result,  a result that matches the gaugino condensate calculations obtained in \cite{Anber:2024mco}: $|\langle\mbox{tr}\left[\lambda\lambda\right]\rangle|^{\scriptsize\mbox{reg}}=16\pi^2\Lambda^3$. Here, we have set $m=0$ in the infinite sums, assuming that the sums after regularization yield in the limit $|m|L\rightarrow 0$ a result continuously connected to the strict limit $m=0$; a finding that we shall verify next.

{\flushleft  {\bf The small mass limit:}} Next, we examine (\ref{bigtracedef}) in the limit $N|m|L\ll 1$. Observe that the combination $\Lambda^3 (m L^4)^{k-1}$ can be rewritten as $\Lambda^3(mL)^{k-1}(\Lambda L)^{3k-3}$, and since we are working in the limit $ N|m|L \ll 1$, $N\Lambda L \ll 1$, and $|m|\ll \Lambda$, the sectors $Q=0, \pm 1/N$ provide the dominant contribution to the partition function and $2$-point correlator, whereas sectors with $k>1$ yield subdominant effects. 

The computation of $Z^T$ greatly simplifies in this limit: using the denominator of (\ref{grand master F 1}), the moduli space volume (\ref{volume of mub}), and (\ref{Dratio last}), we readily obtain
\begin{eqnarray}\label{ZTinQ=1sec}
Z^T=N {\cal F}_0\big\vert_{Reg.} \left[1 +32\pi^2 V |m| (1 + {|m|^2 L^2 \over c \Delta}) \Lambda^3  \cos\left(\frac{\theta_{\scriptsize\mbox{eff}}}{N}\right)\right]\,,
\end{eqnarray}
where $\theta_{\scriptsize\mbox{eff}}\equiv \theta+N\mbox{arg}\,m$ and  ${\cal F}_0\big\vert_{Reg.}$ from (\ref{f0sector2})

The $2$-point function (\ref{bigtracedef}) also involves infinite sums, as is evident from (\ref{Tzero}, \ref{Tone1}, \ref{Tone2}, \ref{Tone3}, \ref{Ttwo}). The contributions of  ${\cal T}_{1 (ii)}$, ${\cal T}_{1 (iii)}$ and ${\cal T}_{2}$, eqns.~(\ref{Tone2}, \ref{Tone3}, \ref{Ttwo}), however, after regulating and subtracting the infinite part, scale  like $|m|^2 L^2$ and we shall ignore them. In ${\cal{T}}_{1 (i)}$, eqn.~(\ref{Tone1}), we keep the $k=1$ and the leading $\sim 1/\Delta$ contribution from $k=-1$ to obtain
\begin{eqnarray}\label{small mV C} \nonumber
&&  \langle \tr \lambda\lambda(0) \rangle^{\scriptsize\mbox{Reg}}\bigg\vert_{|m| LN  \ll 1}   \\
&&\simeq {N {\cal F}_0\big\vert_{Reg.} \over  Z^T} \left\{ \frac{g^2 m^*}{\sqrt{V}} \sum_{\bm p \in \mathbb Z_N^2\neq 0, k_\mu\in \frac{2\pi\mathbb Z}{L_\mu}}\frac{1}{\sqrt{V} M_{\bm p,k}^2}\bigg\vert_{Reg.}  +16\pi^2\Lambda^3 (1 + {(|m|L)^2 \over c \Delta}) e^{i\frac{\theta}{N}} +16\pi^2\Lambda^3 {(m^* L)^2 \over c \Delta} e^{-i \frac{\theta}{N}}\right\}\,. \nonumber \\
\end{eqnarray}
We further note that after regulating the $Q=0$ sector contribution scales like ${g^2 m^*\over L^2} \ll \Lambda^3$ and can be dropped in the weak-coupling, small mass limit.\footnote{We notice that, in order to drop this term, we are making the stronger assumption that $|m| L \ll (\Lambda L)^3$ than simply $|m|L \ll \Lambda L$.} Thus, also recalling $Z^T$ from (\ref{ZTinQ=1sec}), we set the overall prefactor to unity, to obtain, only keeping the leading small-$m$ correction of order $|m|^2 L^2 \over c \Delta$
 \begin{eqnarray}\label{small mV C 1} \nonumber
&&  \langle \tr \lambda\lambda(0) \rangle^{\scriptsize\mbox{Reg}}\bigg\vert_{|m| LN  \ll 1}   \simeq    16\pi^2\Lambda^3 (1 + {(|m|L)^2 \over c \Delta}) e^{i\frac{\theta}{N}} +16\pi^2\Lambda^3 {(m^* L)^2 \over c \Delta} e^{-i \frac{\theta}{N}} . \nonumber \\
\end{eqnarray} 
Identical calculations of the $2$-point function of the dotted fermions yield 
\begin{eqnarray}\label{bar small mV C 1} \nonumber
&&  \langle \tr \bar\lambda\bar\lambda(0) \rangle^{\scriptsize\mbox{Reg}}\bigg\vert_{|m| LN  \ll 1}   \simeq    16\pi^2\Lambda^3 (1 + {(|m|L)^2 \over c \Delta}) e^{-i\frac{\theta}{N}} +16\pi^2\Lambda^3 {(m  L)^2 \over c \Delta} e^{i \frac{\theta}{N}} . \nonumber \\
\end{eqnarray} 
Clearly, both expressions are covariant under the $U(1)_{spurious}$ transformation of eqn.~(\ref{spuriousU1}). 
\subsection{Physical observables and CP violation}

Particularly important observables includes the condensate and pseudo-condensate, both of which are most naturally expressed in terms of the Majorana field $\Psi$ defined in (\ref{Majorana def}) and (\ref{cond1}). We straightforwardly obtain, for the scalar condensate:\footnote{The parameter $c$ here should be taken to be the one appropriate to $k=1$, i.e. $c = 4 \pi/(N-1)$, see (\ref{energyshifts1}).}
\begin{eqnarray}\nonumber \label{scalar condensate}
 \langle\mbox{tr}\left[\bar\Psi\Psi\right] \rangle^{Reg.}\bigg\vert_{|m| LN  \ll 1} &\simeq&
    32\pi^2\Lambda^3  \left(1+  {|m|^2 L^2 \over c \Delta} \right)\cos\left(\frac{\theta}{N}\right)   + 16\pi^2\Lambda^3 \left( {L^2 m^{* \; 2}  \over c \Delta} e^{- i {\theta \over N}}+ { L^2 m^2  \over c \Delta} e^{i {\theta\over N}}\right)  \,,\nonumber \\
\end{eqnarray}
and, for the pseudo-scalar condensate:
\begin{eqnarray}\nonumber \label{pseudo scalar condensate}
 &&-i\langle\mbox{tr}\left[\bar\Psi\gamma_5\Psi\right] \rangle^{Reg.}\bigg\vert_{|m| LN  \ll 1}  =\left( i \langle \tr \lambda\lambda(0) \rangle  - i  \langle \tr \bar\lambda\bar\lambda(0) \rangle^{\scriptsize\mbox{Reg}}\right)\bigg\vert_{|m| LN  \ll 1}  \\
  &&\simeq  -  32\pi^2\Lambda^3  \left(1+ {|m|^2 L^2   \over c \Delta}\right)\sin\left(\frac{\theta}{N}\right)
 +i 16 \pi^2\Lambda^3      \left( {L^2 m^{*\;2} \over c \Delta} e^{- i {\theta \over N}} -{L^2 m^{2} \over c \Delta} e^{  i {\theta \over N}}\right)   \,.
\end{eqnarray}

We can also calculate the following quantity to the leading oder in $|m|NL$:
\begin{eqnarray}\label{energy1} 
\delta {\cal E}&=& - m\langle \mbox{tr}\left[\lambda\lambda\right] \rangle ^{\scriptsize \mbox{reg}} - m^*\langle \mbox{tr}\left[\bar\lambda\bar\lambda\right] \rangle ^{\scriptsize \mbox{reg}} =  - 32\pi^2 \Lambda^3  |m| (1+ 2 {|m|^2  L^2   \over c \Delta})\cos\left(\frac{\theta_{\scriptsize \mbox{eff}}}{N}\right) ,
\end{eqnarray}
where $\theta_{\scriptsize \mbox{eff}}\equiv \theta + N {\rm arg}(m)$. We call it $\delta {\cal E}$ because,  in the infinite volume limit, $\delta {\cal E}$ would be the contribution of the condensates to the vacuum energy: $\delta {\cal E}= - 32\pi^2 \Lambda^3 |m| \cos\left(\frac{\theta_{\scriptsize \mbox{eff}}}{N}\right)$ (however, in the small $\T^4$ where the last equation was obtained, all energy eigenstates contribute).
Clearly, $\delta{\cal{E}}$ is   invariant under (\ref{spuriousU1}) and  has the same $\theta_{\scriptsize \mbox{eff}}$ dependence as the results obtained on $\R^4$ \cite{Konishi:1996iz,Evans:1996hi}. 

We note that $\delta {\cal{E}} =0$, to leading order in $|m|$, at $\theta_{\scriptsize \mbox{eff}} = \pi$ and $N=2$. This can be seen as a ``topological interference'' effect noted in \cite{Unsal:2012zj}: note that the order $|m|$ contribution to $\delta{\cal{E}}$ comes from the $m$ and $m^*$ terms in eqn.~(\ref{soft mass 2}). Applying the selection rule (\ref{selection rule}) then shows that at $\theta_{\scriptsize \mbox{eff}}= \pi$, the contributions of instantons of charge $\pm 1/2$ to $\delta{\cal{E}}$ come with opposite phases $\pm i$ and cancel for $N=2$ (this cancellation is a reflection of the fact that, at $\theta_{\scriptsize \mbox{eff}}=\pi$ there is an exact double degeneracy of the electric flux eigenstates in pure YM theory, due to the parity-center symmetry anomaly, see eqn.~(\ref{vacuum split})).

\section{SYM$^*$ in the semiclassical, yet not calculable, limit of  $\R \times \T^3$}
  
 \label{sec:rtimest3}
 
It is of interest  to also study a different limit, where semiclassical ideas are still expected to hold, albeit without the precise analytical control of the small $L  \Lambda N$, $m LN$ limit enjoyed on the $\T^4$. This is the limit of small $\T^3$ and infinite time ($x_1$), i.e. $\R \times \T^3$.  

The $L_1$-large limit is interesting because it allows one to isolate the ground state energy of the theory on $\T^3$ and study how the soft-breaking affects it. The goal of this section is to calculate---as we shall see, we end up with only a semiclassical estimate, rather than a precise calculation---the energy differences between the minimum values of the energy in two electric flux sectors in SYM$^*$, for small $m$.

Thus, we  imagine taking $L_1 \rightarrow \infty$, while keeping the spatial $\T^3$ small, such that weak coupling on $\T^3$ is justified, i.e. taking $L_{2,3,4} \Lambda N \ll 1$ (or $L \Lambda N \ll 1$, with $L \sim L_{2} \sim L_3 \sim L_4$ a common measure of the $\T^3$ size).
 We revert to the original expression for $Z^T$, eqn.~(\ref{full pf hamiltonian}), and assume that $\hat H_m$ is a small perturbation to $\hat H_{SYM}$ so that 
 the matrix elements of $\hat H_m$ are much smaller than $1/(LN)$, the energy differences between the unperturbed $m=0$ energy levels; parametrically, this implies  $m LN \ll 1$.

   To begin, we note that the $N$ classical zero energy ground states of SYM in  the small-$\T^3$ theory (with a $n_{34}=1$ twist) are center-symmetry images of the trivial vacuum, which we label by $|0\rangle$:
  \begin{eqnarray}\label{kvacua}
  |k \rangle = \hat T_2^k |0\rangle~, ~ k=0,...,N-1~.
  \end{eqnarray}
 The  vacuum $|0\rangle$ corresponds to the classical zero-energy configuration $A=0$, where $A$ is the spatial component of the $SU(N)$ gauge field (recall that the twist removes the zero modes of all fields allowing only a discrete set of $N$ locally pure-gauge configurations, which have classically zero energy). A state in the Hilbert space, denoted by $|k\rangle$, is built around each of these classically zero-energy states. In canonical quantization,\footnote{Canonical quantization on $\T^3$, in the $A_1=0$ (recall that $x_1$ is our time direction) gauge and with a spatial twist is described in \cite{Cox:2021vsa}. For earlier discussion and perturbative calculations in pure YM theory on a twisted small $\T^3$, see \cite{GonzalezArroyo:1987ycm,Daniel:1990iz}.} it has a wave functional $\Psi_k[A] = \langle A |k\rangle$, with the expectation value of the operator $\hat A$ equal to the classical value, $\langle \hat A \rangle_{\Psi_k} = A^{(k)}= i T_2^{k} d T_2^{-k}$, where $T_2(x_2, x_3, x_4)$ is the improper gauge transformation which generates center symmetry transformations in the $x_2$ direction.
  
The expression for  $\langle \hat A \rangle_{\Psi_k}$ given above shows that the  $| k \rangle$ states (\ref{kvacua}) are distinguished by the expectation value of the fundamental Wilson loop winding in the $x_2$ direction, 
  \begin{eqnarray}\label{wloop34}
  W_2 \equiv \langle k |\hat W_2 |k \rangle = e^{i {2\pi \over N} k}, ~ \text{where} ~ \hat W_2 = {1 \over N} \tr_{F} {\cal{P}}e^{i \oint \hat A_2 dx^2}.
  \end{eqnarray}
Next, we introduce the $N$ electric flux states,\footnote{For brevity,  in this section we use the label $e$ to denote  the $N$ values of the flux in the $x_2$ direction, $e_2$. It should not be confused with labels of the fluxes in the $x_{3,4}$ directions.} eigenstates of $\hat T_2$, as a discrete Fourier transform of the $| k \rangle$ states: \begin{eqnarray}\label{electricflux}
 | e \rangle = {1 \over \sqrt{N}} \sum\limits_{k=0}^{N}  e^{- i {2 \pi \over N} e k} |k \rangle~, ~\text{such  that} ~ \hat T_2 | e\rangle = |e \rangle e^{i {2 \pi \over N} e}~, ~ e = 0,..., N-1,
 \end{eqnarray}
 where the second equality follows from $\hat T_2 |k\rangle = |k+1\rangle$, as per (\ref{kvacua}). As we already mentioned, classically the $N$ states, $|e =0,...,N-1\rangle$,  have zero energy, while all other states have energy of order $1/(NL)$.
 We expect that this degeneracy (which is due to chiral symmetry and is exact only in SYM) will be broken once a soft mass is introduced.

At $m$ small enough, $m LN \ll 1$, we expect that the lowest energy states of the SYM$^*$ theory are close to the states $|e\rangle$. More precisely, we expect that the SYM$^*$ minimum energy state in the  flux sector $e$ has significant overlap with the perturbative $|e \rangle$ state. Given that, 
as $L_1 \rightarrow \infty$, we have:
 \begin{eqnarray}
 \langle e| (-)^F e^{- L_1 \hat H} | e \rangle\bigg\vert_{L_1 \rightarrow \infty} = c_e e^{- L_1 E_e}, \; E_e - {\text{lowest energy in $e_2 = e$ flux sector}}, \label{energy0}
 \end{eqnarray}
where $c_e$ is the projection of the perturbative state $|e\rangle$ on the lowest-energy eigenstate of the SYM$^*$ Hamiltonian $\hat H$ in the $e_2 = e$ flux sector (assuming, without loss of generality that the state is bosonic). We denote its energy by $E_e$. 

 Inverting the relation (\ref{energy0}), we have that (dropping ${1 \over L_1} \ln c_e$ term after the first line, since the overlap factor of $\T^3$ states should not depend on the time extent), using (\ref{electricflux}) as well as (\ref{kvacua}):
\begin{eqnarray}
E_e &=& -{1 \over L_1} \ln \langle e | (-)^F e^{- L_1 \hat H} | e \rangle\bigg\vert_{L_1 \rightarrow\infty} + {1 \over L_1} \ln c_e \bigg\vert_{L_1 \rightarrow\infty} ~~ (e = 0, \ldots N)\nonumber \\
&=&  -{1 \over L_1} \ln {1 \over N} \sum\limits_{k,k'=0}^{N-1} e^{i {2 \pi \over N} e (k'-k)} \langle k'| (-)^F e^{- L_1 \hat H}|k \rangle \nonumber \\
&=& -{1 \over L_1} \ln {1 \over N} \sum\limits_{k,k'=0}^{N-1} e^{i {2 \pi \over N} e (k'-k)} \langle 0| (-)^F e^{- L_1 \hat H} \hat T_2^{k - k'} |0 \rangle
 \label{energyflux}
\\
&=&  -{1 \over L_1} \ln\left( \langle 0| (-)^F e^{- L_1 \hat H} |0 \rangle + {1 \over N} \sum\limits_{k,k'=0, \; k \ne k'}^{N-1} e^{i {2 \pi \over N} e (k'-k)} \langle 0| (-)^F e^{- L_1 \hat H} \hat T_2^{k - k'}|0 \rangle \right)\bigg\vert_{L_1 \rightarrow \infty}~,\nonumber
\end{eqnarray}
where we used the fact that there are $N$ terms with $k'=k$.
Thus, we find that, in the $L_1 \rightarrow \infty$ limit:
\begin{eqnarray}
E_e + {1 \over L_1}\ln \langle 0| (-)^F e^{- L_1 \hat H} |0 \rangle  &=& - {1 \over L_1} \ln\left( 1 + {1 \over N} \sum\limits_{k,k'=0, \; k \ne k'}^{N-1} e^{i {2 \pi \over N} e (k'-k)}{ \langle 0| (-)^F e^{- L_1 \hat H}  \hat T_2^{k - k'}|0 \rangle  \over \langle 0| (-)^F e^{- L_1 \hat H} |0 \rangle }\right)~. \nonumber \\
&& \label{energy2}
\end{eqnarray}
Next, we notice that  $q \equiv k'-k = \pm 1, \pm 2,...\pm N-1$ in the sum entering in (\ref{energy2}) and that there are $N-|q|$ terms with $k'-k= q$.
Thus, 
\begin{eqnarray}
 E_e &+&  {1 \over L_1}\ln \langle 0| (-)^F e^{- L_1 \hat H} |0 \rangle \nonumber \\\nonumber
 &=& - {1 \over L_1} \ln \left[ 1 +   \sum\limits_{q=1}^{N-1} {N-q \over N} \left(e^{i {2 \pi \over N} e q}\;{ \langle 0| (-)^F e^{- L_1 \hat H}  \hat T_2^q|0 \rangle  \over \langle 0| (-)^F e^{- L_1 \hat H} |0 \rangle } + e^{ - i {2 \pi \over N} e q}\;{ \langle 0| (-)^F e^{- L_1 \hat H}  \hat T_2^{-q}|0 \rangle  \over \langle 0| (-)^F e^{- L_1 \hat H} |0 \rangle }\right)\right]  \\
&\equiv&   - {1 \over L_1} \ln \left[ 1 +   \sum\limits_{q=1}^{N-1} {N-q \over N} \left(e^{i {2 \pi \over N} e q}\; \Xi_q + e^{ - i {2 \pi \over N} e q}\; \Xi_{-q}\right)\right], 
 \label{energy3} 
\end{eqnarray}
where we defined 
\begin{eqnarray}\nonumber
\Xi_q &=& {\langle 0| (-)^F e^{- L_1 \hat H} \hat T_2^q|0 \rangle \over \langle 0| (-)^F e^{- L_1 \hat H} |0\rangle}\bigg\vert_{L_1 \rightarrow \infty}  \label{xidef} \\
& \simeq& { \int [{\cal{D}} A {\cal{D}}\lambda]^{A(x_1 = 0)=i T_2^q d T_2^{-q}, \; A(x_1 = + L_1 ) =0 }_{\{n_{34}=1\}} \; e^{- S_{SYM} - S_{m}} \over \int [{\cal{D}}A {\cal{D}} \lambda]^{A(x_1 =0) =  A(x_1 = L_1) =0}_{\{n_{34}=1\}}  \; e^{- S_{SYM} - S_{m}}}\bigg\vert_{L_1 \rightarrow \infty, L \Lambda N \ll 1}\,. 
\end{eqnarray}
In the last equation above, we converted the matrix elements   to the path integral representation. Upon doing so, in the second line above we dropped the convolution with the wave functionals $\Psi_0[A]$ of the initial and final state $|0\rangle$ (recall the discussion after  (\ref{kvacua})), in effect replacing them with $\delta$-functionals, taking the initial and final values of $A$ in the path integral to be equal to the classical values in the $|0\rangle$ vacuum. We expect that dropping these factors will have an effect that is absorbed in the---at any rate, incalculable, at present, see below---pre-exponential factor in $\Xi_q$.  In the numerator of (\ref{xidef}), we also took into account the fact that the insertion of $\hat T_2^q$ twists the boundary condition of the fields at $x_1=0$ (this implements the $n_{12}=-q$ twist). We also indicated that the boundary conditions on the spatial $\T^3$ are twisted by $n_{34}=1$.

\begin{figure}[h] 
   \centering
   \includegraphics[width=3.5in]{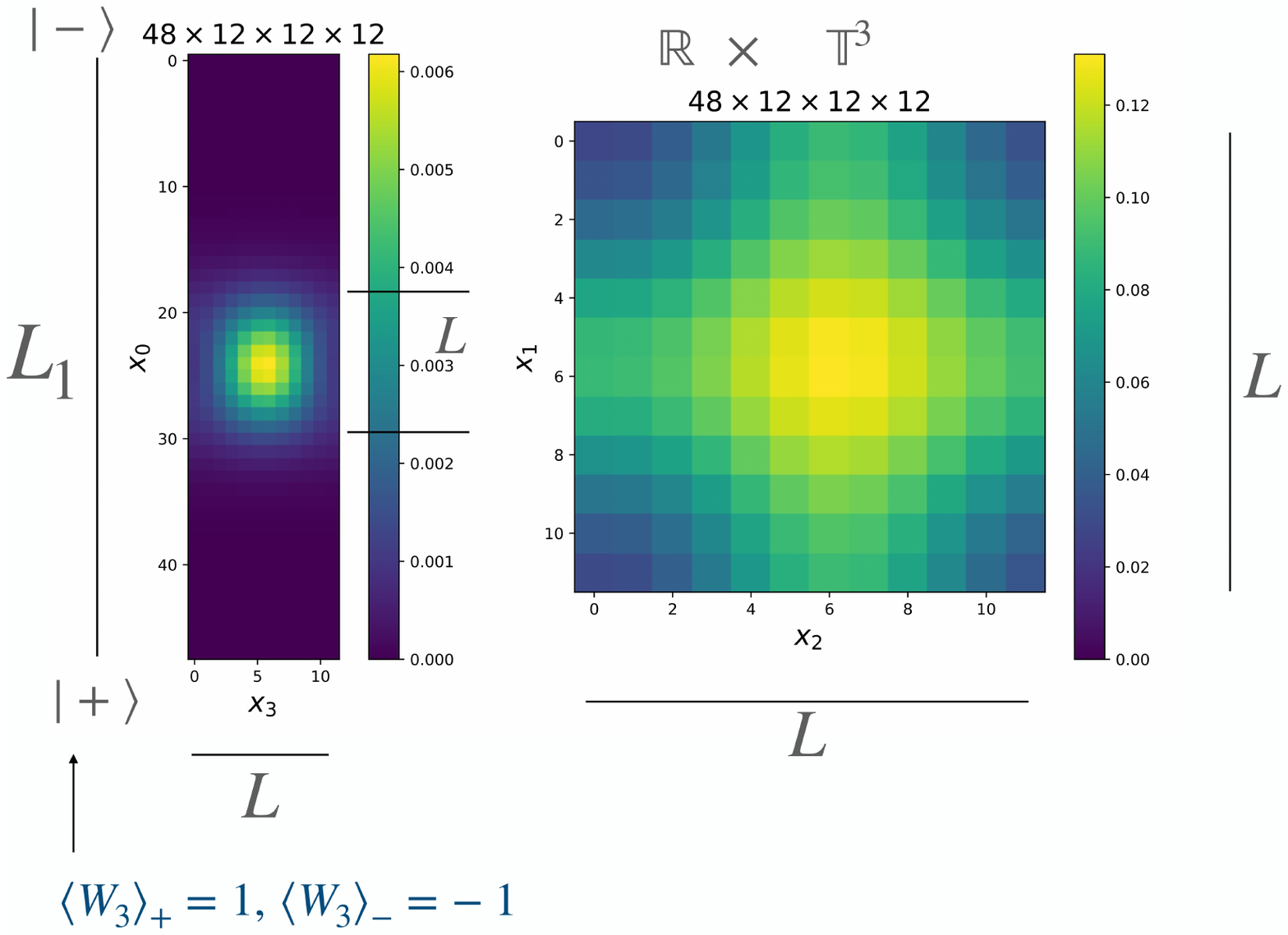} 
  \caption{The action density of a charge-$1/2$ instanton in $SU(2)$ YM theory on an asymmetric $\T^4$  with $L_1 \gg L$  ($L_1=48, L=12$) on the lattice with a unit twist in the small $\T^3$ and a unit twist in the mixed space-time direction. The plots are taken from \cite{Wandler:2024hsq}.  Notice the different axes labeling on the plot, namely $x_0^{plot} \rightarrow x_1$, $x_3^{plot} \rightarrow x_2$, $x_2^{plot} \rightarrow x_3$, $x_1^{plot} \rightarrow x_4$ (thus,  $W_3$ on the plot should be understood as $W_2$).
   The points the plot illustrates are: {\it i.)} that the action is localized in a region of size $L$ and spacetime volume $L^4$ and {\it ii.)}  that the  solution interpolates between the two $|k\rangle$ vacua (\ref{kvacua}, \ref{wloop34}) of the $SU(2)$ theory, the $|k=0\rangle$, $W_2 = 1$, and the $|k=1\rangle$, $W_2=-1$, as described in the text.  
 The dilute gas picture advocated in \cite{RTN:1993ilw,Gonzalez-Arroyo:1995ynx} uses these instantons as the basic blocks. Thus,  in the $L_1 \rightarrow \infty$ limit, one sums over arbitrary even numbers of these instantons, if the boundary conditions in the path integral are untwisted, i.e. if one is computing a transition amplitude $|0\rangle \rightarrow |0\rangle$ (or $|1 \rangle \rightarrow |1 \rangle$). On the other hand, if the transition amplitude is twisted, as it would be in the numerator of $\Xi_1$ of eqn.~(\ref{xidef2}), and interpolates from  $|0\rangle \rightarrow |1\rangle$ (or v.v.), one sums over an arbitrary odd number of such instantons. This picture, adapted here to $SU(N)$, leads to eqn.~(\ref{energy5}) for the splitting of the energies of the electric flux sectors. } 
   \label{instanton1}
\end{figure}

{\flushleft{Let us now make the following comments:}}
\begin{enumerate}
\item The $\ln \langle 0| (-)^F e^{- L_1 \hat H} |0 \rangle$ term appearing on the r.h.s. of (\ref{energy3}) vanishes in SYM but not in SYM$^*$. While it can  receive semiclassical corrections from integer-$Q$ sectors, including perturbative corrections proportional to $|m|^2$, the point is that it does not depend on $e$ and does not affect the energy splittings between the $N$ flux sectors.

\item It is clear that the numerator of $\Xi_q$, eqn.~(\ref{xidef}), receives contributions from  topological sectors with $Q={q\over N} + \Z$, while the denominator receives perturbative contributions plus contributions of sectors of arbitrary integer topological charge, including zero.

In this regard, we recall that the SYM$^*$ partition function $Z^T$ defined in (\ref{FULL PF}), ignoring the issue of regularization, can be written as a sum over the numerators of $\Xi_q$ from (\ref{xidef}):
\begin{eqnarray}
Z^T=\sum_{q=0}^{N-1}  \int [{\cal{D}} A {\cal{D}}\lambda]^{A(x_1 = 0)=i T_2^q d T_2^{-q}, \; A(x_1 = + L_1 ) =0 }_{\{n_{34}=1\}} \; e^{- S_{SYM*} +i\theta Q} \,.
\end{eqnarray}

Now, in the small $\T^3$ limit ($L \Lambda N \ll 1$) one expects to be able to semiclassically evaluate the path integrals $\Xi_q$ contributing to the electric flux energies (\ref{energy3}). 
However, the $L_1 \gg L$ asymmetric limit of a four-torus is not one where the saddle points of topological charge ${q\over N} + \Z$ are known analytically. 
There is evidence, based on minimizing the lattice action (for $SU(2)$ and $SU(3)$) on $\T^4$ with twists $n_{34}=1, n_{12}=-q$ that in the limit of an asymmetric $\T^4$, in topological sectors with  $Q=q/N + \Z$, the saddle points can be represented as a gas of charge $1/N$ objects, as we now describe.\footnote{ Independent of the lattice evidence, one can show, both on $\R \times \T^3_{n_{34}}$ and $\R^2 \times \T^2_{n_{34}}$, with {\it nonzero twist $n_{34}$,} in either $\T^2 $ or $\T^3 $ (as indicated by subscript), that finite-action Euclidean solutions on $\R \times \T^3_{n_{34}}$ and $\R^2 \times \T^2_{n_{34}}$ fall into topological classes labelled by fractional topological charges $Q= {p \over N^\prime} + n$, with  $p = 0,...,N^\prime-1$ and $n \in \Z$,  with  $N^\prime \equiv {N \over {\rm{gcd}}(n_{34},N)}$. Thus, the numerical solution shown on figure~\ref{instanton1} becomes a minimum action solution on $\R \times \T^3_{n_{34}}$ (for additional study of the ``infinite''-time limit of the lattice solution, see \cite{Wandler:2024hsq}).}

Each such object occupies  space-time volume $L^4$ and has topological charge $1/N$. We refer to figure~\ref{instanton1} for a  picture of the action density of the $Q=1/2$ solution in $SU(2)$ YM theory obtained on the lattice, which illustrates the nature of these $Q=1/N$ solutions in the $L_1 \gg L$ limit.  Based on this evidence,\footnote{Further support is also provided by the recent analytical treatment of \cite{Hayashi:2024yjc,Guvendik:2024umd,Hayashi:2024psa,Hayashi:2025doq}, utilizing an asymmetric limit of $\T^3$ with twist.} one can proceed to evaluate $\Xi_q$ using a dilute gas approximation, where these charge $1/N$ objects are the fundamental ``constituents" saturating the path integral. Each of these objects, centered at $x_1^*$ in the time coordinate, interpolates   between values of the Wilson loop $W_2 = e^{i {2\pi \over N} k}$ at $x_1 - x_1^* \ll - L$ to $W_2 =e^{i {2\pi \over N} (k+1)}$ at  $x_1 - x_1^* \gg  L$, i.e., recalling (\ref{wloop34}), between two neighboring $k$-vacua. 

The other feature used in argument for the dilute gas approximation, the splitting of a $Q=q/N$  instanton  into $q$ $Q=1/N$ localized objects in the small-$\T^3$, large-time limit, can be seen, e.g.~in figure 11 of \cite{Anber:2025yub} (for $N=3$, $q=2$ the dissociation can be seen already on an asymmetric lattice of size $(32, 4, 12, 12)$). 
\item Before we continue, we stress that evaluating the path integrals appearing in (\ref{xidef}, \ref{energy3}) by a dilute gas of the charge-$1/N$ objects described above is an  assumption, as far as sectors with $q>1$ is concerned.\end{enumerate}
 To recapitulate, our final expression for the minimum energy in each $e_2 = e$ flux sector is:
\begin{eqnarray}
 E_e &=&  \bar E  - {1 \over L_1} \ln \left[ 1 +   \sum\limits_{q=1}^{N-1} {N-q \over N} \left(e^{i {2 \pi \over N} e q}\; \Xi_q + e^{ - i {2 \pi \over N} e q}\; \Xi_{-q}\right)\right], ~ e=0, \ldots N-1, \nonumber \\ \label{energy4}
\end{eqnarray}
where $\bar E$ denotes the $e$-independent piece in (\ref{energy3}) and $\Xi_q$ is defined in (\ref{xidef}), from which we reproduce only the path integral expression:
\begin{eqnarray}
\Xi_q   \simeq { \int [{\cal{D}} A {\cal{D}}\lambda]^{A(x_1 = 0)=i T_2^q d T_2^{-q}, \; A(x_1 = + L_1 ) =0 }_{\{n_{34}=1\}} \; e^{- S_{SYM} - S_{m}} \over \int [{\cal{D}}A {\cal{D}} \lambda]^{A(x_1 =0) =  A(x_1 = L_1) =0}_{\{n_{34}=1\}}  \; e^{- S_{SYM} - S_{m}}}\bigg\vert_{L_1 \rightarrow \infty, L \Lambda \ll 1} .\nonumber \\
 \label{xidef2}
 \end{eqnarray}
 We now assume that the basic instantons have charge $\pm {1 \over N}$ and fugacity given by
\begin{eqnarray}\label{fugacity}
\zeta L^{-4}\; e^{\pm i {\theta \over N}} \;e^{- {8 \pi^2 \over g^2 N}} = \zeta L^{-4}\; e^{\pm i {\theta \over N}} \;e^{- S_0} ,~ \text{where} ~ S_0 \equiv {8 \pi^2 \over g^2 N},
\end{eqnarray}
where $\zeta$ is an unknown dimensionless coefficient, including $g^2 \equiv g^2 (L)$ dependence, and we have introduced the notation $S_0$ for the action of a charge-$1/N$ instanton.
Thus, we have that  (taking $m$ real and small) the leading contribution to $E_e$ of (\ref{energy4}) is given by\footnote{Recall that all $\Xi_q$ vanish if $m=0$, keeping all fluxes degenerate.}
\begin{eqnarray}
e^{i {2 \pi e\over N}} \Xi_1 + e^{-i {2 \pi e\over N}} \Xi_{-1}= \zeta  {L_1 L^3 \over L^4} \; m L \cos\left( {2 \pi e+\theta \over N}\right) e^{-  S_0}(1  + {\cal{O}}(e^{- NS_0}))~.
\end{eqnarray}
The equation inside the logarithm  in (\ref{energy4}) has to exponentiate after proper evaluation which we will not attempt here,\footnote{A more detailed semiclassical evaluation of (\ref{energy4}), including higher orders, would be an exercise of interest, for, among others, resurgence theory; see \cite{Dunne:2016nmc} for review.} hence, we simply expand the logarithm, keeping the above contribution only, to obtain\footnote{One notes that, at $\theta = \pi$, there is a two fold degeneracy of flux states, $E_e = E_{N-1-e}$, consistent with the parity-center anomaly of SYM$^*$ for real $m$ (for even-$N$, all flux states, while for odd-$N$, all states with $e \ne {N-1\over 2}$, are doubly degenerate). We also stress that the consistency of (\ref{energy5}) with the center-parity anomaly is not an artifact of our keeping the $\Xi_1$ term only: all other terms in (\ref{energy4}) depend on $e$ only in the combination
$e^{i q {2 \pi e + \theta \over N}}$, implying that $E_e$  at $\theta = \pi$  are invariant under  $e \leftrightarrow N-1-e$ (thus interchanging $\Xi_q$ and $\Xi_{-q}$). } 
\begin{eqnarray}\label{vacuum split}
 E_e &\simeq &  \bar E  - \zeta'  m e^{- S_0} \cos\left( {2 \pi e+\theta \over N}\right) + \ldots, ~ e=0,\ldots N-1.  \label{energy5}
\end{eqnarray}
This  equation is a desirable outcome: it shows that the minimum energies in each electric flux sector are split at order $m$. Unfortunately, an analytic calculation of the coefficient $\zeta'$ is currently out of reach, as already alluded to in the title of this section. We also note that an equation identical to (\ref{vacuum split}) can be obtained by soft-supersymmetry breaking methods directly on $\R^4$, for example as in \cite{Konishi:1996iz}.

{\flushleft{\bf Relation to older studies on $\R \times \T^3$:}} Let us now comment on the relation between our analysis of SYM$^*$ on $\R \times \T^3$ and older studies of pure YM theory in the literature \cite{RTN:1993ilw,Gonzalez-Arroyo:1995ynx}, which may be not as familiar as they deserve to be. 

{\flushleft{\bf 1.}} The splitting of the perturbatively degenerate electric flux energies in $SU(2)$ pure YM theory on $\R \times \T^3$ at $\theta=0$ due to nonperturbative effects was studied a long time ago \cite{RTN:1993ilw} using lattice simulations and semiclassical ideas. The semiclassical expression obtained in a dilute gas approximation was fitted to lattice data  to determine the incalculable pre-exponential factor $\zeta'$. Adapted to our notation and our choice of $\T^3$ twist, their result is obtained from (\ref{energy5}), by taking  $N=2$ and $\theta=0$: $E_1 - E_0 = 2  \zeta' m e^{- S_0}$. Their fitting parameter was, in effect, $\zeta' m$---we stress that SYM$^*$ is in the universality class of pure YM, hence the fact that the equations are related shouldn't come as surprise. In addition, the $L$-dependence of the energy split, obtained by replacing $e^{-S_0} \sim (\Lambda L)^{11\over 3}$ appears to fit well the lattice data, as shown in \cite{RTN:1993ilw,Gonzalez-Arroyo:1995ynx}.

For completeness, let us also describe how electric flux differences are measured on the lattice  \cite{RTN:1993ilw,Gonzalez-Arroyo:1995ynx} (here, we generalize their discussion to $SU(N)$ and note that the discussion for SYM$^*$ is identical to that for pure YM). To disentangle the energetics in the softly broken theory, 
consider now the following correlator in the SYM$^*$ theory:
\begin{eqnarray}\label{wilsoncorrelator1}
\langle (W_2^{\dagger}(x_1=T))^k (W_2(x_1=0))^k \rangle &\equiv&  \Tr (-1)^F e^{- {(L_1-T) } H} (W_2^{\dagger}(0))^k e^{- T H}  ( W_2(0)^k)  \over   \Tr (-1)^F e^{-L_1 H} \nonumber  \\
\end{eqnarray}
in the limit $L_1 - T \rightarrow \infty$, $T \rightarrow \infty$. In this limit,  only the lowest energy states contribute to the correlator. We insert two complete set of states, schematically $1 \equiv \sum\limits_{E,e_2} |E, e_2\rangle \langle E, e_2|$ in the numerator of (\ref{wilsoncorrelator1}). To proceed, let us denote $|E_0,   0\rangle$ the
lowest energy state in the SYM$^*$ theory, assuming that it corresponds to vanishing $e_2 = 0$. The operator $W_2^k$ changes the flux by $k$ units. Let us also denote the lowest energy state in the $e_2=k$ sector by $|E_k, k\rangle$. Thus, in the large $L_1$, $T$ limits, we obtain
\begin{eqnarray}\label{wilsoncorrelator2}
 \langle (W_2^{\dagger}(x_1=T))^k (W_2(x_1=0))^k \rangle\bigg\vert_{L_1, T, L_1 - T \rightarrow \infty} &=& {e^{- E_0 (L_1 - T)} e^{- T E_k} |\langle E_0, 0|(W_2^\dagger)^k|E_k,k\rangle|^2 \over e^{- E_0 L_1}} \nonumber \\
 &=&e^{- T(E_k - E_0)}  |(W_2^k)_{0,k}|^2,
 \end{eqnarray}
 where $(W_2^k)_{0,k} = |\langle E_k, k|(W_2^\dagger)|E_0,0\rangle|$ is the matrix element of $W_2^k$ between the lowest energy states in the $e_2=k$ and $e_2 = 0$ sectors.
 This matrix element should not scale with $T$, hence fitting to the exponential fall off determines $E_k - E_0$. 
 
{\flushleft{\bf 2.}} A  final remark regarding older work is that a {\it sketch} of the electric flux energies (given by our eqn.~(\ref{vacuum split})) for pure $SU(6)$ YM theory on a small $\T^3$ as a function of $\theta \in [0, 2\pi)$ appears in Ch.~3, Fig.~3 of van Baal's Ph.D. thesis \cite{vanBaal:1984ra} (unpublished), where the figure is attributed to 't Hooft.

{\bf {\flushleft{Acknowledgments:}}} We thank Rajamani Narayanan for discussions about the summation over twisted sectors.  M.A. thanks the University of Toronto for the warm hospitality during the completion of part of this work.  M.A. is supported by STFC through grant ST/X000591/1.   E.P. is supported by a Discovery Grant from NSERC.

\appendix
\section{Discrete symmetries}
\label{Discrete symmetries}
This appendix discusses the charge-conjugation and parity symmetries, which are essential concepts in $\cal{CP}$-violating processes. 

\subsection{Discrete symmetries in the two-component notation}
The Euclidean action of the SYM$^*$ theory (\ref{symaction2}) is rewritten for use below by taking the derivative act symmetrically:
\begin{eqnarray}
\label{symaction2.1}
S_{\scriptsize\mbox{SYM}^*} &=& \frac{1}{g^2} \int\limits_{\T^4} \tr_\Box \left[ \frac{1}{2} F_{\mu\nu} F_{\mu\nu}  -  \bar\lambda_{\dot\alpha} \bar\sigma_\mu^{\dot\alpha \alpha} (\partial_\mu \lambda_\alpha + i [A_\mu, \lambda_\alpha])  \right. \nonumber \\
&&\left.  \qquad\qquad -  \lambda^{\alpha} \sigma_{\mu\; \alpha \dot\alpha} (\partial_\mu \bar\lambda^{\dot\alpha} + i [A_\mu, \bar\lambda^{\dot\alpha}]) +m\lambda^\alpha\lambda_\alpha+m^*\bar\lambda_{\dot\alpha}\bar\lambda^{\dot\alpha} \right] \,.
\end{eqnarray}
{\flushleft{\bf \underline{Charge conjugation}}} maps the fundamental to its conjugate antifundamental representation of $SU(N)$, thus it acts as:\footnote{Here, $T^{a*}$ is the complex conjugate of the $T^a$ generator. It  can equivalently be written as the transpose, recalling that we work with Hermitean generators.}
\begin{eqnarray}
\label{chargeconjugation}
{\cal{C}}:~~~ A_\mu &\rightarrow& - A_\mu^* \equiv - A_\mu^a T^{a*}, \nonumber \\
 \lambda_\alpha^a T^a & \rightarrow& \lambda_\alpha^a T^{a*}, \\
 \bar\lambda_{\dot\alpha}^a T^a & \rightarrow& \bar\lambda_{\dot\alpha}^a T^{a*}. \nonumber 
 \end{eqnarray}
That this is a symmetry of (\ref{symaction2.1}), for real or complex $m$, follows from $\tr T^a T^b = \tr T^{a*} T^{a*}$ and $[T^{a*}, T^{b*}] = - i f^{abc} T^{c*}$ (while $[T^{a}, T^{b}] =  i f^{abc} T^{c}$, where $f^{abc}$ are the real structure constants). We notice that the gaugino bilinear $\tr \lambda\lambda$ is invariant under ${\cal{C}}$.
Thus, ${\cal{C}}$  is unbroken by $\langle \tr \lambda\lambda\rangle \ne 0$ in SYM theory on $\R^4$  .

However, at finite volume,
the twisted boundary conditions  break the ${\cal{C}}$ symmetry, because the ${\cal{C}}$ transformed field fails to obey the twisted boundary conditions, except in an $SU(2)$ theory, where ${\cal{C}}$ is part of the  gauge group. The quickest way to see that the twists break ${\cal{C}}$ is on the lattice: a plaquette $p$, twisted by a nontrivial two-form $\Z_N$ gauge background $b_p = 0,...,N-1$ (implementing the twisted BC), contributes to the action a term proportional to $e^{i{ 2\pi \over N} b_p} \tr U_p + e^{-i{ 2\pi \over N} b_p} (\tr U_p)^*$, where $\tr U_p$ is the trace of the fundamental Wilson loop around  $p$. As  ${\cal{C}}$ from (\ref{chargeconjugation}) interchanges $U_p$ and $U_p^*$, clearly, for $N \ne 2$ (and, for even $N$, $b_p \ne {N \over 2}$) the action fails to be ${\cal{C}}$-invariant. 

Thus, with nonzero twists, ${\cal{C}}$ is only a symmetry in the infinite volume limit.

{\flushleft{\bf \underline{Parity}}} reverses the direction of three  of the coordinates, which  can be chosen at will in our Euclidean setup. In the $\bar\sigma_\mu = (- i \vec{\sigma}, 1)$ basis we are using it is most straightforward to describe the action of parity   on $\vec{x} = (x_1, x_2, x_3)$. The choice of which three coordinates to reflect is   a matter of convenience and a corresponding transform can be written for  any choice, by changing the basis (in the end ${\cal{P}}$ has the same action on the gaugino bilinear as given in (\ref{parity3}), while the action on the  components of the currents $\bar\lambda^a \bar\sigma_\mu \lambda^b$ is appropriately modified from that given in (\ref{parity3})). 

The action on the gauge field is:
\begin{eqnarray}
\label{parity1}
{\cal{P}}: ~~~ A_4^a(\vec{x}, x_4) &\rightarrow& ~~A_4^a(- \vec{x}, x_4), \nonumber \\
A^a_i(\vec{x}, x_4) &\rightarrow & - A^a_i(-\vec{x}, x_4), ~i=1,2,3.
\end{eqnarray}
On the spinors, the action of parity is as follows:\footnote{\label{footnote:martin}To avoid any confusion, the transformation should be read (omitting the spacetime argument and the group index) as: $\lambda_1 \rightarrow i \bar\lambda^{\dot{1}} = i \bar\lambda_{\dot{2}}$, $\lambda_2 \rightarrow i \bar\lambda^{\dot{2}} = - i \bar\lambda_{\dot{1}}$, etc.}
\begin{eqnarray}
\label{parity2}
{\cal{P}}: ~ \lambda^a_\alpha (\vec{x}, x_4) &\rightarrow& ~~i \bar\lambda^{a \; \dot\alpha}(- \vec{x}, x_4), \nonumber \\
\bar\lambda^a_{\dot\alpha} (\vec{x}, x_4) &\rightarrow & - i \lambda^{a \; \alpha}(-\vec{x}, x_4). 
\end{eqnarray} 

While our discussion of ${\cal{P}}$ can be considered self-contained, we note that  the parity transform can be inferred by using the expressions for  the states and spinor wave functions of a  two-component Weyl spinor with Majorana mass from \cite{Dreiner:2008tw}. One demands that parity map states $|\vec{p}, s\rangle$,  with spin ($s=\pm 1/2$) and momentum $\vec{p}$,  to states $|\text{--}\vec{p}, s\rangle$ and analytically continues the resulting Minkowski field transformations to Euclidean signature, giving (\ref{parity2}).\footnote{One can pursue this route to also find the time reversal transforms, demanding now that both momentum and angular momentum of single-particle states are reversed; we do not need to pursue this here.}

From the above transformation rules, it is easy to see that the   gaugino bilinears entering  (\ref{symaction2.1}) transform as:
\begin{eqnarray}
\label{parity3}
{\cal{P}}: ~~~~\bar\lambda^a_{\dot\alpha} \bar\sigma_4^{\dot\alpha \alpha} \lambda^b_\alpha (\vec{x}, x_4)&\rightarrow&  - \bar\lambda^b_{\dot\alpha} \bar\sigma_4^{\dot\alpha \alpha} \lambda^a_\alpha (-\vec{x}, x_4), \\
\bar\lambda^a_{\dot\alpha} \bar\sigma_i^{\dot\alpha \alpha} \lambda^b_\alpha (\vec{x}, x_4)&\rightarrow& ~~\bar\lambda^b_{\dot\alpha} \bar\sigma_i^{\dot\alpha \alpha} \lambda^a_\alpha (-\vec{x}, x_4),~ i = 1,2,3, \nonumber \\
\tr \lambda^\alpha \lambda_\alpha (\vec{x}, x_4) & \rightarrow & ~~\tr \bar\lambda_{\dot\alpha} \bar\lambda^{\dot\alpha} (-\vec{x}, x_4). \nonumber
\end{eqnarray}
Bilinears where $\bar\sigma_\mu$ is replaced by $\sigma_\mu$ transform analogously, due to $\bar\lambda \bar\sigma_\mu \lambda = - \lambda \sigma_\mu \bar\lambda$.

The form of the  gaugino bilinear   transformations  (\ref{parity3}), along with those of the gauge field  (\ref{parity1}),   imply that the action density of eqn.~(\ref{symaction2.1}) at $\vec{x}$ is the same as the action density at $-\vec{x}$, provided $m=m^*$, hence ${\cal{P}}$ is a symmetry  for real $m$. In particular, in SYM theory on $\R^4$, the gaugino bilinear expectation value in the $k$-th vacuum, $\langle \lambda \lambda \rangle \sim e^{i {2 \pi \over N} k}$, breaks parity for $k\ne 0$ ($P$ is also unbroken in vacua with $k={N\over 2}$ for even $N$).

 In finite volume, in contrast to ${\cal{C}}$, not all twists on $\T^4$ break ${\cal{P}}$. With the choice of $x_i$ ($i=1,2,3$) as parity-reflected coordinates, twists $n_{ij}$ in the spatial planes are consistent with the parity action: the infinite-volume ${\cal{P}}$-transforms of (\ref{parity1}, \ref{parity2}), require a  modification at finite volume to ensure that the fields and their parity transforms obey identical spatial twisted boundary conditions. While the technical details are in  \cite{Cox:2021vsa},  heuristically  this can be argued for by noting that spatial twists are ``discrete magnetic flux'' backgrounds and that magnetic fields preserve parity. Mixed space-time twists $n_{i4}$, however, violate ${\cal{P}}$. This is because in a Hamiltonian picture $n_{i4}$ corresponds to inserting $\hat T_i^{n_{i4}}$ in the partition function. Here, $\hat T_i$ are  the center symmetry generators in the $i$-th direction, which do not commute with ${\cal{P}}$:  under parity, $\hat T_i \rightarrow \hat T_i^{-1}$ (at $\theta =0$), or, equivalently, parity reverses the ``discrete electric flux.''
 
\subsection{Discrete symmetries in the four-component Majorana-spinor notation}

We shall also use the $4$-component  Majorana spinors $\Psi$ and $\bar\Psi$. The Majorana spinors are handy when computing  ${\cal{C}}{\cal{P}}$-odd correlators and are also more convenient to compare with lattice results, since they are used in lattice simulations of SYM, as in the recent \cite{Butti:2022sgy,Bonanno:2024bqg,Bonanno:2024onr}.

The Majorana spinors are defined in terms of the Weyl fermions $\lambda$ and $\bar \lambda$, suppressing momentarily the color index (on which the transposition below does not act), as:\footnote{This is, essentially, the notation of \cite{Butti:2022sgy}, see Sec.~2 there.}
\begin{eqnarray}\label{Majorana def1}
\Psi=\left[\begin{array}{cc}\lambda_{\alpha}\\\bar\lambda^{\dot\alpha} \end{array}\right]\,,\quad \bar\Psi=\left[\begin{array}{cc} \lambda^{\alpha} & \bar\lambda_{\dot\alpha}\end{array}\right] = \left[\begin{array}{cc} \lambda_{\beta} & \bar\lambda^{\dot\beta}\end{array}\right] \cdot \left[\begin{array}{cc}- \epsilon^{\beta \alpha} & 0\cr 0 & -\epsilon_{\dot\beta \dot\alpha} \end{array}\right] \equiv \Psi^t \;C, ~~C = \left[\begin{array}{cc}- \epsilon^{\beta \alpha} & 0\cr 0 & -\epsilon_{\dot\beta \dot\alpha} \end{array}\right]. \nonumber \\
\end{eqnarray}
The use of the $4$-component spinor also necessitates the use of the Euclidean $\gamma$ matrices which are given by:
\begin{eqnarray}\label{gamma def1}
\gamma_\mu=\left[\begin{array}{cc} 0& \sigma_\mu\\\bar\sigma_\mu &0 \end{array}\right]\,, \quad 
\gamma_5=\gamma_1\gamma_2\gamma_3\gamma_4=\left[\begin{array}{cc} -I_{2\times 2}& 0\\0&I_{2\times 2} \end{array}\right]\,.
\end{eqnarray}
The fermionic  terms in the action (\ref{symaction2.1}) take  the form (with $\Psi = \Psi^a T^a$):
\begin{eqnarray}\label{fermion majorana action}
{\cal L}_f=-{1\over g^2} \tr \bar\Psi \gamma^\mu D_\mu \Psi + {m \over g^2} \; \tr \bar\Psi\; {1 - \gamma_5 \over 2} \;\Psi + {m^* \over g^2}\;  \tr \bar\Psi\; {1+ \gamma_5 \over 2}\; \Psi 
\end{eqnarray}
where we used 
\begin{eqnarray}\label{fermion majorana mass}
\tr \bar\Psi \Psi &=& \tr \lambda\lambda + \tr \bar\lambda \bar\lambda \nonumber \\
\tr\bar\Psi \gamma_5 \Psi &=& - \tr \lambda\lambda + \tr \bar\lambda \bar\lambda 
\end{eqnarray}
For completeness, we now list the ${\cal{C}}$ and ${\cal{P}}$ transformations of the Majorana spinors (\ref{Majorana def1}), which follow from the two-component transformations listed above. We have that:
\begin{eqnarray}
\label{chargeconjugation1}
{\cal{C}}:~~~ A_\mu &\rightarrow& - A_\mu^* \equiv - A_\mu^a T^{a*}, \nonumber \\
 \Psi^a T^a & \rightarrow& \Psi^a T^{a*}, \\
 \bar\Psi^a T^a & \rightarrow& \bar\Psi^a T^{a*},\nonumber 
 \end{eqnarray}
while under parity (we give only the spinor transform; the gauge field transform is in (\ref{parity1})), 
\begin{eqnarray}
\label{parity4}
{\cal{P}}:~~~ \Psi(\vec{x}, x_4)& \rightarrow&P \; \Psi (- \vec{x}, x_4) \\
 \bar\Psi(\vec{x}, x_4) & \rightarrow& \bar\Psi(-\vec{x}, x_4) \; P^\dagger \nonumber ,
 \end{eqnarray}
 where the definition of the matrix $P$ and its properties are:
\begin{eqnarray} \label{parity5}
P  &\equiv& \left[\begin{array}{cc} 0 & i I_{2 \times 2} \\   i I_{2 \times 2} & 0\end{array}\right],\; P^\dagger P=1,~ P^2 = - 1, \nonumber \\  P^\dagger \gamma^i P &=& - \gamma^i \; \text{for} \; i=1,2,3,~   P^\dagger \gamma^4 P =  \gamma^4.
 \end{eqnarray}
 
 {\flushleft{\underline{\bf ${\cal{C}}$, ${\cal{P}}$,  and the condensates: }}}
Using the above relations under ${\cal P}$ and ${\cal C}$, we may now check the transformation laws of the condensate $\bar \Psi\Psi$ as well as the pseudo-scalar condensate $\bar\Psi\gamma_5 \Psi$ under the combined ${\cal C} {\cal P}$ operation:\footnote{While we refer to the $\cal{CP}$ transformation of (\ref{scalarcondensate}, \ref{pseudoscalarcondensate}), we recall that both the scalar and pseudoscalar condensates are, in fact, ${\cal{C}}$-invariant.}
\begin{eqnarray}\label{scalarcondensate}
\bar \Psi\Psi=\lambda^\alpha\lambda_\alpha+\bar\lambda_{\dot\alpha}\bar\lambda^{\dot\alpha}  \xrightarrow[]{{\cal C}{\cal P}} \bar\lambda_{\dot\alpha}\bar\lambda^{\dot\alpha}+\lambda^\alpha\lambda_\alpha=\bar \Psi\Psi\,,
\end{eqnarray}
and
\begin{eqnarray}\label{pseudoscalarcondensate}
\bar \Psi\gamma_5\Psi=-\lambda^\alpha\lambda_\alpha+\bar\lambda_{\dot\alpha}\bar\lambda^{\dot\alpha}  \xrightarrow[]{{\cal C}{\cal P}} -\bar\lambda_{\dot\alpha}\bar\lambda^{\dot\alpha}+\lambda^\alpha\lambda_\alpha=-\bar \Psi\gamma_5\Psi\,.
\end{eqnarray}
Thus, as expected, $\bar \Psi\Psi$ behaves as a scalar, while $\bar \Psi\gamma_5\Psi$ is a pseudo-scalar. A theory with $\langle\bar \Psi\gamma_5\Psi\rangle\neq 0$ breaks $\cal{CP}$.

\section{Small-$m$ expansion of $Z^T$ and $\langle \lambda\lambda \rangle$ from the Hamiltonian} 

\label{sec:smallmhamiltonian}
The Hamiltonian formalism is not well-suited to performing actual calculations, as renormalization and regularization are most easily done in a path-integral framework. However, the Hamiltonian formulation aids in the interpretation of various results, as we now discuss. We want to compute and interpret the leading---at small $m LN$---contribution to $Z^T$, as well as to the expectation values (\ref{vevhamiltonian}). 
Taking into account the properties of ${\cal{H}}^{SYM}_{n_{34}=1}$ described above, and using $\hat H_m = \int_{\T^3} d^3 x( - m (\hat \lambda)^2 - m^* (\hat \lambda^\dagger)^2)$, to leading  and subleading order  in $|m|$ we find:\footnote{For $m=0$ and without any operator insertions, the partition function $Z^T = 1$, equaling $1/N$-the Witten index---because of supersymmetry, the contribution of $E>0$ bosonic and fermionic states cancels, while the projection only takes the $e_2=0$ ground state contribution. }
 \begin{eqnarray} \label{soft mass 2}\nonumber
 Z^T &=& 1 + {L_1 m  } \int_{\T^3} d^3 x\; {1 \over N} \sum\limits_{k, E, e_2} (-1)^F \langle E, e_2|  (\hat\lambda(x))^2 e^{-L_1\hat H_{SYM}} \hat T_2^{k} | E, e_2 \rangle   \nonumber \\
 && + {L_1 m^* } \int_{\T^3}  d^3 x \;{1 \over N} \sum\limits_{k, E, e_2} (-1)^F \langle E, e_2|  (\hat\lambda(x)^\dagger)^2 e^{- L_1 \hat H_{SYM}} \hat T_2^{k} | E, e_2 \rangle \nonumber
 \end{eqnarray}
 \begin{eqnarray}\nonumber
 && + {L_1^2 \over 2  }    m^2 \int_{\T^3}  d^3 x d^3 y \;  {1 \over N}\sum\limits_{k, E, e_2} (-1)^F  \langle E, e_2|   (\hat\lambda(x))^2(\hat\lambda(y))^2 e^{- L_1 \hat H_{SYM}} \hat T_2^{k} | E, e_2 \rangle    \label{soft mass 11} \\\nonumber
  &&+ {L_1^2 \over 2  }    (m^*)^2 \int_{\T^3}  d^3 x d^3 y  \; {1 \over N}  \sum\limits_{k, E, e_2} (-1)^F\langle E, e_2|   (\hat\lambda(x)^\dagger)^2(\hat\lambda(y)^\dagger)^2 e^{- L_1 \hat H_{SYM}} \hat T_2^{k} | E, e_2 \rangle\\ \nonumber
   &&+ {L_1^2 \over 2  }    (m m^*) \int_{\T^3}  d^3 x d^3 y \; {1 \over N}  \sum\limits_{k, E, e_2} (-1)^F \langle E, e_2|   \left( (\hat\lambda(x))^2(\hat\lambda(y)^\dagger)^2 + (\hat\lambda(x)^\dagger)^2(\hat\lambda(y) )^2 \right)\\
   &&\quad\quad\quad\quad\quad\quad\quad\quad\quad\quad\quad\quad \times e^{- L_1 \hat H_{SYM}} \hat T_2^{k} | E, e_2 \rangle + {\cal{O}}(|m|^3)\,.
 \end{eqnarray}
Clearly, a similar small-$|m|$ expansion holds for the expectation values (\ref{vevhamiltonian}) and one only has to insert the appropriate $\hat{\cal{O}}$ operator, see (\ref{soft mass 1}) below.
 
 We now note that the chiral-center anomaly and the chiral transformation of $\lambda$ of eqn.~(\ref{anomaly1}) imply relations between the matrix elements that appear above. These are valid for any set of $|E, e_2\rangle$ degenerate states, so for brevity we only keep the $|e_2\rangle$ label below. Thus, for any powers $2p$, $2q$ of insertions of $\lambda$ and $\bar\lambda$, we find
 \begin{eqnarray}
\langle e_2 |  \hat X^{-1} \;(\hat X \hat\lambda^{2 p} (\hat\lambda^\dagger)^{2 q} \hat X^{-1})\; \hat X  | e_2 \rangle &=& e^{ i {2 \pi \over N}(p-q)}  \langle e_2 -1|  \lambda^{2 p} (\lambda^\dagger)^{2 q}   | e_2 - 1\rangle~\implies \nonumber \\
~\langle e_2 |   \lambda^{2 p} (\lambda^\dagger)^{2 q}   \hat   | e_2 \rangle &=&  e^{ i {2 \pi \over N}(p-q)e_2}  \langle  e_2=0|  \hat\lambda^{2 p} (\hat\lambda^\dagger)^{2 q}   | e_2=0\rangle~.
 \end{eqnarray}
This further implies that, at every energy level, the sum over fluxes   entering (\ref{soft mass 2}), using (\ref{T2action}) becomes (here, the Kronecker delta in (\ref{selection rule}) equals unity if $k = q - p\; (\text{mod} N)$ and zero otherwise):
 \begin{eqnarray}\label{selection rule}
  \sum\limits_{e_2=0}^{N-1} \langle e_2 |   \hat\lambda^{2 p} (\hat\lambda^\dagger)^{2 q} \hat T_2^{k}  \hat   | e_2 \rangle &=&   \langle e_2 = 0|  \hat\lambda^{2 p} (\hat\lambda^\dagger)^{2 q}   |   e_2 = 0\rangle\; \sum_{e_2=0}^{N-1}  e^{ i {2 \pi \over N}(p-q + k)e_2} \nonumber \\
 &=&  N \delta_{k, q-p (\text{mod} N)}\langle  e_2 = 0|  \hat\lambda^{2 p} (\hat\lambda^\dagger)^{2 q}   | e_2 =  0\rangle ~.
 \end{eqnarray}
 Thus, the center-chiral anomaly implies   a selection rule on the nonvanishing expectation values of operators like $\lambda^{2 \ell}$ (and c.c.) calculated from 
 (\ref{soft mass 2}). 
 
Before we continue, let us make a comment. One might be tempted to also sum over $k$, as the sum ${1 \over N} \sum\limits_{k=0}^{N-1}$ appears   in every term in (\ref{soft mass 2}). Applying this extra sum to both sides of (\ref{selection rule})  leads to the result  
\begin{eqnarray}
\sum\limits_{k=0}^{N-1} N^{-1}  \sum\limits_{e_2=0}^{N-1} \langle e_2 |   \hat\lambda^{2 p} (\hat\lambda^\dagger)^{2 q} \hat T_2^{k}  \hat   | e_2 \rangle = \langle  e_2 = 0|  \hat\lambda^{2 p} (\hat\lambda^\dagger)^{2 q}   | e_2 =  0\rangle,
\end{eqnarray}thus ending up with a matrix element in the $e_2=0$ state (as already stressed, due to the projector) and with no selection rule on $p, q$. While this is correct, we stress that the path integral representation of a matrix element 
between fixed-$e_2$ states involves a sum over distinct (fractional) topological sectors---as the two are discrete Fourier transforms of each other.\footnote{See the discussion near eqn. (\ref{electricflux}).} Since every  topological sector has its own path integral representation, eqn.~(\ref{selection rule}) for fixed $k$ is, indeed,  useful  in the path integral framework and the semiclassical expansion.

 Now, how do we calculate  (\ref{soft mass 2}) and  corresponding expectation values? We begin by taking the most conservative point of view that $m LN \ll 1$ and $\Lambda L N \ll 1$---the small-$\T^4$ limit 
where semiclassics and the small-$m$ expansion are certainly under control. 
Thus, we replace the various Hilbert space traces appearing in the small-$m$ expansion of $Z_T$ (\ref{soft mass 2}), which all have  the form,
\begin{equation}\label{level-kcontribution}
\sum\limits_{ E, e_2} (-1)^F  \langle E, e_2|  \int_{\T^3} d^3 x (\hat\lambda)^{2 p} (\hat \lambda^\dagger)^{2 q} e^{-L_1\hat H} \hat T_2^{k} | E, e_2 \rangle,\end{equation} 
(or  c.c.), with Euclidean path integrals on the small $\T^4$. The term (\ref{level-kcontribution}) corresponds to a path integral with twists $n_{34}=1$ and $n_{21} =  k$. 
In the small-$L \Lambda N$ semiclassical limit, the leading contribution to each of the terms in the small-$m$ expansion will come from the  value of $k$ allowed by the selection rule (\ref{selection rule}). The  minimum amount of semiclassical suppression is then determined by noting that terms $\sim \hat T_2^k$ are suppressed at least by $e^{- {8 \pi^2 \over g^2 N}p}$, where $p = {\rm{min}}_{n \in \Z} |n - {k \over N}|$.

Thus, even without doing any calculation we can immediately argue that the leading semiclassical contributions to the order $m$ and $m^*$ sums  in (\ref{soft mass 2}) come from $k=N-1$ and $k=1$, respectively (which come with the minimal semiclassical suppression factor, $p=1$). On the other hand, the selection rule (\ref{selection rule}) implies that the order $m^2$ and $(m^*)^2$ terms require $p=2$ so we neglect those terms. The order $m m^*$ term, however, allows $k=0$, which includes an unsuppressed perturbative contribution.  Thus, we  obtain:
 \begin{eqnarray}
Z^T &\simeq&  (1 + |m|^2 L^2 c_0) + c \; ( L m e^{- {8 \pi^2 \over N g^2}} e^{i {\theta \over N}} + L m^* {e^{-{8 \pi^2 \over N g^2}}} e^{ - i {\theta \over N}}) + {\cal{O}}(|m|^3, e^{- {16 \pi^2 \over N g^2}}) \\
&=&  (1 + |m|^2 L^2 c_0) + c' L^4 |m| \Lambda^3 \; \cos {\theta + N \text{arg}\, m \over N}+ {\cal{O}}(|m|^3, e^{- {16 \pi^2 \over N g^2}}),  \nonumber~ \label{ZTsmallmL}
  \end{eqnarray}
  where on the last line we absorbed all undetermined factors in the coefficients $c_0,  c'$. This result, including a calculation of $c_0,c'$ will be obtained using path integrals, by a detailed small-$L_\mu$ semiclassical calculation in the rest of the paper, see (\ref{energy1}).\footnote{Note that $Z_T$ is invariant under $U(1)_{spurious}$ of eqn.~(\ref{spuriousU1}).} 

Note that the order $m$ terms in the partition function $Z_T$ of (\ref{ZTsmallmL})  above do not allow an interpretation as a contribution to the ground state energy due to supersymmetry breaking. This is because in the small torus all energy states contribute to the Hilbert space traces. Later on, see section \ref{sec:rtimest3}, we consider the limit  of large Euclidean time, which is, in principle, also semiclassical (yet is not analytically calculable) where such an extraction is possible.

We can also estimate the scaling of the expectation values of $\langle \hat\lambda^2 \rangle$ and $\langle (\hat\lambda^\dagger)^2 \rangle$, computed in the same $m L N \ll 1$ limit. We simply insert the corresponding operator in each of the terms in the small-$m$ expansion for $Z_T$ in (\ref{soft mass 1}). Consider 
 $\langle \hat\lambda^2 \rangle$ for definiteness, for which we find, for the numerator of (\ref{vevhamiltonian}) with ${\cal{O}}$ replaced by $\hat \lambda^2(x_1=0)$, applying the selection rule (\ref{selection rule}) and indicating the value of $k$ selected by an arrow  in every term:
\begin{eqnarray}
&&N Z_T \langle \hat\lambda^2(0) \rangle =  \tr_{{\cal{H}}^{SYM}_{n_{34}=1}}\left( (-1)^F e^{- L_1 \hat H_{SYM}} \; \hat\lambda^2(0)  \; \hat T_2^{k}\right) \qquad \qquad {\leftarrow k=N-1 \text{} } \nonumber \\ &&+ L_1 m \int_{\T^3_x}  \sum\limits_{k, E, e_2} (-1)^F \langle E, e_2| (\hat\lambda)^2(0)  (\hat\lambda(x))^2 e^{-L_1\hat H_{SYM}} \hat T_2^{k} | E, e_2 \rangle  ~~~~~  {\leftarrow k=N-2, \text{ drop}} \nonumber 
\end{eqnarray}
\begin{eqnarray}
 &&+ L_1 m^*  \int_{\T^3_x}  \; \sum\limits_{k, E, e_2} (-1)^F \langle E, e_2|  (\hat\lambda)^2(0)(\hat\lambda(x)^\dagger)^2 e^{- L_1 \hat H_{SYM}} \hat T_2^{k} | E, e_2 \rangle  ~~ {\leftarrow k=0 \text{  }} \nonumber \\
 &&+ {L_1^2 \over 2}    m^2 \int_{\T^3_{x} \times \T^3_y}   \;  \sum\limits_{k, E, e_2} (-1)^F  \langle E, e_2|   (\hat\lambda)^2(0)(\hat\lambda(x))^2(\hat\lambda(y))^2 e^{- L_1 \hat H_{SYM}} \hat T_2^{k} | E, e_2 \rangle ~~{\leftarrow k=N-3, \text{ drop}} \nonumber \\
&&+ {L_1^2 \over 2}    (m^*)^2 \int_{\T^3_{x} \times \T^3_y}    \;   \sum\limits_{k, E, e_2} (-1)^F\langle E, e_2|   (\hat\lambda)^2(0)(\hat\lambda(x)^\dagger)^2(\hat\lambda(y)^\dagger)^2 e^{- L_1 \hat H_{SYM}} \hat T_2^{k} | E, e_2 \rangle~~{ \leftarrow k=1 \text{  }}  \nonumber\end{eqnarray}
\begin{eqnarray}
 &&+ {L_1^2 }    (m m^*) \int_{\T^3_{x} \times \T^3_y}   \;   \sum\limits_{k, E, e_2} (-1)^F \langle E, e_2|   (\hat\lambda)^2(0) (\hat\lambda(x))^2(\hat\lambda(y)^\dagger)^2 e^{- L_1 \hat H_{SYM}} \hat T_2^{k} | E, e_2 \rangle ~~{\leftarrow k=N-1 \text{}} \nonumber\\
 &&+ {\cal{O}}(|m|^3)~.    \label{soft mass 1}  
 \end{eqnarray}
 
 We now use the fact that
 \begin{equation}
 \tr_{{\cal{H}}^{SYM}_{n_{34}=1}}\left( (-1)^F e^{- L_1 \hat H_{SYM}} \; \hat\lambda^2(0)  \; \hat T_2^{N-1 }\right) \equiv  N  \langle \hat \lambda^2 \rangle_{\R^4, \; SYM} = N 16 \pi^2 \Lambda^3 
\end{equation}
 and proceed by keeping the lowest topological charge contribution in each case, recalling that $\sim \hat T_2^k$ contributions 
 are suppressed by $e^{- {8 \pi^2 \over g^2 N}p}$, where $p = {\rm{min}}_{n \in \Z} |n - {k \over N}|$.
 To write the result, we introduce 
  new undetermined constants $c_1,c_2, c_3$: 
\begin{eqnarray} \label{bilinearhamiltonian}
Z_T \; \langle  \hat\lambda^2 \rangle &=& 16 \pi^2 \Lambda^3(1 + c_1 |m|^2 L^2) +c_2  \;  {m^* \over L^2}+ c_3\;  (m^* L)^2  ~ \Lambda^3 e^{- i {\theta \over N}},\nonumber \\
Z_T \; \langle  (\hat\lambda^\dagger)^2  \rangle &=&  16 \pi^2 \Lambda^3(1 + c_1 |m|^2 L^2)  +c_2 \;    {m  \over L^2}+ c_3 \;  (m  L)^2  ~ \Lambda^3 e^{  i {\theta \over N}}~,
\end{eqnarray}
  where in each case we kept the leading term in the semiclassical and small-$m$ expansion only. The coefficient $c_2$ is perturbative, proportional to $g^2$, while $c_1, c_3$  require a one-(fractional)-instanton calculation.

 The expressions (\ref{bilinearhamiltonian})  for $Z_T \langle  \hat\lambda^2 \rangle$, $Z_T \langle  (\hat\lambda^\dagger)^2\rangle $ obtained from the Hamiltonian formalism match the corresponding leading terms in (\ref{small mV C 1}) and (\ref{bar small mV C 1}) obtained by a  path-integral semiclassical calculation, where  a discussion of how $c_1,c_2, c_3$ are determined can be found. Here, they were obtained using the small-$m$ expansion,  the selection rule following from the chiral-center anomaly in SYM, and the  $ \Lambda NL \ll 1$ semiclassical power-counting.

\section{The propagator in a general $Q = {k \over N}$ self-dual background}
\label{Systematics of the propagator}

The goal of this Appendix is to compute the propagators in the fractional instanton background, needed in the computation of the correlation function (\ref{the full things}). Typically, computing the propagator in the instanton background is a complex task. However, 
 as we discuss below, this task is made significantly simpler within the framework of the leading-order $\Delta$-expansion. Before providing the details, let us discuss the general idea behind calculating the propagators.
 
 Our starting point is the fermion Lagrangian (\ref{fermion majorana action}) written using the Majorana spinor. The corresponding equation of motion is 
 \begin{eqnarray}
 \left(-\slashed D+\left[\begin{array}{cc} m I_2&0\\0&m^* I_2 \end{array}\right]\right)\Psi=0\,,
 \end{eqnarray}
 and $\slashed D=\gamma_\mu D_\mu$. 
 In writing this equation, we stressed the $4\times 4$ dimensional spinor space, while we have suppressed the $N\times N$ indices of the internal space. This makes the treatment compact, while we shall discuss the internal spaces at the due time. 
The corresponding eigenvalue problem---one needs to solve to find the Green's function---is
 \begin{eqnarray}\label{EVPSI}
 -i\slashed D\Sigma_n=\omega_n \Sigma_n\,,
 \end{eqnarray}
 where $\slashed D=D_\mu\gamma_\mu$ and $\Sigma_n$ is a $4$-component spinor. We may also operate on the l.h.s of (\ref{EVPSI}) with $\slashed D$ to rewrite the eigenvalue problem in the form
 \begin{eqnarray}\label{slashed slahed d}
 \slashed D \slashed D\Sigma_n \equiv \left[\begin{array}{cc} D \bar{D} &0 \cr 0 & \bar{D} D\end{array} \right] \Sigma_n = -\omega_n^2\Sigma_n\,, \: \text{where} \; D \equiv \sigma_\mu D_\mu, \bar D \equiv \bar\sigma_\mu D_\mu~.
 \end{eqnarray}
   Notice here that $ -i\slashed D$ is hermitian, and thus, the eigenvalues $\omega_n$ are real. Further, for every mode with $\omega_n  > 0$, there exists a mode with eigenvalue $- \omega_n$, due to anticommutativity of $\slashed D$ with $\gamma_5$.\footnote{Equivalently, in lattice gauge theory terminology, the spectrum of the antihermitean but $\gamma_5$-hermitean ($\slashed D^\dagger = \gamma_5 \slashed D \gamma_5$) Dirac operator is on the imaginary axis  and is symmetric upon reflection across the real axis. }  The set of eigenmodes $\Sigma_n$---with the appropriate BCS---constitutes a complete set of states one uses to express the Green's function. 
   We further note the well-known relations
   \begin{eqnarray}\label{relations 1}
   D \bar D &=& I_{2} D_\mu D_\mu + i F_{\mu\nu} \sigma^{\mu\nu},\nonumber \\
  \bar D D &=& I_{2} D_\mu D_\mu ~,
   \end{eqnarray}
aka the ``Weitzenb\" ock formulae,''  written  here for a self-dual background $(\bar\sigma^{\mu\nu} F_{\mu\nu} = 0$).

To construct the Green's function, we now continue by considering fermions defined on general self-dual non-trivial background of topological charge $Q=k/N$. 
 We assume that the background is generic (as in our detuned $\T^4$ with small $\Delta \ne 0$), such that the  Dirac operator has $2k$ zero modes in the undotted sector, ${1-\gamma_5 \over 2} \Psi = \Psi$, i.e. $\lambda_\alpha$ zero modes, and there are no zero modes in the $\bar\lambda_{\dot\alpha}$ sector---thus, as per the second line of (\ref{relations 1}), the spectrum of the adjoint Laplace operator is negative definite. For use below, we denote the  $2k$ undotted zero modes, obeying (\ref{EVPSI}) with $\omega_n=0$, as\footnote{We stress that $p,\beta$ are indices denoting the zero mode, while $\alpha$ is the spinor index.}
 \begin{eqnarray}\label{kzeromodes1}
 \Sigma_{p,\beta}^{(0)} = \left[\begin{array}{c} \psi_{\alpha\,  p,\beta}^{(0)} \\ 0 \end{array}\right], ~p=1,...,k, ~\beta=1,2,
 \end{eqnarray}
  using a composite index $(p,\beta)$ to label the $2k$ zero modes (this is motivated by their explicit expression on the detuned $\T^4$).
  
We let $\Sigma^{(+)}_n$ and $\Sigma^{(-)}_n$ be the positive and negative eigenmodes, with positive and negative eigenfrequencies, $ \omega_{n} $ and $- \omega_n $, respectively.\footnote{\label{omegasignfootnote}Starting from eqn.~(\ref{post and neg}), we explicitly separate the positive and negative eigenvectors of $-i \slashed D$. Hence, from now on $\omega_n >0$.} Then, we have that $\Sigma_n^{(-)}$ is expressed in terms of $\Sigma_n^{(+)}$:
\begin{eqnarray}\label{post and neg}
\Sigma_n^{(+)}=\left[\begin{array}{c} \psi_{\alpha\, n} \\ \bar\psi^{\dot\alpha}_n \end{array}\right]\,,\quad \Sigma_n^{(-)}=\gamma_5 \Sigma_n^{(+)}=\left[\begin{array}{c} -\psi_{\alpha\, n} \\ \bar\psi^{\dot\alpha}_n \end{array}\right]\,,
\end{eqnarray}
The original eigenvalue equation $-i\slashed D\Sigma_n^{(+)}=\omega_n \Sigma_n^{(+)}$ implies that we can write the nonzero modes undotted wave functions $\psi_{\alpha n}$ in terms of the dotted ones $\bar\psi_{n}^{\dot\alpha}$:
\begin{eqnarray}
\psi_{\alpha n}=-\frac{i}{\omega_n}\sigma_{\mu \; \alpha \dot\alpha} D_\mu\bar\psi_n^{\dot\alpha} \equiv -\frac{i}{\omega_n} (D \psi_n)_\alpha .
\end{eqnarray}

Further, as per eqns.~(\ref{slashed slahed d}, \ref{relations 1}), $\bar\psi_n^{\dot\alpha}$ is an eigenvector of the adjoint covariant Laplace operator with eigenvalue $- \omega_n^2$. Thus, we 
can write $\bar\psi_{n}^{\dot\alpha}$ as the product of a two-component spinor $\bar\zeta^{(s)}_{\dot\alpha}$ and an ordinary $\mathbb C$-function $\phi_n (x)$ in the adjoint representation 
\begin{eqnarray}\label{psidotphi}
\bar\psi_{n, s}^{\dot\alpha}(x)= \bar\zeta_{}^{\dot\alpha(s)}\phi_n(x)\,,
\end{eqnarray}
where we explicitly add the index $s=1,2$ to the index $n$, to account for the fact that there are two linearly independent choices of spinors $\bar\zeta^{\dot\alpha(s)}$.
Here, 
 $\phi_n$ is the eigenvector of the adjoint representation Laplace operator:\begin{equation}
D_\mu D^\mu \phi_n = - \omega_n^2 \phi_n  , \text{where},  \forall n, ~ \omega_n^2 >0. \label{adjoint1}
\end{equation}

As already stated, the adjoint representation Laplace operator is assumed to not have zero modes on the detuned $\T^4$. Thus,  the functions $\phi_n$ with nonzero $\omega_n^2$ are a complete set of functions  in the space of adjoint representation functions $\phi(x)$ obeying the boundary conditions $\phi(x + L_\mu) = \Omega_\mu(x) \phi(x) \Omega_\mu^\dagger(x)$. Without loss of generality, we shall impose a hermiticity condition $\phi^\dagger(x) = \phi(x)$, i.e. consider real adjoint fields in (\ref{adjoint1}).\footnote{In fact,  we are forced to consider Hermitean $\phi_n$ obeying (\ref{completeness1}) and (\ref{normphin1}) when we consider the path integral formulation in terms of eigenmodes for {\it general} (i.e.~not explicitly specified) self-dual backgrounds. This is because our action involves the fields $\bar\Psi$ and $\Psi$, which are not hermitean conjugates of each other, but are related as in (\ref{Majorana def1}). See also Sec. B.2 in \cite{Anber:2022qsz} for a similar treatment.}
 Thus, they obey the completeness relation  \begin{eqnarray}
\sum\limits_n  \varphi_n(x) \otimes \varphi_n(y) = \delta_{x,y}~\label{completeness1}~.
\end{eqnarray}
Here, we use $\delta_{x,y}$ to denote the delta function in the space of adjoint scalar functions with the boundary conditions on the $\T^4$ specified by $\Omega_\mu$; we stress that $\delta_{x,y}$  carries also two sets of adjoint indices  that we do not write explicitly.\footnote{It thus obeys, in view of (\ref{completeness1}, \ref{normphin1}), the delta-function relation $\int_{\T^4} d y \delta_{x,y} \phi(y) = \phi(x)$.}
In addition, we  normalize $\phi_n(x)$ as
\begin{eqnarray}\label{normphin1}
\int_{\mathbb T^4}\tr \phi_n (x)\phi_m(x)= \delta_{mn}\,.
\end{eqnarray}
The spinors $\bar\zeta_{}^{\dot\alpha(s)}$, for $s=1,2$, are two constant independent spinors which we take explicitly as:
\begin{eqnarray}\label{constant spinors}
 \bar\zeta^{\dot\alpha (s)} \equiv \delta^{\dot\alpha s}, ~\text{for}~ s=1,2, \dot\alpha = 1,2.
 \end{eqnarray}
We now go back to our four component spinors $\Sigma_{n,s}^{(\pm)}$ of eqn.~(\ref{post and neg}), but with the extra index $s$ added, as explained after eqn.~(\ref{psidotphi}) (recall that these spinors correspond to nonzero $- i \slashed D$ eigenvalues $\pm  \omega_n $, respectively), and express them in terms of the Laplacian eigenfunctions  $\phi_n$, also recalling footnote \ref{omegasignfootnote}:
\begin{eqnarray}\label{post and neg3}
\Sigma_{n,s}^{(+)}=\left[\begin{array}{c} -\frac{i}{ \omega_n } \sigma_{\mu \;\alpha s}  D_\mu \phi_n \\ \delta^{\dot\alpha s} \phi_n \end{array}\right]\,,\quad \Sigma_{n,s}^{(-)}=\left[\begin{array}{c}  \frac{i}{ \omega_n } \sigma_{\mu \;\alpha s}  D_\mu \phi_n \\ \delta^{\dot\alpha s} \phi_n  \end{array}\right]\,,
\end{eqnarray}
Next, we the relation between the spinors $\Psi$ and $\bar\Psi$ of eqn.~(\ref{Majorana def1}), $\bar\Psi =  (\lambda_\beta, \; \bar\lambda^{\dot\beta})\left( \begin{array}{cc} - \epsilon^{\beta\alpha} & 0 \cr 0 & - \epsilon_{\dot\beta \dot \alpha} \end{array}\right)$,  to define the functions $\bar\Sigma_{n,s}^{(\pm)}$:
\begin{eqnarray}
\bar\Sigma_{n,s}^{(+)}=\left[   \frac{i}{\omega_n} \sigma_{\mu \;\beta s} \epsilon^{\beta \alpha}  D_\mu \phi_n, \;   \epsilon_{ \dot\alpha s} \phi_n  \right]\,,\quad \bar\Sigma_{n,s}^{(-)}=\left[ - \frac{i}{ \omega_n } \sigma_{\mu \;\beta s} \epsilon^{\beta\alpha}  D_\mu \phi_n, \;   \epsilon_{ \dot\alpha s}\phi_n  \right]\,, \nonumber \\
\label{sigmabar}
\end{eqnarray}
Likewise,  from (\ref{kzeromodes1}), we define the function $\bar\Sigma_{0,p}$: 
 \begin{eqnarray}\label{kzeromodes1bar}
\bar\Sigma^{(0)}_{p,\beta} = \left[ - \psi^{(0)}_{\gamma p,\beta} \epsilon^{\gamma\alpha},  0  \right] =  \left[ \psi^{(0) \; \alpha}_{p,\beta},  0  \right], ~p=1,...,k~.
 \end{eqnarray}

In order to define the path integral, we  expand $\Psi$ and $\bar\Psi$ in terms of the complete set of eigenfunctions of (\ref{EVPSI}) (eqns.~(\ref{post and neg3}, \ref{sigmabar}, \ref{kzeromodes1}, \ref{kzeromodes1bar})):\footnote{To avoid confusion, we stress that the $SU(N)$ group indices are carried by the Laplacian eigenvectors $\phi_n$, and therefore by the functions $\Sigma^{(0), (\pm)}$ defined in terms of $\phi_n$. Thus, the equation for, e.g.~$\Psi$, should really read:
$$
\Psi_{ij}  = \xi^0_{p,\beta} \Sigma^{(0)}_{p, \beta \; ij} + \xi_{n,s}^+ \Sigma_{n,s  \; ij}^{(+)} + \xi_{n,s}^- \Sigma_{n,s  \; ij}^{(-)}
$$
where $i,j = 1,...N$ are $SU(N)$ indices. The Grassmann variables carry the indices numbering the Laplacian eigenvectors $\phi_n$ and the functions $\Sigma_{n,s}$, i.e.~$n$ and $s=1,2$. (In the simplest case of a free Laplacian with periodic boundary conditions, as in section \ref{sec:cartansuktimesu1}, every element of the adjoint is an eigenvector of the Laplacian, hence the index $n$ would then, in addition to the momentum label, also include the eigenvector label $ij$. For a general background, however, the eigenvector indices $n$ are distinct from the $SU(N)$ indices.) In what follows, we do not explicitly write the group indices in (\ref{expansion1}). }
\begin{eqnarray}
\label{expansion1}
\Psi &=& \xi^0_{p,\beta} \Sigma^{(0)}_{p, \beta} + \xi_{n,s}^+ \Sigma_{n,s}^{(+)} + \xi_{n,s}^- \Sigma_{n,s}^{(-)},~\nonumber \\
\bar\Psi &=& \xi^0_{p, \beta} \bar\Sigma^{(0)}_{p,\beta} + \xi_{n,s}^+ \bar\Sigma_{n,s}^{(+)} + \xi_{n,s}^- \bar\Sigma_{n,s}^{(-)},
\end{eqnarray}
where $\xi_{p,\beta}$, $\xi_{n,s,\pm}$ are Grassmann variables and a sum over  $p=1,...,k$, $s=1,2$, and $n$ is assumed in each line.

Before we continue we note  various useful identities. First, we note that:
\begin{eqnarray}\label{identity1}
\int\limits_{\T^4} \tr \bar\Sigma^{(0)}_{p,\beta} \Sigma_{n,s}^{(\pm)} \sim \int\limits_{\T^4}  \tr  \psi_{p,\beta}^{(0) \; \alpha} \sigma_{\mu \; \alpha s} D_\mu \phi_n = 0, ~~ \text{likewise}~\int\limits_{\T^4} \tr  \bar\Sigma_{n,s}^{(\pm)} \Sigma^{(0)}_{p} = 0,
\end{eqnarray}
 because $\psi_{ p,\beta}^{(0) \; \alpha}$ and $\sigma_{\mu \; \alpha s} D_\mu \phi_n$ are eigenfunctions of $D \bar D$ with different eigenvalues (the vanishing also follows by direct integration by parts). 
 We normalize the zero-modes (\ref{kzeromodes1}) as follows:\footnote{The zero modes obeying (\ref{identity1.5}) are given, to order $\Delta^0$, in eqn.~(\ref{zeromodesnormalized}). }
 \begin{eqnarray}\label{identity1.5}
\int\limits_{\T^4}\tr  \bar\Sigma^{(0)}_{p,\beta} \Sigma^{(0)}_{q, \beta'}  =\int\limits_{\T^4} \tr  \psi^{(0) \; \alpha}_{ p,\beta}   \psi^{(0)}_{q,\beta', \alpha} = \delta_{pq} \epsilon_{\beta\beta'}~, 
\end{eqnarray}
recalling that $p,q=1,...,k$ and $\beta, \beta' = 1,2$.
 
 Next,  we consider the $\bar\Sigma_{n,s}^{(+)} \Sigma_{m,s'}^{(-)}$ inner product. Rearranging the $\sigma$-matrices, and integrating by parts, we find, from (\ref{post and neg3}, \ref{sigmabar}):
\begin{eqnarray}\label{identity2}
\int\limits_{\T^4}\tr  \bar\Sigma_{n,s}^{(+)} \Sigma_{m,s'}^{(-)} &=&   \epsilon_{s' \dot\gamma} {(\bar\sigma_{\nu} \sigma_\mu)^{\dot\gamma}_{ \; \;  s}  \over \omega_n \omega_m} \int\limits_{\T^4}\tr   \phi_n D_\mu D_\nu \phi_m  +\epsilon_{s's} \int\limits_{\T^4}\tr  \phi_n \phi_m~\stackrel{?}= 0~.\end{eqnarray}
To show its vanishing, we note that  in view of (\ref{normphin1}), the integral in the second term is $\delta_{nm}$. 
The first term is $(\bar\sigma_\nu \sigma_\mu)^{\dot\gamma}_{\;\;s} \int\limits_{\T^4}\tr   \phi_n D_\mu D_\nu \phi_m = \delta^{\dot\gamma}_{\;\; s}   \int\limits_{\T^4} \tr  \phi_n D_\mu D_\mu \phi_m + (\bar\sigma_{\nu\mu})^{\dot\gamma}_{\;\; s}  \int\limits_{\T^4} \tr  \phi_n [D_\mu, D_\nu] \phi_m = - \delta^{\dot\gamma}_s \omega_n^2 \delta_{nm}$, where we used the self-duality of the background, $\bar\sigma_{\nu\mu} F_{\nu \mu}=0$, the Laplace equation (\ref{adjoint1}), and orthonormality of the $\phi_n$ basis (\ref{normphin1}). Plugging this back in (\ref{identity2}) shows that, indeed $\int\limits_{\T^4} \tr \bar\Sigma_{n,s}^{(+)} \Sigma_{m,s'}^{(-)}=0$. 

The vanishing of the inner product $ \bar\Sigma_{n,s}^{(-)} \Sigma_{m,s'}^{(+)}$ follows by literally repeating the same steps, thus
\begin{eqnarray}\label{identity3}
\int\limits_{\T^4}\tr  \bar\Sigma_{n,s}^{(-)} \Sigma_{m,s'}^{(+)} = 0~.\end{eqnarray}  
 
 We are left to determine 
 \begin{eqnarray}\label{identity4}
\int\limits_{\T^4}\tr  \bar\Sigma_{n,s}^{(+)} \Sigma_{m,s'}^{(+)} &=&  - \epsilon_{s' \dot\gamma} {(\bar\sigma_{\nu} \sigma_\mu)^{\dot\gamma}_{ \; \;  s}  \over \omega_n \omega_m} \int\limits_{\T^4} \tr  \phi_n D_\mu D_\nu \phi_m  +\epsilon_{s's} \int\limits_{\T^4}\tr  \phi_n \phi_m .\end{eqnarray}
We note that (\ref{identity4}) only differs from (\ref{identity2}) by the sign of the first term. Thus, instead of cancelling, the two contributions add up and we obtain:
 \begin{eqnarray}\label{identity4.5}
\int\limits_{\T^4}\tr  \bar\Sigma_{n,s}^{(+)} \Sigma_{m,s'}^{(+)} &=&   2   \epsilon_{s's} \delta_{nm} .\end{eqnarray}
Likewise, we find:
 \begin{eqnarray}\label{identity5}
\int\limits_{\T^4} \tr \bar\Sigma_{n,s}^{(-)} \Sigma_{m,s'}^{(-)} &=&   2   \epsilon_{s's} \delta_{nm} .\end{eqnarray}
To calculate the action for $\Im m \ne 0$, we shall also need the various integrals with $\gamma_5$ inserted. As $\gamma_5$ simply changes the sign of the first term appearing in  (\ref{identity2}) and  (\ref{identity4}), it is easy to see that the results are as follows 
\begin{eqnarray}
\label{identitywithgamma5}
\int\limits_{\T^4}\tr  \bar\Sigma^{(0)}_{p,\beta} \gamma_5 \Sigma^{(0)}_{q, \beta'}  & =& -   \delta_{pq} \epsilon_{\beta \beta'}~, \nonumber \\
\int\limits_{\T^4}\tr  \bar\Sigma_{n,s}^{(-)} \gamma_5 \Sigma_{m,s'}^{(-)} &=& 0,~ \int\limits_{\T^4}\tr  \bar\Sigma_{n,s}^{(+)} \gamma_5 \Sigma_{m,s'}^{(+)} = 0,  \\
\int\limits_{\T^4}\tr  \bar\Sigma_{n,s}^{(+)} \gamma_5 \Sigma_{m,s'}^{(-)} &=& 
\int\limits_{\T^4}\tr  \bar\Sigma_{n,s}^{(-)} \gamma_5 \Sigma_{m,s'}^{(+)}  =   2 \delta_{n,m} \epsilon_{s's} \nonumber~,
\end{eqnarray}
which also follow from (\ref{identity2}, \ref{identity3}, \ref{identity5}) by recalling $\gamma_5 \Sigma^{(-)} = \Sigma^{(+)}$ and $\gamma_5 \Sigma^{(+)} = \Sigma^{(-)}$.

Armed with (\ref{identity1}, \ref{identity2}, \ref{identity3}, \ref{identity4.5}, \ref{identity5},\ref{identitywithgamma5}) we substitute (\ref{expansion1}), use (\ref{EVPSI}),
and obtain the fermion action (\ref{fermion majorana action}) as a bilinear in the Grassmann variables, with a sum over all repeated indices understood:
\begin{eqnarray} 
&&  g^2 \int\limits_{\T^4} {\cal{L}}_f = \int\limits_{\T^4}\left( - \tr \bar\Psi \slashed D \Psi + \Re m\; \tr \bar\Psi \Psi - i \Im m \; \tr \bar\Psi \gamma_5 \Psi \right)  \\
  &=& \xi^0_{p \beta} \xi^0_{p \gamma} \epsilon_{\beta\gamma} m +  \xi_{n,s}^+ \xi_{n, s'}^+  \epsilon_{ss'} 2\left[ i \omega_n - \Re m\right] - \xi_{n,s}^- \xi_{n, s'}^-  \epsilon_{ss'}  2 \left[i \omega_n + \Re m\right] 
 + (\xi_{n,s}^+ \xi_{n, s'}^- + \xi_{n,s}^- \xi_{n, s'}^+) \epsilon_{ss'}  \left[2 i  \Im m\right] \nonumber ~.
 \end{eqnarray}
 This can be rewritten in a manner convenient to explicitly do the Grassmann integrals with $e^{-S_f} = e^{- \int_{\T^4} {\cal{L}}_f}$:
 \begin{eqnarray}\label{actionfermionmodes}
&&- \int\limits_{\T^4} {\cal{L}}_f \\ &&
=    \xi^0_{p 1} \xi^0_{p 2} {2 m\over g^2} -   \xi_{n,1}^+ \xi_{n, 2}^+  {4 \over g^2} \left[- i \omega_n + \Re m\right] - \xi_{n,1}^- \xi_{n, 2}^-  {4\over g^2}  \left[i \omega_n + \Re m\right] 
 + (\xi_{n,1}^+ \xi_{n, 2}^- + \xi_{n,1}^- \xi_{n, 2}^+)  \left[{4 \over g^2} i  \Im m\right],  \nonumber
\end{eqnarray}
with summations over $p=1,...,k$ and  $n$, the eigenmodes of the Laplacian (\ref{adjoint1}) understood.

To keep track of any sign ambiguity, we define the path integral measure without yet explicitly specifying the order of integration over the Grassmann variables:\begin{eqnarray}\label{measure}
d \Psi \equiv \prod\limits_{p=1}^k d^2 \xi^0_{p} \; \prod_{n} d^4 \xi_{n},
\end{eqnarray}
where $d^2 \xi^0_p$ represents the measure of  integration over $\xi^0_{p,1}$ and $\xi^0_{p,2}$ and $d^4 \xi_{n}$---over $\xi_{n,1(2)}^\pm$.
This already allows us to calculate the unregulated fermion determinant, which we label ${\cal{D}}_k^{f}(m)$
\begin{eqnarray}\label{zfermion}
&&{\cal{D}}_k^{f}(m)  \\
&& \equiv \int d\Psi \; e^{- \int\limits_{\T^4} {\cal{L}}_f} = \left({2 m \over g^2}\right)^k \prod_n \left({16 \over g^4} (\omega_n^2 + |m|^2)\right) \prod_{p=1}^k \underbrace{\int d^2 \xi^0_p \xi^0_{p1} \xi^0_{p2}}_{\eps_p} \prod_{n} \underbrace{\int d^4 \xi_{n} \xi_{n,1}^+ \xi_{n,2}^+ \xi_{n,1}^- \xi_{n,2}^-}_{\eps_n}, \nonumber \\
\end{eqnarray}
where we define $\eps_p$ and $\eps_n$, each of which can be $\pm1$, depending on our definition of $d^2 \xi^0_p$ and $d^4 \xi_n$. Thus, we have
\begin{eqnarray}
\label{dfermion}
{\cal{D}}_k^{f}(m) =  \left({2 m \over g^2}\right)^k \prod_n \left({16\over g^4} (\omega_n^2 + |m|^2)\right) \prod_{p=1}^k \eps_p  \prod_{n} {\eps_n}.
\end{eqnarray}

The fermion two-point function $\langle \Psi(x) \bar\Psi(y)\rangle$ is now also calculable. We express it in terms of the quantities $\eps_p$ and $\eps_n$ defined in (\ref{zfermion}). 
We denote by $\langle ... \rangle_{unnorm.}$ a correlation function computed with the measure $d \Psi   \exp{(- \int{\cal{L}}_f)}$, as per (\ref{actionfermionmodes}, \ref{measure}),  {\it without} division by ${\cal{D}}_k^{f}(m)$. Then, from (\ref{expansion1}), we find, 
\begin{eqnarray}\label{propagator 0}
&& \langle \Psi(x) \otimes \bar\Psi(y)\rangle_{unnorm.}  \\
&&=  \langle (\xi^0_{p,\beta} \Sigma^{(0)}_{p, \beta}(x)+ \xi_{n,s}^+ \Sigma_{n,s}^{(+)}(x)+ \xi_{n,s}^- \Sigma_{n,s}^{(-)}(x)) \otimes (
 \xi^0_{q, \beta'} \bar\Sigma^{(0)}_{q,\beta'}(y)+ \xi_{n',s'}^+ \bar\Sigma_{n',s'}^{(+)}(y) + \xi_{n',s'}^- \bar\Sigma_{n',s'}^{(-)}(y) )\rangle. \nonumber 
 \end{eqnarray}
Then,
noting that (\ref{actionfermionmodes}) implies that nonzero correllators have $s,s'=1,2$ (or $2,1$) and are diagonal in the mode index $n$, we find: 
\begin{eqnarray}\label{propagator 1}
&&\langle \Psi(x) \otimes \bar\Psi(y)\rangle_{unnorm.}  \\
 &&= \langle \xi^0_{p,1}\xi^0_{q, 2}\rangle \left[\Sigma^{(0)}_{p, 1}(x) \otimes\bar\Sigma^{(0)}_{q,2}(y)-\Sigma^{(0)}_{p, 2}(x) \otimes\bar\Sigma^{(0)}_{q,1}(y)\right]    \nonumber \\
 &&\; +    \langle \xi_{n,1}^+ \xi_{n,2}^+ \rangle \left[\Sigma_{n,1}^+(x) \otimes\bar\Sigma_{n,2}^+(y) - \Sigma_{n,2}^+(x) \otimes\bar\Sigma_{n,1}^+(y)\right] + \langle \xi_{n,1}^- \xi_{n,2}^- \rangle\left[\Sigma_{n,1}^-(x)\otimes \bar\Sigma_{n,2}^-(y) - \Sigma_{n,2}^-(x) \otimes\bar\Sigma_{n,1}^-(y)\right] \nonumber  \\
 &&\; + \langle \xi_{n,1}^+ \xi_{n,2}^- \rangle \left[\Sigma_{n,1}^+(x)\otimes \bar\Sigma_{n,2}^-(y) - \Sigma_{n,2}^-(x) \otimes\bar\Sigma_{n,1}^+(y)\right] + \langle \xi_{n,1}^- \xi_{n,2}^+ \rangle \left[\Sigma_{n,1}^-(x) \otimes\bar\Sigma_{n,2}^+(y) - \Sigma_{n,2}^+(x) \otimes\bar\Sigma_{n,1}^-(y)\right]  \nonumber \end{eqnarray}
Next, from (\ref{actionfermionmodes}), (\ref{measure}), and (\ref{zfermion}), we find the zero-mode correlators:\footnote{To avoid confusion about the dimensionality of the propagator, recall that our functions $\Sigma^{(0), \pm}$ have mass dimension $2$, as per e.g.~(\ref{identity4.5}). Another possible confusion is to recall that there are no $\bar\lambda$ zero modes; if they were included---as they would be there for an e.g. $k=0$ background with periodic B.C., the form of $Z_f$ changes by extra $m^*$ terms.}
\begin{eqnarray}
\label{zerocorrelator}
\langle \xi^0_{p,1}\xi^0_{q, 2}\rangle =    {\cal{D}}_k^{f}(m) \delta_{pq}  {g^2 \over   2} {m^* \over |m|^2},
\end{eqnarray}
 where all factors of $\epsilon_{p,n}$ are inside ${\cal{D}}_k^{f}(m)$ of (\ref{dfermion}).

For the nonzero modes, we find, similarly, that for the same-sign correlators the $\epsilon_{p,n}$ factors are all inside (\ref{dfermion}):
\begin{eqnarray}
\label{samesigncorrelator}
\langle \xi^+_{n,1}\xi^+_{n,2}\rangle &=& {\cal{D}}_k^{f}(m) {g^2 \over 4} {-i \omega_n - \Re m \over \omega_n^2 + |m|^2},  ~\langle \xi^-_{n,1}\xi^-_{n,2}\rangle  =  {\cal{D}}_k^{f}(m) {g^2 \over 4} {  i \omega_n - \Re m \over \omega_n^2 + |m|^2}, \nonumber \\
\end{eqnarray}
as well as for the opposite-sign correlators,
\begin{eqnarray}
\label{oppositesigncorrelator}
\langle \xi^+_{n,1}\xi^-_{n,2}\rangle &=& {\cal{D}}_k^{f}(m) {g^2 \over 4} { -i \Im m \over \omega_n^2 + |m|^2}, ~ 
\langle \xi^-_{n,1}\xi^+_{n,2}\rangle = {\cal{D}}_k^{f}(m) {g^2 \over 4} {  -i \Im m  \over \omega_n^2 + |m|^2},
\end{eqnarray}

In order to read off the components of the  propagator from (\ref{propagator 1}) (again omitting the $\otimes$ sign),
\begin{eqnarray}
\langle \Psi(x)\otimes \bar\Psi(y) \rangle = \left(\begin{array}{cc} \langle \lambda_\alpha(x)\otimes \lambda^\beta(y) \rangle &\langle  \lambda_\alpha(x) \otimes\bar\lambda_{\dot\beta}(y) \rangle \cr \langle \bar\lambda^{\dot\alpha}(x)\otimes \lambda^\beta(y) \rangle & \langle \bar\lambda^{\dot\alpha}(x) \otimes\bar\lambda_{\dot\beta}(y) \rangle\end{array} \right)~,
\end{eqnarray}
the last piece of information, after (\ref{samesigncorrelator},\ref{oppositesigncorrelator}), needed are  the products of $\Sigma, \bar\Sigma$ functions entering  (\ref{propagator 1}), still keeping as much generality as possible. 

To this end, we recall the expressions for $\Sigma^\pm$ and $\bar\Sigma^\pm$ from (\ref{post and neg3}) and (\ref{sigmabar}). Then, we find for the inner products appearing in (\ref{propagator 1})
\begin{eqnarray}
&&\left[\Sigma_{n,1}^+(x) \otimes\bar\Sigma_{n,2}^+(y) - \Sigma_{n,2}^+(x) \otimes\bar\Sigma_{n,1}^+(y)\right] \\
&&= \left(\begin{array}{cc} (\sigma_{\mu \; \alpha 1} \sigma_{\nu \gamma 2}- \sigma_{\mu \; \alpha 2} \sigma_{\nu \gamma 2})\epsilon^{\gamma \beta} {D_\mu \phi_n(x) \otimes D_\nu \phi_n(y) \over \omega_n^2}& (\sigma_{\mu\; \alpha 1} \epsilon_{2 \dot\beta}- \sigma_{ \mu\; \alpha 2} \epsilon_{1 \dot\beta} ) {i D_\mu \phi_n(x) \otimes \phi_n(y) \over \omega_n}\cr (\delta^{\dot\alpha 1} \sigma_{\nu \; \gamma 2}- \delta^{\dot\alpha 2} \sigma_{\nu \; \gamma 1})\epsilon^{\gamma\beta} {i \phi_n(x) \otimes D_\nu \phi_n(y) \over \omega_n}&  (\delta^{\dot\alpha 1} \epsilon_{\dot\beta 2}- \delta^{\dot\alpha 2} \epsilon_{\dot\beta 1}) \phi_n(x) \otimes \phi_n(y)  \end{array}\right), \nonumber 
\end{eqnarray}
With some minor matrix manipulations, this can be cast into the friendlier form:\begin{eqnarray}\label{sigmaplus1}
\left[\Sigma_{n,1}^+(x) \otimes\bar\Sigma_{n,2}^+(y) - \Sigma_{n,2}^+(x) \otimes\bar\Sigma_{n,1}^+(y)\right] 
= \left(\begin{array}{cc} -(\sigma_{\mu} \bar\sigma_{\nu})_\alpha^{\; \;\beta} {D_\mu \phi_n(x) \otimes D_\nu \phi_n(y) \over \omega_n^2}& ~\sigma_{\mu\; \alpha \dot\beta}\; {i D_\mu \phi_n(x)\otimes \phi_n(y) \over \omega_n}\cr - \bar\sigma_\nu^{\; \dot\alpha \beta}\; {i \phi_n(x) \otimes D_\nu \phi_n(y) \over \omega_n}&  -\delta^{\dot\alpha}_{\dot\beta}\; \phi_n(x)\otimes \phi_n(y)  \end{array}\right). \nonumber \\
\end{eqnarray}
We now notice that in the $11$ entry above, we can use
\begin{eqnarray}\label{identity 12}
\sigma_\mu\bar\sigma_\nu D_\mu \phi_n(x) \otimes D_\nu \phi_n(y) = D_\mu \phi_n(x) \otimes D_\mu \phi_n(y) +   \sigma_{\mu\nu}(D_\mu \phi_n(x) \otimes D_\nu \phi_n(y)- D_\nu \phi_n(x) \otimes D_\mu \phi_n(y)),\nonumber \\
\end{eqnarray} but at this point, without explicit expressions for $\phi_n$, there is no general argument for the antisymmetric term vanishing (however,  the antisymmetric term vanishes for $x =y$, which is sufficient for the computation of the gaugino condensate).

Next, we turn to the same product as (\ref{sigmaplus1}), but involving $\Sigma^-$ instead. We notice that  because of (\ref{post and neg3}) and (\ref{sigmabar}), it has the same diagonal entries as (\ref{sigmaplus1}), but opposite-sign off-diagonal entries. Thus, we can immediately write \begin{eqnarray}
\label{sigmaminus1}
 \left[\Sigma_{n,1}^-(x) \otimes  \bar\Sigma_{n,2}^-(y) - \Sigma_{n,2}^-(x)  \otimes \bar\Sigma_{n,1}^-(y)\right] 
= \left(\begin{array}{cc} -(\sigma_{\mu} \bar\sigma_{\nu})_\alpha^{\; \;\beta} {D_\mu \phi_n(x) \otimes  D_\nu \phi_n(y) \over \omega_n^2}&~ - \sigma_{\mu\; \alpha \dot\beta}\; {i D_\mu \phi_n(x)  \otimes \phi_n(y) \over \omega_n}\cr  \bar\sigma_\nu^{\; \dot\alpha \beta}\; {i \phi_n(x)  \otimes D_\nu \phi_n(y) \over \omega_n}&  -\delta^{\dot\alpha}_{\dot\beta}\; \phi_n(x) \ \otimes\phi_n(y)  \end{array}\right). \nonumber \\
\end{eqnarray}
 Finally, we turn to the mixed products multiplying the $\Im m \ne 0$ correlators in (\ref{oppositesigncorrelator}, \ref{propagator 1}). In view of the equality of the correllators, $\langle \xi_{n,1}^+ \xi_{n,2}^- \rangle =\langle \xi_{n,1}^- \xi_{n,2}^+ \rangle  $, as per (\ref{oppositesigncorrelator}), we need only the expression below:
 \begin{eqnarray}\label{sigmamixed}
\left[\Sigma_{n,s}^+(x) \otimes  \bar\Sigma_{n,s'}^-(y) \epsilon^{ss'} + \Sigma_{n,s}^-(x) \otimes  \bar\Sigma_{n,s'}^+(y)\epsilon^{ss'} \right] = \left(\begin{array}{cc} 2 (\sigma_\mu \bar\sigma_\nu)_{\alpha}^{\;\; \beta}\; {D_\mu \phi_n(x) \otimes  D_\nu \phi_n(y) \over \omega_n^2}& 0\cr0 &  - 2 \delta^{\dot\alpha}_{\dot\beta} \; \phi_n(x) \otimes  \phi_n(y)  \end{array}\right).\nonumber \\
\end{eqnarray}
 We also need the outer product of the zero mode wavefunctions that enter (\ref{propagator 1}). From  (\ref{kzeromodes1}, \ref{kzeromodes1bar}), we find:
 \begin{eqnarray}\label{zeromodeproduct}
 \left[\Sigma^{(0)}_{p, 1}(x)  \otimes \bar\Sigma^{(0)}_{q,2}(y)-\Sigma^{(0)}_{p, 2}(x)  \otimes \bar\Sigma^{(0)}_{q,1}(y)\right] =  \left(\begin{array}{cc} \sum\limits_{p=1}^k \left( \psi^{(0)}_{\alpha p, 1}(x)  \otimes  \psi^{(0) \beta}_{\;\;\; p,2}(y) - \psi^{(0)}_{\alpha p, 2}(x)  \otimes  \psi^{(0) \beta}_{\;\;\; p,1}(y)\right) &~ 0\cr 0& ~0\end{array}\right). \nonumber \\
 \end{eqnarray}
 
Finally, substituting into (\ref{propagator 1}) the expression for the Grassmann correllators (\ref{zerocorrelator}, \ref{samesigncorrelator}, \ref{oppositesigncorrelator}) and those for the outer products of the wave functions (\ref{sigmaplus1}, \ref{sigmaminus1}, \ref{sigmamixed}, \ref{zeromodeproduct}),     we obtain the expression:
\begin{eqnarray}\label{propagator 2}
 \langle \Psi(x)  \otimes  \bar\Psi(y) \rangle_{unnorm.} &\equiv& \left(\begin{array}{cc} \langle \lambda_\alpha(x)  \otimes \lambda^\beta(y) \rangle &\langle  \lambda_\alpha(x) \otimes  \bar\lambda_{\dot\beta}(y) \rangle \cr \langle \bar\lambda^{\dot\alpha}(x) \otimes  \lambda^\beta(y) \rangle & \langle \bar\lambda^{\dot\alpha}(x) \otimes  \bar\lambda_{\dot\beta}(y) \rangle\end{array} \right)  \\
&=& {\cal{D}}_k^f(m){g^2 \over 2} \left\{ \left(\begin{array}{cc}  { m^* \over |m|^2}  \sum\limits_{p=1}^k \left( \psi^{(0)}_{\alpha p, 1}(x) \otimes  \psi^{(0) \beta}_{\;\;\; p,2}(y) - \psi^{(0)}_{\alpha p, 2}(x)  \otimes \psi^{(0) \beta}_{\;\;\; p,1}(y)\right) &~ 0\cr 0& ~0\end{array}\right)  \right.\nonumber  \\
&&\left. ~~~~~+  \sum_n \left(\begin{array}{cc}  {m^* \over \omega_n^2 + |m|^2 } \; (\sigma_\mu \bar\sigma_\nu)_\alpha^{\; \; \beta} \;  {D_\mu \phi_n(x)  \otimes  D_\nu \phi_n(y) \over \omega_n^2}&  {\sigma_{\mu \; \alpha \dot\beta} \over \omega_n^2 + |m|^2}\; {D_\mu \phi_n(x)  \otimes  \phi_n(y)}\cr - { \bar\sigma_{\nu}^{\dot\alpha \beta}  \over \omega_n^2 + |m|^2} \; { \phi_n(x)  \otimes D_\nu \phi_n(y)}&  {m \over \omega_n^2 + |m|^2} \; \delta^{\dot\alpha}_{\dot\beta}\; \phi_n(x)  \otimes \phi_n(y)  \end{array}\right) \right\}. \nonumber
\end{eqnarray}
Comments:
\begin{enumerate}
\item Eqn.~(\ref{propagator 2}) is the general expression propagator in a $Q=k/N$ self dual background under the assumption of  $2k$  dotted fermion zero modes saturating the index (this can be relaxed, see below). We recall again that the ${\cal{D}}_k^f(m)$ factor, given in (\ref{dfermion}), is due to the fact that the correlator is assumed to be unnormalized. Furthermore, all Grassmann ordering ambiguity is contained in this factor (as explicitly outlined in (\ref{zfermion})).

\item The propagator is expressed in terms of sums over the zero and nonzero modes in the instanton background:
  In writing the zero mode contribution to the propagator, we allowed the zero mode wave-functions to be $x$-dependent, as is the case at the subleading orders in the $\Delta$-expansion in the fractional instanton background.

\item The sum over nonzero modes is expressed in terms of $\phi_n$, the nonzero mode eigenfunctions of the adjoint scalar Laplace operator  (\ref{adjoint1}), in a self-dual background of charge $Q=k/N$, normalized as in (\ref{normphin1});  thus $\phi_n$ are taken Hermitean. Self-duality of $F_{\mu\nu}$ was essential to find the action in the diagonal form (\ref{actionfermionmodes}) (it was used e.g. in deriving (\ref{identity2})).

 Due to the nature of the adjoint path integral over $\Psi$, with $\bar\Psi = \Psi^t C$, no complex conjugates of $\phi_n$ appear in the path integral and in the propagators (as in Sec. B.2 in \cite{Anber:2022qsz}). 
 The orthonormality of $\phi_n$ (\ref{normphin1})  was essential in the derivation of the action (\ref{actionfermionmodes}), and the consequent expression for the propagator, but completeness in the form (\ref{completeness1}) was not used. Thus, should one study a background that also has dotted zero modes (which would change the Laplacian completeness condition) these can simply be added to the above propagator. 
 
   \item All products of wave functions and their derivatives appearing in (\ref{propagator 2}) should be understood as outer products,  explicitly the l.h.s. and r.h.s. should be understood as
 \begin{eqnarray}\label{outer 1}
 \langle \lambda_\alpha(x) \otimes  \lambda^\beta(y) \rangle &\rightarrow& \langle \lambda_{ij \; \alpha}(x) \lambda_{kl}^\beta(y) \rangle, \nonumber \\
 D_\mu \phi_n(x) \otimes \phi_n(y) &\rightarrow& (D_\mu \phi_{n})_{ij}(x)\; \phi_{n \; kl}(y),~ \text{etc.}
 \end{eqnarray}
Throughout,  we suppressed the adjoint indices,  $i,j,k,l = 1,...N$; these can be restored in (\ref{propagator 2}), following (\ref{outer 1}).

These indices can be further split into $SU(\ell)$ $(C,B,..)$ and $SU(k)$ $(C',B',...)$ indices. This is necessary when one develops explicit expressions for $\phi_n$ and the propagators by solving for the fractional instanton Laplacian eigenfunctions in the $\Delta$-expansion, as we describe below.

\item All expressions in (\ref{propagator 2}) are valid in an exactly self-dual background, assumed to be ``generic,'' i.e. such that the Laplacian (\ref{adjoint1}) has no zero modes. We have no exact expression for such a background. However, within the $\Delta$ expansion, we know the background as  a series expansion in $\Delta$, see (\ref{perturbed A}). 
Thus,  in (\ref{propagator 2}), all terms should be understood using the same $\Delta$-expansion: the eigenvalues of the Laplacian $\omega_n$, its eigenfunctions $\phi_n$, the background entering the covariant derivative ($D_\mu = \partial + i [A_\mu,..]$), as well as the zero mode wavefunctions $\psi_{\alpha p,i}^{(0)}$ are all given as an expansion in $\Delta$.

In the following sections, we evaluate the terms in the propagator to order $\Delta^0$. However, we note that there are $k$ eigenvalues of the Laplacian of order $\Delta^1$, whose leading contribution to the propagator is of order $1\over \Delta$ in the $|m|L \ll \sqrt{\Delta}$ limit (see discussion near (\ref{smallmvsdelta}) below), which we also evaluate. 

 \end{enumerate}

\section{The $Q={k \over N}$ propagator in the diagonal $SU(k) \times SU(\ell) \times U(1)$ sector}
\label{sec:propagatorsuktimesu1}

We now use the general formula for the propagator in the self-dual $Q=k/N$ background, eqn.~(\ref{propagator 2}),  to construct the propagators around the instanton in the $\Delta$-expansion. To this order, we need to find expressions for the eigenvalues and eigenvectors of the covariant adjoint Laplacian (\ref{adjoint1}).  

In this section,  we concentrate on adjoint components $\phi_{ij}$   lying entirely within the ``diagonal,'' i.e.~the $SU(k) \times SU(\ell) \times U(1)$ part of the  $SU(N)$-adjoint. These components of the  adjoint fields   do not couple to the order-$\Delta^0$ background (\ref{naked expressions}), since, being proportional to $\omega$ (\ref{omega}), it commutes with the fields inside $SU(k) \times SU(\ell) \times U(1)$. Therefore, all $k^2$ diagonal components (we add the one along the $U(1)$ generator) obey the free Laplace equation, eqn.~(\ref{adjoint1}) with $D_\mu = \partial_\mu$

However, the diagonal fields are subject to the boundary conditions (\ref{BCS lambda A}, \ref{BCS lambda a}). In order to find a basis of free fields obeying these BCS, it is convenient to split the diagonal fields into $SU(k) \times U(1)$ components (i.e. ones along the traceless part of the $k\times k$ part of the $SU(N)$ matrix plus the ones proportional to $\omega$) and $SU(\ell)$ components (i.e. in the traceless part of the $\ell\times \ell$ part of $SU(N)$. 

\subsection{The $SU(k) \times U(1)$ diagonal components}
\label{sec:diagonalsuktimesu1}
 
{\flushleft{We}} first recall that the free laplacian on $\T^4$ with periodic BC, has $2^4$ real, normalized as in (\ref{normphin1}), eigenfunctions of the Laplacian for every $\omega_n^2$, i.e.    for each choice of ``momenta'' $n_{1,2,3,4} > 0$:
\begin{eqnarray}\label{definitionofF} 
\phi_{(n_\mu, f_\mu)} &=& {4 \over \sqrt{V}} \prod_\mu f_{\mu}(n_\mu x_\mu)~, \\
f_\mu (n_\mu x_\mu) &=& \cos{2 \pi n_\mu x_\mu\over L_\mu}, ~\text{or}~ \sin{2 \pi n_\mu x_\mu\over L_\mu}~.   \nonumber  
\end{eqnarray} All these have the same $\omega_n^2 = \sum_\mu ({2 \pi n_\mu \over L_\mu})^2$. That these $\phi_{(n,f_\mu)}$ are correctly normalized follows from recalling that $\int\limits_{0}^L dx \cos({2 \pi n x\over L}) \cos({2 \pi m x\over L}) = {L \over 2} \delta_{nm}$, with an identical expression for the integral with $\cos \rightarrow \sin$, while $\sin$ and $\cos$ are orthogonal.

A look at the BCS (\ref{BCS lambda A}) shows that $\phi_{B'C'}$ is periodic in all   directions but $x_2$. On the other hand, the $\phi$-component along the $U(1)$ generator $\omega$ is periodic in all directions. Thus, the only complication that arises is  for the $B'\ne C'$ components of the adjoint $\phi_{B'C'}$, which are not periodic in the $x_2$ direction.  

\subsubsection{The Cartan components of the diagonal $SU(k) \times U(1)$: the propagator of the ${\cal{O}}(\Delta^0)$ nonzero modes}

\label{sec:cartansuktimesu1}

Thus,  for each of the diagonal components $\phi_{B'B'}$ we find the following orthonormal basis of $2^4$ eigenfunctions for each eigenvalue  $\omega_n^2$ of the Laplacian:\footnote{Below, for brevity we keep all $n_\mu > 0$, but it is easy to see that  our considerations generalize to relaxing this to $\sum\limits_\mu n_\mu n_\mu >0$, i.e. allowing some but not all $n_\mu$ to vanish, thus including all nonzero $\omega_n^2$.}
\begin{eqnarray}\label{diagonalsukmodes}
\phi_{B'B' \; (n_\mu, f_\mu)}(x) &=&   {2^{4 \over 2}\over \sqrt{V}} \prod\limits_{\mu =1}^4 f_\mu(x_\mu)~, ~ \text{for every} ~ B', ~ 2^4~ \text{eigenfunctions} \nonumber \\
\omega_n^2 &=&  \sum\limits_{\mu=1}^4 ({2 \pi n_\mu \over L_\mu})^2~, ~ n_{1,2,3,4} > 0,~~ (\text{we rename} ~\phi_{B'B'} \rightarrow \phi_b, \text{see below}).
\end{eqnarray}
Recall that there are $k$ zero eigenvalues of the order-$\Delta^0$ Laplacian, which lie in the $SU(k) \times U(1)$ subspace. These are lifted at subleading order in $\Delta$. The contribution of these eigenvalues and the corresponding eigenfunction to the propagator will be computed separately, see the following section \ref{sec:liftingzeromodes}.

In order to  include  the $U(1)$ mode in the $\omega$-direction (\ref{omega}) in a proper manner, 
let now $\bm H_{(k)}$ be the Hermitean  $SU(k)$ Cartan generators (which we complete into $N \times N$ matrices by simply adding appropriate zeros). Let us also define  the $U(1)$ generator along $\omega$ as $\frac{\omega}{2\pi\sqrt{k(N-k)}}$. Then combine these into a new basis of $SU(k) \times U(1)$ Cartan generators, obeying the relations given below
\begin{eqnarray}\label{suktimesu1cartan}
\bm {\tilde H}\equiv \left(\frac{\omega}{2\pi   \sqrt{N k(N-k)}}, \bm H_{(k)}\right),~~\mbox{tr}\left[\tilde H^{b_1} \tilde H^{b_2}\right]=\delta_{b_1b_2}, ~~ b_1, b_2=1,2,..,k.
\end{eqnarray} 
Using this basis, we can group the $SU(k) \times U(1)$ components of the adjoints $\phi$ as fields along the $SU(k)$ Cartan directions and the $U(1)$ $\phi_\omega$ into one set of fields $\tilde\phi_{i}$, where $i=1,2,..,k$. In view of the Cartan normalization given above, we simply replace $\phi_{B'B'} \rightarrow \tilde\phi_{i}$ to define the $k$ eigenmodes of the Laplacian given in (\ref{diagonalsukmodes}). Likewise, the fields $\lambda, \bar\lambda$ have components along the $SU(k)\times U(1)$ Cartan generators (\ref{suktimesu1cartan}) which we denote by $\tilde\lambda_i$ and whose propagators will be read off---after first working out the contribution of $\phi_i$ to (\ref{propagator 2}). Explicitly, we define
the gaugino components along the $SU(k) \times U(1)$ Cartan generators (\ref{suktimesu1cartan})
\begin{eqnarray}\label{lambdacartan}
\lambda(x) = \sum\limits_{b=1}^k \tilde\lambda_b \tilde{H}^b + \text{off diagonal}
\end{eqnarray}
and in what follows, find their propagator from (\ref{propagator 2}).

To obtain the  propagator $\langle \lambda_b \lambda_{b'}\rangle$, from (\ref{propagator 2}), one has, for every $\omega_n^2$, to sum over all $2^4$ different choices of wave functions and their derivatives appearing in (\ref{propagator 2}). The $\sum_n$ in the nonzero-mode contribution to the propagator can thus be written as a sum over the values of $\omega_n^2$, i.e. a sum over $n_\mu >0$, and a sum over the $2^4$ various choices of $f_\mu$ for a fixed 
$\omega_n^2 = (2 \pi)^2 \sum_\mu {n_\mu^2/L_\mu^2}$. 

 We start with the sum over $f_\mu$ appearing in the $22$ element of the propagator for a given $\omega_n^2$:
 \begin{eqnarray}
\sum_{f_\mu} \phi_{(n_\mu, f_\mu)}(x) \phi_{(n_\mu, f_\mu)}(y) &=& {16 \over V} \sum_{\{f\} = \{\cos, \sin\}} \prod_\lambda f_\lambda(x_\lambda) f_\lambda(y_\lambda) \nonumber \\
&=& {16 \over V}\prod_\lambda \left(\sum_{f_\lambda= \{\cos(...), \sin(...)\}}  f_\lambda(x_\lambda) f_\lambda(y_\lambda)\right)
 \end{eqnarray}
 We now note that for every $\lambda = 1,...,4$, 
\begin{eqnarray}\label{sum1} \sum_{f_\lambda(...)= \{\cos({2 \pi n_\lambda ...\over L_\lambda}), \sin{2 \pi n_\lambda ...\over L_\lambda})\}}  f_\lambda(x_\lambda) f_\lambda(y_\lambda) = {1 \over 2}\left( e^{i {2\pi n_\lambda \over L_\lambda}( x_\lambda-y_\lambda)} + e^{-i {2\pi n_\lambda \over L_\lambda}( x_\lambda-y_\lambda)}\right),\end{eqnarray}
meaning that we can, instead of summing over positive values of $n_\mu$ only, sum over positive and negative values. Thus, the $\bar\lambda(x) \bar\lambda(y)$-propagator from (\ref{propagator 2}) can be equivalently written as:
\begin{eqnarray}\label{22element 1}
\langle \bar\lambda_b^{\dot\alpha}(x) \bar\lambda_{b' \dot\beta}(y) \rangle_{unnorm.} = \delta_{bb'} {\cal{D}}_k^f(m) \; {g^2 \over 2 V} \sum\limits_{p_\mu \in {2 \pi\over L_\mu} \Z}' { 
\delta_{\dot\beta}^{\dot\alpha} m \over p_\mu^2 + |m|^2} e^{i p_\mu (x_\mu - y_\mu)}~,
\end{eqnarray} 
where the prime on the sum denotes omission of $p_\mu=0$.

Next, consider the sums appearing in the $12$ and $21$ matrix elements in (\ref{propagator 2}) (where for definiteness and to not overcrowd the notation) we took the derivative in the $x_1$-direction:
 \begin{eqnarray}
&&\sum_{f_\mu} {\partial_1 \phi_{(n_\mu, f_\mu)}(x) \phi_{(n_\mu, f_\mu)}(y)} \nonumber \\&=& {16 \over V} \sum_{\{f\} = \{\cos, \sin\}} {\partial_1 f_1(x_1) f_1(y_1)} \prod_{\lambda=2,3,4} f_\lambda(x_\lambda) f_\lambda(y_\lambda) \nonumber \\
&=& {1 \over V} {i 2 \pi n_1 \over L_1}(e^{i {2 \pi n_1 \over L_1}(x_1-y_1)} - e^{-i {2 \pi n_1 \over L_1}(x_1-y_1)})   \prod_{\lambda=2,3,4} \left( e^{i {2\pi n_\lambda \over L_\lambda}( x_\lambda-y_\lambda)} + e^{-i {2\pi n_\lambda \over L_\lambda}( x_\lambda-y_\lambda)}\right) ~.\nonumber \\
 \end{eqnarray}
Here, we used (\ref{sum1}) in the $\lambda=2,3,4$ directions, as well as (giving also the equation for the case when the derivative acts on $y_1$)
\begin{eqnarray}\label{derivative 22} 
 \sum_{f_1(...)= \{\cos({2 \pi n_1 ...\over L_1}), \sin{2 \pi n_1 ...\over L_1})\}} \partial_1 f_1(x_1) f_1 (y_1) &=& {1 \over 2}{i 2 \pi n_1 \over L_1}(e^{i {2 \pi n_1 \over L_1}(x_1-y_1)} - e^{-i {2 \pi n_1 \over L_1}(x_1-y_1)}) \nonumber \\
  \sum_{f_1(...)= \{\cos({2 \pi n_1 ...\over L_1}), \sin{2 \pi n_1 ...\over L_1})\}}  f_1(x_1) \partial_1 f_1 (y_1) &=&- {1 \over 2}{ i 2 \pi n_1 \over L_1}(e^{i {2 \pi n_1 \over L_1}(x_1-y_1)} - e^{-i {2 \pi n_1 \over L_1}(x_1-y_1)}) \nonumber \\
\end{eqnarray} 
Clearly this holds with $x_1$ replaced  any other direction and allows us to extend the summation over $n_\mu >0$ to both positive and negative $n_\mu$, giving for the $12$ and $21$ propagators in (\ref{propagator 2}) the expressions:
\begin{eqnarray}\label{12and21element}
\langle  \lambda_{b \alpha}(x) \bar\lambda_{b' \; \dot\beta}(y) \rangle_{unnorm.} &=&\delta_{bb'} {\cal{D}}_k^f(m) \; {g^2 \over 2 V} \sum\limits_{p_\mu \in {2 \pi\over L_\mu} \Z}' { i 
\sigma_{\mu \; \alpha \dot\beta}   p_\mu \over p_\mu^2 + |m|^2} e^{i p_\mu (x_\mu - y_\mu)}~,
 \nonumber \\
 \langle \bar\lambda^{\dot\alpha}_b(x) \lambda_{b'}^\beta(y) \rangle_{unnorm.} &=&\delta_{bb'}   {\cal{D}}_k^f(m) \; {g^2 \over 2 V} \sum\limits_{p_\mu \in {2 \pi\over L_\mu} \Z}'  {  i  \bar\sigma_\nu^{\dot\alpha\beta}  p_\nu \over p_\mu^2 + |m|^2 } e^{i p_\mu (x_\mu - y_\mu)}~,
\end{eqnarray} 
 
 Finally, we consider the $11$ element, $(\sigma_\mu \bar\sigma_\nu)_\alpha^{\; \; \beta} \;  {D_\mu \phi_n(x) \; D_\nu \phi_n(y) \over \omega_n^2}$, for which, in view of (\ref{identity 12}) we only need to consider $\sum_\mu {D_\mu \phi_n(x) \; D_\mu \phi_n(y) \over \omega_n^2}$ as well as the antisymmetric combination appearing in the second term of (\ref{identity 12}). We start with the latter
 \begin{eqnarray}
&&\sum_{f_\mu} {\partial_1 \phi_{(n_\mu, f_\mu)}(x) \partial_2 \phi_{(n_\mu, f_\mu)}(y)} - {\partial_2 \phi_{(n_\mu, f_\mu)}(x) \partial_1 \phi_{(n_\mu, f_\mu)}(y)} \nonumber \\
&=& {16 \over V} \left(\sum\limits_{f_1}( \partial_1 f_1(x_1) f_1(y_1)) (\sum_{f_2} f_2 (x_2) \partial_2 f_2(y_2)) - ( \sum_{f_1} f_1(x_1) \partial_1 f_1(y_1)) (\sum_{f_2} \partial_2 f_2 (x_2)  f_2(y_2))\right)  \nonumber  \\
&& ~ \times \prod_{\lambda=3,4} ( \sum_{f_\lambda}  f_\lambda(x_\lambda) f_\lambda(y_\lambda)) \nonumber \\
&=& 0, 
\end{eqnarray}
where the vanishing of the second line follows from applying (\ref{derivative 22}). Thus, we now focus on $\sum_\mu {D_\mu \phi_n(x) \; D_\mu \phi_n(y) \over \omega_n^2}$, from which we compute one term, taking $\mu=1$ and momentarily omitting the $1/\omega_n^2$ factor:
\begin{eqnarray}\label{twoderivative}
&&\sum_{f} \partial_1 \phi_{n_\mu, f_\mu} (x) \; \partial_1 \phi_{n_\mu, f_\mu}(y)  \\
&=& {16 \over V} \sum_{\{f\} = \{\cos, \sin\}} {\partial_1 f_1(x_1)  \partial_1 f_1(y_1)} \prod_{\lambda=2,3,4} f_\lambda(x_\lambda) f_\lambda(y_\lambda) \nonumber \\
&=& {1 \over V} \left({ 2 \pi n_1 \over L_1}\right)^2 (e^{i {2 \pi n_1 \over L_1}(x_1-y_1)} + e^{-i {2 \pi n_1 \over L_1}(x_1-y_1)})   \prod_{\lambda=2,3,4} \left( e^{i {2\pi n_\lambda \over L_\lambda}( x_\lambda-y_\lambda)} + e^{-i {2\pi n_\lambda \over L_\lambda}( x_\lambda-y_\lambda)}\right) ~,\nonumber  
\end{eqnarray}
where the identity used to go from the second to the third line is easily verified. 
Replacing $1 \rightarrow \mu$, summing over $\mu$, remembering the zero mode wave functions (\ref{zeromodesnormalized}) as well as the $1\over \omega_n^2$ factor, we find that
the $11$-propagator in (\ref{propagator 2}) is:
\begin{eqnarray} \label{11element}
  \langle \lambda_{b\;\alpha}(x) \lambda_{b'}^\beta(y) \rangle_{unnorm.} &=&\delta_{bb'}\; \delta_\alpha^\beta\; {\cal{D}}_k^f(m) {g^2 \over 2 V}    \left({ m^* \over |m|^2}   +   \sum\limits_{p_\mu \in {2 \pi\over L_\mu} \Z}' {m^*  \over p_\mu^2 + |m|^2} e^{i p_\mu (x_\mu - y_\mu)}\right), \nonumber \\
\end{eqnarray}
where we also remembered the zero mode wavefunctions (\ref{zeromodesnormalized}).

To summarize, here we found the propagators along the Cartan generators of the diagonal $SU(k)\times U(1)$. They are given in (\ref{11element}, \ref{12and21element}, \ref{22element}). 
We stress
that to these propagators, we have to add the contribution of the lowest  eigenvalue of the Laplacian, which is order $\Delta$.

\subsubsection{The Cartan components of the diagonal $SU(k) \times U(1)$: the lifting of the ${\cal{O}}(\Delta^0)$  zero modes at order $\Delta$}

\label{sec:liftingzeromodes}

The covariant Laplacian (\ref{adjoint1}), in the ${\cal{O}}(\Delta^0)$ background, has $k$ zero modes: these are constant modes in the Cartan directions of $SU(k) \times U(1)$.\footnote{As our results in the various sections here show, upon decomposing the Laplacian into different $SU(N)$ components, the $SU(\ell)$ as well as the off-diagonal $k \times \ell$ and $\ell \times k$ blocks have no zero  modes at order $\Delta^0$.} Similar to the definition of the Cartan components of the gaugino  (\ref{lambdacartan}), we define these unperturbed  hermitean zero modes, $\phi^{(0)}$, using the $\bm{\tilde H}$ basis  of (\ref{suktimesu1cartan}):
\begin{eqnarray}\label{laplacianzeromodes}
\phi^{(0) b} ={1 \over \sqrt{V}} \; \tilde H^b, ~b=1,...k, ~\text{with}~ \int\limits_{\T^4} \tr \phi^{(0) b} \phi^{(0) a} = \delta^{ab}.
\end{eqnarray}
 We also recall that $b=1$ corresponds to the $U(1)$-generator and $b=2,...,k$---to the $SU(k)$ Cartan generators from (\ref{suktimesu1cartan}) and that the basis is $N\times N$ dimensional (with the $SU(k)$ Cartan generators extended by appropriately adding zeros).
  
 At next order in $\Delta$,  the zero modes of the Laplacian---and, by eqns.~(\ref{slashed slahed d}) and (\ref{relations  1}), of the dotted fermions---are lifted. Here, we calculate the shift of the lowest eigenvalues of Laplacian away from zero to leading nontrivial order in $\Delta$. Such an estimate was carried out in \cite{Anber:2022qsz} in the $SU(2)$ case, and we generalize it here for $SU(N)$ and arbitrary $k$. 
 
 We use first-order perturbation theory  to determine the shift of the lowest eigenvalue of the operator $- D_\mu D_\mu$ away from zero. 
 To this order, we need to  find the eigenvalues of the $k \times k$ matrix 
 \begin{eqnarray}
E^{ba} &\equiv& \int\limits_{\T^4} \tr \phi^{(0) b} (- D_\mu D_\mu)_{pert.} \phi^{(0) a} =    \int\limits_{\T^4} \tr (D_\mu)_{pert.} \phi^{(0)  b} (D_\mu)_{pert.} \phi^{(0) a} \nonumber \\
 &=& - \int\limits_{\T^4} \tr [A_\mu^{pert.}, \phi^{(0)b}][A_\mu^{pert.}, \phi^{(0)a}]
 \end{eqnarray}
We used the definition of the covariant derivative, see eqn.~(\ref{EOM}). The  leading-order $\Delta$-expansion of the background (\ref{perturbed A}) shows that the background $A_\mu$ is perturbed by, to order $\sqrt{\Delta}$, by:
\begin{eqnarray} \label{perturbed A2}
A_\mu^{pert.}=  \sqrt{\Delta}\left[\begin{array}{cc}0 & w_\mu\\ w_\mu^{\dagger}&0\end{array}\right]   
 + {\Delta}\left[\begin{array}{cc}{\cal{S}}^{(k)}_\mu & 0\\  0&{\cal{S}}^{(\ell)}_\mu\end{array}\right] + {\cal{O}}(\Delta^{3 \over 2})\,,
\end{eqnarray}
using the $k$-$\ell$ block decomposition of  $SU(N)$ as in (\ref{perturbed A}) (the leading order perturbation $w_\mu$ is in the $k \times \ell$ and $\ell \times k$ components of the $N\times N$ matrix). Then, we find, to order $\Delta$:
\begin{eqnarray} \label{energyshifts}
E^{11} &=& {2 N  \Delta \over k (N-k)}\sum\limits_{C'=1}^k \sum\limits_{D=1}^\ell {1\over V} \int\limits_{\T^4}  (w_\mu^\dagger)_{DC'} (w_\mu)_{C'D}  \nonumber \\
E^{1b}&=& {2 \Delta \over k (N-k)} \sum\limits_{C'=1}^k (\tilde H^b )_{C'C'}   \sum\limits_{D=1}^\ell {1\over V}  \int\limits_{\T^4} (w_\mu)_{C'D} (w_\mu^\dagger)_{DC'}, ~ \text{for}~b=2,...,k,  \\
E^{ab} &=& {2} \Delta \sum\limits_{C'=1}^k(\tilde H^a \tilde H^b)_{C'C'}   \sum\limits_{D=1}^\ell {1\over V}  \int\limits_{\T^4} (w_\mu)_{C'D} (w_\mu^\dagger)_{DC'}, ~ \text{for}~a,b=2,...,k~.  \nonumber 
\end{eqnarray}

 To proceed, we have to evaluate the integral entering all energy shifts above. Using the formulae\footnote{For the reader who wants to reproduce (\ref{integral1}), we offer a guide to the relevant equations in \cite{Anber:2023sjn}: the functions $(w_\mu)_{C'D}$ are defined in eqns.~(4.20, 4.21) there, while $\Phi_{C'C}$ is in (3.21). For $r=k$, one finds that the coefficients ${\cal{C}}_4^A = 0$, while   ${\cal{C}}_2^A$ are determined from eqn.~(5.3). The integral (\ref{perturbed A2}) is then computed yielding the quoted result.} from \cite{Anber:2023sjn}, we find, $\forall C' =1,\ldots k$:
\begin{eqnarray} \label{integral1}
\sum\limits_{D=1}^\ell {1\over V}  \int\limits_{\T^4} (w_\mu)_{C'D} (w_\mu^\dagger)_{DC'} &=& {2 \pi \over k N}{1 \over \sqrt{V}}~ \rightarrow  {2 \pi \over k N}{1 \over L^2},    
\end{eqnarray}
where we replaced $\sqrt{V}$ by $L^2$ for brevity. This  then gives for (\ref{energyshifts}) the diagonal result
\begin{eqnarray}\label{energyshifts1}
E^{11} &=& {4 \pi \over k (N-k)} {\Delta \over L^2}~, \nonumber   \\
E^{1b} &=& 0~,\\
E^{ab} & =& {4 \pi \over k N} {\Delta\over L^2} \; \tr \tilde H^a \tilde H^b =  \delta^{ab}  {4 \pi \over k N} {\Delta\over L^2}, \;
\text{for} \; a,b = 2,...,k~.\nonumber
\end{eqnarray}  

Because of the diagonality of the above matrix elements, there is no need to change the unperturbed basis of eigenvectors, and we take the leading-order wave functions to be simply the ones in (\ref{laplacianzeromodes}), $\phi^{(0) b}$.
From (\ref{propagator 2}), we then find that
the contribution of the order-$\Delta$ eigenvalues affects the $\bar\lambda-\bar\lambda$ propagator. Adding to
(\ref{22element}) the contribution of the eigenvalues (\ref{energyshifts1}), we find \begin{eqnarray}\label{22element}
&&\langle \bar\lambda_b^{\dot\alpha}(x) \bar\lambda_{b' \dot\beta}(y) \rangle_{unnorm.}  =  \delta_{bb'} {\cal{D}}_k^f(m) \; {g^2 \over 2 V} \sum\limits_{p_\mu \in {2 \pi\over L_\mu} \Z}' { 
\delta_{\dot\beta}^{\dot\alpha} m \over p_\mu^2 + |m|^2} e^{i p_\mu (x_\mu - y_\mu)}~   \\
&& + \delta_{bb'}  \delta_{b1}  {\cal{D}}_k^f(m) \; {g^2 \over 2 V} \delta^{\dot\alpha}_{\dot\beta}  {m \over {4 \pi \over k(N-k)} {\Delta\over L^2}  +  |m|^2 }~+\delta_{bb'} \sum\limits_{a=2}^k \delta_{ba}{\cal{D}}_k^f(m) \; {g^2 \over 2 V} \delta^{\dot\alpha}_{\dot\beta}  {m \over {4 \pi \over kN} {\Delta\over L^2}  +  |m|^2 }. \nonumber  
\end{eqnarray} 

We next recall that the SYM result is reproduced by taking $m \rightarrow 0$ while keeping $\Delta$ strictly positive, hence, parametrically $\Delta \gg (m L)^2$. Thus, including the coefficient, in what follows we study the small soft mass limit where $|m|L$ obeys
 \begin{eqnarray}\label{smallmvsdelta}
{|m|L \over \sqrt{c \Delta}} \ll 1~, ~ c = {4 \pi \over k N}~.
 \end{eqnarray}
In this limit, further assuming\footnote{For brevity only: this assumption allows us to write the terms on the last line in (\ref{22element}) as one.} that $k \ll N$, we rewrite the propagator (\ref{22element}) in the simpler form
 \begin{eqnarray}\label{22element2}
 \langle \bar\lambda_b^{\dot\alpha}(x) \bar\lambda_{b' \dot\beta}(y) \rangle_{unnorm.}  &=&  \delta_{bb'} {\cal{D}}_k^f(m) \; {g^2 \over 2 V} \sum\limits_{p_\mu \in {2 \pi\over L_\mu} \Z}' { 
\delta_{\dot\beta}^{\dot\alpha} m \over p_\mu^2 + |m|^2} e^{i p_\mu (x_\mu - y_\mu)}~   \\
&& +  \delta_{bb'}   {\cal{D}}_k^f(m) \; {g^2 \over 2 V}\; \delta^{\dot\alpha}_{\dot\beta} \; {m L^2 \over c \Delta } (1 - {|m|^2 L^2 \over c \Delta} + \ldots). \nonumber  
\end{eqnarray} 
Let us now make some comments on the limit (\ref{smallmvsdelta}) and the expressions for the propagators and eigenvalues in the $Q={k/N}$ background that we use throughout this paper:
\begin{enumerate}
\item From now on, we shall ignore terms suppressed in the limit (\ref{smallmvsdelta}) and keep only the leading term in the propagator (\ref{22element2}), proportional to $
\sim {m L^2 \over c \Delta }$.
\item We also note that all Laplacian eigenvalues receive order-$\Delta$ corrections. This includes the  $p_\mu^2 \ne0$  eigenvalues of the free Laplacian whose contributions to the propagator are on the first line in (\ref{22element2}). These corrections are in principle computable via perturbation theory, but this is an arduous task shall not attempt here. Thus, in the limit (\ref{smallmvsdelta}), we   ignore the additive $|m|^2$ terms in the denominator of the first line in (\ref{22element2}) and replace the propagator (\ref{22element2}) by:
\begin{eqnarray}\label{22element3}
 \langle \bar\lambda_b^{\dot\alpha}(x) \bar\lambda_{b' \dot\beta}(y) \rangle_{unnorm.}  &=&  \delta_{bb'}  {\cal{D}}_k^f(m) \; {g^2 \over 2 V} \; \delta_{\dot\beta}^{\dot\alpha} \; \left({m  L^2 \over c \Delta } + \sum\limits_{p_\mu \in {2 \pi\over L_\mu} \Z}' { 
 m \; e^{i p_\mu (x_\mu - y_\mu)} \over p_\mu^2  } \right), \nonumber \\
\end{eqnarray} 
recalling that the sum excludes $p_\mu^2 =0$. 
\begin{enumerate}
\item We also note that the sum in the second term in the propagator (\ref{22element3}), while divergent and in need of regularization (we discuss this later), scales as $m L^2$, and is thus parametrically suppressed compared to the first term, due to the fact that $\Delta \ll 1$.

\item We stress that the comment here applies to all other propagators in the $Q={k/N}$ background computed in the $\Delta$ expansion: the $|m|^2$ contributions in the denominators are to be ignored in the $(|m| L)^2 \ll \Delta$ limit.
\end{enumerate}
\item Finally, we note that, in contrast with  the $22$ element (\ref{22element2}), the $11$, $12$, and $21$ elements of the  propagator (\ref{propagator 2}) do not receive a similar $ {1 \over \Delta}$ contribution, since the leading-order unperturbed eigenvectors (\ref{laplacianzeromodes}) obey $\partial_\mu \phi^{(0) b} = 0$.

There are, however, other contributions to the $11, 12$ and $21$ elements  of the propagator  (\ref{propagator 2}) from the order-$\Delta^0$ zero eigenvalues of the Laplacian (lifted at order $\Delta$), but which do not scale as $1 \over \Delta$ and are  in principle computable. 
To see this,  we note that the background entering the covariant derivatives  in the $11$, $12$, and $21$ elements of (\ref{propagator 2}) is shifted at order $\sqrt{\Delta}$. The eigenvectors (\ref{laplacianzeromodes}) are also shifted at order-$\Delta$, in a manner computable in perturbation theory. Both these shifts conspire to cancel the $1\over \Delta$ contribution from $\omega_n^2 = 0 + {\cal{O}}(\Delta)$ and lead to order-$1$ contributions to the propagators. In the spirit of keeping only the leading $|m|^2 L^2 \over c \Delta$ terms in all our expansions, we shall not need to compute these corrections here.
\end{enumerate}

\subsubsection{The propagator of the non-Cartan components of the diagonal $SU(k) \times U(1)$} 
\label{sec:noncartansuktimesu1}

We now continue with finding the free Laplacian eigenfunctions with $B' \ne C'$, obeying the BCS (\ref{BCS lambda A}). For every choice of an above-the-diagonal element $B',C'$ (such that $B' > C'$; recall that there are $k(k-1)/2$ such choices) there  are Hermitean eigenfunctions of the Laplacian which  involve both the $\phi_{B'C'}$ and $\phi_{C'B'}$ components, with  all other components vanishing.\footnote{As opposed to the Cartan sector, in the off-diagonal part of the $SU(k)$ matrix, hermiticity can be satisfied with complex eigenfunctions, explicitly given in (\ref{noncartansuktimesu1}).} The eigenfunctions are labeled by a set of integers $n_\mu$ as well as a choice of $B' > C'$. Thus, the eigenfunction of the Laplacian $\phi_{(n_\mu, B' > C')}$ is a $k\times k$ hermitean matrix with only two nonzero off-diagonal entries, which are complex conjugates to each other. There are two such linearly independent Hermitean matrices for every choice of $B' > C'$, corresponding to the fact that the Hermitean matrices (schematically, showing only the two nonzero entries inside the $k \times k$ matrix)
$$
\left(\begin{array}{cc} 0 & f \cr f^* & 0\end{array}\right) ~\text{and} \left(\begin{array}{cc} 0 & i f \cr -i f^* & 0\end{array}\right) 
$$
are linearly independent under addition with real coefficients  (a restriction needed in order to preserve the Hermiticity of the sum of two arbitrary hermitean matrices).

In a somewhat baroque but explicit notation, we use $(\phi_{(n_\mu, B' > C', \alpha)})_{D'E'}$ to label the $D'E'$ element of the $k \times k$ matrix representing the Hermitean eigenfunction of the free Laplacian:
\begin{eqnarray}\label{noncartansuktimesu1}
&&(\phi_{(n_\mu, B' > C', \alpha)})_{D'E'} \\
&&={1\over\sqrt{2}}  {\alpha \over \sqrt{V}}   e^{  i 2 \pi (n_\mu + \delta_{\mu 2} {B'-C'\over k}) {x_\mu \over L_\mu}} \delta_{B'D'} \delta_{C' E'} + {1\over\sqrt{2}}    {\alpha^* \over \sqrt{V}}  e^{ - i 2 \pi (n_\mu + \delta_{\mu 2} {B'-C'\over k}) {x_\mu \over L_\mu}}   \delta_{B'E'} \delta_{D' C'}, \nonumber \\
&&\text{with eigenvalue} ~\omega_n^2 =  \sum\limits_{\mu=1}^4 \left(2 \pi \;{n_\mu+ \delta_{\mu 2} {B'-C'\over k} \over L_\mu}\right)^2~,~~ n_\mu \in \Z, ~ \alpha = \{1, i\}~,\nonumber
\end{eqnarray}
where the  normalization factor is worked out in (\ref{normphin 11}) below (and solely for brevity we did not indicate the $B'C'$-dependence when denoting the eigenvalue by $\omega_n^2$).
 Clearly the boundary conditions (\ref{BCS lambda A}) are obeyed. Hermiticity is also manifest, as $(\phi_{(n_\mu, B'>C', \alpha)})_{D'E'}^* = (\phi_{(n_\mu, B'>C', \alpha)})_{E'D'}$. The fact that the normalization condition (\ref{normphin1}) holds follows by explicit  calculation  from (\ref{noncartansuktimesu1}). Mindful about the $B'>C'$ and $E'>F'$ condition, we obtain for the inner product:\begin{eqnarray} \label{normphin 11}
&& \int_{\T^4} \tr \phi_{(n_\mu, B'> C', \alpha)} \phi_{(m_\mu, E'>F', \beta)}= \int_{\T^4} \sum\limits_{D',G'=1}^k (\phi_{(n_\mu, B'>C', \alpha})_{D'G'}(x) (\phi_{(m_\mu, E'> F',\beta)})_{G'D'}(x)  \nonumber \\
&&= {1 \over V} \delta_{B'E'} \delta_{C'F'}  \int_{\T^4}d^4 x ( {\alpha^* \beta \over 2}  e^{-  \sum\limits_{\mu=1}^4  i 2 \pi (n_\mu-m_\mu) {x_\mu \over L_\mu}} + {\alpha \beta^* \over 2}  e^{  \sum\limits_{\mu=1}^4  i 2 \pi (n_\mu-m_\mu) {x_\mu \over L_\mu}} ) \nonumber \\
&&=\delta_{B'E'} \delta_{C'F'}   \prod_\mu \delta_{n_\mu m_\mu}~{\alpha \beta^* + \alpha^* \beta \over 2} = \delta_{B'E'} \delta_{C'F'}    \delta_{\alpha \beta} \prod_\mu\delta_{n_\mu m_\mu}, 
 \end{eqnarray}
remembering that $\alpha$ and $\beta$ are either $1$ or $i$ and using $\delta_{\alpha\beta}=1$ if $\alpha=\beta$ and zero otherwise. In eqn.~(\ref{normphin 11}), we have thus showed that satisfying (\ref{normphin1})  requires the overall ${1\over\sqrt{2}}  $ factor in (\ref{noncartansuktimesu1}), and that the different eigenfunctions are indeed orthogonal.

 Before we continue, we also note that since each eigenfunction (\ref{noncartansuktimesu1}) only involves the $\phi_{B'C'}$ and $\phi_{C'B'}$ components with some given $B'\ne C'$, the only  nonzero propagators are  of the form 
$\langle \lambda_{B'C'}(x) \lambda_{C' B'}(y) \rangle$, etc., as follows from (\ref{propagator 2}) and the outer product definition (\ref{outer 1}), and as we now show in detail.
We begin  by  evaluating the contribution of the eigenfunctions with $B' > C'$ (summed over $n_\mu$, $\alpha=1,i$, and over $B'>C'$, i.e. over all eigenvalues of the Laplacian) to the $22$ element of  (\ref{propagator 2}), remembering that it is understood as an outer product, i.e. carries two sets of adjoint indices as in (\ref{outer 1}). On one hand, this matrix element gives the unnormalized propagator
\begin{eqnarray}
\label{lambdabarsquared1}
\langle \bar\lambda^{\dot\alpha}_{D'E'}(x) \bar\lambda_{\dot\beta \; F'G'}(y) \rangle~,
\end{eqnarray}
where we recall  that here $D' \ne E'$ and $F' \ne G'$, because we are computing the off-diagonal $SU(k)$ elements in this section.
By (\ref{propagator 2}) the $D' \ne E'$, $F' \ne G'$ matrix element (\ref{lambdabarsquared1}) is equal to the sum over all eigenfunctions (\ref{noncartansuktimesu1}), i.e. over all $n_\mu$ and $B'>C'$:
\begin{eqnarray} \label{result22}
&& {g^2 \over 2} {\cal{D}}_k^f(m) \; 
 \sum\limits_{{n_\mu, B'>C', \alpha}} { m\delta^{\dot\alpha}_{\dot\beta} \over |m|^2 + \omega_n^2 } \; (\phi_{(n_\mu, B'>C',\alpha)})_{D'E'}(x)(\phi_{(n_\mu, B'>C',\alpha)})_{F'G'}(y) \nonumber \\
&&={g^2 \over 4V} {\cal{D}}_k^f(m) \sum\limits_{p_\mu = {2 \pi n_\mu \over L_\mu}, B'>C'}  
  {m\delta^{\dot\alpha}_{\dot\beta} \over |m|^2 +  (p_\mu  + \delta_{\mu 2} {2 \pi \over L_2}{B'-C'\over k})^2 }\\
 && \times  \left(
 e^{  i   (x_\mu + y_\mu)(p_\mu + \delta_{\mu,2} {2 \pi (B'-C')\over k L_2})}  \delta_{B'D'} \delta_{C'E'} \delta_{B'F'}  \delta_{C'G'} (\sum_\alpha \alpha^2) \right. \nonumber \\
 && \left. ~~+e^{i   (x_\mu - y_\mu)(p_\mu + \delta_{\mu,2} {2 \pi (B'-C')\over k L_2})}  \delta_{B'D'} \delta_{C'E'} \delta_{B'G'} \delta_{C'F'} (\sum_\alpha |\alpha|^2)  \right. \nonumber \\
&& \left.~~ + e^{- i   (x_\mu - y_\mu)(p_\mu + \delta_{\mu,2} {2 \pi (B'-C')\over k L_2})} \delta_{B'E'} \delta_{C'D'} \delta_{B'F'} \delta_{C'G'}   (\sum_\alpha |\alpha|^2)  \right. \nonumber \\
&& \left. ~~+ e^{- i   (x_\mu + y_\mu)(p_\mu + \delta_{\mu,2} {2 \pi (B'-C')\over k L_2})} 
 \delta_{B'E'} \delta_{C'D'} \delta_{B'G'} \delta_{C'F'} (\sum_\alpha (\alpha^*)^2\right)    \nonumber 
\end{eqnarray}
 Recalling that $\alpha = 1, i$, the terms with summing $\alpha^2$ or $(\alpha^*)^2$ drop out, while the others give a factor of $2$. The Kronecker symbols remove the sum over group indices, but we have to be conscious about the $B' > C'$ condition. Proceeding carefully, we rewrite (\ref{result22}) as:
\begin{eqnarray} \label{result23}
&&{g^2 \over 2V} {\cal{D}}_k^f(m) \times \\
&& \left[ \sum\limits_{p_\mu = {2 \pi n_\mu \over L_\mu}}  
  { m\delta^{\dot\alpha}_{\dot\beta} \over |m|^2 +  (p_\mu  + \delta_{\mu 2} {2 \pi \over L_2}{D'-E'\over k})^2 }  
  \left( e^{  i    x_\mu(p_\mu + \delta_{\mu,2} {2 \pi (D'-E')\over k L_2}) - i y_\mu(p_\mu - \delta_{\mu,2} {2 \pi (F'-G')\over k L_2}) } \delta_{D'G'}\delta_{E'F'} \theta_{D'E'}   \right)  \right. \nonumber \\
  &&\left.+  \sum\limits_{p_\mu = {2 \pi n_\mu \over L_\mu}}  
  { m\delta^{\dot\alpha}_{\dot\beta} \over |m|^2 +  (p_\mu  - \delta_{\mu 2} {2 \pi \over L_2}{D'-E'\over k})^2 }  
  \left( e^{-  i    x_\mu(p_\mu - \delta_{\mu,2} {2 \pi (D'-E')\over k L_2}) + i y_\mu(p_\mu + \delta_{\mu,2} {2 \pi (F'-G')\over k L_2}) } \delta_{D'G'} \delta_{E'F'} \theta_{E'D'}  \right)\right],  \nonumber \\
\end{eqnarray}
where we defined the ``theta-function'' $\theta_{E'D'} = 1$ if $E' > D'$ and $\theta_{E'D'}=0$  if $E'<D'$.
 We now note that the two lines have the same Kronecker symbols   but opposite theta-functions). We can combine  these two lines together, allowing us to drop the theta-function, upon changing the sign of $p_\mu$ in the second line, giving the following expression for the non-Cartan components of the $\bar\lambda\bar\lambda$ propagator 
 \begin{eqnarray}
\label{lambdabarsquared2}
&&\langle \bar\lambda^{\dot\alpha}_{D'E'}(x) \bar\lambda_{\dot\beta \; F'G'}(y) \rangle~ \nonumber \\
&& = {g^2 \over 2V} {\cal{D}}_k^f(m)  \\
&& \times \sum\limits_{p_\mu = {2 \pi n_\mu \over L_\mu}}  
  {m\delta^{\dot\alpha}_{\dot\beta} \over |m|^2 +  (p_\mu  + \delta_{\mu 2} {2 \pi \over L_2}{D'-E'\over k})^2 }  
    e^{  i    x_\mu(p_\mu + \delta_{\mu,2} {2 \pi (D'-E')\over k L_2}) - i y_\mu(p_\mu - \delta_{\mu,2} {2 \pi (F'-G')\over k L_2}) } \delta_{D'G'}\delta_{E'F'}~.    \nonumber
\end{eqnarray}
After this exercise, finding the $11$ element of   (\ref{propagator 2}) becomes straightforward. It is easily seen  that only the symmetric part of $\sigma_\mu \bar\sigma_\nu$ survives (recall (\ref{identity 12})) with the derivatives canceling the $\omega_n^2$ factor in the denominator and leading to
\begin{eqnarray}\label{lambdasquared2}
&&\langle \lambda_{\alpha \; D'E'}(x) \lambda^\beta_{F'G'}(y) \rangle \nonumber\\
&&={g^2 \over 2V} {\cal{D}}_k^f(m) \\
&&\times  \sum\limits_{p_\mu = {2 \pi n_\mu \over L_\mu}}  
{ m^*  \delta_\alpha^\beta \over |m|^2 +(p_\mu  + \delta_{\mu 2} {2 \pi \over L_2}{D'-E'\over k})^2  }  e^{  i    x_\mu(p_\mu + \delta_{\mu,2} {2 \pi (D'-E')\over k L_2}) - i y_\mu(p_\mu - \delta_{\mu,2} {2 \pi (F'-G')\over k L_2}) } \delta_{D'G'}\delta_{E'F'}~.  \nonumber 
\end{eqnarray}
The kinetic part of the propagator can similarly be calculated from the $12$ part of (\ref{propagator 2}).

\subsection{The propagator of the $SU(\ell)$   components}

\label{appx:suellpropagator}

 Finally, we turn to finding the propagators (\ref{propagator 2}) in the $SU(\ell)$ space. We need to find the eigenfunctions of the free Laplacian (\ref{adjoint1}), subject to the BCS given by (\ref{BCS lambda a}). For now, we generically denote these by  $\phi$,   now $SU(\ell)$-algebra elements (hermiticity will be imposed later), periodic in the $x_1$ and $x_2$ directions. Using the explicit form of the transition functions, eqn.~(\ref{the set of transition functions for Q equal r over N, general solution}), the BCS in the $x_3$ and $x_4$ directions can be rewritten as
\begin{eqnarray}\label{cond 34}
\phi(x+L_3\hat e_3)= P_\ell \phi(x) P_\ell^{-1}\,,\quad \phi(x+L_4\hat e_4) = Q_\ell\phi(x)Q_\ell^{-1}\,,
\end{eqnarray} 
where $P_\ell$ and $Q_\ell$ are the $SU(\ell)$ shift and clock matrices that satisfy $P_\ell Q_\ell=Q_\ell P_\ell e^{i\frac{2\pi}{\ell}}$. We decompose $\phi(x)$ using the basis for $SU(\ell)$ generators constructed using the $P_\ell$ and $Q_\ell$ matrices \cite{Gonzalez-Arroyo:1982hyq,GonzalezArroyo:1987ycm}:
\begin{eqnarray}\label{exp Jp}
\phi(x)=\sum_{\bm p=(p_3,p_4)\neq (0,0)}\phi_{\bm p}(x)J_{\bf p}\,,~ \phi \in su(\ell), 
\end{eqnarray}
where $p_3,p_4$ each run from $0$ to $\ell-1$ (i.e.~$\bm p = (p_3,p_4)\in \Z_\ell^2$), with $\bm p\neq \bm 0$, and $J_{\bf p}\in su(\ell)$ is given by\footnote{The phase factors in the definition of $J_{\bm p}$ are chosen such that (\ref{JPELL1}) and (\ref{J identity}) hold.}
\begin{eqnarray}\label{JPELL1}
J_{\bm p}=e^{-i\frac{\pi p_3p_4}{\ell}}Q_\ell^{-p_3}P_\ell^{p_4}\,,~ J_{\bm p}^\dagger =  e^{- i {\pi p_3 p_4 \over \ell}} Q_\ell^{p_3} P_\ell^{- p_4} \equiv J_{- \bm p},\end{eqnarray}
satisfying the normalization condition
\begin{eqnarray}\label{J identity}
\mbox{tr}\left[J_{\bm p}J_{\bm p'}\right]=\mbox{tr}\left[\bm 1_{\ell\times \ell}\right]\delta_{\bm p,-\bm p'}=\ell\delta_{\bm p,-\bm p'}\, (\text{or} ~\mbox{tr}\left[J_{\bm p}J^\dagger_{\bm p'}\right] = \ell \delta_{\bm p, \bm p'}). 
\end{eqnarray}
 
With the decomposition (\ref{exp Jp}), we find that  the BCS (\ref{BCS lambda a}, \ref{cond 34}) translate into the following conditions on the modes $\phi_{\bm p}(x)$:
\begin{eqnarray}\label{phisuellbc}
\phi_{\bm p}(x+L_3\hat e_3)&=&e^{-i\frac{2\pi p_3}{\ell}}\phi_{\bm p}(x)\,, \quad \phi_{\bm p}(x+L_4\hat e_4)=e^{-i\frac{2\pi p_4}{\ell}}\phi_{\bm p}(x)\,, \nonumber \\
\phi_{\bm p}( x + L_1 \hat e_1) &=& \phi_{\bm p}( x)\, , \qquad  \qquad \; \phi_{\bm p}( x + L_2 \hat e_2) = \phi_{\bm p}( x)~.
\end{eqnarray}
Thus, we conclude that the eigenfunctions of the free Laplacian (\ref{adjoint1}) on $\T^4$, in the $SU(\ell)$ subspace, i.e. subject to (\ref{phisuellbc}) are labelled by an integer  $n_\mu \in \Z$ plus the Lie algebra label $\bm p \in \Z_\ell^2, \bm p \ne \bm 0$:
\begin{eqnarray}\label{phiell1}
 \phi_{\bm p, n_\mu}  &=&  e^{- i 2 \pi {x_\mu \over L_\mu} (n_\mu  + \delta_{\mu 3} \frac{p_3}{\ell } + \delta_{\mu 4} \frac{p_4}{\ell}) } J_{\bm p}, n_\mu \in \Z,   p_3 \in [0,\ell-1], p_4 \in [0, \ell -1], (p_3,p_4) \ne (0,0),\ \nonumber \\
\omega_n^2 &\rightarrow& \omega_{\bm p, n_\mu}^2 = \sum\limits_{\mu=1}^4 \left({2 \pi   \over L_\mu}(n_\mu + \delta_{\mu 3} {p_3 \over \ell} + \delta_{\mu 4} {  p_4 \over \ell })\right)^2, \end{eqnarray}
 with $J_{\bm p}$ from (\ref{JPELL1}). Eqn.~(\ref{phiell1}) gives the complex eigenfunctions of the adjoint Laplacian  with eigenvalue $\omega_{\bm{p}, n_\mu}^2$, obeying the BCS (\ref{cond 34}) and periodic in $x_{1,2}$. It is easy to see, using (\ref{J identity}),  and integrating over $\T^4$, that they are orthogonal in the complex norm $\int \tr \phi^\dagger_{\bm p, n_\mu} \phi_{\bm p', n'_\mu} \sim \delta_{\bm p, \bm p'} \delta_{n,n'}$. However, n order to define our path integral, we need to find a hermitean basis, with norm (\ref{normphin1}), a task we take up below.
 
 For the purpose of counting independent solutions, imposing hermiticity, and figuring out the normalization, we stress that in what follows we keep the range of $(p_3,p_4)$ as indicated in (\ref{phiell1}) above, i.e. in their fundamental domain. Allowing $p_3, p_4$ to ``wrap around'' (i.e. take values outside the $[0,\ell-1]$ range) can be compensated by integer shifts of $n_\mu$. 

Next, we also note that the eigenspace of the Laplacian with eigenvalue  $\omega_{\bm p, n_\mu}^2$ is degenerate.
To this end, we first observe that eigenfunctions $\phi_{\bm p, n_\mu}$ with $n_1$ and $-n_1$, keeping all other labels the same, have the same eigenvalue, likewise for $n_2 \rightarrow - n_2$. The degeneracy in the $1,2$ directions can be taken into account by taking the real eigenfunctions from (\ref{definitionofF}), $f_\mu = (\sin, \cos)$, and redefining the eigenfunctions (\ref{phiell1}) by making the $x_{1,2}$-dependent parts real. Thus we define 
\begin{eqnarray}\label{phiell11}
 \phi_{p_3, p_4, n_\mu, f_1, f_2}  &=&  f_1(n_1 x_1) f_2 (n_2 x_2)
 e^{- i 2 \pi {x_3 \over L_3} (n_3 +  \frac{p_3}{\ell }) } e^{-  i 2 \pi {x_4 \over L_3} (n_4 +  \frac{p_4}{\ell}) } J_{p_3, p_4}, \end{eqnarray}
and restrict $n_1, n_2 \ge 0$.\footnote{Noting that now $n_{1,2}=0$ is allowed, while of course if $f_\mu = \sin$, the wave function for $n_\mu=0$ will vanish.} 

One further observes that there is also a degeneracy upon reflecting $n_3$: one needs to take $n_3 \rightarrow - n_3 - 1$ simultaneously with $p_3 \rightarrow \ell - p_3$. 
Let us call the image of (\ref{phiell1}) under this reflection  $\phi_{p_3, p_4, n_\mu, f_1, f_2}'$:
\begin{eqnarray}
\label{phiell12}
 \phi_{p_3, p_4, n_\mu, f_1, f_2}'  \equiv f_1(n_1 x_1) f_2 (n_2 x_2)
 e^{ i 2 \pi  {x_3 \over L_3} (n_3 +  \frac{p_3}{\ell }) } e^{-  i 2 \pi {x_4 \over L_3} (n_4 +  \frac{p_4}{\ell}) } J_{\ell-p_3,p_4}     \end{eqnarray}
 Now, if we restrict $n_3 \ge 0$ and allow all $p_3 \in [0, \ell-1]$ ($p_3$ is integer), the two sets of functions (\ref{phiell1}) and (\ref{phiell12}) account for all states degenerate due to the above $n_3$ reflection. This is because $n_3 \rightarrow - n_3 -1$, being a ``reflection of the $n_3$-axis around $n_3 = -1/2$,'' maps all $n_3 \ge 0$ states (with some $p_3$) to $n_3 < 0$ states (with a different $p_3$).
 In other words, all $n_3 <0$ states for any $p_3$ are obtained upon reflection of $n_3 \ge 0$ states.

Finally, a reflection of $n_4$, $n_4 \rightarrow - n_4$ also leaves $\omega_{\bm p, n_\mu}$ invariant if it is accompanied by 
$p_4 \rightarrow \ell - p_4$. Thus, we shall also restrict $n_4 \ge 0$, while allowing all $p_3$, but we have to account for the images under the $n_4$-reflection of both (\ref{phiell1}) and (\ref{phiell12}). We thus end with sixteen functions (accounting for the four choices of $f_{1,2}$ in each of the four functions below)\
\begin{eqnarray}\label{phiIdefinition1}
\phi^1_{p_3, p_4, n_\mu, f_1, f_2}  &=&  f_1(n_1 x_1) f_2 (n_2 x_2)
 e^{- i 2 \pi {x_3 \over L_3} (n_3 +  \frac{p_3}{\ell }) } e^{-  i 2 \pi {x_4 \over L_3} (n_4 +  \frac{p_4}{\ell}) } J_{p_3, p_4},  \\
 \phi_{p_3, p_4, n_\mu, f_1, f_2}^2  &=&  f_1(n_1 x_1) f_2 (n_2 x_2)
 e^{- i 2 \pi {x_3 \over L_3} (n_3 +  \frac{p_3}{\ell }) } e^{  i 2 \pi {x_4 \over L_3} (n_4 +  \frac{p_4}{\ell}) } J_{p_3, \ell - p_4}, 
 \nonumber \\
  \phi_{p_3, p_4, n_\mu, f_1, f_2}^3  &=& f_1(n_1 x_1) f_2 (n_2 x_2)
 e^{ i 2 \pi  {x_3 \over L_3} (n_3 +  \frac{p_3}{\ell }) } e^{-  i 2 \pi {x_4 \over L_3} (n_4 +  \frac{p_4}{\ell}) } J_{\ell-p_3,p_4},  \nonumber \\
   \phi_{p_3, p_4, n_\mu, f_1, f_2}^4  &=& f_1(n_1 x_1) f_2 (n_2 x_2)
 e^{ i 2 \pi  {x_3 \over L_3} (n_3 +  \frac{p_3}{\ell }) } e^{  i 2 \pi {x_4 \over L_3} (n_4 +  \frac{p_4}{\ell}) } J_{\ell-p_3,\ell - p_4},\nonumber  \end{eqnarray}
 where now we restrict
 \begin{equation}\label{domain 1}
 n_\mu \ge 0, \forall \mu, ~\text{and} ~ p_3  \in [0, \ell-1], p_4  \in [0, \ell-1], (p_3,p_4) \ne (0,0),
\end{equation}
where we remind the reader that $p_{3,4}$ are integer valued.
These are the analogues of the $2^4$ functions $\prod_\mu e^{\pm i k_\mu x_\mu}$ for a free particle, whose real linear combinations give rise to the $2^4$ real functions $\prod_\mu f_\mu$ of (\ref{definitionofF}). 
The difference here is  the nontrivial embedding in the gauge group from the $J_{\bm p}$ factors.

It is clear from (\ref{phiIdefinition1}) that the factors multiplying $J_{\bm p}$ in $\phi^1$ and $\phi^4$ are $h.c.$ to each other, as are the factors multiplying $J_{\bm p}$ in $\phi^2$ and $\phi^3$. To study the behaviour of the $J$-factors upon hermitean conjugation, we note that
  (\ref{J identity}) and (\ref{JPELL1}) imply that
\begin{eqnarray}\label{signsJ}
J_{\ell - p_3, \ell-p_4} &=& (-1)^{\ell - p_3 - p_4} J_{-p_3, - p_4} = (-1)^{\ell - p_3 - p_4} J_{p_3,p_4}^\dagger, \\ 
J_{p_3, \ell-p_4} &=& (-1)^{p_3} J_{p_3, -p_4}= (-1)^{p_3} J_{-p_3, p_4}^\dagger = (-)^{p_3 + p_4} J_{\ell -p_3, p_4}^\dagger~. \nonumber
\end{eqnarray}
These relations, substituted into (\ref{phiIdefinition1}) imply that, up to signs, $\phi^1 \sim (\phi^4)^\dagger$, $\phi^2 \sim (\phi^3)^\dagger$. Explicitly, 
\begin{eqnarray}\label{conjugate1}
  \phi_{p_3, p_4, n_\mu, f_1, f_2}^3  &=&   (-)^{p_3 + p_4} (\phi^2_{p_3, p_4, n_\mu, f_1, f_2})^\dagger,  \nonumber \\
   \phi_{p_3, p_4, n_\mu, f_1, f_2}^4  &=& (-)^{\ell - p_3 - p_4} (\phi^1_{p_3, p_4, n_\mu, f_1, f_2} )^\dagger \nonumber ~.
 \end{eqnarray}
Thus, taking  sums and difference of $\phi^1$ and $\phi^{4}$, as well as of $\phi^2$ and $\phi^3$, we can obtain a basis of hermitean $2^4$ hermitean or antihermitean functions. 
To proceed, we first simplify the notation and define the functions:
\begin{eqnarray} \label{gdefinition}
g_\mu(x_\mu, n_\mu, p_\mu) \equiv e^{- i 2 \pi {x_\mu \over L_\mu} (n_\mu + {p_\mu \over \ell})}~, ~\mu = 3,4. \end{eqnarray}
Then, we redefine the functions $\phi^{1, 2, 3, 4} \rightarrow \psi^{1,2,3,4}$ by phase factors so that the hermitean conjugation (\ref{conjugate1}) works without signs. Explicitly, we take the functions $\psi^{i}$ are
\begin{eqnarray}
\psi^1_{p_3, p_4, n_\mu, f_1, f_2} &=& (-)^{\ell - p_3 - p_4}f_1 f_2 g_3 g_4 \; J_{p_3,p_4} \\
\psi^4_{p_3, p_4, n_\mu, f_1, f_2} &=& f_1 f_2 g_3^* g_4^* \; J_{\ell-p_3, \ell-p_4} = (\psi^1_{p_3, p_4, n_\mu, f_1, f_2})^\dagger \nonumber \\
\psi^2_{p_3, p_4, n_\mu, f_1, f_2} &=& (-)^{p_3 + p_4} f_1 f_2 g_3 g_4^* J_{p_3, \ell - p_4} \nonumber \\
\psi^3_{p_3, p_4, n_\mu, f_1, f_2} &=&   f_1 f_2 g_3^* g_4 \; J_{\ell- p_3,  p_4} = (\psi^2_{p_3, p_4, n_\mu, f_1, f_2})^\dagger~, \nonumber
\end{eqnarray}
where verifying the ${h.c.}$ relations follows immediately from (\ref{signsJ}). 
 We continue by introducing the hermitean eigenfunction basis, $\Psi^{a, \alpha}$, with $a = 1,2$ and $\alpha \in \{1, i\}$. 
 These functions are defined in terms of $\psi^1$ and $\psi^2$ and their hermitean conjugates as follows:
\begin{eqnarray} \label{PhiDefinitions}
\Phi^{1,\alpha}_{p_3, p_4, n_\mu, f_1, f_2} &=& c_1 (\alpha \psi^1_{p_3, p_4, n_\mu, f_1, f_2} + \alpha^* (\psi^1_{p_3, p_4, n_\mu, f_1, f_2})^\dagger) = c_1 f_1 f_2 \left[ \alpha (-)^{\ell- p_3 - p_4}  g_3 g_4 J_{p_3, p_4} + \alpha^*  g_3^* g_4^* J_{\ell-p_3, \ell - p_4}\right] \nonumber \\
&=& c_1 f_1 f_2 (-)^{\ell - p_3 - p_4} \left[ \alpha   g_3 g_4 J_{p_3, p_4} + \alpha^*  g_3^* g_4^*  J_{ p_3,  p_4}^\dagger \right] \nonumber \\
\Phi^{2,\alpha}_{p_3, p_4, n_\mu, f_1, f_2}  &=& c_2 (\alpha \psi^2_{p_3, p_4, n_\mu, f_1, f_2} + \alpha^* (\psi^2_{p_3, p_4, n_\mu, f_1, f_2})^\dagger))=
c_2 f_1 f_2 \left[  \alpha (-)^{p_3 + p_4}  g_3 g_4^* J_{p_3, \ell- p_4} + \alpha^*   g_3^* g_4 J_{\ell-p_3,  p_4}\right] \nonumber \\
 &=&
c_2 f_1 f_2 \left[  \alpha   g_3 g_4^* J_{\ell- p_3,   p_4}^\dagger + \alpha^*   g_3^* g_4 J_{\ell-p_3,  p_4}\right] ,  ~ \alpha = (1, i),  
 \end{eqnarray}
 We used (\ref{signsJ}) on the way\footnote{In particular, the last expression for $\Phi^{1, \alpha}$ implies that the $(-)^{\ell + p_3 + p_4}$ sign can be absorbed into $c_1$; we assume this in what follows.} and stress that the range of indices $p_3, p_4, n_\mu$ are in  (\ref{domain 1}) and $f_1, f_2$ are the $\sin$ and $\cos$  functions of (\ref{diagonalsukmodes}). We stress again that the expressions (\ref{PhiDefinitions}) are simply the analogue of the $2^4$ real eigenfunctions  of the free Laplacian with periodic BCS, the only difference being their nontrivial embedding into the gauge group, owing to the presence of a 't Hooft twist. 
 
Now, consider the inner product in order to determine the normalization $c_{1,2}$ of the Hermitean functions:
\begin{eqnarray}\label{normalizationPhi1 1}
\int\limits_{\T^4} \tr \Phi^{a, \alpha}_{p_3, p_4, n_\mu, f_1, f_2}  \Phi^{b, \beta}_{p'_3, p'_4, n'_\mu, f'_1, f'_2} 
\end{eqnarray} 
Before we actually calculate the norm, we make the following comments:
\begin{enumerate}
\item
It is immediately clear that the inner product vanishes, upon only integrating over $\T^4$ and taking (\ref{domain 1}) into account, unless $f_1=f_1'$ and $f_2 = f_2'$, as well as $n_\mu  = n_\mu'$, for all $\mu$, and $p_{3}=p_3'$, $p_4 = p_4'$ (this is just to say that eigenstates with different eigenvalues of the Laplacian $\omega_{\bm p, n_\mu}^2$ (\ref{phiell1}) are orthogonal). 
\item We also note that the inner product vanishes unless $a=b$. This is because the products with $a \ne b$ always involve integrals of $g_3 g_3$, $g_4 g_4$ (and their $h.c.$), which vanish, recalling (\ref{gdefinition}). 
\end{enumerate}
We are thus left to calculate the norms (\ref{normalizationPhi1}) for $a=b$.
For each $\Phi^{a, \alpha}$, the form given in terms of $J$ and $J^\dagger$ 
is the most useful to compute the traces. We begin, using (\ref{J identity})\footnote{Note that $\tr J_{p_3,p_4}^2$, which is nonzero for $p_3 = p_4 = \ell/2$, does not contribute because of the $x_3$, $x_4$ integrals.} and the integrals involving $f_1, f_2$, see paragraph after (\ref{definitionofF}), with
\begin{eqnarray}\label{normalizationPhi1}
\int\limits_{\T^4} \tr \Phi^{1, \alpha}_{p_3, p_4, n_\mu, f_1, f_2}  \Phi^{1, \beta}_{p_3, p_4, n_\mu, f_1, f_2} &=&c_1^2 \ell  (\alpha \beta^* + \alpha^* \beta) \int\limits_{\T^4} f_1^2 f_2^2 g_3 g_3^* g_4 g_4^* \nonumber \\
&=& \delta_{\alpha \beta} c_1^2 \; 2 \ell    V \left\{ \begin{array}{cc} {1\over 4}, & \text{if} \; n_1 \ne 0, n_2 \ne 0, \cr
{1 \over 2}, & \text{if} \; n_1 =0, n_2 \ne 0, \cr
{1 \over 2}, & \text{if} \; n_1 \ne 0, n_2 = 0, \cr
1, & \text{if} \; n_1 = 0, n_2 = 0. \end{array} \right. 
\end{eqnarray}
The expression for the norm of $\Phi^{2, \alpha}$ is identical. Thus we conclude that:
 \begin{eqnarray}\label{normalizationC}
c_{1}^2 = c_2^2 ={  {2}^{1- \delta_{n_1,0} - \delta_{n_2,0}} \over  \ell V} ~.
 \end{eqnarray}
 
 Before we continue to find the propagator from (\ref{propagator 2}), we summarize the final form of the functions $\Phi^{a, \alpha}$:
 \begin{eqnarray} \label{PhiDefinitions2}
\Phi^{1,\alpha}_{p_3, p_4, n_\mu, f_1, f_2} &=&  { \sqrt{2}^{1- \delta_{n_1,0} - \delta_{n_2,0}} \over \sqrt{\ell V}} f_1 f_2  \left[ \alpha   g_3 g_4 J_{p_3, p_4} + \alpha^*  g_3^* g_4^*  J_{ p_3,  p_4}^\dagger \right], \nonumber \\
\Phi^{2,\alpha}_{p_3, p_4, n_\mu, f_1, f_2}  &=&  { \sqrt{2}^{1- \delta_{n_1,0} - \delta_{n_2,0}} \over \sqrt{\ell V}} f_1 f_2 \left[  \alpha   g_3 g_4^* J_{\ell- p_3,   p_4}^\dagger + \alpha^*   g_3^* g_4 J_{\ell-p_3,  p_4}\right] ,   
 \end{eqnarray}
 where $f_1, f_2$ are from (\ref{definitionofF}), $g_3, g_4$ are from (\ref{gdefinition}), $\alpha = 1$ or $=i$, and $J_{\bm p}$ are from (\ref{JPELL1}). 
 
Now, turning to (\ref{propagator 2}), for a given eigenvalue $\omega^2_{\bm p, n_\mu}$ (\ref{phiell1}), to find the $22$ component of the propagator, we have to evaluate, skipping the overall $\delta^{\dot\alpha}_{\dot\beta} {\cal D}^f_k(m) {g^2 \over 2} {m \over \omega^{2}_{\bm p, n_\mu} + |m|^2}$ factor, the following sum over $a$ and $\alpha$ and $f_1$, $f_2$, keeping $p_3, p_4, n_\mu$ fixed:
 \begin{eqnarray}\label{partial 1} 
&& \sum\limits_{a=1}^2 \sum\limits_{\alpha = 1, i} \sum_{f_{1,2} = (sin, cos)} \Phi^{a, \alpha}(x) \otimes \Phi^{a, \alpha}(y) = 
 2 c_1^2   \sum\limits_{f_{1,2}= (sin, cos)} f_1(x_1) f_2(x_2) f_1(y_1) f_2 (y_2)   \\
&&\qquad  \times\left(  g_3(x_3) g_4(x_4) g_3^*(y_3) g_4^*(y_4) J_{p_3,p_4} \otimes J^\dagger_{p_3,p_4} +   g_3(x_3)   g_4^* (x_4) g_3^*(y_3) g_4(y_4) J^\dagger_{\ell - p_3, p_4} \otimes J_{\ell-p_3,p_4}  \right. \nonumber \\ 
&&\left. \qquad \; ~ +   g_3^*(x_3) g_4^*(x_4) g_3 (y_3) g_4 (y_4) J^\dagger_{p_3,p_4} \otimes J_{p_3,p_4} +  g_3^*(x_3)
  g_4 (x_4) g_3 (y_3) g_4^* (y_4)J_{\ell - p_3, p_4} \otimes J^\dagger_{\ell - p_3, p_4}   \right), \nonumber 
 \end{eqnarray}
where the overall factor of $2$ occurs because of the sum over $\alpha = 1, i$ )the last line is from $\Phi^2$ and the second to last line is from $\Phi^1$). 
The sum over $f_\mu$ can be dealt with (\ref{sum1}), which also works for $n_1$ or $n_2$ are zero, because the r.h.s. of (\ref{sum1}) gives unity (and recall that the different normalization for this case is taken care of by $c_{1}$, eqn.~(\ref{normalizationC})). We recall that 
\begin{eqnarray}\label{recall eqn1} 
 \sum\limits_{f_{1,2}= (sin, cos)} f_1(x_1) f_2(x_2) f_1(y_1) f_2 (y_2) =  
 {1 \over 4}  \prod\limits_{\mu=1}^2  \left( e^{ i {2\pi n_\lambda \over L_\lambda}( x_\lambda-y_\lambda)} + e^{-i {2\pi n_\lambda \over L_\lambda}( x_\lambda-y_\lambda)}  \right)  ,
\end{eqnarray}   allowing the sum over $n_{1,2}$ to be extended to negative values as well.

The way the   products of functions $g_3$, $g_4$ and their conjugates appear also makes it clear that the sum over $n_3, n_4$ can be extended to positive and negative values; however, we have to unravel the group structure. To this end, we consider the expression in brackets, the last two lines in (\ref{partial 1}), term-by-term. 

The first term  in (\ref{partial 1}) is
\begin{eqnarray}\label{first term}
&&   g_3(x_3) g_4(x_4) g_3^*(y_3) g_4^*(y_4) J_{p_3,p_4} \otimes J^\dagger_{p_3,p_4} \nonumber \\
&& \nonumber \\
&&
= e^{- i 2 \pi(n_3 + {p_3 \over \ell}) x_3  - i 2 \pi(n_4 + {p_4 \over \ell}) x_4  }J_{p_3,p_4} \otimes  e^{  i 2 \pi(n_3 + {p_3 \over \ell}) y_3    + i 2 \pi(n_4 + {p_4 \over \ell})  y_4  } J_{-p_3,-p_4},   
 \end{eqnarray}
 we recall that $n_{3, 4} \ge 0$ and $p_4$ is in the fundamental domain. 
 
 The second term  in (\ref{partial 1}) is:
 \begin{eqnarray}\label{second term}
&&   g_3(x_3) g_4^* (x_4) g_3^*(y_3) g_4 (y_4) J^\dagger_{\ell-p_3,p_4} \otimes J_{\ell - p_3,p_4} \nonumber \\ &&\nonumber \\
&&= e^{- i 2 \pi(n_3 + {p_3 \over \ell}) x_3     + i 2 \pi(n_4 + {p_4 \over \ell}) x_4  }J_{p_3 - \ell,- p_4} \otimes  e^{  i 2 \pi(n_3 + {p_3 \over \ell}) y_3     - i 2 \pi(n_4 + {p_4 \over \ell}) y_4  }  J_{\ell -p_3,p_4}~.\nonumber
 \end{eqnarray}
 Here, of course, we also have $n_{3,4} \ge 0$. We now  change variables $n_4 = - n_4' -1$, $p_4 = \ell - p_4'$ and obtain
  \begin{eqnarray}\label{second term 2}
&&   g_3(x_3) g_4^* (x_4) g_3^*(y_3) g_4 (y_4) J^\dagger_{\ell-p_3,p_4} \otimes J_{\ell - p_3,p_4} \nonumber \\ &&\nonumber \\
&&= e^{- i 2 \pi(n_3 + {p_3 \over \ell}) x_3     - i 2 \pi(n_4' + {p_4' \over \ell}) x_4  }J_{p_3 - \ell,  p_4'-\ell} \otimes  e^{  i 2 \pi(n_3 + {p_3 \over \ell}) y_3    + i 2 \pi(n_4' + {p_4' \over \ell}) y_4  }  J_{\ell -p_3,\ell-p_4'} \nonumber \\
&& = e^{- i 2 \pi(n_3 + {p_3 \over \ell}) x_3     - i 2 \pi(n_4' + {p_4' \over \ell}) x_4  }J_{p_3 ,  p_4' } \otimes  e^{  i 2 \pi(n_3 + {p_3 \over \ell}) y_3    + i 2 \pi(n_4' + {p_4' \over \ell}) y_4  }  J_{  -p_3, -p_4'}~,
 \end{eqnarray}
 where in the last line we used the definition of $J_{\bm p}$  (\ref{JPELL1}), noting that the change of the phase factors in the two $J_{\bm p}$ factors upon going from the second to the third line cancel out. We note that now $n_4' < 0$ and $p_4'= \ell - p_4$ is in the fundamental domain and that the last line in (\ref{second term 2}) has the same form as (\ref{first term}), but includes negative values of $n_4'$ but $n_3 \ge 0$. 
   
 The third term   in (\ref{partial 1}) is:
  \begin{eqnarray}\label{third term}
&&   g_3^*(x_3) g_4^* (x_4) g_3 (y_3) g_4 (y_4) J^\dagger_{p_3,p_4} \otimes J_{p_3,p_4}  \nonumber \\ &&\nonumber \\  &&=e^{  i 2 \pi(n_3 + {p_3 \over \ell})x_3   + i 2 \pi(n_4 + {p_4 \over \ell})x_4 }J_{- p_3  ,- p_4} \otimes e^{ - i 2 \pi(n_3 + {p_3 \over \ell})y_3   - i 2 \pi(n_4 + {p_4 \over \ell}) y_4 } J_{ p_3,p_4} \nonumber  \end{eqnarray}
We now change both $n_3 = - n_3' -1$, $p_3 = \ell - p_3'$ and $n_4 = - n_4' - 1$, $p_4 = \ell - p_4'$ to obtain, performing the similar changes in the $J_{\bm p}$ factors, 
  \begin{eqnarray}\label{third term 2}
&&   g_3^*(x_3) g_4^* (x_4) g_3 (y_3) g_4 (y_4) J^\dagger_{p_3,p_4} \otimes J_{p_3,p_4}  \nonumber \\ &&\nonumber \\  &&=e^{ - i 2 \pi(n_3' + {p_3' \over \ell})x_3   - i 2 \pi(n_4' + {p_4' \over \ell})x_4 }J_{p_3'  , p_4'} \otimes e^{   i 2 \pi(n_3' + {p_3' \over \ell})y_3  + i 2 \pi(n_4' + {p_4' \over \ell}) y_4 } J_{- p_3',-p_4'}.    \end{eqnarray}
 We now note that (\ref{third term 2} is of the same form as (\ref{first term}), but with $n_3' <0$, $n_4'<0$ and $p_3', p_4'$ (equal to $\ell - p_3, \ell-p_4$, respectively) in the fundamental domain. 
 
Finally, the fourth term  in (\ref{partial 1}) is:
   \begin{eqnarray}\label{fourth term}
&&   g_3^*(x_3) g_4  (x_4) g_3 (y_3) g_4^* (y_4) J_{\ell-p_3,p_4} \otimes J^\dagger_{\ell-p_3,p_4} \nonumber \\&&\nonumber \\  &&= e^{  i 2 \pi(n_3 + {p_3 \over \ell})x_3    - i 2 \pi(n_4 + {p_4 \over \ell})x_4 }J_{\ell - p_3, p_4} \otimes e^{  -i 2 \pi(n_3 + {p_3 \over \ell})y_3   + i 2 \pi(n_4 + {p_4 \over \ell}) y_4 } J_{ p_3-\ell,p_4}, \nonumber 
 \end{eqnarray}
 it is clear that we now perform $n_3 = - n_3' -1$, $p_3 = \ell - p_3'$ to find
    \begin{eqnarray}\label{fourth term 2}
&&   g_3^*(x_3) g_4  (x_4) g_3 (y_3) g_4^* (y_4) J_{\ell-p_3,p_4} \otimes J^\dagger_{\ell-p_3,p_4} \\&&\nonumber \\  &&= e^{-  i 2 \pi(n_3' + {p_3' \over \ell})x_3    - i 2 \pi(n_4 + {p_4 \over \ell})x_4 }J_{p_3', p_4} \otimes e^{  i 2 \pi(n_3' + {p_3' \over \ell})y_3   + i 2 \pi(n_4 + {p_4 \over \ell}) y_4 } J_{ -p_3',-p_4}, \nonumber 
 \end{eqnarray}
 which clearly has the same form as (\ref{first term}) but includes $n_3' <0$ but $n_4\ge0$.
 
 Thus, combining the four terms, including also (\ref{recall eqn1}) allows us to extend the range of all $n_\mu$ to $n_\mu \in \Z$. 
 Before combining everything, we note that  the same changes that we made in the four terms above that led to (\ref{second term 2}, \ref{third term 2}, \ref{fourth term 2}) have to also be made in $\omega_{\bm p, n_\mu}$ of (\ref{phiell1}), which, as recalled just above (\ref{partial 1}), enters the propagator. Thus, we find that (changing the sign of $n_1, n_2$ in the process) the propagator in the $SU(\ell) \in SU(N)$ modes of the fermions (obeying $\sum_{B=1}^\ell \bar\lambda_{BB} = 0)$ in the fractional instanton background is given by:
 \begin{eqnarray}\label{lambdabarsuell1}
&& \langle \bar\lambda^{\dot\beta}_{BC}(x)  \bar\lambda_{\dot\alpha \;DE}(y)\rangle  
  \\
 &&= \delta_{\dot\alpha}^{\dot\beta}  \; {\cal D}^f_k(m) \; {g^2 \over 2 \ell V}  \nonumber \\
 && \times \sum\limits_{k_\mu = {2 \pi n_\mu \over L_\mu}, n_\mu \in \Z} \sum\limits_{(p_3,p_4) \in \Z_\ell^2} {m \;  e^{- i (x_\mu - y_\mu) (k_\mu + \delta_{\mu 3} {2 \pi p_3 \over \ell L_3} + \delta_{\mu 4} {2 \pi p_3 \over \ell L_3})}  \over |m|^2 +  \sum\limits_{\mu = 1}^4 \left(k_\mu + \delta_{\mu 3} {2 \pi p_3 \over \ell L_3} + \delta_{\mu 4} {2 \pi p_3 \over \ell L_3}\right)^2}  \; (J_{p_3, p_4})_{BC}  (J_{- p_3, - p_4})_{DE}, \nonumber 
 \end{eqnarray}
 where $J_{\bm p}$ is from (\ref{JPELL1}).
We note that an identical expression for $\langle \lambda_{\alpha \; BC}(x)  \lambda(x)^\beta_{DE }(y) \rangle$ propagator is identical to (\ref{lambdabarsuell1}), with the replacement $m \rightarrow m^*$ and $\delta^{\dot\alpha}_{\dot\beta} \rightarrow \delta_\alpha^\beta$.

 \section{The $Q={k \over N}$ propagator in the off-diagonal $k \times \ell$ and $\ell \times k$ sector}
 \label{sec:offdiagonalktimesell}
 
Here, we study the eigenvalues of the order-$\Delta^0$ Laplacian (\ref{adjoint1}) in the $k \times \ell$ (and the c.c. $\ell \times k$ blocks). These off-diagonal blocks are the only ones that couple to the background (\ref{naked expressions}), the constant field strength in the direction of the $U(1)$ generator $\omega$ (\ref{omega}); for use below, we write the background gauge field as $A_\mu = \omega A^\omega_\mu$.
The adjoint Laplacian operator acting on wave functions in the $k \times \ell$ block is, proportional to the unit matrix\footnote{For vanishing holonomies; including them does not spoil the diagonal nature of (\ref{laplacian ktimesell}) and can be done in a trivial manner.} and is given by  
\begin{eqnarray}\label{laplacian ktimesell}
D^2&=&\left(\partial_\mu+i2\pi N A_\mu^\omega\right)^2=\Box+i4\pi N A_\mu^\omega\partial_\mu-4\pi^2N^2 A_\mu^\omega A_\mu^\omega\\
&=&\partial_1^2+\partial_3^2+\left(\partial_2-i2\pi \frac{x_1}{L_1L_2}\right)^2+\left(\partial_4-i2\pi \frac{x_3}{\ell L_3L_4}\right)^2\,, \nonumber
\end{eqnarray}
and we used $\partial_\mu A_\mu^\omega=0$. We are looking for solutions of the equations
\begin{eqnarray} \label{laplacian k times ell equation}
D^2 \Phi_{C'C \; n} = - \omega^2_n \Phi_{C'C \; n}~,
\end{eqnarray}
obeying the BCS (\ref{BCS lambda beta}). The Laplacian acting on functions in the $\ell \times k$-block inside $SU(N)$ is identical to (\ref{laplacian ktimesell}) but with a minus sign in the covariant derivative, and BCS complex conjugate to (\ref{BCS lambda beta}).
In \cite{Anber:2023sjn}, it was shown that there are no normalizable zero modes of (\ref{laplacian k times ell equation}), (this is because (\ref{laplacian k times ell equation}), by (\ref{relations 1}), is the equation for the dotted fermions  studied there);  this result also follows from our more general discussion below.

\subsection{Boundary conditions, eigenvectors of the covariant Laplacian, and norm}  To continue, we note that there is a subtlety, related to the BCS (\ref{BCS lambda beta}) for the case $\ell > 1$, that we have to address. To this end, we go back to the boundary conditions in the form (\ref{conditions on gauge field}), with transition functions (\ref{the set of transition functions for Q equal r over N, general solution}),  which have to be obeyed by any hermitean eigenfunction of the laplacian (\ref{laplacian ktimesell}). We write the Hermitean Laplacian eigenvectors in terms of  $N\times N$ matrices, decomposed into $k\times k$, $k\times \ell$, etc., blocks. The  part on which the Laplacian acts as in (\ref{laplacian k times ell equation}) is $\Phi$, a $k \times \ell$ matrix. 
The Hermitean  eigenvectors $\phi_n$ (recall (\ref{adjoint1})) are, therefore:
\begin{eqnarray}\label{eigenvectors ktimesell hermitean}
\phi_n^\alpha \rightarrow \left(\begin{array}{cc} 0 & \alpha \Phi_n \cr \alpha^* \Phi_n^\dagger & 0\end{array} \right)~, ~ \alpha = 1 ~\rm{or} ~i.
\end{eqnarray}
 We write the BCS for $\phi_n^\alpha$ in the general form
\begin{eqnarray}\label{BCSD4}
\left(\begin{array}{cc} 0 &\alpha \Phi_n \cr \alpha^* \Phi^\dagger & 0\end{array} \right)(x + \hat{e}_\mu L_\mu) = \Omega_\mu(x)  \left(\begin{array}{cc} 0 & \alpha \Phi_n \cr \alpha^* \Phi_n^\dagger & 0\end{array} \right)(x) \;\Omega_\mu^\dagger (x).
\end{eqnarray}
Since every element of $\Phi$ is  acted upon in the same way by the Laplacian (\ref{laplacian ktimesell}), thus the $k\times \ell$ matrix eigenvector $\Phi_n$ obeys:
\begin{eqnarray} \label{laplacian equation 2}
D^2 \Phi_{ n} = - \omega^2_n \Phi_{ \; n}~.
\end{eqnarray}
The normalization condition  (\ref{normphin1})  for the Hermitean $N\times N$ eigenvectors $\left(\begin{array}{cc} 0 &  \alpha\Phi_n \cr \alpha^* \Phi_n^\dagger & 0\end{array} \right)$ is given in terms of $\Phi_{n}, \Phi_{m}$ as:
\begin{eqnarray}\label{normphi1234}
 \int\limits_{\T^4} \tr_{N \times N}\; \phi_n^\alpha \phi_m^\beta =\delta_{\alpha,\beta} \delta_{nm} \rightarrow \int\limits_{\T^4} \tr_{k \times k} (\alpha \beta^* \Phi_n \cdot \Phi^\dagger_m + \beta\alpha^* \Phi_m \cdot \Phi^\dagger_n) = \delta_{\alpha, \beta}\delta_{nm}~.
\end{eqnarray}

We now turn to the BCS (\ref{BCSD4}) written in terms of the $k \times \ell$ matrix $\Phi$:
\begin{eqnarray}
\Phi(x+\hat{e}_1 L_1) &=& (-1)^{k-1} e^{i 2 \pi {x_2 \over L_2} }\; \Phi(x),  \nonumber \\
\Phi(x+\hat{e}_2 L_2) &=& Q_k \; \Phi(x),  \nonumber \\
\Phi(x+\hat{e}_3 L_3) &=& e^{i 2 \pi {x_4 \over \ell L_4} }\; \Phi(x) \;P_\ell^{-1},  \label{PhikbyellBC} \\
\Phi(x+\hat{e}_4 L_4) &=& \Phi(x) \; Q_\ell^{-1}, ~~~~~ (\rm{where}~~ ||\Phi||_{C'C} = \Phi_{C'C}). \nonumber 
\end{eqnarray}
Recall that  $Q_\ell$ is the diagonal ``clock'' matrix whose elements are different powers of the $\ell$-th root of unity, and that $P_{\ell}$ is the ``shift'' matrix, which has 
 the form given after the commutation relation, eqn.~(\ref{clockshift}).  Our point now is that while all components of $\Phi$ are acted upon identically by the Laplacian (\ref{laplacian ktimesell}), the BCS (\ref{PhikbyellBC}) in the $x_3$ direction mix components with different second index $C$, for $\ell >1$ (recall also that despite the noncommutativity of $P_\ell$ and $Q_\ell$, $\Phi(x)$ is single valued on the $\T^4$ due to the $x_{4}$-dependent phases).

{To recap,  the BCS (\ref{PhikbyellBC}), imply  that every row of the $k\times \ell$ matrix $\Phi$ (whose matrix elements are $\Phi_{C'C}$)   maps  to itself upon traversing the torus  and thus offers a separate solution of the eigenvalue equation, as the BCS are diagonal in the index $C'$. However, the BCS relate the elements inside each row of the matrix, because the action of $P_\ell^{-1}$ in the $x_3$ BC cyclically permutes the $\ell$ row elements. Thus, every eigenvector of the Laplacian, $\Phi$, obeying the Laplace equation (\ref{laplacian equation 2}) and the boundary conditions (\ref{PhikbyellBC})  will  involve all $\ell$ elements of the chosen row.\footnote{Thus, the twisted BCS on the $\T^4$, which relate all components in a given row, imply that the number of  degrees of freedom in the $k\times \ell$ part of the $N \times N$ matrix is smaller than $k \ell$: there are only $k$ independent Grassmann degrees of freedom.}

Explicitly, we add the $C'$ index (which labels degenerate eigenvectors) to the eigenvalue index $n$ of the $k\times \ell$  matrix representing the eigenvector (\ref{eigenvectors ktimesell hermitean}): $\Phi_n \rightarrow \Phi_{n \; C'}$:
 \begin{eqnarray}\label{eigenvectormatrix1}
 \Phi_{n \; C'} = \left(\begin{array}{ccccc}0 & 0 & 0& ...&0 \cr
 ... & ... & ... & ... & ... \cr
  0 & 0 & 0 & ... &0 \cr
 \Phi_{n \; C'1} & \Phi_{n \; C'2} & \Phi_{n\; C'3} &...& \Phi_{n\; C'\ell} \cr
 0 & 0 & 0&...&0  \cr ... & ... & ... & ... & ... \end{array} \right)~,
 \end{eqnarray}
 showing  that an eigenvector involves all $\ell$ nonzero entries in the $C'$-th row. The inner product is determined from (\ref{normphi1234}). 

\subsection{The general eigenvector consistent with the twisted boundary conditions} We now turn to determining the spectrum and eigenvectors of (\ref{laplacian equation 2}).  
 The first step is to   find the most general Fourier expansion of $\Phi_{C'C}(x)$, $C=1,...,\ell$, for a fixed $C'$ (as per the above discussion) consistent with the BCS (\ref{BCS lambda beta})  or (\ref{PhikbyellBC}) on $\T^4$.\footnote{The reader can skip to the final result,  eqn.~(\ref{generalphiBCS}), and check that the BCS are obeyed.}

 We begin by considering the $C$-th element in the chosen row and note that the $x_2$ and $x_4$ BCS (\ref{BCS lambda beta}) (equivalently, the first and last row in (\ref{PhikbyellBC})) imply that 
 $$
 \Phi_{C'C}(x) = \sum\limits_{m,p \in \Z} e^{i {2 \pi x_2 \over L_2}(m+ {2 C'  - 1 -k \over 2k}) + i {2 \pi x_4 \over L_4}(p - {2 C - 1 - \ell \over 2\ell})} \Phi_{m,p, C'C}(x_1, x_3).
 $$
Further, the $x_1$ BC from (\ref{BCS lambda beta}) or (\ref{PhikbyellBC})) implies that the Fourier component $\Phi_{m,p, C'C}(x_1, x_3)$ obeys
 $$
 \Phi_{m,p,C' C} (x_1 + L_1, x_3) = \gamma_k^{-k} \Phi_{m-1,p, C'C}(x_1, x_3)~.
 $$
Hence, for any $m$, we have  $\Phi_{m,p,C',C}(x_1,x_3) = \gamma_k^{- m k} \Phi_{m=0,p,C',C}(x_1 - m L_1, x_3)$. Thus all $x_2$-Fourier components labelled by $m$ can be related to the $m=0$ one.
Then, dropping the now superfluous $m=0$ subscript, 
we can write for the general $\Phi_{C'C}$ obeying the BCS:
\begin{eqnarray}\label{PhiFourier2}
 \Phi_{C'C}(x) = \sum\limits_{m,p \in \Z} e^{i {2 \pi x_2 \over L_2}(m+ {2 C'  - 1 -k \over 2k}) + i {2 \pi x_4 \over L_4}(p - {2 C - 1 - \ell \over 2\ell}) - i m \pi (1-k)} \; \Phi_{p, C'C}(x_1- m L_1, x_3),
 \end{eqnarray}
 This expression can be modified slightly, while still obeying the BCS, to make it periodic w.r.t. $C' \rightarrow C'+k$, upon multiplication by a phase and a shift of the $x_1$ argument:
 \begin{eqnarray}\label{PhiFourier21}
 \Phi_{C'C}(x) = \sum\limits_{m,p \in \Z} e^{i {2 \pi x_2 \over L_2}(m+ {2 C'  - 1 -k \over 2k}) + i {2 \pi x_4 \over L_4}(p - {2 C - 1 - \ell \over 2\ell}) - i (m  + {C'\over k}) \pi (1-k)} \; \Phi_{p, C}(x_1- (m + {C' \over k}) L_1, x_3). \nonumber \\
 \end{eqnarray}
 while we also omit the subscript $C'$ from the Fourier mode,  $\Phi_{p, C',C}\rightarrow \Phi_{p, C}$ since all the dependence on $C'$ is already accounted for.
 
{\flushleft{F}}inally, the $x_3$ BC in  (\ref{BCS lambda beta}) or (\ref{PhikbyellBC})) determines, remembering that $[C+1]_\ell = C+1$, for $1 \le C \le \ell$, and $[C+1]_\ell  = 1$ for $C=\ell$,
 \begin{eqnarray}\label{x3bc0}
 \Phi_{p, C}(x_1, x_3+L_3) = \left\{ \begin{array}{cl} \gamma_\ell^{-1} \Phi_{p,  C+1}(x_1,x_3), &  ~1 \le C \le \ell -1, \cr
 \gamma_\ell^{-1} \Phi_{p-1, 1}(x_1,x_3), &~C=\ell ~.\end{array}\right.
 \end{eqnarray}
 The recursion relations (\ref{x3bc0}) between the Fourier coefficients $\Phi_{p,C',C}$ involve both the Fourier index $p \in \Z$ and the $SU(\ell)$ index $C=1,...,\ell$. They imply that all $x_4$-direction Fourier components,  labelled by $p \in \Z$,  combine with those of different values of $C$, and that any $p,C$-Fourier component can be expressed in terms of one chosen  single function,\footnote{As in the discussion that led to (\ref{PhiFourier2}), the function $\Phi$ on the r.h.s. of (\ref{x3bc2}) can be taken one of the $\Phi_{p,C}$ Fourier modes, e.g. $\Phi_{p=0, C=1}$.}  denoted $\Phi(x_1, x_3)$, with an appropriately shifted $x_3$ argument. Explicitly, 
 \begin{eqnarray}\label{x3bc2}
 \Phi_{p, C}(x_1, x_3) = \gamma_\ell^{C-1-p\ell} \Phi(x_1, x_3 + (C - p \ell)L_3).
 \end{eqnarray}

 Thus, substituting (\ref{x3bc2}) in (\ref{PhiFourier2}), we find that the general Fourier expansion of the $\Phi_{C'C}(x_1,x_3)$ component of the eigenvector (\ref{eigenvectormatrix1}) obeying the BCS (\ref{BCS lambda beta}) or (\ref{PhikbyellBC}) has the form 
 \begin{eqnarray}
 \label{generalphiBCS}
  \Phi_{C'C}(x) &=& \sum\limits_{m,p \in \Z} e^{i {2 \pi x_2 \over L_2}(m+ {2 C'  - 1 -k \over 2k}) + i {2 \pi x_4 \over L_4}(p - {2 C - 1 - \ell \over 2\ell}) - i (m + {C' \over k}) \pi (1-k) + i \pi {1 - \ell \over \ell}(C-1-p\ell)} \nonumber \\
  && ~~~~~ \times \; \Phi(x_1- (m+ {C' \over k}) L_1, x_3 + (C - p \ell) L_3)~, \end{eqnarray}
 determined by a single $x_{2,4}$-Fourier coefficient, the function $\Phi(x_1, x_3)$. This expression for $\Phi_{C'C}$ is now manifestly periodic w.r.t. both $C$ and $C'$, i.e. invariant under $C' \rightarrow C'+k$ and $C \rightarrow C + \ell$.

  Now we come to the Laplacian eigenvalue problem (\ref{laplacian k times ell equation}). We substitute the Fourier series (\ref{generalphiBCS}) into the eigenvalue equation (\ref{laplacian k times ell equation}), undo the Fourier sum, and perform a few trivial shifts of variables. In terms of the Fourier coefficient $\Phi_{ n}(x_1,x_3)$ in (\ref{generalphiBCS})\footnote{To avoid confusion, we stress that $\Phi_{ n}(x_1,x_3)$ is the Fourier component $\Phi (x_1,x_3)$ appearing in  (\ref{generalphiBCS}), with an  eigenvalue index $n$ attached.} the eigenvalue equation (\ref{laplacian k times ell equation}) has the form of the Schr\" odinger equation of two simple harmonic oscillators (SHOs) in the $x_1$ and $x_3$ directions:
  \begin{eqnarray}\label{secondorderphi}
&& \left[- {1\over 2} \partial_1^2 + {\Omega^2 \over 2} \left(x_1 + L_1{1+ k \over 2k}\right)^2   - {1\over 2} \partial_3^2 +{\tilde\Omega^2 \over 2} \left(x_3 - L_3  {1 + \ell \over 2 } \right)^2 \right]\Phi_{ n}(x_1, x_3)  =  {\omega_n^2 \over 2} \; \Phi_{ n}(x_1,x_3)~,\nonumber \\\end{eqnarray}
The SHO frequencies in the $x_1$ and $x_3$ directions are
\begin{eqnarray} \label{omegas}
\Omega =  {2 \pi \over L_1 L_2}~, ~\tilde\Omega = {2 \pi \over \ell L_3 L_4}~, 
\end{eqnarray}
where we note that 
\begin{eqnarray} \label{omegasequal}
\Omega = \tilde\Omega,  \; \text{due to} \; L_1 L_2 = \ell L_3 L_4,
\end{eqnarray}
using  the self-duality condition, $\Delta=0$ from (\ref{def of Delta}), from the leading-order in the $\Delta$ expansion.

\subsection{The eigenvalues and the explicit form of the eigenvectors} We next  determine the spectrum of the Laplacian, i.e. the values of $\omega_n^2$ that yield normalizable solutions, with an inner product given by  the $\T^4$-integral on the l.h.s. of eqn.~(\ref{normphi1234}). 

Thus, we consider two hermitean eigenvectors (\ref{eigenvectors ktimesell hermitean}), labelled by $(\alpha, C', n)$ and $(\beta, D', m)$, ($\alpha, \beta = \{1, i\}$). We recal the $k\times \ell$ form of the eigenvector (\ref{eigenvectormatrix1}) in terms of (\ref{generalphiBCS}) and find their inner product (\ref{normphi1234}):\footnote{To go to the second line, we noted that the integrals over $x_2$ and $x_4$ reduce the four Fourier sums to two.}
\begin{eqnarray}\label{normphi123}
&&\int\limits_{\T^4} \tr_{k \times k} ( \alpha \beta^* \Phi_{n \; C'} \cdot \Phi^\dagger_{m \; D'} + \beta \alpha^* \Phi_{m \; D'} \cdot \Phi^\dagger_{n \; C'}) = \delta_{C'D'} \sum\limits_{C=1}^\ell \int\limits_{\T^4} ( \alpha \beta^*\Phi_{n C'C} \Phi_{m C'C}^* +   \beta \alpha^*\Phi_{m C'C} \Phi_{n C'C}^*) \nonumber \\
  && =   L_2 L_4\; \delta_{C'D'} \sum\limits_{C=1}^\ell \sum\limits_{p,m \in \Z} \int\limits_{ 0}^{L_3} dx_3 \int\limits_{0}^{L_1} dx_1  \; (\alpha \beta^* \Phi_{C'\; n}(x_1', x_3') \Phi_{C'\; m}^*(x_1', x_3') + {\rm{h.c.}})\bigg\vert_{x_1'=x_1 - m L_1, x_3' = x_3 + (C - p \ell) L_3} \nonumber \\
  && =   L_2 L_4 \;\delta_{C'D'}  \int\limits_{-\infty}^\infty dx_1 \int\limits_{ -\infty}^{\infty}
 d x_3 \;  (\alpha \beta^* \Phi_{C'\; n}(x_1, x_3) \Phi_{C'\; m}^*(x_1, x_3) + \beta \alpha^* \Phi_{C'\; n}(x_1, x_3)^* \Phi_{C'\; m}(x_1, x_3)). \nonumber \\
  \end{eqnarray}
To obtain the last line, we used  the sum over $m$   to  extend the $x_1$ integral over the entire real line and the combined sums over $p$ and $C$ to extend the $x_3$ integral to the real line.

The net result is that  we have shown that the eigenvalue problem (\ref{laplacian k times ell equation}), subject to the BCS (\ref{BCS lambda beta}), reduces to the  standard quantum mechanics problem of two SHOs in the ($x_1, x_3$)-plane, normalizable on the $x_1, x_3$ plane, as per (\ref{normphi123}). The ``Shr\" odinger equation" is given by eqn.~(\ref{secondorderphi}), where the oscillators have  frequencies (\ref{omegas}, \ref{omegasequal}). The normalizable wave functions give rise to the eigenvalues $n$ labelled by $\ell_{(1) },\ell_{(3) }$ which determine the eigenvalue of the Laplacian:
\begin{eqnarray}
\label{EVSktimesell}
\omega_n^2 \rightarrow \omega^2_{\ell_{(1) },\ell_{(3) }} = {4 \pi \over L_1 L_2} (\ell_{(1) }+ \ell_{(3) } + 1), ~\ell_{(1) }, \ell_{(3) } = 0, 1, 2,....
\end{eqnarray}
The eigenfunctions of (\ref{secondorderphi}) are given in terms of the standard unit-normalized eigenfunctions of the one-dimensional Harmonic oscillator, $h_{\ell}(x)$, with eigenvalue $\Omega (\ell + {1 \over 2})$, $\ell=0,1,2,...$,  explicitly given by:
\begin{eqnarray}\label{phicell}
&& \Phi_{C' n} \rightarrow \Phi_{C', \ell_{(1) },\ell_{(3) }}(x_1, x_3)  =  \tilde c  \; h_{ \ell_{(1)}}(x_1)\; h_{  \ell_{(3)}} (x_3), \\ && h_\ell(x) = {1\over \sqrt{2^\ell \ell!}} \left({\Omega \over \pi}\right)^{1\over 4} e^{- {\Omega x^2 \over 2}} H_\ell(\sqrt{\Omega} x), \;\text{and} \; \int\limits_{-\infty}^{\infty} dx h_\ell(x) h_{\ell'}(x) = \delta_{\ell, \ell'}~,\nonumber 
\end{eqnarray}
where $H_\ell(x)$ is the $\ell$-th Hermite polynomial.
The normalization $ \tilde c$ must be chosen so that the norm (\ref{normphi123}) equals unity,\footnote{It is clear that eigenvectors with the same $\ell_{(1) },\ell_{(3) }$ but with $\alpha \ne \beta$ are orthogonal.} namely:\begin{eqnarray}\label{equationfornorm}
\tilde c = {1 \over \sqrt{2 L_2 L_4}}.
\end{eqnarray}

To summarize, our final expression for the $C$-th component of the eigenvector (\ref{eigenvectormatrix1}), the $k \times \ell$ matrix $\Phi_{C'  \ell_{(1) },\ell_{(3) }}$ with nonzero $C'$-th row, labelled by $(C', \ell_{(1) },\ell_{(3)})$ is, substituting (\ref{phicell}) and the normalization (\ref{equationfornorm}) into (\ref{generalphiBCS})\footnote{Here, we added the (harmless) extra phase factor $e^{i {1 + k \over 2 k}  \pi (1-k)}$, to restore agreement with the similar equations for Dirac zero modes in \cite{Anber:2023sjn}.}
 \begin{eqnarray}
 \label{finalphiBCS}
  \Phi_{C'C \; \ell_{(1) },\ell_{(3) }}(x) &=&~{1 \over \sqrt{2 L_2 L_4}} \sum\limits_{m,p \in \Z} e^{i {2 \pi x_2 \over L_2}(m+ {2 C'  - 1 -k \over 2k}) + i {2 \pi x_4 \over L_4}(p - {2 C - 1 - \ell \over 2\ell}) - i (m+ {C'\over k} - {1 + k \over 2 k}) \pi (1-k) + i \pi {1 - \ell \over \ell}(C-1-p\ell)} \nonumber \\
  && ~~~~~~~~ \times \; h_{\ell_{(1)}}(x_1- L_1 (m+{C'\over k} - {1 + k \over 2k})) \; h_{\ell_{(3)}}(x_3 + L_3 (C - p \ell -  {1+ \ell \over 2}))~,  \nonumber \\
\end{eqnarray}
 where $h_{\ell}(x)$, as per (\ref{phicell}) is the $\ell$-th unit normalized eigenstate of the SHO, and the normalization  is such that the norm (\ref{normphi123}) equals unity. Explicitly, the norm integral (\ref{normphi123}) is
 $$
 2 \int\limits_{\T^4} d^4 x \sum\limits_{C=1}^\ell   \Phi_{C'C \; \ell_{(1) },\ell_{(3) }}(x)  \Phi^*_{C'C \; \ell_{(1) },\ell_{(3) }}(x)  = 1.
 $$
One extra step, useful  to obtain expressions similar to our other propagators, is to define a  dimensionless $\varphi_{C'C \; \ell_{(1) },\ell_{(3) }}$ as follows:
\begin{eqnarray}
\label{morefinalphiBCS}
 \Phi_{C'C \; \ell_{(1) },\ell_{(3) }}(x) = {1 \over \sqrt{2V}}\; \varphi_{C'C \; \ell_{(1) },\ell_{(3) }}(x)~,
\end{eqnarray}
where
\begin{eqnarray}\label{varphiktimesellnorm}
 \int\limits_{\T^4} d^4 x \sum\limits_{C=1}^\ell   \varphi_{C'C \; \ell_{(1) },\ell_{(3) }}(x)  \varphi^*_{C'C \; \ell_{(1) },\ell_{(3) }}(x) = V.
\end{eqnarray}

At the end, we also include the instanton moduli from (\ref{r over N abelian sol}) (these were previously omitted when writing (\ref{laplacian ktimesell}), but  have to be included since the propagators have to be integrated over the moduli space), which we now write as
\begin{eqnarray}\label{moduli11}
\hat\phi_\mu^{C'} = {2 \pi \over L_\mu} \bm{a}_\mu \cdot \bm{\nu}_{C'} - {2 \pi N z_\mu \over L_\mu}~,
\end{eqnarray}
where $\bm{\nu}_{C'}$ are the $k$ weights of the fundamental representation of $SU(k)$. 
Thus, we write the final expression for the eigenvector $\varphi_{C'C ;\ell_{(1) },\ell_{(3) }}$ as:
  \begin{eqnarray}
 \label{finalvarphiBCS}
  \varphi_{C'C \; \ell_{(1) },\ell_{(3) }}(x) &=&~{ \sqrt{V \over L_2 L_4}} e^{- i (x_3 \hat\phi_3^{C'} + x_1 \hat\phi_1^{C'})} \sum\limits_{m,p \in \Z} e^{i ({2 \pi x_2 \over L_2} + L_1 \hat\phi_1^{C'}) (m+ {2 C'  - 1 -k \over 2k}) + i ({2 \pi x_4 \over L_4} + \ell L_3 \hat\phi_3^{C'}) (p - {2 C - 1 - \ell \over 2\ell})}\nonumber \\
  &&  \times~e^{ - i (m+ {C'\over k} - {1 + k \over 2 k}) \pi (1-k) + i \pi {1 - \ell \over \ell}(C-1-p\ell)} \; h_{\ell_{(1)}}(x_1- {L_1 L_2 \hat\phi_2^{C'} \over 2 \pi} - L_1 (m+{C'\over k} - {1 +k \over 2 k})) \nonumber \\
  && \times ~ \; h_{\ell_{(3)}}(x_3 -  {\ell L_3 L_4 \hat\phi_4^{C'} \over 2 \pi}+ L_3 (C - p \ell - {1 + \ell \over 2}))~.  \nonumber \\
\end{eqnarray}
  This expression for $\varphi_{C'C ;\ell_{(1) },\ell_{(3) }}$ is used to construct the propagator.

\subsection{Final form of the propagator in the $k\times\ell$ and $\ell\times k$ subspace} To compute the $22$ component of the propagator  from (\ref{propagator 2}) we need to evaluate the sum over eigenvectors corresponding to the same eigenvalue (\ref{EVSktimesell}), i.e. labelled by 
($\alpha, C',  \ell_{(1)}, \ell_{(3)}$):
\begin{eqnarray}\label{sum22elementktimesell}
\sum_{C', \alpha} \; (\phi^\alpha_{C' \ell_{(1)} \ell_{(3)} }(x))_{ij} \; (\phi^\alpha_{C' \ell_{(1)} \ell_{(3)}}(y))_{kl}
\end{eqnarray}
From (\ref{eigenvectors ktimesell hermitean}) and (\ref{eigenvectormatrix1}), we can explicitly write:\footnote{For brevity only, we use a  slightly idiosyncratic notation. The indices $i,j$ are $SU(N)$ indices and the index $C=1...\ell$, while $C' = 1...k$; when we write $\delta_{iC} \delta_{jC'}$ we really mean that the $SU(N)$ index $i$ is in the $SU(\ell)$ part and the index $j$ is in the $SU(k)$ part; this should not cause confusion because the eigenvector $\phi^\alpha$ (\ref{eigenvectors ktimesell hermitean}) has no $k\times k$ or $\ell\times\ell$ components.}
\begin{eqnarray}\label{phimatrixktimesell}
(\phi^\alpha_{C' \ell_{(1)} \ell_{(3)} })_{ij} = \alpha \delta_{i C'}  \delta_{jC} \Phi_{C' C \ell_{(1)} \ell_{(3)} } + \alpha^*  \delta_{iC} \delta_{jC'}  \Phi_{C' C \ell_{(1)} \ell_{(3)} }^*~.
\end{eqnarray}
Using (\ref{phimatrixktimesell}), we find, recalling that $\sum\limits_{\alpha=1,i} \alpha^2 = 0$, $\sum\limits_{\alpha=1,i} |\alpha|^2 = 2$:
\begin{eqnarray}
&&\sum_{C', \alpha} \; (\phi^\alpha_{C' \ell_{(1)} \ell_{(3)} }(x))_{ij} \; (\phi^\alpha_{C' \ell_{(1)} \ell_{(3)}}(y))_{kl} \\
&& = 2  \left(  \delta_{i C'} \delta_{jC} \Phi_{C'C \ell_{(1)} \ell_{(3)}}(x) \; \delta_{kD} \delta_{lC'} \Phi^*_{C'D \ell_{(1)} \ell_{(3)}}(y) +   \delta_{i C} \delta_{jC'} \Phi^*_{C'C \ell_{(1)} \ell_{(3)}}(x) \; \delta_{kC'} \delta_{lD} \Phi_{C'D \ell_{(1)} \ell_{(3)}}(y)\right) \nonumber \\
&&={1 \over V} \left(  \delta_{i C'} \delta_{jC} \varphi_{C'C \ell_{(1)} \ell_{(3)}}(x) \; \delta_{kD} \delta_{lC'} \varphi^*_{C'D \ell_{(1)} \ell_{(3)}}(y) +   \delta_{i C} \delta_{jC'} \varphi^*_{C'C \ell_{(1)} \ell_{(3)}}(x) \; \delta_{kC'} \delta_{lD} \varphi_{C'D \ell_{(1)} \ell_{(3)}}(y)\right), \nonumber
\end{eqnarray}
where in the last line we used (\ref{morefinalphiBCS}) to rewrite the sum using the functions $\varphi_{C'C \ell_{(1)} \ell_{(3)}}$ with normalization (\ref{varphiktimesellnorm}).

From this equation and the $22$ component of the propagator (\ref{propagator 2}), we immediately find the following nonzero $\langle \bar\lambda^{\dot\alpha} (x) \bar\lambda_{\dot\beta}(y) \rangle $ propagators in the off-diagonal $k\times\ell$ and $\ell\times k$ components of the $SU(N)$ adjoint:
\begin{eqnarray}\label{barlambdaktimesellpropagator}
 \langle \bar\lambda^{\dot\alpha}_{C'C}(x) \bar\lambda_{\dot\beta \; DC'}(y) \rangle =   \delta^{\dot\alpha}_{\dot\beta} \; {\cal{D}}_k^f(m)\; {g^2 \over 2V} \sum\limits_{\ell_{(1)}, \ell_{(2)} = 0}^\infty {m \;  \varphi _{C'C \ell_{(1)} \ell_{(3)}}(x) \;  \varphi^*_{C'D \ell_{(1)} \ell_{(3)}}(y)  \over \omega_{\ell_{(1)}, \ell_{(3)} }^2 + |m|^2}  ~,
\end{eqnarray}
where $ \omega_{\ell_{(1)}, \ell_{(3)} }^2 =  {4 \pi \over L_1 L_2} (\ell_{(1) }+ \ell_{(3) } + 1)$ and 
$\varphi _{C'C \ell_{(1)} \ell_{(3)}} = \sqrt{2 V} \Phi _{C'C \ell_{(1)} \ell_{(3)}}$, with $\Phi_{C'C \ell_{(1)} \ell_{(3)}}$ given in
(\ref{finalphiBCS}), and $\varphi$ normalized as in (\ref{varphiktimesellnorm}). We stress that 
 combinations of the $k\times \ell$ indices other than those in (\ref{barlambdaktimesellpropagator}) give vanishing contribution.

To compute the $11$ element of (\ref{propagator 2}), the $\langle \lambda_\gamma(x) \lambda^\beta(y) \rangle$  propagator in the off-diagonal $k\times \ell$ and $\ell \times k$ subspace, we need:
\begin{eqnarray}\label{sum11elementktimesell}
&& \sum_{C', \alpha} \; \sigma_{\mu \gamma \dot\gamma} D_\mu (\phi^\alpha_{C' \ell_{(1)} \ell_{(3)} }(x))_{ij} \; \bar\sigma_\nu^{\dot\gamma \beta}  D_\nu (\phi^\alpha_{C' \ell_{(1)} \ell_{(3)}}(y))_{kl}
\end{eqnarray}
where the derivative of $\phi^\alpha_{C' \ell_{(1)} \ell_{(3)} }$ is 
\begin{eqnarray}\label{phimatrixktimesell11}
&&D_\mu (\phi^\alpha_{C' \ell_{(1)} \ell_{(3)} })_{ij} = \alpha  \delta_{i C'}  \delta_{jC} D_\mu \Phi_{C' C \ell_{(1)} \ell_{(3)} } + \alpha^*  \delta_{iC} \delta_{jC'}  (D_\mu \Phi_{C' C \ell_{(1)} \ell_{(3)} })^*,
\end{eqnarray}
where the covariant derivatives are explicitly given by\footnote{It is understood that in (\ref{phimatrixktimesell11}), the derivative action on $\Phi_{C' C}$ includes the corresponding $\hat\phi^{C'}$ from (\ref{covariantsubstitution}).}
\begin{eqnarray}\label{covariantsubstitution}
&&~D_1 \rightarrow \partial_1 + i \hat\phi_1^{C'},\; D_2 \rightarrow \partial_2 - i {2 \pi \over L_1 L_2} x_1 + i \hat\phi_2^{C'}, \\
&&~  D_3 \rightarrow \partial_3+ i \hat\phi_3^{C'}, \;D_4 \rightarrow \partial_4 - i{2 \pi \over \ell L_3 L_4} x_3 + i \hat\phi_4^{C'}. \nonumber  
\end{eqnarray}

As in the other components of the propagator, the sum over $\alpha$ in (\ref{sum11elementktimesell}) leaves only terms of the form 
\begin{eqnarray}\label{munu1}
 \sigma_{\mu \gamma \dot\gamma} D_\mu \Phi_{C' C \ell_{(1)} \ell_{(3)} }(x) \; \bar\sigma_\nu^{\dot\gamma \beta} (D_\nu  \Phi_{C' D \ell_{(1)} \ell_{(3)} }(y))^*
\end{eqnarray}
remaining on the r.h.s. of (\ref{sum11elementktimesell}), omitting the Kronecker delta symbols involving $i,j,k,l$.

In terms of the product (\ref{munu1}), we can write the nonvanishing $k\times \ell$ and $\ell \times k$ components of the $\langle \lambda_\gamma(x) \lambda^\beta(y) \rangle$ propagator, with the derivatives substituted from (\ref{covariantsubstitution}):
\begin{eqnarray} \label{lambdaktimesellpropagator}
 \langle \lambda_{\gamma \; C'C} (x) \lambda^\beta(y)_{D C'}\rangle  = {\cal{D}}_k^f(m)\; {g^2 \over 2V}  \sum\limits_{\ell_{(1)}, \ell_{(2)} = 0}^\infty {m^* \;   \sigma_{\mu \gamma \dot\gamma} D_\mu  \varphi _{C'C \ell_{(1)} \ell_{(3)}}(x) \;  \bar\sigma_\nu^{\dot\gamma \beta}  D_\nu^* \varphi^*_{C'D \ell_{(1)} \ell_{(3)}}(y)  \over \omega_{\ell_{(1)}, \ell_{(2)} }^2(\omega_{\ell_{(1)}, \ell_{(3)} }^2 + |m|^2)} ~,\nonumber\\
\end{eqnarray}
Here, as in (\ref{barlambdaktimesellpropagator}), 
$ \omega_{\ell_{(1)}, \ell_{(3)} }^2 =  {4 \pi \over L_1 L_2} (\ell_{(1) }+ \ell_{(3) } + 1)$ and 
$\varphi _{C'C \ell_{(1)} \ell_{(3)}} = \sqrt{2 V} \Phi _{C'C \ell_{(1)} \ell_{(3)}}$, with $\Phi_{C'C \ell_{(1)} \ell_{(3)}}$ given in
(\ref{finalphiBCS}), and $\varphi$ normalized as in (\ref{varphiktimesellnorm}). We stress that 
 combinations of the $k\times \ell$ indices other than those in (\ref{barlambdaktimesellpropagator}) give vanishing contribution. The form of the off-diagonal ($k\times \ell$/$\ell\times k$) propagator $\langle \lambda_\gamma(x) \lambda^\beta(y) \rangle$  given in (\ref{lambdaktimesellpropagator}) will be sufficient for our calculations.

 \section{The propagator in the $Q=0$ sector with a single twist $n_{34}=1$}
 \label{sec:singletwistpropagator}
 
 All the work to find this propagator was already performed when studying the $SU(\ell)$ part in section \ref{appx:suellpropagator}. This is because the here the BCS for all $SU(N)$ adjoint components (rather than just for the $SU(\ell)$ part)  are twisted in the $x_3$ and $x_4$ directions and 
 periodic in $x_1$ and $x_2$. 
 
Thus, the propagators  in the $Q=0$ sector with a single unit twist $n_{34}=1$ have a form identical to (\ref{lambdabarsuell1}) but with $\ell \rightarrow N$, $p_3,p_4 \in [1, N-1]$ (both $p_{3,4}$ take integer values), and the matrices $J$ now referring to $SU(N)$, i.e. given by the same $J_p$ as (\ref{JPELL1}) but with $\ell \rightarrow N$. Naturally, the indices on the fermions in (\ref{lambdabarsuell1})  are allowed to run over all $N$ values.

\section{Expressions for open Wilson  lines to order $\Delta^0$}
\label{Expressions of Wilsons lines}

In the $SU(k)$ space, the nonvanishing components of the open Wilson lines ${\cal W}_\mu$ (\ref{Wilson 1 main}) entering the gauge invariant correllators of section \ref{Guage-invariant observables} are:
\begin{eqnarray}\label{abelian wilson gr1}\nonumber
||{\cal W}_{1 C'B'}|| (x) &=& \exp\left[i 2\pi x_1\left(-\ell \frac{z_1}{L_1}I_k + \frac{\bm a_1 \cdot \bm H_{(k)}}{L_1}\right)\right],\\\nonumber
||{\cal W}_{2 C'B'}|| (x) &=& \exp\left\{i 2\pi x_2\left[ -\ell\left( \frac{z_2}{L_2} + \frac{x_1}{N L_1 L_2}\right)I_k + \frac{\bm a_2 \cdot \bm H_{(k)}}{L_2}\right]\right\},\\\nonumber
||{\cal W}_{3 C'B'}|| (x) &=& \exp\left[i 2\pi x_3\left(-\ell \frac{z_3}{L_3}I_k + \frac{\bm a_3 \cdot \bm H_{(k)}}{L_3}\right)\right],\\
||{\cal W}_{4 C'B'}|| (x) &=& \exp\left\{i 2\pi x_4\left[ -\ell\left( \frac{z_4}{L_4} + \frac{x_3}{N\ell L_3 L_4}\right)I_k + \frac{\bm a_4 \cdot \bm H_{(k)}}{L_4}\right]\right\}.
\end{eqnarray}
Recall that ${\bm H}_{(k)}\equiv (H_{(k)}^1,...,H_{(k)}^{k-1})$ are the $SU(k)$ Cartan generators obeying $\tr \left[{ H}_{(k)}^{a}  { H}_{(k)}^{b} \right]= \delta^{ab}$, $a,b=1,...,k-1$. They can be expressed as, ${ H}_{(k)}^b$=
diag$({ \nu}^b_1,  { \nu}^b_2,..., { \nu}^b_k)$, where $\bm\nu_1,...,\bm\nu_k$ are the weights of the fundamental representation of $SU(k)$. These are $(k-1)$-dimensional vectors that obey $\bm \nu_{B'} \cdot \bm \nu_{C'} = \delta_{B'C'} - {1 \over k}$, where $B',C'=1,..,k$. 

Similarly, in the $SU(\ell)$ space, the nonvanishing components are:
\begin{eqnarray}\label{abelian wilson gr2}\nonumber
||{\cal W}_{1 CB}|| (x) &=& \exp\left[i 2\pi x_1\left(k \frac{z_1}{L_1}I_\ell\right)\right],\\\nonumber
||{\cal W}_{2 CB}|| (x) &=& \exp\left\{i 2\pi x_2\left[ k\left( \frac{z_2}{L_2} + \frac{x_1}{N L_1 L_2}\right)I_\ell\right]\right\},\\\nonumber
||{\cal W}_{3 CB}|| (x) &=& \exp\left[i 2\pi x_3\left(k \frac{z_3}{L_3}I_\ell\right)\right],\\
||{\cal W}_{4 CB}|| (x) &=& \exp\left\{i 2\pi x_4\left[ k\left( \frac{z_4}{L_4} + \frac{x_3}{N\ell L_3 L_4}\right)I_\ell\right]\right\}.
\end{eqnarray}

We may also express ${\cal W}_\mu$ in terms of the basis $\bm {\tilde H}\equiv \left(\frac{\omega}{2\pi\sqrt{k(N-k)}}, \bm H_{(k)}\right)$. Defining
\begin{eqnarray}\nonumber
\tilde {\bm a_1}&\equiv&\left(-2\pi\sqrt{k(N-k)}z_1,\bm a_1\right)\,,\\\nonumber
\tilde {\bm a_2}&\equiv&\left(-2\pi\sqrt{k(N-k)}\left(z_2+\frac{x_1}{NL_1}\right),\bm a_2\right)\,,\\\nonumber
\tilde {\bm a_3}&\equiv&\left(-2\pi\sqrt{k(N-k)}z_3,\bm a_3\right)\,,\\
\tilde {\bm a_4}&\equiv&\left(-2\pi\sqrt{k(N-k)}\left(z_4+\frac{x_3}{N\ell L_3}\right),\bm a_4\right)\,,
\end{eqnarray}
we can write
\begin{eqnarray}\label{Wilson lines in tilde H coor}
{\cal W}_{\mu} (x)=\exp\left[i2\pi  \tilde {\bm a_\mu}\cdot \tilde{\bm H}\frac{x_\mu}{L_\mu}\right]\,.
\end{eqnarray}

The open Wilson lines ${\cal W}_\mu(x)$ are not gauge invariant.  For Wilson lines that wind around the torus, one can define gauge-invariant Wilson lines as
\begin{equation}\label{wilsondef}
 W_\mu^{}[A](x)=\mbox {tr}\left[e^{i \int_0^{L_\mu }\hat A_\mu(x)}\Omega_\mu^{}(x)\right]\,.
 \end{equation}
To show that $W_\mu$ is gauge invariant, we first note that the Wilson lines ${\cal W}_\mu(x)$ and the transition functions\footnote{The transformation of the transition functions ensure consistency with (\ref{conditions on gauge field}).} $\Omega_\mu(x)$ transform under a gauge transformation $U(x)$ as
\begin{eqnarray}
{\cal W}'_{\mu}(x)&=&U(x, x_\mu= 0){\cal W}_{\mu}(x) U^\dagger(x, x_\mu=L_\mu)\,,\nonumber \\
\Omega'_\mu(x)&=& U(x, x_\mu=L_\mu) \Omega_\mu(x) U^\dagger(x, x_\mu= 0)
\end{eqnarray}
which ensures gauge invariance of (\ref{wilsondef}). 
The gauge-invariant Wilson loops that wind around the four cycles, $ W_\mu^{}[A](x)=\mbox {tr}\left[e^{i \int_0^{L_\mu }\hat A_\mu(x)}\Omega_\mu^{}(x)\right]$, are:
\begin{eqnarray}\label{Wilson 1}
\nonumber
W_1 &=&(-1)^{(k-1)} e^{-i2\pi (N-k)\left(z_1-\frac{x_2}{N L_2}\right)}\left[\sum\limits_{C'=1}^k e^{i 2\pi  \bm a_1\cdot \bm \nu_{C'}}\right]+(N-k)e^{i2\pi k \left(z_1-\frac{x_2}{N L_2}\right)}\,, \\\nonumber
W_2&=& e^{-i2\pi (N-k)\left(z_2+\frac{x_1}{N L_1}\right)} \; \left[\sum\limits_{C'=1}^k  e^{i 2\pi  (\bm a_2 - {\bm \rho \over k})\cdot \bm\nu_{C'} }\right] + (N-k)e^{i2\pi  k \left(z_2+\frac{x_1}{N L_1}\right)}\,,\\\nonumber
W_3&=& e^{-i2\pi (N-k)\left(z_3-\frac{x_4}{N \ell L_4}\right)}\left[\sum\limits_{C'=1}^k  e^{i 2\pi  \bm a_3 \cdot \bm\nu_{C'} }\right] + (N-k)\; e^{i2\pi  k \left(z_3-\frac{x_4}{N \ell L_4}\right)}\,\gamma_\ell \, \delta_{\ell,1}\,,\\
W_4&=& e^{-i2\pi (N-k) \left(z_4+\frac{x_3}{N \ell L_3}\right)}\left[\sum\limits_{C'=1}^k  e^{i 2\pi   \bm a_4 \cdot \bm\nu_{C'} }\right]+(N-k)\; e^{i2\pi k \left(z_4+\frac{x_3}{N \ell L_3}\right)} \gamma_\ell \; \; \delta_{\ell,1}\,.
\end{eqnarray}
%

\section{$\zeta$-function regularization}
\label{zeta function regularization}

In this appendix, we briefly discuss the regularization of the sums in the sectors with $Q=k/N$ and $Q=0$. We shall use the $\zeta$-function regularization technique.  

We start with the $Q=k/N$-sector sum 
  \begin{eqnarray}{\cal S}=\sum_{k_\mu\in\frac{2\pi \mathbb Z_\mu}{L_\mu},\,k_\mu k_\mu\neq0 }\frac{m^2}{m^2+k_\mu k_\mu}\,.\end{eqnarray} Here, we are taking the mass to be real. In the bulk of the paper, the mass term that appears in the denominator is the absolute value $|m|^2$, while the mass term in the numerator is $m^2$, meaning that it can take complex values.  However, generalizing the results of this appendix to a general complex mass can easily be achieved by tracking the phase of the mass that appears in the numerator. Also, we shall regularize the sums for arbitrary value of $m$, but at the end we investigate the sum in the leading order in $mL_\mu$. One of the main purpose of this appendix is to show that the regularized sum behaves as expected; in particular, we shall show that ${\cal S}\sim m^2L_\mu^2$ in the limit $mV^{1/4}\ll1$, and thus, in the strict limit $m=0$, the sum vanishes. 

  We rewrite ${\cal S}$ as
\begin{eqnarray}
{\cal S}=2^4{\cal S}_4+2^3{\cal S}_3+2^2{\cal S}_2+2{\cal S}_1\,,
\end{eqnarray}
where 
\begin{eqnarray}\label{allsums}\nonumber
{\cal S}_4&=&\sum_{n_1,.., n_4=1}^\infty\frac{m^2}{m^2+\left(\frac{2\pi n_1}{L_1}\right)^2+\left(\frac{2\pi n_2}{L_2}\right)^2+\left(\frac{2\pi n_3}{L_3}\right)^2+\left(\frac{2\pi n_4}{L_4}\right)^2}\,,\\\nonumber
{\cal S}_3&=&\sum_{n_1,.., n_3=1}^\infty\frac{m^2}{m^2+\left(\frac{2\pi n_1}{L_1}\right)^2+\left(\frac{2\pi n_2}{L_2}\right)^2+\left(\frac{2\pi n_3}{L_3}\right)^2}+3\,\mbox{other permutations}\,,\\\nonumber
{\cal S}_2&=&\sum_{n_1,n_2=1}^\infty\frac{m^2}{m^2+\left(\frac{2\pi n_1}{L_1}\right)^2+\left(\frac{2\pi n_2}{L_2}\right)^2}+5\,\mbox{other permutations}\,,\\
{\cal S}_1&=&\sum_{n_1}^\infty\frac{m^2}{m^2+\left(\frac{2\pi n_1}{L_1}\right)^2}+3\,\mbox{other permutations}\,.
\end{eqnarray}
We are interested in the limit $\mbox{Lim}_{m\rightarrow 0}\,{\cal S}_{1,2,3,4}$, which we shall study in what follows.

To this end, we define
\begin{eqnarray}
E_Q^{c^2}(s; a_1,a_2,...,a_Q)\equiv \sum_{n_1,n_2,..,n_Q=1}^{\infty}\left(c^2+a_1^2n_1^2+a_2^2n_2^2+...+a_Q^2n_Q^2\right)^{-s}\,.
\end{eqnarray}
 First, we consider the sum
\begin{eqnarray}
E_1^{c^2}(s;1)\equiv\sum_{n=1}^\infty(n^2+c^2)^{-s}\,,
\end{eqnarray}
which is given by \cite{Elizalde:1994gf}
\begin{eqnarray}\label{Expressions E}\nonumber 
E_1^{c^2}(s;1)&=&-\frac{1}{2}c^{-2s}+\frac{\sqrt{\pi}}{2\Gamma(s)}|c|^{1-2s}\left[\Gamma\left(s-\frac{1}{2}\right) +4\sum_{p=1}\left(\pi p |c|\right)^{s-\frac{1}{2}}K_{s-1/2}\left(2\pi p |c|\right)\right]\,.\\
\end{eqnarray}
Particularly important series are those with $s=1$ and $s=1/2$. The series $E_1^{c^2}(1;1)$ is convergent, with the sum given by
\begin{eqnarray}
E_1^{c^2}(1;1)=\frac{-1+c\pi \coth c\pi}{2c^2}\,,
\end{eqnarray}
and we have
\begin{eqnarray}
\mbox{lim}_{|c|\rightarrow 0} c^2E_1^{c^2}(1;1)\approx \frac{\pi^2 c^2}{6}\,.
\end{eqnarray}
The series $E_1^{c^2}(1/2;1)$ has a pole at $s=1/2$, removing it we end up with
\begin{eqnarray}
E_1^{c^2}(1/2;1)=-\frac{1}{2|c|}-\frac{1}{2}\left(\gamma+2\log|c|+\psi\left(\frac{1}{2}\right)\right)+2\sum_{p=1}^\infty K_0(2\pi p |c|)\,.
\end{eqnarray}
This series converges well for values of $|c|\gtrsim 1$. However, the series does not converge well in the limit $|c|\rightarrow 0$. This is, however, the limit we are interested in. Nevertheless, it is possible to study the series numerically to find that
\begin{eqnarray}
\mbox{lim}_{|c|\rightarrow 0}E_1^{c^2}(1/2;1)\approx \gamma\,,
\end{eqnarray}
where $\gamma=0.5772$ is Euler's constant, 
a result that is $|c|$-independent. Therefore, we have
\begin{eqnarray}
\mbox{lim}_{|c|\rightarrow 0} c^2E_1^{c^2}(1/2;1)\approx \gamma c^2\,.
\end{eqnarray}
Next, we consider the sum
\begin{eqnarray}
E_2^{c^2}(s; a_1,a_2)\equiv\sum_{n_1,n_2=1}^\infty\left(c^2+a_1^2n_1^2+a_2^2n_2^2\right)^{-s}\,.
\end{eqnarray}
Applying (\ref{Expressions E}) twice we obtain
\begin{eqnarray}\nonumber
&&E_2^{c^2}(s; a_1,a_2)=-\frac{a_2^{-2s}}{2}E_1^{c^2/a_2^2}(s;1)+\frac{a_2^{1-2s}}{a_1}\frac{\sqrt{\pi} \Gamma\left(s-\frac{1}{2}\right)}{2\Gamma(s)}E_1^{c^2/a_2^2}\left(s-\frac{1}{2};1\right)\\\nonumber
&+&2\frac{a_1^{-2s}\sqrt \pi}{\Gamma(s)}\sum_{n_2,p=1}(\pi p)^{s-1/2}\left[\frac{c^2}{a_1^2}+\frac{a_2^2n_2^2}{a_1^2}\right]^{-s/2+1/4}K_{s-1/2}\left\{2\pi p \sqrt{\frac{c^2}{a_1^2}+\frac{a_2^2n_2^2}{a_1^2}} \right\}\,.\\
\end{eqnarray}
We are interested in $E_2^{c^2}(1; a_1,a_2)$ in the limit $|c|\rightarrow 0$. From our previous discussions,  the first two terms give a constant. Also, notice that the third term in the expression of $E_2^{c^2}(1; a_1,a_2)$ yields a constant in this limit, since the double sum starts at $n_2=p=1$. Thus, we find
\begin{eqnarray}\nonumber
&&\mbox{lim}_{|c|\rightarrow 0}\, c^2E_2^{c^2}(1; a_1,a_2)=\\
&&c^2\left(-\frac{\pi^2}{12 a_2^2}+\frac{\pi \gamma}{2a_1a_2}+2\frac{\sqrt \pi}{a_1^{2}}\sum_{n_2,p=1}(\pi p)^{1/2}\left(\frac{a_2n_2}{a_1}\right)^{-1/2}K_{1/2}\left(2\pi p \frac{a_2n_2}{a_1}\right)\right)\,.
\end{eqnarray}

We can further repeat the exercise for $E_3^{c^2}(s; a_1,a_2,a_3)$ defined via
\begin{eqnarray}
E_3^{c^2}(s; a_1,a_2,a_3)\equiv\sum_{n_1,n_2,n_3=1}^\infty\left(c^2+a_1^2n_1^2+a_2^2n_2^2+a_3^2a_3^2\right)^{-s}\,.
\end{eqnarray}
We find
\begin{eqnarray}\nonumber
E_3^{c^2}(s; a_1,a_2,a_3)&=&\frac{a_3^{-2s}}{4}E_1^{c^2/a_3^2}(s;1)-\frac{a_3^{1-2s}}{a_2}\frac{\sqrt{\pi} \Gamma\left(s-\frac{1}{2}\right)}{4\Gamma(s)}E_1^{c^2/a_3^2}\left(s-\frac{1}{2};1\right)\\\nonumber
&-&\frac{a_3^{1-2s}}{a_1}\frac{\sqrt{\pi} \Gamma\left(s-\frac{1}{2}\right)}{4\Gamma(s)}E_1^{c^2/a_3^2}\left(s-\frac{1}{2};1\right)+\frac{a_3^{-2s+2}}{a_1a_2}\frac{\pi\Gamma(s-1)}{4\Gamma(s)}E_1^{c^2/a_3^2}\left(s-1;1\right)\\\nonumber
&-&\frac{a_2^{-2s}\sqrt \pi}{\Gamma(s)}\sum_{n_3,p=1}(\pi p)^{s-1/2}\left[\frac{c^2}{a_2^2}+\frac{a_3^2n_3^2}{a_2^2}\right]^{-s/2+1/4}K_{s-1/2}\left\{2\pi p \sqrt{\frac{c^2}{a_2^2}+\frac{a_3^2n_3^2}{a_2^2}} \right\}\\\nonumber
&+&\frac{a_2^{1-2s} \pi}{a_1\Gamma(s)}\sum_{n_3,p=1}(\pi p)^{s-1}\left[\frac{c^2}{a_2^2}+\frac{a_3^2n_3^2}{a_2^2}\right]^{-s/2+1/2}K_{s-1}\left\{2\pi p \sqrt{\frac{c^2}{a_2^2}+\frac{a_3^2n_3^2}{a_2^2}} \right\}\\\nonumber
&+&2\frac{a_1^{-2s}\sqrt \pi}{\Gamma(s)}\sum_{n_2,n_3,p=1}(\pi p)^{s-1/2}\left[\frac{c^2}{a_1^2}+\frac{a_2^2n_2^2}{a_1^2}+\frac{a_3^2n_3^2}{a_1^2}\right]^{-s/2+1/4}\\
&&\times K_{s-1/2}\left\{2\pi p \sqrt{\frac{c^2}{a_1^2}+\frac{a_2^2n_2^2}{a_1^2}+\frac{a_3^2n_3^2}{a_1^2}} \right\}\,.
\end{eqnarray}
Setting $s = 1$, we observe that in the limit $c \rightarrow 0$,  
$\lim_{c \rightarrow 0} c^2 E_1^{c^2/a_3^2}(0;1) \sim c^4.$
Hence, the contribution from $E_1^{c^2/a_3^2}(0;1)$ is subleading and can be neglected at leading order in $c$. Consequently, just as with $E_2^{c^2}(s; a_1, a_2)$ and $E_1^{c^2}(s; a_1)$, we find that  
$\lim_{c \rightarrow 0} c^2 E_3^{c^2}(s; a_1, a_2, a_3) \sim c^2$
to leading order in $c$.

This systematic procedure can be extended to compute $E_4^{c^2}(s; a_1, a_2, a_3, a_4)$ in a similar fashion. After straightforward yet tedious computations, we find
\begin{eqnarray}\nonumber
&&E_4^{c^2}(s; a_1,a_2,a_3,a_4)={\cal R}_1+{\cal R}_2+{\cal R}_3+{\cal R}_4\\\nonumber
&&-\frac{a_2^{-2s}\sqrt \pi}{a_4^{-2s}\Gamma(s)}\sum_{n_3,n_4,p=1}(\pi p)^{s-1/2}\left[\frac{c^2}{a_2^2}+\frac{a_3^2n_3^2}{a_2^2}+\frac{a_4^2n_4^2}{a_2^2}\right]^{-s/2+1/4}K_{s-1/2}\left\{2\pi p \sqrt{\frac{c^2}{a_2^2}+\frac{a_3^2n_3^2}{a_2^2}+\frac{a_4^2n_4^2}{a_2^2}} \right\}\\\nonumber
&&+\frac{a_2^{1-2s} a_4^{2s} \pi}{a_1\Gamma(s)}\sum_{n_3,n_4,p=1}(\pi p)^{s-1}\left[\frac{c^2}{a_2^2}+\frac{a_3^2n_3^2}{a_2^2}+\frac{a_4^2n_4^2}{a_2^2}\right]^{-s/2+1/2}K_{s-1}\left\{2\pi p \sqrt{\frac{c^2}{a_2^2}+\frac{a_3^2n_3^2}{a_2^2}+\frac{a_4^2n_4^2}{a_2^2}} \right\}\\\nonumber
&&+2\frac{a_1^{-2s}a_4^{2s}\sqrt \pi}{\Gamma(s)}\sum_{n_2,n_3,n_4,p=1}(\pi p)^{s-1/2}\left[\frac{c^2}{a_1^2}+\frac{a_2^2n_2^2}{a_1^2}+\frac{a_3^2n_3^2}{a_1^2}+\frac{a_4^2n_4^2}{a_1^2}\right]^{-s/2+1/4}\\
&&\times K_{s-1/2}\left\{2\pi p \sqrt{\frac{c^2}{a_1^2}+\frac{a_2^2n_2^2}{a_1^2}+\frac{a_3^2n_3^2}{a_1^2}+\frac{a_4^2n_4^2}{a_1^2}} \right\}\,,
\end{eqnarray}
where
\begin{eqnarray}\nonumber
{\cal R}_1&=&-\frac{a_4^{-2s}}{8}E_1^{c^2/a_4^2}(s;1)+a_3^{-2s}\frac{\sqrt{\pi} \Gamma\left(s-\frac{1}{2}\right)}{8\Gamma(s)}E_1^{c^2/a_4^2}\left(s-\frac{1}{2};1\right)\\\nonumber
&+&\frac{\sqrt \pi}{2\Gamma(s)}a_3^{-2s}\sum_{n_4,p=1}\left[\frac{c^2}{a_4^2}+n_4^2\right]^{1/4-s/2}(\pi p)^{s-1/2}K_{s-1/2}\left\{2\pi p \sqrt{\frac{c^2}{a_4^2}+n_4^2} \right\}\,,\\\nonumber
{\cal R}_2&=&\frac{\sqrt \pi \Gamma(s-1/2)}{8\Gamma(s)}\frac{a_4^{-2s+1}}{a_2}E_1^{c^2/a_4^2}\left(s-\frac{1}{2};1\right)-\frac{\pi\Gamma(s-1)}{8\Gamma(s)}\frac{a_4^{2-2s}}{a_2a_3}E^{c^2/a_4^2}(s-1;1)\\\nonumber
&&-\frac{ \pi}{2\Gamma(s)}\frac{a_3^{1-2s}}{a_2}\sum_{n_4,p=1}\left[\frac{c^2}{a_3^2}+\frac{a_4^2n_4^2}{a_3^2}\right]^{1/2-s/2}(\pi p)^{s-1}K_{s-1}\left\{2\pi p \sqrt{\frac{c^2}{a_3^2}+\frac{a_4^2n_4^2}{a_3^2}} \right\}\,,\\
\end{eqnarray}
and
\begin{eqnarray}\nonumber
{\cal R}_3&=&\frac{\sqrt \pi \Gamma(s-1/2)}{8\Gamma(s)}\frac{a_4^{-2s+1}}{a_1}E_1^{c^2/a_4^2}\left(s-\frac{1}{2};1\right)-\frac{\pi\Gamma(s-1)}{8\Gamma(s)}\frac{a_4^{2-2s}}{a_1a_3}E^{c^2/a_4^2}(s-1;1)\\\nonumber
&&-\frac{ \pi}{2\Gamma(s)}\frac{a_3^{1-2s}}{a_1}\sum_{n_4,p=1}\left[\frac{c^2}{a_3^2}+\frac{a_4^2n_4^2}{a_3^2}\right]^{1/2-s/2}(\pi p)^{s-1}K_{s-1}\left\{2\pi p \sqrt{\frac{c^2}{a_3^2}+\frac{a_4^2n_4^2}{a_3^2}} \right\}\,,\\\nonumber
{\cal R}_4&=&-\frac{\pi \Gamma(s-1)}{8\Gamma(s)}\frac{a_4^{-2s+2}}{a_1a_2}E_1^{c^2/a_4^2}(s-1;1)+\frac{\pi^{3/2}\Gamma\left(s-\frac{3}{2}\right)}{8\Gamma(s)}\frac{a_4^{-2s+3}}{a_1a_2a_3}E_1^{c^2/a_4^2}\left(s-\frac{3}{2};1\right)\\\nonumber
&&+\frac{a_3^{-2s+2}}{a_1a_2}\frac{\pi^{3/2} }{2\Gamma(s)}\sum_{n_4,p=1}\left[\frac{c^2}{a_3^2}+\frac{a_4^2n_4^2}{a_3^2}\right]^{1-s/2}(\pi p)^{s-3/2}K_{s-3/2}\left\{2\pi p \sqrt{\frac{c^2}{a_3^2}+\frac{a_4^2n_4^2}{a_3^2}} \right\}\,.\\
\end{eqnarray}
Repeating the same analysis as before, we conclude that $\lim_{c \rightarrow 0} c^2 E_4^{c^2}(s; a_1, a_2, a_3,a_4) \sim c^2$.

The key takeaway is that  ${\cal S} \sim {\cal O}((mV^{1/4})^2)$ to leading order in $mV^{1/4}$.

Next, we turn to the sum in $ Q=0$ sector
\begin{eqnarray}
{\cal S}_{(Q=0)}=\sum_{k_\mu=\frac{2\pi \mathbb Z}{L_\mu}, \bm p\equiv (p_3,p_4)\neq 0}\frac{m}{m^2+M_{\bm p,k}^2}\,,
\end{eqnarray}
where
\begin{eqnarray}
M_{\bm p,k}^2=\left[k_1^2+k_2^2+\left(k_3+\frac{2\pi p_3}{NL_3}\right)^2+\left(k_4+\frac{2\pi p_4}{NL_4}\right)^2\right]\,, \quad p_3,p_4=1,2,..,N-1\,.
\end{eqnarray}
We can write ${\cal S}_{(Q=0)}$ as
\begin{eqnarray}
{\cal S}_{(Q=0)}=2^4{\cal S}_{(Q=0)4}+2^3{\cal S}_{(Q=0) 3}+2^2{\cal S}_{(Q=0)2}+2{\cal S}_{(Q=0)1}\,,
\end{eqnarray}
where
\begin{eqnarray}\label{Dsums q0}\nonumber
{\cal S}_{(Q=0)4}&=&\sum_{n_1,.., n_4=1}^\infty\frac{m}{m^2+\left(\frac{2\pi n_1}{L_1}\right)^2+\left(\frac{2\pi n_2}{L_2}\right)^2+\left(\frac{2\pi n_3}{NL_3}\right)^2+\left(\frac{2\pi n_4}{NL_4}\right)^2}\,,\\\nonumber
{\cal S}_{(Q=0)3}&=&\sum_{n_1,.., n_3=1}^\infty\frac{m}{m^2+\left(\frac{2\pi n_1}{L_1}\right)^2+\left(\frac{2\pi n_2}{L_2}\right)^2+\left(\frac{2\pi n_3}{NL_3}\right)^2}+3\,\mbox{other permutations}\,,\\\nonumber
{\cal S}_{(Q=0)2}&=&\sum_{n_1,n_2=1}^\infty\frac{m}{m^2+\left(\frac{2\pi n_1}{L_1}\right)^2+\left(\frac{2\pi n_2}{L_2}\right)^2}+4\,\mbox{other permutations}\,,\\
{\cal S}_{(Q=0)1}&=&\sum_{n_1}^\infty\frac{m}{m^2+\left(\frac{2\pi n_1}{L_1}\right)^2}+1\,\mbox{other permutation}\,.
\end{eqnarray}
Comparing (\ref{Dsums q0}) with (\ref{allsums}), we see that the sums ${\cal S}_{(Q=0)2}$ and ${\cal S}_{(Q=0)1}$ in (\ref{Dsums q0}) do not include the permutations that set both $p_1=p_2=0$. We remind the reader that the condition $\bm p\neq0$ is vital to ensure that the algebra of the fermions is in $su(N)$. Moreover, comparing the sums  (\ref{allsums}) with  (\ref{Dsums q0}), we find that the latter can be obtained from the former by sending $L_3\rightarrow N L_3$ and $L_4\rightarrow N L_4$, keeping $L_1,L_2$ intact, as well as eliminating one power of the mass in the numerators of  (\ref{allsums}). At the end, we conclude that ${\cal S}_{(Q=0)} \sim {\cal O}(m V^{1/2})$ to leading order in $mV^{1/4}$. The exact numerical value of the coefficients can be readily extracted from $E_Q^{c^2}(1,a_1,..,a_Q)$.

In the remainder of this appendix, we compute the leading-order correction to the partition function in the $Q = 0$ sector, beyond the Witten index $N$, as a small mass is introduced. In this sector, the partition function is given by the ratio between the fermion and the gauge-boson determinants:
\begin{eqnarray}\label{start discu of prod}
Z_{Q=0}=N\frac{\prod_{k_\mu, \bm p\neq \bm 0}\left[|m|^2+M_{\bm p,k}^2\right]^{\frac{1}{2}}}{\prod_{k_\mu, \bm p\neq \bm 0}\left[M_{\bm p,k}^2\right]^{\frac{1}{2}}}\,.
\end{eqnarray}
Consider the product
\begin{eqnarray}
P=\prod_{k_\mu, \bm p\neq \bm 0}\left[|m|^2+M_{\bm p,k}^2\right]^{\frac{1}{2}}\rightarrow 2\log P=\sum_{k_\mu,\bm p\neq \bm 0}\log\left[|m|^2+M_{\bm p,k}^2\right]\,,
\end{eqnarray}
from which we find
\begin{eqnarray}
2\frac{\partial \log P}{\partial |m|^2}=\sum_{k_\mu,\bm p\neq \bm 0}\frac{1}{|m|^2+M_{\bm p,k}^2}=\frac{{\cal S}_{(Q=0)}}{|m|}\,,
\end{eqnarray}
and in the limit $|m|V^{1/4}\ll 1$
\begin{eqnarray}
2\frac{\partial \log P}{\partial |m|^2}\sim 2 cV^{1/2}\rightarrow P=d\exp\left [c |m|^2V^{1/2}\right]\,,
\end{eqnarray}
where $c, d$ are constants. The constant $c$ can be determined exactly from the regularized sums obtained above, while $d$ is a constant of integration.
Then, in the limit $|m|V^{1/4}\ll 1$, we find
\begin{eqnarray}
Z_{Q=0}=N\exp\left [c |m|^2V^{1/2}\right]\approx N+Nc|m|^2 V^{1/2}\,.
\end{eqnarray}
 This result matches the calculations in the Hamiltonian formalism.

  \bibliography{SYMCPrefs.bib}
  
  \bibliographystyle{JHEP}
  \end{document}